\documentclass[11pt,letterpaper]{article}
\pdfoutput=1
\usepackage[dvipsnames]{xcolor}
\usepackage{amsmath,amsfonts,amssymb,amsthm,amssymb,bbm,bm}
\usepackage{pdfsync}

\usepackage{graphicx,import}
\usepackage[latin1]{inputenc}
\usepackage{hyperref}
\usepackage{empheq}
\usepackage[all]{xy}
\usepackage{stmaryrd}
\usepackage{rotating}
\usepackage{color}  
\usepackage{slashed}
\usepackage{cancel}
\usepackage{array, makecell} %
\usepackage{comment}
\usepackage{tikz-cd}
\usepackage{hhline}
\usepackage{comment}
\usepackage{mathrsfs}
\usepackage{enumitem}
\usepackage{MnSymbol}
\usepackage{booktabs}
\usepackage{subcaption}
\usepackage{cite}
\numberwithin{equation}{section}

\newcommand{\p}{\partial}

\newcommand{\bit}{\begin{itemize}}
\newcommand{\eit}{\end{itemize}}
\newcommand{\bd}{\begin{description}}
\newcommand{\ed}{\end{description}}

\newcommand{\bc}{\begin{center}}
\newcommand{\ec}{\end{center}}

\newcommand{\be}{\begin{equation}}
\newcommand{\ee}{\end{equation}}
\newcommand{\bea}{\begin{eqnarray}}
\newcommand{\eea}{\end{eqnarray}}
\newcommand{\bs}{\begin{subequations}}
\newcommand{\es}{\end{subequations}}

\newcommand{\sgn}{\mathrm{sgn}}

\def\p{\partial}

\def\bz{\bar{z} }

\begin{document}

\begin{titlepage}

\unitlength = 1mm
\ \\
\vskip 1cm
\begin{center}

{\LARGE \textsc{On bulk reconstruction in Lorentzian AdS \vspace{12pt}\\
and its flat space limit}}
\date{}

\vspace{0.8cm}
N\'uria Navarro and Ana-Maria Raclariu
\vspace{0.7cm}\\
\textit{Department of Mathematics, King's College London, }\vspace{5pt}\\
\textit{Strand, London, WC2R 2LS, United Kingdom}
\vspace{40pt}

\begin{abstract}
\vspace{10pt}

We revisit the reconstruction of a free quantum field in 4-dimensional Lorentzian Anti-de-Sitter (AdS$_4$) spacetime in terms of primary operators in the boundary 3d CFT (CFT$_3$). We show that the positive and negative energy subspaces of solutions to the Klein--Gordon equation in AdS can be spanned with bulk-to-boundary propagators with appropriate time orderings. As a result, free scalar fields on a codimension-1 bulk hypersurface $\Sigma_{\tau}$ can be expressed in terms of operators integrated over boundary regions in the past or future of $\Sigma_{\tau}$ with kernels given by time-ordered or anti-time-ordered propagators. We present various equivalent representations for the bulk field in terms of either CFT primaries or their shadows.  We show from both a representation theoretic perspective and by direct computation of various flat space limits that our construction is the AdS analog of the canonical quantization of a free scalar in flat space. Depending on the choice of $\Delta$ and the location of the boundary insertions one obtains a decomposition of the scalar in either a plane wave basis ($\Delta \rightarrow \infty$) or a Carrollian basis (fixed $\Delta$). Finally, we show  that the free scalar in AdS$_4$ can be alternatively decomposed in terms of wavefunctions associated with principal series representations of dimension $\delta = 1 + i\lambda$ of an $\mathfrak{so}(3,1)$ subalgebra of the AdS$_4$ isometry algebra. We demonstrate that the latter become, in the limit of large AdS radius and for fixed $\delta$, scalar conformal primary wavefunctions in flat space. 

\end{abstract}
\vspace{0.5cm}
\end{center}
\vspace{0.8cm}

\end{titlepage}
\tableofcontents

\newpage

\section{Introduction}

The holographic principle posits that gravitational theories, namely theories with a dynamical metric, admit an equivalent description in terms of quantum field theories on a fixed background. Top-down \cite{Maldacena:1997re,Aharony:1999ti, Gubser:1998bc, Witten:1998qj,Aharony:2008ug} and bottom up \cite{Banks:1998dd,deHaro:2000vlm,Skenderis:2002wp,Ryu:2006bv, Hamilton:2006az, Harlow:2011ke} realizations of the holographic principle in asymptotically Anti-de-Sitter (AdS) spacetimes provided a concrete dictionary relating gravity to conformal field theories (CFT) in one lower number of dimensions. 

More recently, proposals aiming to generalize this correspondence to theories of gravity with other asymptotics have emerged, notably asymptotically de Sitter (dS) \cite{Strominger:2001pn, Maldacena:2002vr,Anninos:2011ui} and flat spacetimes \cite{deBoer:2003vf, Pasterski:2016qvg, Raclariu:2021zjz,Pasterski:2021rjz, Donnay:2023mrd,Bagchi:2025vri}. In all these cases, the absence of an asymptotic time-like boundary appears to preclude a dual description relying on a lower-dimensional local, unitary CFT.\footnote{See however \cite{Anninos:2023epi,Anninos:2024wpy, Anninos:2024xhc,Silverstein:2024xnr,Anninos:2025zgr} for very recent progress in our understanding of finite-distance timelike boundaries in gravity and their implications for holography.} In dS spacetime, all proposals to date draw, at least to some extent, lessons from AdS/CFT. For instance, the perturbative computation of cosmological correlators in dS leverages the analytic continuation of holographic observables, including AdS-Witten diagrams \cite{McFadden:2009fg,Bzowski:2023nef}, while non-perturbative aspects are expected to be captured by solutions to a (generalized) bootstrap problem for the Euclidean conformal group \cite{Sleight:2019hfp,Baumann:2020dch,Baumann:2022jpr}. On the other hand, the static patch proposal \cite{PhysRevD.15.2738,Anninos:2011af,Anninos:2012qw} shares similarities with AdS holography for Rindler subregions. 

So far, AdS/CFT has been less of a guiding principle in formulating a holographic correspondence in asymptotically flat spacetimes. Instead, the existing proposals in this context were motivated by a careful analysis of asymptotic symmetries and their relation to scattering amplitudes, which revealed that Poincar\'e symmetry is promoted to an infinite-dimensional asymptotic symmetry algebra  containing a Virasoro subalgebra \cite{Barnich:2010eb,He:2014laa,Kapec:2014opa,Kapec:2016jld}. This led to the proposal that the flat space S-matrix is encoded in a set of correlation functions of operators of a codimension-2 celestial conformal field theory (CCFT) \cite{Pasterski:2016qvg, Pasterski:2017kqt} or codimension-1 Carrollian field theory \cite{Bagchi:2010zz,Bagchi:2016bcd,Donnay:2022wvx} where these symmetries act naturally. The study of subleading terms in the low energy or large-distance asymptotic expansions of gravitational observables further revealed an infinite tower of higher spin  CCFT operators obeying a $w_{1+\infty}$ algebra \cite{Guevara:2021abz,Strominger:2021mtt,Freidel:2021ytz,Adamo:2021lrv}. It is still largely unclear to what extent all these symmetries may constrain (quantum) gravitational dynamics -- see however \cite{Bittleston:2024efo,Fernandez:2024qnu,Charanya:2026pnh} for recent progress on using these symmetries to bootstrap amplitudes in self-dual backgrounds. 

At least from a bottom up perspective, holography in asymptotically flat spacetimes appears to share few, if any, properties with AdS/CFT. On the other hand, it has been observed since the early days of AdS/CFT that flat space scattering amplitudes should emerge from AdS/CFT observables in the limit of vanishing cosmological constant \cite{Susskind:1998vk,Giddings:1999jq,Giddings:1999qu}. Various prescriptions have since been shown to allow for perturbative scattering amplitudes to be extracted from careful limits of Euclidean CFT correlators in either position space \cite{Fitzpatrick:2011jn,Raju:2012zr,Maldacena:2015iua,Komatsu:2020sag,vanRees:2022zmr}, momentum space \cite{Marotta:2024sce} or Mellin \cite{Penedones:2010ue,Fitzpatrick:2011ia} representations, after suitable analytic continuation. Unfortunately, the relation between the different approaches is not always obvious (see \cite{Li:2021snj} for an attempt at reconciling the different prescriptions), and extracting loop scattering amplitudes from AdS-Witten diagrams or conformal correlators is technically rather challenging, since the loop integrals probe the whole AdS spacetime and not just a locally flat patch thereof.

More recently, the connection between flat space holography and limits of AdS/CFT have been explored either via Carrollian limits of CFT correlators \cite{deGioia:2022fcn,Bagchi:2023fbj,deGioia:2024yne,Alday:2024yyj,Lipstein:2025jfj}, or by considering a foliation of Minkowski spacetime with Euclidean (A)dS slices \cite{Iacobacci:2022yjo,Sleight:2023ojm,Melton:2023bjw, Melton:2025ecj}. Notably, it was shown in \cite{deGioia:2023cbd,deGioia:2025mwt}, that all known 2d celestial symmetries emerge in a Carrollian limit of standard CFT$_3$ correlation functions involving a conserved current or the stress tensor. The bulk (AdS) origin of these symmetry enhancements remains so far rather unclear. 
Our paper takes a very first step in establishing this by revisiting the relation between the bulk-boundary correspondences for a free scalar in AdS$_{4}$ and for a free field in 4-dimensional Minkowski spacetime (Mink$_4$). In particular, we construct different bases that span the Hilbert space of a free scalar in AdS$_4$ and relate them to CFT$_3$ primary operators and their shadows. The associated wavefunctions are determined by the representation theory of SO$(3,2)$ and a Lorentz SO$(3,1)$ subalgebra thereof. From a representation theoretic perspective, these constructions yield precise AdS analogues of asymptotic scattering states in Mink$_4$ in either plane wave, Carrollian or conformal primary bases. We show that the latter are obtained from the former in a flat space limit, which realizes the In\"on\"u--Wigner contraction of the $\mathfrak{so}(3,2)$ algebra to 4d Poincar\'e. Related observations have appeared recently in \cite{Berenstein:2025tts}. We review the standard flat space story briefly before outlining our main results.

The main observables in $(d+1)$-dimensional asymptotically flat spacetimes (AFS), are scattering amplitudes. These are matrix elements of a unitary map -- the S-matrix -- that encodes all interactions, typically assumed to be localized in space and time.\footnote{This assumption notoriously fails in theories with massless particles, but this doesn't mean that the S-matrix is not useful in constructing observables in this case.} The matrix elements are usually evaluated in a basis of incoming/outgoing asymptotic states assumed to transform in positive energy representations of the Poincare algebra. A natural basis for scalar single-particle states is obtained from the expansion of a free field in a basis of plane waves
\begin{equation}
\label{eq:plane-waves}
\Phi(x) = \int \frac{d^d \vec{p}}{2\omega (2\pi)^d} \left( e^{-i p \cdot x} a^\dagger(\vec{p})+e^{i p \cdot x} a(\vec{p})  \right),
\end{equation}
by acting with the creation operator on the vacuum,
\begin{equation}
\label{eq:pos-en}
|p\rangle = a^{\dagger}(\vec{p}) |0\rangle.
\end{equation}
Here the vacuum state $|0\rangle$ is defined by
\begin{equation}
  a(\vec{p}) |0\rangle =0
  \end{equation}
and $p$ are $4$-momenta subject to the on-shell condition $p^2 = -m^2$. The on-shell condition ensures that as $\vec{p}$ varies, \eqref{eq:pos-en} span an irreducible, positive energy representation of the Poincar\'e algebra \cite{Weinberg:1995mt}. Asymptotic states are then multi-particle states, namely elements of the Fock space constructed in terms of \eqref{eq:pos-en}. 

The expansion \eqref{eq:plane-waves} is clearly not unique, as it relies on the choice of basis of solutions to the Klein--Gordon equation. In this case, the operators $a^{\dagger}(\vec{p}), ~a(\vec{p})$ live in momentum space. One could equivalently express the bulk field in terms of a linear superposition of operators living on the null boundary of AFS. For massless particles $m^2 = 0$ of energy $\omega$, the momenta $p$ can be parameterized in terms of points on cross-sections of $\mathscr{I}^{\pm}$ with unit normal $\Omega$ 
\begin{equation}
p = \omega \hat{q}(\Omega) = \omega \left(1, \Omega \right)
\end{equation} 
and 
the desired expansion is obtained by observing that \cite{Pasterski:2017kqt,Banerjee:2018gce,Donnay:2022wvx}
\begin{equation}
\label{eq:Carroll-bases}
\Psi^{\pm}_{\Delta}(u_0,\hat{q}; x) \equiv \int_0^{\infty} d\omega  \omega^{\Delta - 1} e^{\pm i \omega u_0 - \epsilon \omega} e^{\pm i \omega \hat{q} \cdot x} =  i^{-\Delta} \frac{\Gamma(\Delta)}{(\mp u_0 \mp \hat{q} \cdot x - i\epsilon)^{\Delta}}
\end{equation}
By virtue of being superpositions of positive/negative energy plane waves, these wavefunctions have positive/negative energy, however they are no longer momentum eigenstates. They are instead wavepackets of definite boost weight $\Delta$.  
For fixed $\Delta$, $u_0 \in (-\infty, \infty)$ and arbitrary $\hat{q}$, \eqref{eq:Carroll-bases} span positive and negative energy irreducible representations of the massless Poincar\'e algebra. 
Choosing $\Delta = d - 1$, we can rewrite \eqref{eq:plane-waves} as
\begin{equation}
\label{eq:Carroll-exp}
\Phi(x) = \frac{1}{2(2\pi)^d} \int_{-\infty}^{\infty} du_0 \int d\Omega \left[ \Psi_{d-1}^+(u_0,\hat{q};x) \mathcal{O}(u_0,\Omega) + \Psi_{d-1}^-(u_0,\hat{q};x) \mathcal{O}^{\dagger}(u_0,\Omega)   \right],
\end{equation}
where the operators $\mathcal{O}$ are related to $a$ by a Fourier transform in $\omega$, namely
\begin{equation}
\label{eq:Carr-op}
    \mathcal{O}(u, \Omega) = \frac{1}{2\pi} \int_0^{\infty} d\omega e^{-i\omega u} a(\omega, \Omega).
\end{equation} Since $u_0$ can be chosen to be a retarded time at $\mathscr{I}^{+}$ or an advanced time at $\mathscr{I}^-$, \eqref{eq:Carroll-exp} relates an on-shell bulk field to operators on the conformal boundary of Mink$_{d+1}$.  

Furthermore, one can directly obtain from \eqref{eq:Carroll-exp} the Kirchhoff--d'Adh\'emar formula in Minkowski space \cite{Penrose,Newman:1976gc,Donnay:2022wvx} which allows for a bulk field to be expressed in terms of its value at a cut of $\mathscr{I}$. This can be seen by noting that 
\begin{equation}
\Psi^{\pm}_{d - 1}(u_0, \hat{q}; x) = (\pm i)^{2 - d} \partial_{u_0}^{d - 2} \Psi_{1}^{\pm}(u_0, \hat{q}; x), 
\end{equation}
integrating by parts and evaluating the $u_0$ integrals. Then \eqref{eq:Carroll-exp} becomes
\begin{equation}
\label{eq:KA-Mink}
\Phi(x) =  \frac{i^d}{2(2\pi)^{d-1}}   \int d\Omega \left[ \partial_{u_0}^{d - 2}\mathcal{O}(u_0,\Omega) + (-1)^{d-2} \partial_{u_0}^{d - 2}\mathcal{O}^{\dagger}(u_0,\Omega)    \right] \left. \right|_{u_0 = - \hat{q}\cdot x}.
\end{equation}
The RHS of this formula can be expressed in terms of the boundary value of the field by noting that in the limit $r \rightarrow \infty$, 
\begin{equation}
\label{eq:bdry-val}
\lim_{r \rightarrow \infty} r^{\frac{d - 1}{2}} \p_u^{\frac{d-1}{2}} \Phi(x) = \frac{ i }{2(2\pi)^{\frac{d-1}{2}}}  \left[ \partial_{u}^{d - 2}\mathcal{O}(u,\Omega) - \partial_{u}^{d - 2}\mathcal{O}^{\dagger}(u,\Omega)    \right].
\end{equation}

One can further decompose \eqref{eq:Carroll-exp} in terms of highest weight representations of the $\mathfrak{so}(d,1)$ Lorentz subalgebra of Poincar\'e. Assuming that $|u_0| < |\hat{q}\cdot x|$, one can achieve this by expanding the wavefunctions \eqref{eq:Carroll-bases} in powers of $u_0$
\begin{equation}
\label{eq:Carroll-wf-taylor}
\Psi_{\Delta}^{\pm}(u_0, \hat{q};x) = \sum_{n = 0}^{\infty} i^{\mp \Delta} \frac{(u_0 \pm i\epsilon)^n}{\Gamma(n + 1)} \frac{ \Gamma(\Delta + n)}{(- \hat{q}\cdot x \mp i\epsilon)^{\Delta + n}} \equiv  \sum_{n = 0}^{\infty} i^{\mp \Delta} \frac{\Gamma(\Delta + n) (u_0 \pm i\epsilon)^n}{\Gamma(n + 1)} \psi_{\Delta + n}^{\pm}(\hat{q};x).
\end{equation}
We recognize in this expansion $\psi_{\Delta + n}^{\pm}(\hat{q}; x)$ which are the (unnormalized) conformal primary wavefunctions of $\mathfrak{so}(d,1)$ \cite{Pasterski:2017kqt}. They span lowest-weight representations of dimension $\Delta + n$ of this Lorentz subalgebra. For $|u_0| > |\hat{q}\cdot x|$, one finds a similar expansion with $n \in -\Delta - \mathbb{N}$. For $d = 3$, these two sets of wavefunctions are precisely dual to the bases of celestial Goldstone and memory/conformally soft modes constructed in \cite{Freidel:2022skz}. It is also manifest from \eqref{eq:Carroll-wf-taylor} that the relation between Carrollian and conformal primary wavefunctions is given by light-transforms with respect to $u_0$ \cite{Bagchi:2023fbj, deGioia:2024yne} analogous to those introduced in \cite{Donnay:2022wvx,Donnay:2022sdg} for Carrollian operators defined by the extrapolate dictionary. In particular, one can project out the incoming or outgoing conformal primary wavefunction by appropriately choosing the $i\epsilon$ prescription in the kernel of the light transform. 
The wavefunctions in \eqref{eq:Carroll-bases} admit an equivalent representation as an integral over principal series wavefunctions
\begin{equation}
\label{eq:principal-series}
\Psi_{\Delta}^{\pm}(u_0, \hat{q};x) = \int_{\lambda \in \mathcal{C}} \frac{d\lambda}{\sin \pi \lambda}\frac{(-1)^{\lambda}\Gamma(\lambda + \Delta)}{\Gamma(\lambda)} i^{\mp \Delta}  (u_0 \pm i\epsilon)^{\lambda} \psi_{\Delta + \lambda}^{\pm}(\hat{q};x),
\end{equation}
where $\mathcal{C}$ is a contour along the imaginary axis. This can be established by deforming the contour to the right complex $\lambda$ plane after suitable regularization \cite{Mitra:2024ugt,deGioia:2025mwt}.

Substituting \eqref{eq:principal-series} into \eqref{eq:Carroll-exp} and taking the integration contour\footnote{Note that for odd $d$, $\psi^{\pm}_{d - 1 + \lambda}$ and $\psi^{\pm}_{\lambda}$ are related by an integer number of derivatives with respect to $u$. The relative normalization cancels the ratio of Gamma functions in \eqref{eq:principal-series} which therefore doesn't introduce additional zeros for this choice of integration contour.} at $\lambda = -\frac{d - 1}{2} + i\mathbb{R}$ for odd values of $d$ we obtain the decomposition of a free massless scalar in terms of conformal primary wavefunctions
\begin{equation}
\label{eq:conformal-primary-exp}
\Phi(x) =  \frac{1}{2(2\pi)^d} \int_{\mathbb{R}} d\lambda \int d\Omega \left[\psi^+_{\frac{d - 1}{2} + i\lambda}(\hat{q};x) O^+_{\frac{d - 1}{2} - i\lambda}(\Omega) +\psi^-_{\frac{d - 1}{2} - i\lambda}(\hat{q};x) O^-_{\frac{d - 1}{2} + i\lambda}(\Omega)  \right)
\end{equation}
where the operators $O$ are related to the operators $\mathcal{O}$ in \eqref{eq:Carroll-exp} by
\begin{equation}
O^{+}_{\frac{d - 1}{2} - i\lambda}(\Omega) =  \frac{(-1)^{ i\lambda - \frac{d-1}{2}} i^{ 1 - d}}{\pi}\Gamma\left(\frac{d - 1}{2} + i\lambda\right) \Gamma\left(\frac{1 + d}{2} - i\lambda\right) \int_{-\infty}^{\infty} du (u + i\epsilon)^{ i\lambda - \frac{d - 1}{2}} \mathcal{O}(u, \Omega)
\end{equation}
and $\mathcal{O}^-_{\frac{d - 1}{2} + i\lambda}(\Omega) = \left(\mathcal{O}^+_{\frac{d - 1}{2} - i\lambda}\right)^{\dagger}$.
They are operators in a $(d-1)$-dimensional ``celestial'' conformal field theory \cite{Pasterski:2017kqt,Kapec:2021eug,Pano:2023slc}.

Since all these results are implied by representation theory of the higher-dimensional Poincar\'e group, it should not be surprising that they admit generalizations to other Lie groups. Of particular interest are SO$(d,2)$ and SO$(d+1,1)$ which are the the asymptotic symmetry groups of Lorentzian $(d+1)$-dimensional (A)dS$_{d+1}$ spacetimes. In this paper we construct the counterparts of \eqref{eq:plane-waves}, \eqref{eq:Carroll-exp} and \eqref{eq:conformal-primary-exp} in the case of Lorentzian AdS spacetimes. This will allow us to obtain a precise relation between the Carrollian/celestial holographic dictionaries and the standard AdS/CFT dictionary. The latter is usually formulated in terms of AdS highest weight wavefunctions, which unfortunately obscures the relation between the two.  As we will explain in detail, our results shed some new light on well-studied topics in this context such as bulk reconstruction \cite{ Hamilton:2005ju, Hamilton:2006az, Hamilton:2006fh,Dong:2016eik} and the flat space limit of AdS/CFT \cite{Susskind:1998vk,Giddings:1999jq,Giddings:1999qu}. 
Although, we focus on the case of a scalar field in AdS$_4$, our results admit generalizations to any spacetime dimension and higher spin fields. Building on this work, we hope to eventually achieve a unified bottom-up realization of the holographic principle in different backgrounds. 

We now summarize the content of our paper, highlighting our main new results. 

{\bf A basis of AdS bulk-to-boundary propagators.} It is well-known that the positive frequency ($\omega = \Delta + 2n + \ell$) subspace of solutions to the Klein--Gordon wave equation in AdS$_4$ is a highest weight irreducible representation of the $3$-dimensional conformal algebra $\mathfrak{so}(3,2)$ of dimension $\Delta$ and vanishing spin. We show that another basis for this irreducible highest weight representation can be constructed from (anti-)time-ordered AdS bulk-to-boundary propagators provided that the boundary points are restricted to specific subregions. To see this, we express the (anti-)time-ordered bulk-to-boundary propagators as a linear superposition of positive/negative frequency wavefunctions
\begin{equation}
\label{eq:time-ordered-btb-intro}
\begin{split}
   G_{\Delta}^{\rm T.O.}(X;P) =  G_{\Delta}^+(X; P) \Theta(\tau - \tau_p) +  G_{\Delta}^-(X;P) \Theta(\tau_p - \tau),
    \end{split}
\end{equation}
where
\begin{equation}
\label{eq:Gpm}
 G^+_{\Delta}(X;P) =  \sum_{k = 0}^{\infty} \sum_{\ell = 0}^{\infty} \pi \sqrt{2 \ell + 1} (-1)^{k}\frac{2^{\Delta}\Gamma(\Delta + \ell + k)}{k! \Gamma(\Delta)\Gamma(\frac{3}{2} + \ell)}  \psi^{+}_{k \ell 0}(\tau - \tau_p, \rho, \Omega \cdot \Omega_p), \quad G^-_{\Delta}(X;P) = G^+_{\Delta}(X;P)^*.
\end{equation}
We derive this formula from first principles starting with the definition
\begin{equation}
    G_{\Delta}^{\rm T.O.}(X; P) = \frac{1}{(-P\cdot X + i\epsilon)^{\Delta}}
\end{equation}
in Section \ref{sec:mode-expansion-btb}. Here $X$ and $P$ are embedding space coordinates denoting bulk and boundary points in AdS. $\tau, \rho$ and $\Omega$ are the global AdS coordinates introduced in \eqref{eq:global-AdS}. 

The mode expansion \eqref{eq:time-ordered-btb-intro} suggests that  $G_{\Delta}^{\rm T.O.}$ with respectively $\tau > \tau_p$ and $\tau < \tau_p$ span the positive and negative frequency components of the space of solutions to the Klein--Gordon (KG) equation. We confirm this by computing the inner products of (anti-)time ordered propagators on a \textit{bulk} constant time surface $\Sigma_{\tau_0}$. We show in Section \ref{sec:inner-prodG} that this is given by 
\begin{equation}
     \label{eq:time-ordered-btb-ip-intro} 
\begin{split}
\langle G^{\rm T.O.}_{\Delta}(X;P_1), &G^{\rm T.O.}_{\Delta}(X;P_2) \rangle_{\tau = \tau_0} = 2^{\Delta +1} \pi^{3/2} \frac{\Gamma(\Delta - \frac{1}{2})}{\Gamma(\Delta)} \\
&\times \left( G_{\Delta}^+(P_2, P_1) \Theta(\tau_0 - \tau_1) \Theta(\tau_0 - \tau_2) - G_{\Delta}^-(P_2, P_1) \Theta(\tau_1 - \tau_0) \Theta(\tau_2 - \tau_0)\right).
\end{split}
\end{equation}
The formula for the anti-time-ordered propagators follows by complex conjugation. 
The RHS of this equation can also be expressed in terms of (anti-)time-ordered propagators -- see eq. \eqref{eq:time-ordered-btb-ip-to}.  The inner products of propagators with boundary insertions above/below $\Sigma_{\tau_0}$ are (up to a sign) equal to CFT$_3$ two-point functions with specified time orderings. They hence form non-orthogonal bases, unlike the standard AdS wavefunctions. We furthermore compute the inner products involving shadow transforms of AdS bulk-to-boundary propagators in Section \ref{sec:shadow-ip-comp}. The results allow us to establish that $G_{\Delta}^{\rm T.O.}$ with boundary points above (below) $\Sigma_{\tau}$ form orthogonal, delta-function normalizable bases for positive (negative) energy subspaces of solutions to the KG equation with respect to the \textit{shadow inner product} in AdS. 

{\bf AdS bulk reconstruction, revisited.} The positive and negative frequency bases of bulk-to-boundary propagators allow us to ``reconstruct'' a free AdS field as an integral over boundary operators. The different choices of positive and negative frequency wavefunctions yield equivalent representations that may involve either primary operators $\mathcal{O}_{\Delta}$, shadows thereof  $\widetilde{\mathcal{O}}_{3 - \Delta}$ or both. For instance, we can write 
\begin{equation}
\label{eq:bulk-rec-intro}
     \Phi(\tau, \rho, \Omega) = \int_{B} d\tau_p d\Omega_p\left( \alpha  G^{\rm A.T.O.}_{3 - \Delta}(\tau, \rho, \Omega; \tau_p, \Omega_p) \mathcal{O}_{\Delta}(\tau_p, \Omega_p) + \alpha^*  G^{\rm T.O.}_{3-\Delta}(\tau, \rho, \Omega; \tau_p, \Omega_p) \mathcal{O}_{\Delta}(\tau_p, \Omega_p) \right),
\end{equation}
where $\alpha$ is a coefficient fixed by consistency with the extrapolate dictionary given in \eqref{eq:a-b}. The boundary integration region is taken to be a time interval of length $\pi$ for $\tau_p > \tau$. Taking the boundary integration region to be an interval of length $\pi$ with $\tau_p < \tau$ amounts to considering the hermitian conjugate of this formula, hence gives an equivalent representation of $\Phi$.  

This formula is similar to the HKLL formula, where the kernels are also related to bulk-to-boundary propagators of dimension $3 - \Delta$ \cite{Hamilton:2006az} (we review this in Section \ref{sec:bulk-rec}). However, unlike in the HKLL case, our result is not just supported in the bulk subregion spacelike separated from the boundary. Since the propagators have specified time orderings, one can uniquely analytically continue the result beyond these subregions. Furthermore, we demonstrate in Section \ref{sec:bulk-rec-AdS} that our reconstruction formula has \textit{normalizable} boundary asymptotics and the boundary value of the field is related to a primary $\mathcal{O}_{\Delta}$ in agreement with the extrapolate dictionary \cite{Harlow:2011ke}. The key identity that allows us to establish this is the relation between the mode expansion of $\mathcal{O}_{\Delta}$ and its shadow $\widetilde{\mathcal{O}}_{3 - \Delta}$, which we compute in \eqref{eq:O-tilde-modes}. 

We show in Section \ref{sec:bulk-rec-AdS} that one may obtain equivalent formulae by letting $\Delta \rightarrow 3 - \Delta$ in \eqref{eq:bulk-rec-intro} either in one or both terms, for instance
    \begin{equation}
\label{eq:exp-fin-intro}
     \Phi(\tau, \rho, \Omega) =  \int_{B} d\tau_p d\Omega_p \left( \alpha G^{\rm A. T.O.}_{\Delta}(\tau, \rho, \Omega; \tau_p, \Omega_p) \widetilde{\mathcal{O}}_{3 - \Delta}^+(\tau_p, \Omega_p) + \alpha^* G^{\rm T.O.}_{\Delta}(\tau, \rho, \Omega; \tau_p, \Omega_p) \widetilde{\mathcal{O}}_{3 - \Delta}^-(\tau_p, \Omega_p) \right).
\end{equation}
This equivalence can be proven by using the inner products of (anti-)time-ordered bulk-to-boundary propagators with the lowest-weight AdS wavefunctions to express each of the terms in \eqref{eq:bulk-rec-intro} and \eqref{eq:exp-fin-intro} in terms of the positive and negative frequency components of the standard mode expansion \eqref{eq:AdS-bulk-field}.

As a byproduct of our bulk reconstruction formulae, we obtain in Section \ref{sec:AdS-KA} an AdS analogue of the Kirchhoff--d'Adh\'emar formula in Minkowski space \eqref{eq:KA-Mink}. This allows for a positive integer $\Delta$ free field in AdS to be reconstructed from its value at a cut of the timelike boundary of AdS -- see eq. \eqref{eq:KA-AdS}.  
We show that in the flat space limit, this formula reduces to
\begin{equation}
 \begin{split}
      \Phi(u, r, \Omega) 
      &= -\frac{(-1)^{\Delta-1} \ell^{-\Delta+1}}{2^{\Delta-1}\pi^{1/2}\Gamma(\Delta - \frac{1}{2})}\int d\Omega_p \p_u^{\Delta} \phi(u_p, \Omega_p)\left. \right|_{u_p = -\ell^{-1} \hat{q}\cdot x}, \quad \Delta \in \mathbb{N},
      \end{split}
 \end{equation}
where $u, r, \Omega$ are retarded coordinates. For $\Delta = 1$ this agrees up to a factor of $2$ with \eqref{eq:KA-Mink}, \eqref{eq:bdry-val} for $d = 3$. While the relation between the AdS and flat space extrapolate dictionaries has been discussed in \cite{deGioia:2022fcn,Bagchi:2023fbj,Bagchi:2023cen,deGioia:2024yne,Alday:2024yyj}, the derivation of the Kirchhoff--d'Adh\'emar formula \eqref{eq:KA-Mink} from a flat space limit of its AdS counterpart is to the extent of our knowledge new. The reconstruction of the bulk metric and geodesics from boundary data in asymptotically AdS spacetimes has been studied in \cite{Caron-Huot:2025she, Caron-Huot:2025hmk}.

{\bf Principal series and $\mathfrak{so}(3,1)$ wavefunctions in AdS$_4$.}
The space of solutions to the KG equation in AdS$_4$ can be decomposed in terms of wavefunctions spanning irreducible representations of an $\mathfrak{so}(3,1)$ subalgebra of $\mathfrak{so}(3,2)$. We construct the associated wavefunctions which are the AdS$_4$ analogs of conformal primary wavefunctions in 4d Minkowski space \cite{Pasterski:2016qvg,Pasterski:2017kqt}.

The first such basis is obtained by foliating the AdS$_4$ subregions determined by the lightcone through the origin with Euclidean (A)dS$_3$ slices.  This construction is the AdS analog of that introduced in \cite{deBoer:2003vf} for Minkowski spacetime. Just as in the flat space case, the AdS$_4$ Laplacian decomposes into a sum of the Laplacian on the (A)dS$_3$ leaves, which we parameterize with coordinates $\hat{x}$, and a differential operator with respect to the orthogonal direction $\alpha$. As a result the solutions factorize into (A)dS$_3$ bulk-to-boundary propagators (which correspond to scalar irreducible representations of $\mathfrak{so}(3,1)$ of dimension $\delta$) and a function $f_{\Delta,\delta}(\alpha)$  obeying a second order ODE, namely
\begin{equation}
\label{eq:hyp-wf-intro}
    \Psi_{\Delta,\delta}(\alpha, \hat{x}; \hat{q}) = f_{\Delta,\delta}(\alpha) G_{\delta}(\alpha, \hat{x}; \hat{q}).
\end{equation}
Here $\hat{q}$ parameterize points on a constant time slice on the AdS boundary and $f_{\Delta,\delta}(\alpha)$ are explicitly determined in \eqref{eq:fa}. 
We show that Sturm-Liouville theory provides a complete basis of solutions to this equation for $\delta = 1 + i\lambda$, $\lambda \in \mathbb{R}_+$ which correspond to $\mathfrak{so}(3,1)$ principal series representations.

An alternative set of globally defined $\mathfrak{so}(3,1)$ wavefunctions is obtained by considering a foliation of AdS$_4$ with null cones. We construct a set of wavefunctions for the AdS$_4$ Laplacian that diagonalize $\mathfrak{so}(3,1)$ boosts with eigenvalue $\delta$,
\begin{equation}
\label{eq:psi-null-intro}
\begin{split}
    \Psi_{\Delta, \delta}^{\rm null}( \tau, r, \Omega; \Omega_q) &= \frac{\Gamma(\Delta - \delta)}{\Gamma(\Delta - \frac{1}{2})} {}_2F_1\left[\frac{1}{2}(\Delta - \delta), \frac{1}{2}(\Delta - \delta + 1); \Delta - \frac{1}{2}; \frac{1}{(\ell \cos\tau - r \sin \tau)^2} \right] \\
    &\times \frac{1}{(r \cos \tau + \ell \sin \tau - r \Omega \cdot \Omega_q)^{\delta}} \frac{1}{(\ell^2 \cos \tau - \ell r \sin \tau + i\epsilon)^{\Delta - \delta}}.
    \end{split}
\end{equation}
We expect these wavefunctions to form an equivalent basis of solutions to the KG equation in AdS$_4$ provided that $\delta = 1 + i\lambda$. We provide two constructions of these wavefunctions in Section \ref{sec:null-cpw-ads}. 
A free field in AdS$_4$ hence admits the decomposition
\begin{equation}
\label{eq:AdS-cpw-dec-intro}
    \Phi(X)  = \int_{\mathbb{R}} d\lambda \int d^{2} \hat{q}  \Psi_{\Delta, 1+ i\lambda}(X;\hat{q}) \mathcal{O}_{1 - i\lambda}(\hat{q}),
\end{equation}
where $\mathcal{O}_{1 - i\lambda}$ are operators transforming in representations of an $\mathfrak{so}(3,1)$ subalgebra of $\mathfrak{so}(3,2)$ of dimension $\delta = 1 - i\lambda$. 
We also study the flat space limits of our formulas and, in particular, demonstrate that \eqref{eq:hyp-wf-intro} and \eqref{eq:psi-null-intro} reduce to massive and massless conformal primary wavefunctions with respect to a Lorentz subalgebra of 4d Poincar\'e.

It will be interesting to study under what conditions $\delta \in \mathbb{Z}$ may form a basis. We expect appropriate generalizations of these wavefunctions to give rise to AdS analogs of conformally soft modes in flat space. The AdS counterparts of the holographic symmetry algebras \cite{Guevara:2021abz,Strominger:2021mtt} and their $\Lambda$ deformations \cite{Taylor:2023ajd} will be discussed elsewhere.

{\bf Bridging AdS and flat space holography.}
Unlike the standard AdS wavefunctions, AdS bulk-to-boundary propagators diagonalize boosts towards a point on the boundary with eigenvalue $\Delta$. They are the $\mathfrak{so}(3,2)$ analogs of the Carrollian wavefunctions for the 4d Poincar\'e algebra \cite{Banerjee:2018gce, Donnay:2022wvx, Bagchi:2023cen}. We show in Section \ref{sec:flat-space-lim} that in the flat space limit $\ell, \Delta \rightarrow \infty$, the time-ordered bulk-to-boundary propagators reduce to plane waves. In particular, we show that if the bulk and boundary points are null separated, one obtains a massless plane wave, while massive plane waves are obtained for analytically continued boundary times. This provides a direct derivation of the prescriptions used in \cite{Komatsu:2020sag} to extract massive scattering amplitudes from CFT correlators.  For fixed $\Delta$, it was already shown in \cite{deGioia:2022fcn, Bagchi:2023fbj, deGioia:2024yne} that bulk-to-boundary propagators become Carrollian wavefunctions. We clarify the relation between these wavefunctions and plane waves, by showing that the former are linear superpositions of positive or negative energy plane waves (depending on the $i\epsilon$ prescription) that diagonalize boosts. As such, they provide alternative bases of wavefunctions for irreducible Poincar\'e representations. This provides a representation theoretic explanation for the modified Mellin transform introduced in \cite{Banerjee:2018gce}. 

These results allow us to show that the bulk reconstruction formulae we propose reduce to their flat space counterparts \eqref{eq:plane-waves} or \eqref{eq:Carroll-bases} depending on whether $\Delta \rightarrow \infty$ or $\Delta$ is kept fixed. In the latter case, we obtain a precise identification between the Carrollian operators \eqref{eq:Carr-op} and primary operators or shadows thereof in CFT$_3$. We show that two point functions involving 3d shadow operators solve the conformal Carroll Ward identities. Together with our computation of the AdS shadow inner products in Section \ref{sec:AdS-bases}, this suggests that CFT correlators involving primary and shadow operator insertions with specific kinematics share similar properties to correlators of electric Carrollian field theories \cite{Bagchi:2025vri}. 

{\bf Relation to previous works. } Equation \eqref{eq:time-ordered-btb-intro} can be obtained indirectly from a limit of an integral representation of  the AdS bulk-to-bulk propagator \cite{Giddings:1999qu}. We explain this relation in Appendix \ref{eq:Giddings}. Similar formulas have appeared in the vast literature on AdS bulk reconstruction (for reviews, see \cite{Harlow:2018fse, Kajuri:2020vxf}). We note however that \eqref{eq:time-ordered-btb-intro} is related to, yet differs from the HKLL kernel constructed in \cite{Hamilton:2006az}, as we discuss in Section \ref{sec:bulk-rec} (this is also emphasized in \cite{Harlow:2018fse}).

To the extent of our knowledge, the bulk inner product of bulk-to-boundary propagators on a fixed bulk slice with arbitrary boundary insertions is new. Of course, the form of the result is expected by conformal invariance \cite{Kaplan1}, but the precise normalization and the effects of the different $i\epsilon$ prescriptions were important in establishing a precise relation between the  Lorentzian AdS and flat space holographic dictionaries. Unlike the bulk inner product, the boundary inner product of bulk-to-boundary propagators is straightforward to compute, see for instance \cite{Melton:2025ecj}. One can use this result together with Stokes' theorem to deduce the \textit{difference} between inner products on distinct bulk slices. Our result \eqref{eq:time-ordered-btb-ip-intro} is compatible with these computations. 

AdS-inspired ideas have been used to revisit the path integral and associated observables in flat space in \cite{Kim:2023qbl,Jain:2023fxc,Kraus:2024gso,Ammon:2025avo}. In particular, \cite{Kraus:2024gso} discusses the flat space limit of bulk wavefunctions obtained by smearing boundary sources with bulk-to-boundary propagators. The boundary limit of these wavefunctions is identified with the non-normalizable boundary values of the bulk fields or equivalently, the sources for boundary operators which are kept fixed in the path integral. Our paper is complementary in that it revisits the reconstruction of the normalizable component of the bulk field and its canonical quantization in terms of boundary operators. It will be interesting to consider the relation between the analytic continuation of the Euclidean path integral considered in \cite{Jain:2023fxc} and the intrinsically Lorentzian prescription proposed here, as well as the relation to the holographic renormalization procedure proposed in \cite{Ammon:2025avo}.

Finally, the more recent literature on the flat space limit of AdS/CFT usually relies on HKLL reconstruction and the properties of the HKLL kernel in various limits (a limit of large radial quantum number has to be taken), to justify the different prescriptions relating CFT correlators to flat space scattering amplitudes. This is nicely summarized in \cite{Li:2021snj}. We found that these approaches obscure the geometric picture, including how the limit impacts the bulk isometries and the bulk-boundary map. Similar observations have been made and, in part, addressed recently in \cite{Berenstein:2025tts}, which our work is in agreement with. We provide some further technical details and a more precise explanation of why and how the different versions of the flat space (Carrollian/celestial) holographic dictionaries (so far developed largely for free fields) are an output of flat space limits of the holographic dictionary in AdS. This paper is accompanied by the letter \cite{Navarro:2025}.

\section{Preliminaries: free fields in AdS$_{d+1}$}

In this section we review the canonical quantization of a free field in Lorentzian $(d+1)$-dimensional Anti-de-Sitter space (AdS$_{d+1}$). A standard basis of solutions to the Klein--Gordon equation  is given by the wavefunctions \eqref{eq:AdS-wf} \cite{Kaplan1, Penedones:2016voo}. The positive energy wavefunctions span unitary lowest-weight representations of $\mathfrak{so}(d,2)$, as briefly outlined in Section \ref{sec:rel-cft}. The decomposition of a free field in this basis is the starting point of bulk reconstruction in AdS$_{d+1}$, which expresses the on-shell bulk field as a linear superposition of boundary operators in a $d$-dimensional conformal field theory (CFT$_d$). We review this procedure first developed by HKLL \cite{Hamilton:2005ju,Hamilton:2006az} and explain some of its shortcomings in Section \ref{sec:bulk-rec}.

\subsection{AdS isometries}
\label{sec:AdS-free-field}

AdS$_{d+1}$ of radius $\ell$ can be described as a hyperbolic surface embedded into $\mathbb{R}^{2,d}$ with coordinates $X^A$ and metric $\eta_{AB}$,
\begin{equation}
    \label{eq:AdS-emb}
  \eta_{AB} X^{A} X^B \equiv  -(X^0)^2 + (\vec{X})^2 - (X^{d+1})^2 = -\ell^2.
\end{equation}
A set of global coordinates in AdS$_{d+1}$ is obtained by parameterizing \cite{Penedones:2016voo, Kaplan1}
\be 
\label{eq:global-AdS}
\begin{split}
    X^0 &= \ell \frac{\sin \tau}{\cos \rho}, \quad  X^{d+1} = \ell \frac{\cos\tau}{\cos \rho},\\
    X^i &= \ell \tan \rho \Omega_i, \quad \Omega \cdot \Omega = 1.\\ 
\end{split}
\ee
Here $\tau$ and $\rho$ are time and radial AdS coordinates, while $\Omega$ is the unit normal to a point on the $(d-1)$-dimensional sphere $S^{d-1}$. In these coordinates, the AdS metric takes the form
\begin{equation}
\label{eq:AdS-metric}
ds^2 = \frac{\ell^2}{(\cos \rho)^2}\left( -d\tau^2 + d\rho^2 + \sin^2 \rho d \Omega^2_{d-1} \right),
\end{equation}
where the metric on $S^{d-1}$ is
\begin{equation}
    d\Omega^2_{d - 1} = \gamma_{\alpha\beta} dx^{\alpha} dx^{\beta}.
\end{equation}
It is clear from \eqref{eq:AdS-emb} that the AdS$_{d+1}$ isometry group is one of the connected components of the Lorentz group SO$(d,2)$ in $\mathbb{R}^{2,d}$. 
The isometry algebra $\mathfrak{so}(d,2)$ is generated by the embedding space Lorentz generators 
\be 
M_{AB} = -i \left(X_A \p_{B} - X_B \p_{A}\right)
\ee
whose commutators are given by 
\be 
[M_{AB}, M_{CD}] = i\left(\eta_{AC} M_{BD} + \eta_{BD} M_{AC} - \eta_{BC} M_{AD} - \eta_{AD} M_{BC} \right).
\ee

It will be useful to reorganize these $\mathfrak{so}(d,2)$ generators in terms of generators of $\mathfrak{so}(d)$ rotations, and boosts associated to the two times $X^0$ and $X^{d+1}$
\be
\begin{split}
M_{ij} &\equiv -i \left(\Omega_i \p_{\Omega_j} - \Omega_j \p_{\Omega_i}\right), \quad i, j = 1, \cdots d,\\
K_i &\equiv i M_{i0} + M_{i d+1}, \\
P_i &\equiv i M_{i0} - M_{i d+1},\\
D &\equiv -M_{0 d+1}.
\end{split}
\ee
Using the Lorentz algebra, it can be shown that 
\be 
\begin{split}
[D, P_i] &= P_i, \hspace{0.34in} [D, K_i] = -K_i, \hspace{1.21in} [P_i, K_j] = 2\delta_{ij} D + 2i M_{ij}\\
[D, M_{ij}] &= 0, \quad\quad [M_{ij}, P_k] = i \left( \delta_{ik} P_j - \delta_{jk} P_i \right), \quad \quad [M_{ij}, K_k] = i \left( \delta_{ik} K_j - \delta_{jk} K_i \right).
\end{split}
\ee
From the bulk perspective, $D$ generates time translations, while $M_{ij}$ generate rotations. From the boundary perspective,
$D$, $P_i$, $K_i$ and $M_{ij}$ correspond to dilatation, translations, special conformal generators and Lorentz generators of a $d$-dimensional conformal field theory (CFT$_d$). The AdS time-translation generator $D$ together with the Cartan elements of $\mathfrak{so}(d)$ span a Cartan subalgebra of $\mathfrak{so}(d,2)$. 
Note that $P_i, K_i$ (for fixed $i$) and $D$ generate an $\mathfrak{sl}(2,\mathbb{R})$ sub-algebra of $\mathfrak{so}(d,2)$.

In the coordinates \eqref{eq:global-AdS}, setting $\ell = 1$, the AdS$_{d+1}$ isometry generators take the form 
\be
\label{eq:AdS-Gens}
\begin{split}
D &= i\frac{\p}{\p \tau}, \quad M_{ij} = -i \left(\Omega_i \nabla_j - \Omega_j \nabla_i \right), \\
K_i &=  -i e^{i\tau} \left[ \Omega_i\left( \cos \rho \p_{\rho} + i\sin \rho \p_{\tau} \right)  + \frac{1}{\sin \rho} \nabla_i \right], \\
P_i &=  i e^{-i\tau} \left[ \Omega_i \left(\cos \rho \p_{\rho} - i\sin\rho \p_{\tau} \right) + \frac{1}{\sin \rho} \nabla_i  \right],
\end{split}
\ee
where $\nabla_i$ is defined by
\begin{equation}
    \nabla_i = \frac{\partial}{\partial \Omega_i} - \Omega_i \Omega^j \frac{\partial}{\partial\Omega_j}.
\end{equation} 
Note that $K^{\dagger}_i = -P_i$. Note also that $P_i$ and $K_i$ get exchanged under $\tau \rightarrow -\tau$. We therefore expect that this transformation exchanges the positive and negative energy AdS wavefunctions, which we will confirm in the following.

\subsection{AdS wavefunctions and canonical quantization}

The standard canonical quantization of a free scalar field in AdS starts with the identification of a basis of solutions to the Klein--Gordon (KG) equation 
\begin{equation}
    \label{eq:AdS-wave-eq}
    (\Box_{AdS} - m^2) \Phi = 0.
\end{equation}
This equation is an eigenvalue equation for a quadratic Casimir of $\mathfrak{so}(d,2)$, and one can show that positive/negative energy solutions with $m^2 = \Delta(\Delta - d)$ that are regular at the origin are in correspondence with lowest/highest-weight representations of $(\Delta, s = 0)$, provided that the eigenvalues of $D$ are quantized. We review the full derivation of this fact in Appendices \ref{app:highest-weight}, \ref{app:scalar} and summarize the the main results below.  

In the global coordinates \eqref{eq:global-AdS}
\begin{equation}
\Box_{AdS} = \cot^2 \rho \Box_{d-1} + (d - 1) \cot \rho \p_{\rho} + \cos^2 \rho \left(\p_{\rho}^2 - \p_{\tau}^2 \right), \quad  \Box_{d-1} = \gamma^{\alpha\beta} \nabla_{\alpha} \nabla_{\beta} .
\end{equation}
A convenient basis of solutions is given by eigenstates of the Cartan subalgebra, namely, wavefunctions obeying
\begin{equation}
\begin{split}
    D \psi_{n, \ell}^{\pm} &= \pm \omega_n \psi_{n, \ell}^{\pm},\\
    \Box_{d - 1} \psi_{n, \ell}^{\pm} & = -\ell(\ell + d- 2) \psi_{n, \ell}^{\pm}.
    \end{split}
\end{equation}
Substituting these into \eqref{eq:AdS-wave-eq} and imposing regularity at the origin, we obtain the following wavefunctions
\begin{equation}
\label{eq:AdS-wf}
\psi_{n\ell J}^{+}(\tau, \rho, \Omega) = e^{- i\omega_n \tau} f_{n\ell}(\rho) Y_{\ell J}(\Omega),\quad\quad \psi_{n\ell J}^{-}(\tau, \rho, \Omega) = e^{ i\omega_n \tau} f_{n\ell}(\rho) Y^*_{\ell J}(\Omega).
\end{equation}
Here
\begin{equation}
\label{eq:omega}
\omega_n = \Delta + 2n + \ell, \quad \Delta(\Delta - d) = m^2,  \quad n, \ell \in \mathbb{N}\ ,
\end{equation}
the radial component of the wavefunction is given by
\begin{equation}
\label{eq:radial}
f_{n\ell}(\rho) = \sin^{\ell} \rho \cos^{\Delta} \rho F_{n\ell}(\rho), \quad F_{n\ell}(\rho) = {}_{2} F_1\left(-n , \ell + \Delta + n;\frac{d}{2} + \ell; \sin^2 \rho \right),
\end{equation}
and $Y_{\ell J}$ are spherical harmonics on $S^{d-1}$. Note that the quantization condition \eqref{eq:omega} leading to \eqref{eq:radial} follows by requiring that the solutions are normalizable. For applications to scattering, it will sometimes be convenient to describe AdS with retarded or advanced Bondi coordinates. The corresponding wavefunctions are related to \eqref{eq:AdS-wf} by a simple coordinate transformation which we discuss in Appendix \ref{app:null-foliation}. For completeness, we also include Appendix \ref{app:Rindler} where we show that the AdS Rindler wavefunctions \cite{Parikh:2012kg,Sugishita:2022ldv}  are also simply obtained from \eqref{eq:AdS-wf} by a complexified coordinate transformation. 

The wavefunctions  \eqref{eq:AdS-wf} form an orthogonal basis with respect to the Klein--Gordon inner product on a constant global time slice in AdS. Given a codimension-1 slice $\Sigma$, the Klein--Gordon inner product is defined by 
\begin{equation}
\label{eq:KG-ip}
\langle \Phi_1, \Phi_2 \rangle \equiv  i \int_{\Sigma} d\Sigma^\mu \left(\Phi_{1}^{\dagger}\partial_{\mu}\Phi_{2}-\Phi_{2}\partial_{\mu}\Phi_{1}^{\dagger}\right),
\end{equation}
obeying 
\begin{equation}
    \label{eq:KG-sym}
    \langle \Phi_1, \Phi_2 \rangle^\dagger=-\langle \Phi_1^\dagger, \Phi_2^\dagger \rangle =\langle \Phi_2, \Phi_1 \rangle.  
\end{equation}
Taking $\Sigma$ to be a surface of constant time $\tau$ and assuming the normal to $\Sigma$ is future-pointing, this becomes
\begin{equation}
\label{eq:KG-tau}
\langle \Phi_1, \Phi_2 \rangle_{\tau = \tau_0} =  i \int_0^{\pi/2} d\rho (\tan\rho)^{d-1} \int_{S^{d-1}} d \Omega  \left(\Phi_{1}^{\dagger}\partial_{\tau}\Phi_{2}-\Phi_{2}\partial_{\tau}\Phi_{1}^{\dagger}\right).
\end{equation}
 Orthogonality of \eqref{eq:AdS-wf} then follows upon observing that their radial component is related to the Jacobi polynomials  
\begin{equation}
\begin{split}
    P_{n}^{\alpha,\beta}(x) \equiv \left(\begin{matrix}
        n + \alpha\\
        n
    \end{matrix} \right){}_2F_1\left(-n, n + \alpha + \beta + 1; \alpha + 1; \frac{1 - x}{2} \right),
    \end{split}
\end{equation}
which are well-known to be orthogonal \cite{jacobipol}
\begin{equation}
\label{eq:Jacobi-ortho}
\begin{split}
    \int_{-1}^1 dx (1 - x)^{\alpha} (1 + x)^{\beta} P_{n}^{(\alpha, \beta)}(x) P_{k}^{(\alpha, \beta)}(x) = \begin{cases}
        0, \quad k \neq n\\
        \dfrac{2^{\alpha + \beta + 1}\Gamma(\alpha + n + 1)\Gamma(\beta + n + 1)}{n! \Gamma(\alpha + \beta + n + 1)(\alpha + \beta + 2n + 1)}, \quad k = n.
    \end{cases}
    \end{split}
\end{equation}

In our case
\begin{equation}
\begin{split}
    \alpha &= \frac{d-2}{2} + \ell, \quad \beta = \Delta - \frac{d}{2}, \quad   x = 1 - 2 \sin^2{\rho}.
    \end{split}
\end{equation}
Also using the orthogonality of spherical harmonics, 
\begin{equation}
\label{eq:orthog-sph-harm}
\int_{S^{d-1}} d \Omega Y_{\ell J}(\Omega) Y^*_{\ell' J'} = \delta_{\ell \ell'} \delta_{J J'}
\end{equation}
it follows immediately that 
\begin{equation}
\label{eq:wf-ip-fin}
\begin{split}
\langle \psi_{n\ell J}^{\pm}, \psi_{k \ell' J'}^{\pm} \rangle &= i \int_0^{\pi/2} d\rho (\tan\rho)^{d-1} \int_{S^{d-1}} d \Omega \left(\psi_{n \ell J}^{\pm *}(\tau, \rho, \Omega) \p_{\tau} \psi^{\pm}_{k \ell' J'}(\tau, \rho, \Omega) - (\psi_{n\ell J}^{\pm*} \leftrightarrow \psi^{\pm}_{n'\ell' J'})\right) \\
&= \pm  N_{n \ell}^d \delta_{nk} \delta_{\ell\ell'}\delta_{JJ'},\\
\langle \psi_{n\ell J}^{\pm}, \psi_{k \ell' J'}^{\pm} \rangle &= 0,
\end{split}
\end{equation}
where the normalization factor is 
\begin{equation}
\begin{split}
N_{n \ell}^d
&= \frac{\Gamma(n + 1)\Gamma(\ell + \frac{d}{2})^2}{\Gamma(n + \ell + \frac{d}{2})} \frac{\Gamma(\Delta + n - \frac{d-2}{2})}{\Gamma(\Delta + \ell  + n)} .
\end{split}
\end{equation}
The wavefunctions \eqref{eq:AdS-wf} are the counterparts of plane waves in $\mathbb{R}^{1,d}$. The space of all such wavefunctions decomposes into subspaces distinguished by the sign of the eigenvalues of $D$. Since $D$ generates time-translations in AdS (see \eqref{eq:AdS-Gens}), these subspaces consist of respectively positive- and negative-energy wavefunctions which are orthogonal and complete with respect to the Klein--Gordon inner product.

It follows that free fields in AdS$_{d+1}$ admit the decomposition \cite{Kaplan1, Penedones:2016voo}
\begin{equation}
\label{eq:AdS-bulk-field}
    \begin{split}
       \Phi(\tau,\rho,\Omega) = \sum_{n \ell J}\left( \psi^+_{n\ell J}(\tau,\rho,\Omega) a_{n \ell J} + \psi^-_{n \ell J}(\tau,\rho,\Omega) a^{\dagger}_{n \ell J} \right).
    \end{split}
\end{equation}
In the quantum theory, $a_{n \ell J}^{\dagger}$ and $a_{n \ell J}$ are promoted to creation and annihilation operators obeying the canonical commutation relations 
\begin{equation}
    [a_{n \ell J}, a_{n' \ell' J'}^{\dagger}] = (N_{n\ell}^d)^{-2} \langle \psi^+_{n\ell J}, \psi^+_{n' \ell' J'} \rangle  = (N_{n\ell}^d)^{-1} \delta_{nn'} \delta_{\ell\ell'} \delta_{JJ'}.
\end{equation}
For fixed $\Delta,$ the Hilbert space of a free field in AdS is spanned by states $ |n, \ell, J\rangle$ obtained by the action of creation operators on the vacuum
\begin{equation}
\label{eq:vacuum}
  |n, \ell, J \rangle =  a_{n \ell J}^{\dagger} |0\rangle, \quad  a_{n \ell J}|0\rangle = 0, \quad \forall~ n, \ell \in \mathbb{N}.
\end{equation}
For positive $\Delta \geq \Delta_0$, these states span a unitary, lowest-weight representation of $\mathfrak{so}(d,2)$ of dimension $\Delta$ and spin 0.\footnote{Note that this representation clearly includes states of non-vanishing angular momentum (equivalently non-vanishing eigenvalue of the $\mathfrak{so}(d)$ Casimir). Such states are obtained by the action of translations on a lowest-weight state of fixed $\Delta$ and $\ell = 0$. Nevertheless, the wavefunctions \eqref{eq:AdS-wf} transform, by virtue of being solution of the KG equation, in scalar lowest-weight representations of $\mathfrak{so}(d,2)$.} The lowest weight state is $|0, 0, 0\rangle$ which follows from the fact that $\psi_{000}^+$ is annihilated by the $K_i$ generators.

\subsection{Relation to unitary CFT}
\label{sec:rel-cft}

 Local operators in CFT$_d$ correspond to highest weight representations of $\mathfrak{so}(d,2)$. To see this, recall that scalar operators $\mathcal{O}_{\Delta}(x)$ on $\mathbb{R}^{1,d-1}$ transform under the action of $\mathfrak{so}(d,2)$ as
 \begin{equation}
 \label{eq:bdry-cft-act}
     \begin{split}
     [D,\mathcal{O}_{\Delta}(x)] &= -i\left(x^{j}\p_{j} + \Delta \right) \mathcal{O}_{\Delta}(x),\\
     [P_{i},\mathcal{O}_{\Delta}(x)] &= -i \p_{i}\mathcal{O}_{\Delta}(x), \\
[K_{i}, \mathcal{O}_{\Delta}(x)] &= - i \left( x^2 \p_{i} - 2 x_{i} x^{j}\p_{j} - 2 x_{i} \Delta \right) \mathcal{O}_{\Delta}(x),\\
[M_{ij}, \mathcal{O}_{\Delta}(x)] &= -i \left( x_{i} \p_{j} - x_{j} \p_{i} \right)\mathcal{O}_{\Delta}(x).
     \end{split}
 \end{equation}
 By construction, $\mathcal{O}_{\Delta}(0)$ is a primary operator, meaning that it diagonalizes the dilatation generator $D$ and is annihilated by the special conformal generators $K$. Via the state operator correspondence, it defines the lowest-weight state in a lowest-weight representation of $\mathfrak{so}(d,2)$. The representation is generated by acting with the translation generators $P_i$ on the lowest-weight state. Clearly, any $\mathcal{O}_{\Delta}(x)$ can be expressed as a linear combination of these states and, conversely, any state in the representation can be obtained from $\mathcal{O}_{\Delta}(x)$ by acting with derivatives and evaluating the result at the origin.  The representation is unitary provided that $\Delta$ is lower bounded (see eg. \cite{Kaplan1}). 

 The lowest-weight representation defined by the operator $\mathcal{O}_{\Delta}(x)$ is isomorphic to the space of positive energy modes of a field of mass $m^2 = \Delta(\Delta - d)$ in AdS$_{d + 1}$. One way to see this is to consider the limit as the renormalized field $\Phi$ approaches the boundary, 
 where we defined the (rescaled) boundary values of the wavefunctions
 \begin{align}
 \label{eq:hat-psi}
     \widehat{\psi}_{n\ell J}^{\pm}(\tau, \Omega) \equiv \lim_{\rho \rightarrow \frac{\pi}{2}}(\cos\rho)^{-\Delta} \psi_{n\ell J}^{\pm}(\tau, \rho, \Omega) =\begin{cases}  e^{- i \omega_n \tau} F_{n\ell}\left(\frac{\pi}{2}\right) Y_{\ell J}(\Omega),\\
     e^{ i \omega_n \tau} F_{n\ell}\left(\frac{\pi}{2}\right) Y^*_{\ell J}(\Omega).\end{cases}
 \end{align}
 The boundary values of the fields \eqref{eq:AdS-bulk-field} define operators of dimension $\Delta$ in a CFT$_d$ on $S^{d-1} \times \mathbb{R}$,
 \begin{equation}
\label{eq:extrapolate0}
\begin{split}
    \mathcal{O}_{\Delta}(\tau, \Omega) &= \lim_{\rho \rightarrow \frac{\pi}{2} - \epsilon} \epsilon^{-\Delta} \Phi(\tau, \rho, \Omega) = \sum_{n = 0}^{\infty} \sum_{\ell = 0}^{\infty} \sum_J\left(\widehat{\psi}^-_{n\ell J}(\tau, \Omega) a_{n\ell J}^{\dagger} + \widehat{\psi}^+_{n\ell J}(\tau, \Omega) a_{n\ell J} \right)\\
    &\equiv  \mathcal{O}_{\Delta}^-(\tau, \Omega) + \mathcal{O}^+_{\Delta}(\tau,\Omega).
    \end{split}
\end{equation}
Here $\mathcal{O}_{\Delta}^-$ and $\mathcal{O}^+_{\Delta}$ respectively create positive  frequency excitations in the ``in'' vacuum state defined in \eqref{eq:vacuum} and its hermitian conjugate ``out'' vacuum. These states can be used to construct AdS analogs of asymptotic incoming and outgoing scattering states in Minkowski space. Provided that all interactions can be localized to a finite AdS region, an S-matrix-like bulk observable can be constructed by evaluating transition elements of the time-evolution operator between these states. The AdS/CFT correspondence asserts that this bulk observable is encoded in correlation functions of primary operators in the boundary CFT. 

It can be shown that the operators in \eqref{eq:extrapolate0} inherit conformal  transformation properties (given by the counterparts of \eqref{eq:bdry-cft-act} on $S^{d-1}\times \mathbb{R}$) from the action of the AdS isometries \eqref{eq:AdS-Gens} on bulk fields (see eg. \cite{Kaplan1}). 
\eqref{eq:extrapolate0} is known as the extrapolate dictionary \cite{Harlow:2011ke}. One can invert this equation using orthogonality of the boundary wavefunctions \eqref{eq:hat-psi} to express the bulk modes in terms of boundary operators. Substituting the result into \eqref{eq:AdS-bulk-field}  yields an expression of the AdS field at a bulk point in terms of boundary operators. This procedure is known as AdS bulk reconstruction and will be reviewed next -- see also \cite{Harlow:2018fse,Kajuri:2020vxf} for extensive reviews.\footnote{While \cite{Kajuri:2020vxf} includes slightly more technical details than \cite{Harlow:2018fse}, we caution the reader that it contains a rather large amount of typos and inaccurate statements. }

\subsection{HKLL \& bulk reconstruction}
\label{sec:bulk-rec}

 We have seen in \eqref{eq:extrapolate0} that given a bulk field in AdS$_{d+1}$, one can construct a primary operator in a CFT$_{d}$. In order to establish an isomorphism between the Hilbert space of a particle in AdS$_{d+1}$ and that of a local operator of dimension $\Delta$ in CFT$_d$, one must construct an inverse of this map, namely, express the bulk field as linear superposition of boundary operators
\begin{equation}
\label{eq:K-int}
    \Phi(\tau, \rho, \Omega) \stackrel{?}{=} \int_{B} d\tau' d\Omega' K(\tau, \rho, \Omega; \tau', \Omega') \mathcal{O}_{\Delta}(\tau', \Omega').
\end{equation}
The kernel or ``smearing function'' $K$ is fixed by requiring that $\Phi$ is a solution to the KG equation with the normalizable asymptotics.
A natural candidate for $K$ is the time-ordered bulk-to-boundary propagator, which in the embedding space takes the form 
\begin{equation}
    G^{\rm T.O.}_{\Delta}(X; P) = \frac{C_{\Delta}}{(-P \cdot X + i\epsilon)^{\Delta}}.
\end{equation}
Here $P$ are null momenta in $\mathbb{R}^{2,d}$ parameterizing points on the AdS$_{d+1}$ boundary,
\begin{equation}
\label{eq:bdry-P}
P = \lim_{\rho \rightarrow \frac{\pi}{2}} \cos\rho \;X
\end{equation}
and $C_{\Delta}$ is a normalization constant. Since the bulk-to-boundary propagator is a solution to \eqref{eq:AdS-wave-eq} for any null vector $P$, this choice ensures that $\Phi$ is also a solution. 
However, $G^{\rm T.O.}_{\Delta}$ (or rather $G^{\rm T.O.}_{d - \Delta}$ by covariance of the boundary integral in \eqref{eq:K-int}) naively cannot provide the desired asymptotics of the bulk field $\Phi$ since its near-boundary expansion takes the form 
\begin{equation}
    \lim_{\rho \rightarrow \frac{\pi}{2} - \epsilon} G^{\rm T.O.}_{\Delta}(X; P) = m_{d - \Delta}\epsilon^{d - \Delta} \delta^{(d)}(P_X - P) + \cdots + n_{\Delta} \epsilon^{\Delta} \frac{1}{(-P\cdot P_X + i\epsilon)^{\Delta}} + \cdots.
\end{equation}
In other words, the bulk-to-boundary propagator consists of a linear combination of normalizable and non-normalizable modes near the boundary, thereby violating the required asymptotics. 

One could remedy this problem by considering a covariant linear combination of the form
\begin{equation}
    \label{eq:G-int}
    \Phi(\tau, \rho, \Omega) \stackrel{?}{=} \alpha \int_{B} d\tau' d\Omega' G_{d - \Delta}(\tau, \rho, \Omega; \tau', \Omega') \mathcal{O}_{\Delta}(\tau', \Omega') + \beta \int_{B} d\tau' d\Omega' G_{\Delta}(\tau, \rho, \Omega; \tau', \Omega') \widetilde{\mathcal{O}}_{d - \Delta}(\tau', \Omega') ,
\end{equation}
with $\alpha$ and $\beta$ chosen such as the field has normalizable asymptotics provided that $G_{\Delta}$ are defined with appropriate  time orderings.  
To the extent of our knowledge, a formula such as \eqref{eq:G-int} has not appeared in the vast literature on bulk reconstruction. Perhaps this is because it involves both the local operator $\mathcal{O}_{\Delta}$ and its shadow $\mathcal{O}_{d - \Delta}$, and hence naively appears to be in tension with boundary locality.\footnote{We thank Dionysios Anninos and Kostas Skenderis for a discussion on this point.} On the other hand, \eqref{eq:G-int} is analogous to the Carrollian expansion of a free field in Minkowski space \cite{Donnay:2022wvx}, which in turn leads to the conformal primary basis expansion in celestial CFT \cite{Pasterski:2017kqt}. A similar expression has also appeared in the de Sitter context in \cite{Xiao:2014uea}. 
Finding the precise definition \eqref{eq:G-int} consistent with canonical/geometric quantization of $\Phi$ in AdS such that it agrees with the extrapolate dictionary and HKLL reconstruction in a subregion of AdS is one of the main subjects of this paper. We will explain in Sections \ref{sec:AdS-bases} and \ref{sec:bulk-rec-AdS} how time-ordered and anti-time-ordered bulk-to-boundary propagators can be used to construct various equivalent representations of the bulk field in terms of boundary operators. We will furthermore show that all formulas reduce to analogous formulas in either the Carrollian or celestial versions of flat space holography in appropriate limits.

In the remainder of this section, we review the first approach, leading to \eqref{eq:K-int} involving only $\mathcal{O}_{\Delta}$ and with a kernel $K$ whose properties differ from those of the bulk-to-boundary propagator. This is known as the HKLL bulk reconstruction \cite{Hamilton:2005ju, Hamilton:2006az, Hamilton:2006fh}. The special case of AdS$_2$ already captures the main features of the HKLL formula, so in the remainder of this section we will restrict to $d = 1$. 

The starting point of \cite{Hamilton:2005ju} is the bulk-to-bulk propagator, which is a solution to the sourced KG equation
\begin{equation}
\label{eq:AdS-wave-eq-source}
    (\Box_{AdS_2} - m^2) K(X, X') = \frac{1}{\sqrt{-g}} \delta^2(X - X').
\end{equation}
The AdS$_2$ metric in global coordinates takes the form
\begin{equation}
    ds^2 = \frac{\ell^2}{(\cos\rho)^2}\left( - d\tau^2 + d\rho^2 \right)
\end{equation}
and the Laplacian is
\begin{equation}
    \Box_{AdS_2} = -\frac{(\cos\rho)^2}{\ell^2}\left(\p_{\tau}^2 + \tan\rho \p_{\rho} - \p_{\rho}^2 \right).
\end{equation}
The first important assumption of HKLL is that $K$ should be only non-vanishing for bulk points that are spacelike separated from the boundary. This leads to the Ansatz
\begin{equation}
\label{eq:bulk-bulk-Ansatz}
    K_{\rm HKLL}(X, X') = f(\sigma) \Theta\left( - (\tau - \tau')^2 + (\rho - \rho')^2\right),
\end{equation}
where $\sigma$ is the invariant interval in AdS$_{2}$
\begin{equation}
    \sigma = -X \cdot X' = \frac{\cos(\tau - \tau') - \sin \rho \sin \rho'}{\cos \rho \cos \rho'}.
\end{equation}
Substituting \eqref{eq:bulk-bulk-Ansatz} into the sourced Klein--Gordon equation \eqref{eq:AdS-wave-eq-source} leads to the following differential equation for $f$
\begin{equation}
\label{eq:KG2}
    (\sigma^2 - 1) f''(\sigma) + 2\sigma f'(\sigma) - \Delta(\Delta - 1) f(\sigma) = 0, \quad f(1) = \frac{1}{4}.
\end{equation}
The theta function in \eqref{eq:bulk-bulk-Ansatz} generates the delta function source term on the RHS of \eqref{eq:AdS-wave-eq-source}, and matching the coefficients of these terms leads to the boundary condition $f(1) = \frac{1}{4}$. Note that $\sigma = 1$ corresponds to the lightcone in AdS defined by $\tau - \tau' = \rho - \rho'$.

\begin{figure}
\begin{center}
\includegraphics[scale=0.4]{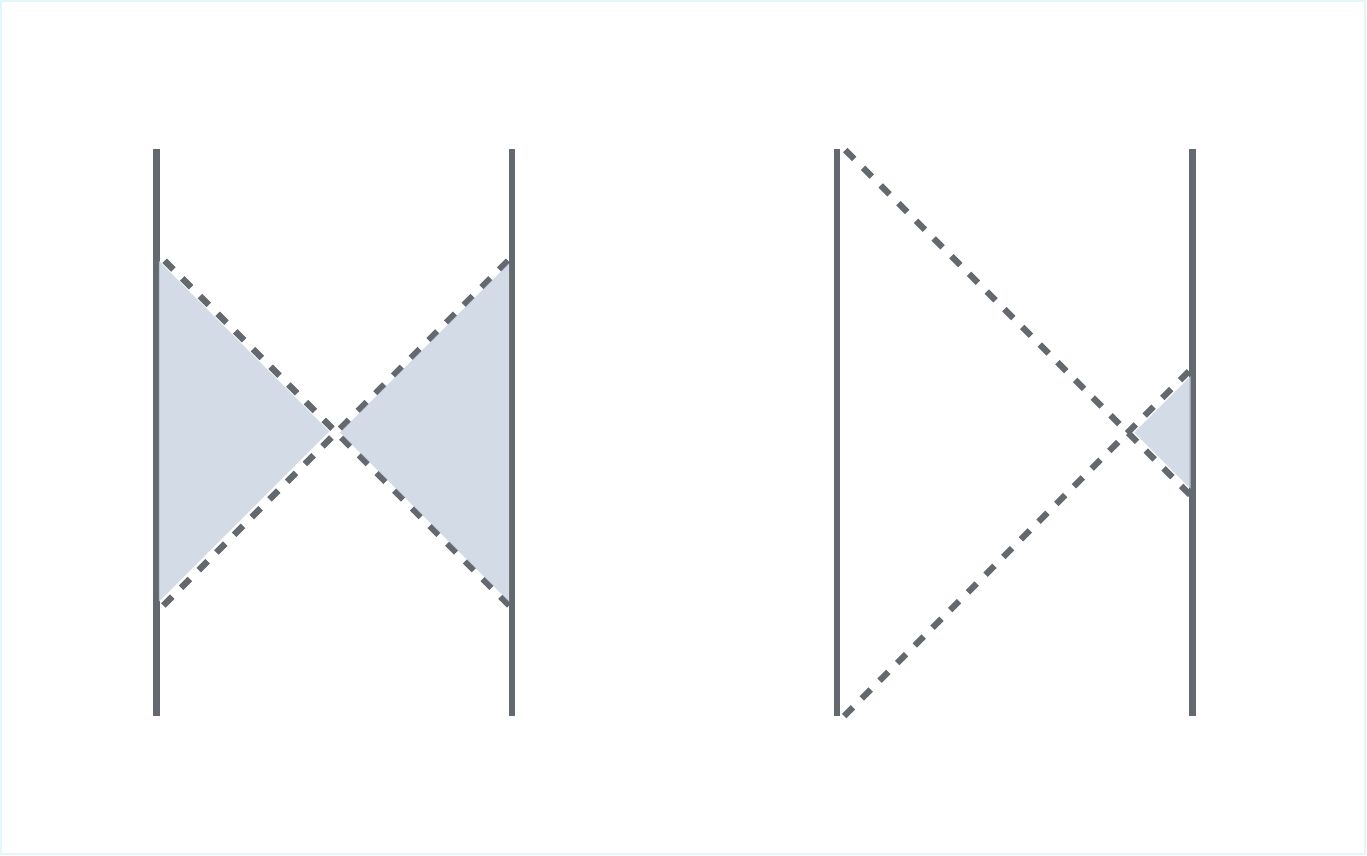}
\end{center}
\caption{Left: The HKLL formula allows for the bulk field in the colored regions to be reconstructed as integrals over operators over their corresponding boundaries. Right: The theta function that restricts the support of the HKLL kernel to the colored diamond becomes delta-function localized near the boundary. This ensures that the bulk field has normalizable near-boundary fall off.}
\label{Fig:1}
\end{figure}

The solutions to \eqref{eq:KG2} are Legendre polynomials 
\begin{equation}
    f(\sigma) = \frac{1}{4} P_{\Delta - 1}(\sigma). 
\end{equation}
The near-boundary behaviour of $f(\sigma)$ is 
\begin{equation}
   f(\sigma) \sim \frac{\Gamma(2 \Delta - 1)}{2^{\Delta + 1} \Gamma(\Delta)^2} \sigma^{\Delta - 1}, \quad \sigma \rightarrow \infty,
\end{equation}
 so the coefficient of the theta function in \eqref{eq:bulk-bulk-Ansatz} is \textit{non-normalizable} near the boundary. Indeed, it is shown in \cite{Hamilton:2005ju} that in the limit as $X'$ is taken to the boundary, namely $\rho' \rightarrow \frac{\pi}{2}$, $f(\sigma)$ becomes proportional to the AdS$_2$ bulk-to-boundary propagator $G_{1 - \Delta}$,
 \begin{equation}
 \label{eq:HKLL-kernel}
 \begin{split}
     K_{\rm HKLL}(\tau, \rho; \tau') &= \lim_{\rho' \rightarrow \frac{\pi}{2}} (2\Delta - 1) (\cos\rho')^{\Delta - 1}K(X,X')\\
     &= \frac{2^{\Delta - 1} \Gamma(\Delta + \frac{1}{2})}{\sqrt{\pi}\Gamma(\Delta)} \left(\frac{\cos \rho}{\cos(\tau - \tau') - \sin \rho} \right)^{1 - \Delta} \Theta\left(\frac{\pi}{2} - \rho  - |\tau - \tau'| \right),
     \end{split}
 \end{equation}
 where the simplification of the theta function is particular to the AdS$_2$ case if one restricts to the bulk region spacelike separated from one component of the boundary  (see the right panel of Figure \ref{Fig:1}). 
The bulk field in this region can then be reconstructed using one of Green's identities as
\begin{equation}
\label{eq:HKLL-reconstruction}
    \Phi(\tau, \rho) = \int d\tau' K_{\rm HKLL}(\tau, \rho; \tau') \phi_R(\tau'),
\end{equation}
where $\phi_R(\tau')$ is the renormalized boundary value of the field. 

 Since $G_{1 - \Delta}$ behaves as $\epsilon^{1 - \Delta}$ near the boundary $\rho \rightarrow \frac{\pi}{2} - \epsilon$, one may worry that the formula \eqref{eq:HKLL-reconstruction} reconstructs the \textit{non-normalizable} component of the field. This naive issue is resolved by the theta function whose support as $\rho \rightarrow \frac{\pi}{2}$ becomes localized around $\tau = \tau'$, 
\begin{equation}
 \Theta\left(\frac{\pi}{2} - \rho  - |\tau - \tau'| \right) = -\left(\rho - \frac{\pi}{2} \right) \delta(\tau - \tau') + \cdots , \quad \rho \rightarrow \frac{\pi}{2}
\end{equation}
and so integrating against a test function
\begin{equation}
   \int d\tau' F(\tau') \left(\frac{\cos \rho}{\cos(\tau - \tau') - \sin \rho} \right)^{1 - \Delta} \Theta\left(\frac{\pi}{2} - \rho  - |\tau - \tau'| \right) \propto F(\tau) (\cos\rho)^{\Delta} + \cdots , \quad \rho \rightarrow \frac{\pi}{2}.
\end{equation}
The expected asymptotics are hence recovered at the expense of having to adjust the boundary integration region depending on the point at which one chooses to reconstruct the bulk field. In particular, one has to integrate over the whole AdS boundary in order to reconstruct the bulk field at the origin. This is in stark contrast to the reconstruction of a free field in Minkowski space where, at least in the massless limit, specifying the boundary value of the field at a \textit{cut} of $\mathscr{I}^{\pm}$ is sufficient to reconstruct the field at the origin \cite{cite-key}. This result leads to the so-called Kirchhoff--d'Adh\'emar formula \cite{Penrose, Newman:1976gc}. Recent progress towards generalizing these ideas to AdS using the Hamilton-Jacobi formalism can be found in \cite{Caron-Huot:2025she, Caron-Huot:2025hmk}.

To summarize, the HKLL formula \eqref{eq:HKLL-kernel} and its higher-dimensional generalizations \cite{Hamilton:2006az, Hamilton:2006fh} provide only a partial solution to the original goal of reconstructing the bulk state space in terms of the boundary one. The HKLL kernel is only defined in a subregion of AdS and it is not clear how to consistently analytically continue the result beyond this patch.  
In Section \ref{sec:bulk-rec-AdS} we will propose precise representations of the form \eqref{eq:G-int} which reconstruct a free field at an arbitrary AdS bulk point. In particular, we will show that the HKLL kernel can be replaced by bulk-to-boundary propagators with specified time-orderings, thereby allowing one to uniquely analytically continue the field beyond the bulk wedge of spacelike separated points from the boundary. We also obtain an analog of the Kirchhoff--d'Adh\'emar formula for $\Delta \in \mathbb{N}$ (and in particular massless) fields in AdS. This shows that for such values of $\Delta$, the bulk field can in fact be reconstructed from its value at a cut of the boundary. 

Furthermore, we will demonstrate that the decomposition of a free field in Minkowski space in terms of plane waves, Carrollian and conformal primary bases, arise from flat space limits of reconstruction formulas in terms of bulk-to-boundary propagators, such as \eqref{eq:G-int} and its counterparts, where the bulk-to-boundary propagators are replaced by wavefunctions with respect to an $\mathfrak{so}(3,1)$ subalgebra of the AdS$_4$ isometry algebra. The latter will be constructed in Section \ref{sec:boost-estates}. For simplicity, we will restrict to the case of 4-dimensional spacetimes ($d = 3$), but the analysis can be easily generalized to any $d$. 

It will be very interesting to revisit ideas building on the HKLL proposal such as entanglement wedge reconstruction \cite{Almheiri:2014lwa,Jafferis:2015del,Dong:2016eik}, as well as bulk reconstruction in more general backgrounds \cite{Hamilton:2005ju, Papadodimas:2012aq} in the light of our results. We hope that the precise relation between the AdS and flat space bulk reconstruction established here in a very simple case will allow us to generalize these interesting features of AdS/CFT to asymptotically flat spacetimes.

\section{Bulk inner products of AdS bulk-to-boundary propagators}
\label{sec:bulk-to-boundary-basis}

In this section we determine the conditions under which bulk-to-boundary propagators with specified time orderings in Lorentzian AdS$_4$ form a basis with respect to the Klein--Gordon inner product \eqref{eq:KG-ip}. We start by computing the inner product of bulk-to-boundary propagators and their Lorentzian shadow transforms on a constant time slice $\Sigma_{\tau}$ in AdS. In Section \ref{sec:AdS-bases} we will use these results to construct different bases of solutions to the KG equation labeled by points on the AdS boundary as opposed to the discrete quantum numbers $n, \ell, m$. Equipped with such a basis, we will show in Section \ref{sec:bulk-rec-AdS} that bulk fields in AdS admit different equivalent representations as linear superpositions of boundary operators or their shadows, generalizing the HKLL construction reviewed in Section \ref{sec:bulk-rec}.

\subsection{Mode expansion of bulk-to-boundary propagator}
\label{sec:mode-expansion-btb}

In global AdS$_{4}$ coordinates \eqref{eq:global-AdS}, the time-ordered scalar bulk-to-boundary propagator takes the form
\begin{equation}
\label{eq:scalar-btb}
    G_{\Delta}^{\rm T.O.}(X; P) \equiv \frac{1}{\left(-P \cdot X  + i\epsilon \right)^{\Delta}} = \frac{(\cos \rho)^{\Delta}}{\left(\cos(\tau - \tau_p) - \sin \rho (\Omega \cdot \Omega_p) + i\epsilon \right)^{\Delta}}.
\end{equation}
It will be convenient to expand this propagator in terms of the basis \eqref{eq:AdS-wf} of AdS wavefunctions. We start by  decomposing  \eqref{eq:scalar-btb} into spherical harmonics. The spherical harmonics, normalized according to  \eqref{eq:orthog-sph-harm}, are given by
\begin{equation}
Y^{\ell}_{~m}(\theta, \varphi) = \sqrt{\frac{(2 \ell + 1)(\ell - m)!}{4\pi (\ell + m)!}} P^{m}_{~\ell}(\cos\theta) e^{i m \varphi}, \quad P^m_{~\ell}(x) = (-1)^m (1 - x^2)^{m/2} \frac{d^m}{dx^m} P_{\ell}(x)\ ,
\end{equation}
where $P_{\ell}$ are the Legendre polynomials
\begin{equation}
P_{\ell}(x) = \frac{1}{2^{\ell} \ell!} \frac{d^{\ell}}{dx^{\ell}} \left[(x^2 - 1)^{\ell} \right].
\end{equation}
We can, without loss of generality, take 
\begin{equation}
\Omega \cdot \Omega_p = \cos\theta.
\end{equation}

We first expand \eqref{eq:scalar-btb} in the $P_{\ell}(\cos \theta)$ basis. We perform this decomposition in Appendix \ref{eq:btb-exp}, which yields\footnote{We have additionally made use of the identity $\int_{-1}^{1} dx P_{\ell}^0(x) P_{\ell'}^0(x) = \frac{2}{2\ell + 1} \delta_{\ell \ell'}$.} 
\begin{equation}
\label{eq:time-ordered-pm}
    G_{\Delta}^{\rm T.O.}(X;P) =  G_{\Delta}^+(X; P) \Theta(\tau - \tau_p) +  G_{\Delta}^-(X;P) \Theta(\tau_p - \tau),\\
\end{equation}
where 
\begin{equation}
\label{eq:Gp-int}
\begin{split}
G^+_{\Delta}(X; P) &\equiv \sum_{\ell = 0}^{\infty} \frac{2 \ell + 1}{2} c_{\ell}(\tau - \tau_p, \rho) P_{\ell}(\Omega \cdot \Omega_p), \\
 c_{\ell}(\tau - \tau_p, \rho) &= \sum_{k = 0}^{\infty} \sqrt{\pi}  (-1)^{k}\frac{2^{\Delta}\Gamma(\Delta + \ell + k)}{k! \Gamma(\Delta)\Gamma(\frac{3}{2} + \ell)}  (\cos\rho)^{\Delta} (\sin\rho)^{\ell} e^{-i(\tau - \tau_p)(\Delta + \ell + 2k)}\\
 &\times {}_2 F_1\left(-k, \Delta + \ell + k; \frac{3}{2} + \ell; \sin^2 \rho\right),
\end{split}
\end{equation}
and 
\begin{equation}
\label{eq:Gpm-cc}
    G^{-}_{\Delta}(X; P) = \left[G^+_{\Delta}(X; P)\right]^*
\end{equation}
 multiplied by the appropriate $\Theta$ functions as in \eqref{eq:time-ordered-pm} are retarded and advanced propagators. See eg. \cite{Sleight:2019hfp} for a summary of the relation between time-ordered, advanced and retarded propagators in (A)dS. 

As expected, the coefficients $c_{\ell}$ in \eqref{eq:Gp-int} can be expressed in terms of the radial wavefunctions \eqref{eq:radial} with $d = 3$, and so
\begin{equation}
\label{eq:mode-exp-btb-fin}
\begin{split}
G^+_{\Delta}(X; P) &= \sum_{k = 0}^{\infty} \sum_{\ell = 0}^{\infty} \sqrt{\pi}(2 \ell + 1) (-1)^{k}\frac{2^{\Delta-1}\Gamma(\Delta + \ell + k)}{k! \Gamma(\Delta)\Gamma(\frac{3}{2} + \ell)}  e^{-i(\tau - \tau_p)(\Delta + \ell + 2k)} f_{k \ell} (\rho) \left. \right|_{d = 3} P_{\ell}(\Omega \cdot \Omega_p)\\
&=  \sum_{k = 0}^{\infty} \sum_{\ell = 0}^{\infty} \pi \sqrt{2 \ell + 1} (-1)^{k}\frac{2^{\Delta}\Gamma(\Delta + \ell + k)}{k! \Gamma(\Delta)\Gamma(\frac{3}{2} + \ell)}  \psi^{+}_{k \ell 0}(\tau - \tau_p, \rho, \Omega \cdot \Omega_p).\\
\end{split}
\end{equation}
We conclude that the time-ordered bulk-to-boundary propagator admits the following expansion in terms of AdS wavefunctions, 
\begin{equation}
\label{eq:time-ordered-btb}
\begin{split}
   G_{\Delta}^{\rm T.O.}(X;P)    &= \sum_{k = 0}^{\infty} \sum_{\ell = 0}^{\infty}  \Big[ b^+_{k\ell} \psi^{+}_{k \ell 0}(\tau - \tau_p, \rho, \Omega \cdot \Omega_p)\Theta(\tau-\tau_{p}) + b^-_{k\ell} \psi^{-}_{k \ell 0}(\tau - \tau_p, \rho, \Omega \cdot \Omega_p)\Theta(\tau_{p}-\tau) \Big]
    \end{split}
\end{equation}
with
\begin{equation}
\label{eq:bklto}
    b_{k\ell}^{+} =\pi \sqrt{2 \ell + 1} (-1)^{k}\frac{2^{\Delta}\Gamma(\Delta + \ell + k)}{k! \Gamma(\Delta)\Gamma(\frac{3}{2} + \ell)}, \quad b_{k\ell}^{-} = (b_{k\ell}^+)^*.
\end{equation}

This expression can be shown to agree with an integral representation of the time-ordered bulk-to-boundary propagator given in eq. A.31 of \cite{Giddings:1999qu}.\footnote{We note that the appendix of \cite{Giddings:1999qu} contains an inconsistency among the normalization of the wavefunctions A.18 and the representation A.31 of the bulk-to-boundary propagator. Our explicit computations imply that eq. A.31 is out by a factor of $\omega_{n\ell}$. } We establish this equivalence in Appendix \ref{eq:Giddings}. 
The mode expansion of the anti-time-ordered propagator is obtained from \eqref{eq:time-ordered-pm} by complex conjugation, namely
\begin{equation}
\label{eq:mode-exp-ato-btb}
     G_{\Delta}^{\rm A.T.O.}(X;P)    = \sum_{k = 0}^{\infty} \sum_{\ell = 0}^{\infty}  \Big[ b^-_{k\ell} \psi^{-}_{k \ell 0}(\tau - \tau_p, \rho, \Omega \cdot \Omega_p)\Theta(\tau-\tau_{p}) + b^+_{k\ell}  \psi^{+}_{k \ell 0}(\tau - \tau_p, \rho, \Omega \cdot \Omega_p)\Theta(\tau_{p}-\tau) \Big].
\end{equation}
One can use \eqref{eq:time-ordered-pm} and its complex conjugate to express $G_{\Delta}^{\pm}$ in terms of $G_{\Delta}^{\rm (A.)T.O.}$, 
\begin{equation}
\label{eq:btb-from-to}
\begin{split}
    G_{\Delta}^+(X;P) &= G_{\Delta}^{\rm T.O.}(X;P) \Theta(\tau - \tau_p) + G_{\Delta}^{\rm A.T.O}(X;P)\Theta(\tau_p - \tau), \\
     G_{\Delta}^-(X;P) &= G_{\Delta}^{\rm A.T.O.}(X;P) \Theta(\tau - \tau_p) + G_{\Delta}^{\rm T.O}(X;P)\Theta(\tau_p - \tau).
    \end{split}
 \end{equation}

\subsection{Bulk inner product}
\label{sec:inner-prodG}

The decomposition \eqref{eq:mode-exp-btb-fin} allows us to evaluate the KG inner product of bulk-to-boundary propagators \eqref{eq:scalar-btb} on a constant $\tau = \tau_0$ slice denoted $\Sigma_{\tau_0}$.  We assume that the normal to $\Sigma_{\tau_0}$ is future pointing. Using the identity
\begin{equation}
\label{eq:wf-ip-main}
\langle \psi^\pm_{n\ell 0}(\tau - \tau_1, \rho, \Omega \cdot \Omega_1),  \psi^\pm_{k \ell' 0}(\tau - \tau_2, \rho, \Omega \cdot \Omega_2)\rangle_{\tau = \tau_0}  =  \pm \sqrt{\frac{4\pi}{2\ell + 1}} N_{n \ell}^{d=3} ~ e^{\mp i\omega_n (\tau_1 - \tau_2)}Y_{\ell 0}(\Omega_1 \cdot \Omega_2)  \delta_{nk} \delta_{\ell \ell'}
\end{equation}
derived in Appendix \ref{sec:AdS-prop}, we straightforwardly find that 
\begin{equation}
\langle G^{\pm}_{\Delta}(X;P_1), G^{\pm}_{\Delta}(X;P_2) \rangle_{\tau = \tau_0} = \pm \sum_{k, \ell = 0}^{\infty} 2^{2\Delta +1} \pi^{5/2} \sqrt{2\ell + 1} \frac{\Gamma(\Delta + \ell + k)\Gamma(\Delta + k -\frac{1}{2})}{k! \Gamma(\Delta)^2 \Gamma(k + \ell + \frac{3}{2})}e^{\mp i\omega_k (\tau_1 - \tau_2)}Y_{\ell 0}(\Omega_1 \cdot \Omega_2) .
\end{equation}
The RHS can be expressed in terms of the wavefunction \eqref{eq:AdS-wf} evaluated on the boundary $\rho = \frac{\pi}{2}$ using the identity
\begin{equation}
{}_2 F_1\left(-k, \ell + \Delta + k; \frac{3}{2} + \ell; 1 \right) = (-1)^k \frac{\Gamma(\frac{3}{2} + \ell) \Gamma(\Delta + k - \frac{1}{2})}{\Gamma(\frac{3}{2}  + \ell + k)\Gamma(\Delta - \frac{1}{2})},
\end{equation}
namely
\begin{equation}
\begin{split}
\langle G^{\pm}_{\Delta}(X;P_1), G^{\pm}_{\Delta}(X;P_2) \rangle_{\tau = \tau_0} &= \pm \sum_{k, \ell = 0}^{\infty} 2^{2\Delta +1} \pi^{5/2} \sqrt{2\ell + 1} (-1)^k \frac{\Gamma(\Delta + \ell + k)\Gamma(\Delta - \frac{1}{2})}{k! \Gamma(\Delta)^2 \Gamma(\frac{3}{2} + \ell)}\\
&\times \widehat{\psi}^{\pm}_{k\ell 0}(\tau_1 - \tau_2, \Omega_1 \cdot \Omega_2),
\end{split}
\end{equation}
where $\widehat{\psi}^\pm_{k\ell 0}$ were defined in (\ref{eq:hat-psi}).
We now recognize that the RHS is proportional to a boundary-to-boundary propagator and therefore
\begin{equation}
\label{eq:Gpm-ip}
    \langle G^{\pm}_{\Delta}(X;P_1), G^{\pm}_{\Delta}(X;P_2) \rangle_{\tau = \tau_0} = \pm 2^{\Delta +1} \pi^{3/2} \frac{\Gamma(\Delta - \frac{1}{2})}{\Gamma(\Delta)} G_{\Delta}^{\pm}(P_1, P_2).
\end{equation}

Putting everything together, using the definition \eqref{eq:time-ordered-btb} of the time-ordered propagator and the inner products of AdS wavefunctions \eqref{eq:wf-ip-fin}, we find
    \begin{equation}
    \label{eq:time-ordered-btb-ip} 
\begin{split}
\langle G^{\rm T.O.}_{\Delta}(X;P_1), &G^{\rm T.O.}_{\Delta}(X;P_2) \rangle_{\tau = \tau_0} = 2^{\Delta +1} \pi^{3/2} \frac{\Gamma(\Delta - \frac{1}{2})}{\Gamma(\Delta)} \\
&\times \left( G_{\Delta}^+(P_1, P_2) \Theta(\tau_0 - \tau_1) \Theta(\tau_0 - \tau_2) - G_{\Delta}^-(P_1, P_2) \Theta(\tau_1 - \tau_0) \Theta(\tau_2 - \tau_0)\right).
\end{split}
\end{equation}
Here we assumed that $\tau \neq \tau_1 \neq \tau_2$, otherwise \eqref{eq:time-ordered-btb-ip} receives additional delta function contributions from $\p_{\tau}$ acting on the step functions in \eqref{eq:time-ordered-btb}.
If the normal to $\Sigma_{\tau_0}$ is past-pointing, the inner product picks up an overall sign. 

Using \eqref{eq:btb-from-to}, we can equivalently write
    \begin{equation}
    \label{eq:time-ordered-btb-ip-to} 
\begin{split}
\langle G^{\rm T.O.}_{\Delta}(X;P_1), &G^{\rm T.O.}_{\Delta}(X;P_2) \rangle_{\tau = \tau_0} = 2^{\Delta +1} \pi^{3/2} \frac{\Gamma(\Delta - \frac{1}{2})}{\Gamma(\Delta)} \\
&\times \bigg[\left( G_{\Delta}^{\rm T.O.}(P_1, P_2)  \Theta(\tau_{12})+G_{\Delta}^{\rm A.T.O.}(P_1, P_2)  \Theta(\tau_{21})\right)\Theta(\tau_{01}) \Theta(\tau_{02}) \\ &-\left( G_{\Delta}^{\rm T.O.}(P_1, P_2)  \Theta(\tau_{21})+G_{\Delta}^{\rm A.T.O.}(P_1, P_2)  \Theta(\tau_{12})\right) \Theta(\tau_{10}) \Theta(\tau_{20})\Bigg].
\end{split}
\end{equation}
Using eqns. \eqref{eq:Gpm}, \eqref{eq:mode-exp-ato-btb} and \eqref{eq:Gpm-ip} we also deduce that 
\begin{equation}
\label{eq:ato-to-ip}
    \begin{split}
\langle G^{\rm A.T.O.}_{\Delta}(X;P_1), &G^{\rm  T.O.}_{\Delta}(X;P_2) \rangle_{\tau = \tau_0} = 2^{\Delta +1} \pi^{3/2} \frac{\Gamma(\Delta - \frac{1}{2})}{\Gamma(\Delta)} \\
&\times \left( G_{\Delta}^+(P_1, P_2) \Theta(\tau_0 - \tau_2) \Theta(\tau_1 - \tau_0) - G_{\Delta}^-(P_1, P_2) \Theta(\tau_2 - \tau_0) \Theta(\tau_0 - \tau_1)\right).
\end{split}
\end{equation}
Note that in contrast to \eqref{eq:time-ordered-btb-ip-to}, this inner product vanishes if the boundary points $P_1$ and $P_2$ are both either in the future or the past of $\Sigma_{\tau_0}$.

\subsection{Mode expansion of shadow transformed bulk-to-boundary propagator}
\label{sec:mode-exp-shadow}

We will also be interested in the mode expansions of shadow transforms of bulk-to-boundary propagators. In Euclidean CFT, the shadow transform maps a primary operator of dimension $\Delta$ to another primary of dimension $3 - \Delta$ via \cite{Simmons-Duffin:2012juh}
\begin{equation}
    \widetilde{\mathcal{O}}_{3 - \Delta}(P) \equiv \mathcal{N} \int d^3 P' \frac{1}{(-P\cdot P')^{3 - \Delta}} \mathcal{O}_{\Delta}(P').
\end{equation}
We thus expect that 
\begin{equation}
    \widetilde{G}_{3 - \Delta}(X; P) \propto \int d^3 P' \frac{1}{(-P\cdot P
    ')^{3 - \Delta}} G_{\Delta}(X; P'),
\end{equation}
and so the mode expansion of the shadow propagator should be obtained, up to normalization, from that in Section \ref{sec:mode-expansion-btb} by setting $\Delta \rightarrow 3 - \Delta$. 

Here we work in the Lorentzian setting, where one has to carefully account for the time-ordering of both the propagators and the shadow kernels. We compute the results obtained with different prescriptions in Appendix \ref{sec:btb-shadow}. It will be convenient to define the shadow integration regions, $B$ and $\bar{B}$, with respect to the reference time $\tau$ associated with the bulk point $X$ at which $G_{\Delta}(X; P)$ is evaluated as follows
\begin{equation}
\label{eq:B-bB}
    B: \tau < \tau_{p'} < \tau + \pi, \quad \quad \bar{B}: -\pi + \tau < \tau_{p'} < \tau.
\end{equation}
Furthermore, we take the point $P$ to lie in the boundary region complementary to the integration region.
We find
\begin{equation}
\label{eq:shadow-G-0}
    \begin{split}
   \int_{B}  \frac{d^3 P'}{(-P\cdot P' - i\epsilon)^{3 - \Delta}} G_{\Delta}^{\rm T.O.}(X; P') = e^{i\pi\left(  \Delta-\frac{3}{2}\right)} \pi^{3/2} 2^{\Delta + 1} \cos \pi \Delta \frac{\Gamma(\Delta - \frac{3}{2})}{\Gamma(\Delta)} G_{3 - \Delta}^{\rm A.T.O.}(X; P), ~~ P \in \bar{B}
    \end{split}
\end{equation}
and similarly in the case where $B$ and $\bar{B}$ are exchanged, with the phase $e^{i\pi\left(  \Delta-\frac{3}{2}\right)}$ replaced with $e^{-i\pi\left(  \Delta-\frac{3}{2}\right)}$ as explained in Appendix \ref{sec:btb-shadow}.

We also find that
\begin{equation}
    \begin{split}
    \int_{B} d^3 P' \frac{1}{(-P\cdot P' + i\epsilon)^{3 - \Delta}} G_{\Delta}^{\rm T.O.}(X; P') = 0, \quad P \in \bar{B},
    \end{split}
\end{equation}
and similarly for $B \leftrightarrow \bar{B}$. We normalize the shadow transform such that 
\begin{equation}
\label{eq:shadow-norm}
    \widetilde{G_{\Delta}^{\rm T.O.}}(X;P)\equiv \frac{e^{-i\pi(\Delta - \frac{3}{2})} \Gamma(\Delta)}{2^{1 + \Delta}\pi^{3/2} \cos \pi \Delta \Gamma(\Delta - \frac{3}{2})}  \int_{B}  \frac{d^3 P'}{(-P\cdot P' - i\epsilon)^{3 - \Delta}} G_{\Delta}^{\rm T.O.}(X; P') = G_{3 - \Delta}^{\rm A.T.O.}(X;P).
\end{equation}
The results involving anti-time-ordered propagators are obtained by complex conjugation. We note that the shadow transform with the appropriate kernel by definition maps points in $B$ to points in $\bar{B}$ and vice versa, and consequently implements an inversion with respect to $\Sigma_\tau$ \cite{Chen:2023tvj, Chen:2024kuq, Jorstad:2023ajr}. Recall also that $X^2 = -\ell^2$ (here $\ell = 1$) so eq. \eqref{eq:shadow-G-0} is consistent with scaling $X \rightarrow \lambda X$ after restoring factors of $\ell$. 

We conclude that the mode expansions of the shadow transforms of AdS bulk-to-boundary propagators are obtained from the mode expansions of (anti-)time-ordered bulk-to-boundary propagators \eqref{eq:time-ordered-btb} and \eqref{eq:mode-exp-ato-btb} by substituting $\Delta \rightarrow 3 - \Delta$.

\subsection{Bulk inner product of shadow transformed bulk-to-boundary propagators}
\label{sec:shadow-ip-comp}

In this section we compute the inner products of $G_{\Delta}^{\rm (A).T.O.}(X;P)$ and  $G_{3 - \Delta}^{\rm(A).T.O.}(X;P)$. Up to normalization, these coincide with the inner products of bulk-to-boundary propagators and their shadow transforms \eqref{eq:shadow-G-0}. By conformal symmetry, for generic $\Delta$ we expect this to be proportional to $\delta^{3}(P_1, P_2)$
\begin{equation}
\label{eq:shadow-ip-guess}
     \langle G^{\rm T.O.}_{\Delta}(X;P_1), G^{\rm T.O.}_{3-\Delta}(X;P_2) \rangle_{\tau = \tau_0}  \propto \delta^3(P_1, P_2).
\end{equation}

The direct evaluation of this inner product is slightly trickier than that of the inner product \eqref{eq:time-ordered-btb}, which was essentially implied by orthogonality of the Jacobi polynomials. Here, we instead need to compute the following integral of radial wavefunctions
\begin{equation}
\label{eq:radial-ip}
I_r(\Delta,3 - \Delta) \equiv \int_0^{\pi/2} d\rho (\sin \rho)^{\ell + \ell'} (\cos\rho)^3 (\tan\rho)^2 F_{n\ell}^{\Delta}(\rho) F_{k\ell'}^{3 - \Delta}(\rho).
\end{equation}
Relating these to Jacobi polynomials
\begin{equation}
F_{n\ell}^{\Delta}(\rho) = \frac{n! \alpha!}{(n + \alpha)!} P_{n}^{(\alpha, \beta)}(x), \quad \alpha = \frac{1}{2} + \ell,~~ \beta = \Delta - \frac{3}{2},~~x = 1 - 2 \sin^2\rho,
\end{equation}
we see that \eqref{eq:radial-ip} is equivalent to the following integral
\begin{equation}
\label{eq:shadow-int-eval}
I_r(\Delta,3 - \Delta) \equiv \frac{1}{4} \int_{-1}^1 dx (1 - x)^{\alpha}  2^{-\alpha} \frac{n! \alpha!}{(n + \alpha)!}\frac{k! \alpha!}{(k + \alpha)!} P_n^{(\alpha, \beta)}(x) P_k^{(\alpha, -\beta)}(x).
\end{equation}
To evaluate this inner product, we use the symmetry properties of the Jacobi polynomials \cite{jacobipol}
\begin{equation}
\label{eq:poly-prop}
\begin{split}
P_n^{(\alpha, \beta)}(z) &= (-1)^n P_n^{(\beta, \alpha)}(-z), \\
P_n^{(-k, \beta)}(z) &= \frac{\Gamma(n + \beta + 1)}{ \Gamma(n + \beta + 1 - k)} \frac{(n - k)!}{n!} \left(\frac{z-1}{2} \right)^k P_{n - k}^{(k, \beta)}(z).
\end{split}
\end{equation}
The latter identity holds for $k \in \mathbb{N}$. \eqref{eq:poly-prop} allows us to express $P_{n}^{(\alpha, -\beta)}$ in terms of $P_{n}^{(\alpha, \beta)}$, namely
\begin{equation}
\begin{split}
P_n^{(\alpha, -\beta)}(z) &= (-1)^n P_n^{(-\beta, \alpha)}(-z) = (-1)^n \frac{\Gamma(n + \alpha + 1)}{\Gamma(n + \alpha + 1 - \beta)} \frac{(n - \beta)!}{n!} \left( \frac{-z-1}{2}\right)^{\beta} P_{n - \beta}^{(\beta, \alpha)}(-z)\\
&=  \frac{\Gamma(n + \alpha + 1)}{\Gamma(n + \alpha + 1 - \beta)} \frac{(n - \beta)!}{n!} \left( \frac{z+1}{2}\right)^{\beta} P_{n - \beta}^{( \alpha, \beta)}(z),
\end{split}
\end{equation}
where in the last equality we applied the first identity in \eqref{eq:poly-prop}.
Using the orthogonality of Jacobi polynomials \eqref{eq:Jacobi-ortho} we can now evaluate \eqref{eq:radial-ip} and consequently the inner product \eqref{eq:shadow-ip-guess}. We spell out the details in Appendix \ref{app:shadow} and quote the result
\begin{equation}
    \label{eq:shadow-bulk-ip}
    \begin{split}
  \langle G^{\rm T.O.}_{\Delta}(X;P_1),& G^{\rm T.O.}_{3-\Delta}(X;P_2) \rangle_{\tau = \tau_0}   =\sum_{n, \ell = 0}^{\infty} \frac{2^4 \pi^{5/2} \sqrt{2 \ell + 1} }{\Gamma(\Delta)\Gamma(3 - \Delta)}  Y_{\ell0}(\Omega_1\cdot \Omega_2)\\
  &\times \left[ e^{i\pi\left(\Delta - \frac{3}{2}\right)} e^{-i\omega_n\tau_{12}} \Theta(\tau_0 - \tau_1)\Theta(\tau_0 - \tau_2) - e^{-i\pi\left(\Delta - \frac{3}{2}\right)} e^{i\omega_n\tau_{12}} \Theta(\tau_1 - \tau_0)\Theta(\tau_2 - \tau_0) \right].
  \end{split}
\end{equation}
This result is obtained provided that $\Delta - \frac{3}{2} \in \mathbb{N}$. Evaluating the same inner product with $\Delta \rightarrow 3 - \Delta$ yields a similar result for $\frac{3}{2} - \Delta \in \mathbb{N}$.
For non-coincident points $P_1, P_2$,  $\mathfrak{so}(3,2)$ covariance fixes $\Delta = \frac{3}{2}$. We will be interested in generic $\Delta$, in which case conformal symmetry also allows for a distributional term in the inner product. We compute this contribution below.

Assuming that $\tau_{12} \in (-\frac{\pi}{2}, \frac{\pi}{2})$, we can use the identities\footnote{For $\tau_{12} \in \mathbb{R}$, the infinite sum yields a Dirac delta comb instead.}
\begin{equation}
\sum_{n = 0}^{\infty} e^{2\pi i n t}  = \frac{i}{2\pi} \frac{1}{t + i\epsilon},\quad
\delta(t) = \sum_{n = -\infty}^{\infty} e^{2\pi i n t}  = \frac{i}{2\pi} \left[\frac{1}{t + i\epsilon} - \frac{1}{t - i\epsilon} \right], 
\end{equation}
and
\begin{equation}
    \sum_{\ell = 0}^{\infty} (2\ell + 1) P_\ell(x) t^\ell= \frac{1 - t^2}{(1 + t^2 - 2 t x)^{3/2}}, \quad Y_{\ell 0}(\Omega_1 \cdot \Omega_2) = \frac{\sqrt{2\ell + 1}}{2 \sqrt{\pi}} P_{\ell}(\Omega_1 \cdot \Omega_2)
\end{equation}
to evaluate the sums. The result is
\begin{equation}
\begin{split}
     \langle G^{\rm T.O.}_{\Delta}(X;P_1), &G^{\rm T.O.}_{3-\Delta}(X;P_2) \rangle_{\tau = \tau_0}  \\
     &= \frac{1}{2}\frac{2^{5/2} \pi^2 }{\Gamma(\Delta)\Gamma(3 - \Delta)} \left[\frac{e^{i\pi(\Delta - \frac{3}{2} )}e^{-i\left( \Delta - \frac{1}{2} \right)\tau_{12}}}{\tau_{12} - i\epsilon}  \frac{\sin(\tau_{12} -i\epsilon) }{(\cos (\tau_{12} - i\epsilon) - \Omega_1 \cdot \Omega_2)^{3/2}} \Theta(\tau_0 - \tau_1) \Theta(\tau_0 - \tau_2)\right.\\
     &\left.- \frac{e^{-i\pi(\Delta - \frac{3}{2} )} e^{i(\Delta - \frac{1}{2})\tau_{12}}}{\tau_{12} + i\epsilon} \frac{\sin(\tau_{12} +i\epsilon) }{(\cos (\tau_{12} + i\epsilon) - \Omega_1 \cdot \Omega_2)^{3/2}}\Theta(\tau_1 - \tau_0) \Theta(\tau_2 - \tau_0)\right].
     \end{split}
\end{equation}

The distributional contribution to this inner product may be extracted from the identity
\begin{equation}
  \lim_{\epsilon \rightarrow 0}  \frac{1}{\tau_{12} \pm i\epsilon} = \mp i \pi \delta(\tau_{12}) + \mathcal{P}\left(\frac{1}{\tau_{12}}\right),
\end{equation}
where $\mathcal{P}$ denotes the principal value. For $\Delta \neq \frac{3}{2}$, we then find that
\begin{equation}
\label{eq:to-sh}
\begin{split}
     \langle G^{\rm T.O.}_{\Delta}(X;P_1), G^{\rm T.O.}_{3-\Delta}(X;P_2) \rangle_{\tau = \tau_0} & \supset  \frac{(2\pi)^{3} }{\Gamma(\Delta)\Gamma(3 - \Delta)} \rm \delta(\tau_{12})\delta(\Omega_{1}\cdot\Omega_{2}-1)\\
     &\times \left[ e^{i\pi(\Delta - \frac{3}{2})}\Theta(\tau_0 - \tau_1)\Theta(\tau_0 - \tau_2) - e^{-i\pi(\Delta - \frac{3}{2})}\Theta(\tau_1 - \tau_0)\Theta(\tau_2 - \tau_0)\right],
     \end{split}
\end{equation}
where we have used the identity 
\begin{equation}
    \lim_{\epsilon \rightarrow 0} \frac{\epsilon}{(2 - 2\Omega_1 \cdot \Omega_2 + \epsilon^2)^{3/2}} = \delta(1 - \Omega_1 \cdot \Omega_2).
\end{equation}

This result can be easily generalized to $\tau_{12} \in \mathbb{R}$. For instance, the distributional contribution to the inner product of the time-ordered shadow propagator with the anti-time-ordered propagator is given by
\begin{equation}
\label{eq:shadow-ip}
\begin{split}
    &\langle G_{\Delta}^{\rm A. T.O.}(X;P_1), G_{3 - \Delta}^{\rm T.O.}(X;P_2)\rangle_{\tau = \tau_0} \supset  \sum_{k \in \mathbb{Z}} \frac{ (2 \pi)^{3} }{\Gamma(\Delta) \Gamma(3 - \Delta)} \\
&\times \left[ e^{i\pi (\Delta k + \Delta - \frac{3}{2})} \Theta(\tau_1 - \tau_0)\Theta(\tau_0 - \tau_2) - e^{-i\pi (\Delta k+ \Delta - \frac{3}{2})} \Theta(\tau_0 - \tau_1)\Theta(\tau_2 - \tau_0)\right] \delta(\tau_{12}+k\pi) \delta(\Omega_1\cdot \Omega_2^k - 1),
\end{split}
\end{equation}
where $\Omega^k$ is antipodally related to $\Omega$ for odd $k$. The principal value contribution appears to violate the expected conformal covariance properties of the KG inner product for $\Delta \neq \frac{3}{2}$, hence it should vanish. We couldn't verify this by direct computation. For boundary points with $\Delta \tau_{12} \sim k \pi$ it can be shown that the principal value contribution is also delta function localized around $\Omega_1 = \Omega_2^k$ and the full inner product is of the form \eqref{eq:shadow-ip} with $\delta(\tau_{12} + k \pi) \rightarrow \frac{1}{\tau_{12} +k \pi \pm i\epsilon}.$ The relation between this result and the structure of Carrollian two-point functions will be discussed in Section \ref{sec:flat-space-lim}.

\subsection{Boundary inner product and consistency with Stokes' theorem}
\label{sec:bdry-ip}

In this section we verify our results  using the near-boundary expansions of bulk-to-boundary propagators and Stokes' theorem. 

Recall that $G^{\pm}_{\Delta}(P,X)$ are solutions  to the wave equation in AdS, namely
\begin{equation}
    \Box_X G^{\pm}_{\Delta}(X; P) = \Delta(\Delta - 3) G^{\pm}_{\Delta}(X;P).
\end{equation}
This immediately implies that
\begin{equation}
\label{eq:greens}
    \int_{AdS_4} d^4\Sigma \left[(G_{\Delta}^{\pm})^*(X;P_1) \Box_X G^{\pm}_{\Delta}(X;P_2) - \Box_X (G_{\Delta}^{\pm})^*(X;P_2) G^{\pm}_{\Delta}(X;P_1) \right]  = 0.
\end{equation}
Applying Stokes' theorem to \eqref{eq:greens} we find
\begin{equation}
\label{eq:boundary-stokes}
    \int_{\partial AdS_4} d^3\Sigma^{\mu}  \left[(G_{\Delta}^{\pm})^*(X;P_1) \partial_{\mu} G^{\pm}_{\Delta}(X;P_2) -  G^{\pm}_{\Delta}(X;P_2) \partial_{\mu} (G_{\Delta}^{\pm})^*(X;P_1) \right] = 0.
\end{equation}
This formula relates the KG inner products on bulk and boundary slices. 

We can now use the results of Section \ref{sec:mode-expansion-btb} to compute the inner products of bulk to boundary propagators on the AdS boundary. We find that 
\begin{equation}
\label{eq:Wightman-ip}
    \langle G_{\Delta}^{\pm}(X; P_1), G^{\pm}_{\Delta}(X; P_2) \rangle \left. \right|_{\rho = \frac{\pi}{2}} = 0,
\end{equation}
in agreement with \eqref{eq:Gpm-ip} being independent on $\tau_0$, while 
\begin{equation}
\label{eq:boundary-ip}
\begin{split}
     \langle G_{\Delta}^{\rm T.O.}(X; P_1), G^{\rm T.O.}_{\Delta}(X; P_2) \rangle \left. \right|_{\rho = ct.} 
     &= - 2^{\Delta + 1}\pi^{3/2} \frac{\Gamma(\Delta - \frac{1}{2})}{\Gamma(\Delta)} \left( G_{\Delta}^{\rm T.O.}(P_1, P_2) + G^{\rm A.T.O.}_{\Delta}(P_1,P_2) \right),\\
     &\qquad \qquad \qquad \qquad \qquad \quad\tau_1, \tau_2 \in [\tau_i, \tau_f],
     \end{split}
\end{equation} 
where the boundary integral is performed over a region $\tau \in [\tau_i, \tau_f]$ and $\Delta \tau = \tau_f - \tau_i < \pi$ with both $P_1$ and $P_2$ assumed to be contained in the integration region. We verify \eqref{eq:Wightman-ip} and \eqref{eq:boundary-ip} in Appendix \ref{app:boundary-ip} using the near-boundary expansion of the time-ordered bulk-to-boundary propagator
\begin{equation}
\label{eq:boundary-exp}
\begin{split}
    \lim_{\rho \rightarrow \frac{\pi}{2} - \epsilon} G^{\rm T.O.}_{\Delta}(X;P) &= \left( -2^{\Delta}\frac{\pi^{3/2}\Gamma(\Delta - \frac{3}{2})}{\Gamma(\Delta)} \epsilon^{3 - \Delta} i \delta(\tau - \tau_p)\delta^{(2)}(\Omega - \Omega_p) + \mathcal{O}(\epsilon^{5 - \Delta})\right) \\
   &+ \left(\frac{\epsilon^{\Delta}}{\left( \cos(\tau - \tau_p) - \Omega \cdot \Omega_p + i\epsilon \right)^{\Delta}} + \mathcal{O}(\epsilon^{\Delta + 2}) \right)
   \end{split}
\end{equation}
derived in Appendix \ref{app:delta-id}.
Eq. \eqref{eq:boundary-exp} can be proven using the Mellin representation of the bulk-to-boundary propagator 
\begin{equation}
\label{eq:Mellin}
G^{\rm T.O.}_{\Delta}(X;P) = \frac{i^{-\Delta}}{\Gamma(\Delta)} \int_0^{\infty} d\alpha \alpha^{\Delta - 1} e^{i \alpha(-P\cdot X + i\epsilon)}
\end{equation}
and the stationary phase approximation 
\begin{equation}
\label{eq:stat-phase}
\int_{\mathbb{R}^n} dx g(x) e^{ik f(x)} = \sum_{x_0} e^{ik f(x_0)} |{\rm det} ({\rm Hess}(f(x_0)))|^{-1/2} e^{i\frac{\pi}{4} {\rm sgn}({\rm Hess}(f(x_0)))} \left( 2\pi/k \right)^{n/2} g(x_0) + \mathcal{O}(k^{-n/2}),
\end{equation}
where $x_0$ are the critical points of $f$ (that is, points for which $\nabla f(x) = 0$), ${\rm Hess}(f)$ is the Hessian matrix of $f(x)$ and ${\rm sgn(Hess}(f))$ is the difference between the number of positive and negative eigenvalues of ${\rm Hess}(f)$.   It is easy to check that \eqref{eq:time-ordered-btb-ip} and \eqref{eq:boundary-ip} satisfy \eqref{eq:boundary-stokes}.

Using similar methods, we find
\begin{equation}
\label{eq:bdr-to-ato}
\begin{split}
     \langle G_{\Delta}^{\rm A.T.O.}(X; P_1), G^{\rm A.T.O.}_{\Delta}(X; P_2) \rangle \left. \right|_{\rho = ct.} 
     &= 2^{\Delta + 1}\pi^{3/2} \frac{\Gamma(\Delta - \frac{1}{2})}{\Gamma(\Delta)} \left( G_{\Delta}^{\rm T.O.}(P_1, P_2) + G^{\rm A.T.O.}_{\Delta}(P_1,P_2) \right)\\
     \langle G_{\Delta}^{\rm T.O.}(X; P_1), G^{\rm A.T.O.}_{\Delta}(X; P_2) \rangle \left. \right|_{\rho = ct.}& = 0, \quad \quad \quad \quad  \tau_1, \tau_2 \in [\tau_i, \tau_f].
     \end{split}
\end{equation}
In Appendix \ref{sec:delta-function-modes} we perform a consistency check of all these formulas using the mode expansion \eqref{eq:mode-exp-btb-fin} of bulk-to-boundary propagators. In particular, we will see explicitly how the delta function appears in the near-boundary expansion of the time-ordered propagator but not of $G^{\pm}(X;P)$. Putting \eqref{eq:time-ordered-btb-ip-to}, \eqref{eq:ato-to-ip} and \eqref{eq:bdr-to-ato} together, it is easy to show that Stokes' theorem  \eqref{eq:boundary-stokes} is again satisfied.

\subsection{Inner product on null plane}
\label{sec:null-plane-ip}

It is interesting to consider the case where the bulk inner product of $G_{\Delta}^{\rm T.O.}(X;P)$ is evaluated on a null surface. As long as no boundary points are crossed as the spacelike slice is deformed to a null slice, we expect the result to be the same as in \eqref{eq:time-ordered-btb-ip}. In this section we show that this inner product receives an additional delta-function contact contribution in the case where the boundary points are contained in the null surface. Such kinematic configurations appear in the flat space/bulk-point limit \cite{Susskind:1998vk, Hijano:2020szl, deGioia:2024yne} so we can not discard them. In this section we show that the inner product of bulk-to-boundary propagators on a null plane in AdS$_4$ is indeed proportional to a delta function in the case where the boundary points are contained in the null plane. We leave the generalization to boundary points away from the null plane to the future.

We can parameterize a null plane through the origin in AdS as follows
\begin{equation}
\label{eq:null-plane}
    X^0 = r q^0,\quad X^i = r q^i, \quad X^{4} = \ell,
\end{equation}
where $q$ is the null vector
\begin{equation}
   q = (1 + z\bz, z + \bz, -i(z - \bz), 1 - z\bz).
\end{equation}
The null plane \eqref{eq:null-plane} intersects the boundary whose points are parameterized by
\begin{equation}
   P= \lim_{r \rightarrow \infty} r^{-1} X^A
\end{equation}
and the time-ordered bulk-to-boundary propagator in these coordinates takes the form
\begin{equation}
    G_{\Delta}^{\rm T.O.}(X(r,z);P(w)) = \frac{1}{\left(2 r|z - w|^2 + i\epsilon \right)^{\Delta}}.
\end{equation}
The KG inner product of bulk-to-boundary propagators on the null plane is then proportional to
\begin{equation}
\label{eq:null}
\begin{split}
    \langle G_{\Delta}^{\rm T.O.}(X;P_1), G_{\Delta}^{\rm T.O.}(X;P_2)\rangle_{\rm null} &\propto \int_{-\infty}^{\infty} dr \int d^2 z \frac{\Delta r^{2-2\Delta}\left((r+i\epsilon)^{-1} - (r - i\epsilon)^{-1}\right)}{\left(2 |z - w_1|^2 - i\epsilon \sgn(r) \right)^{\Delta}\left(2 |z - w_2|^2 + i\epsilon \sgn(r) \right)^{\Delta}}\\
    &\propto \Delta \epsilon^{2 - 2\Delta} \int d^2 z \frac{1}{\left(2 |z - w_1|^2 \right)^{\Delta}\left(2 |z - w_2|^2 \right)^{\Delta}}\\
    &\propto \delta^{(2)}(w_1 - w_2), \quad \Delta = 1.
    \end{split}
\end{equation}
In the last line we made use of the conformal integral formulas \cite{Osborn:2012vt}. Note that the result is  divergent for $\Delta > 1$ and vanishing for $\Delta < 1$. It is consistent with the expected behavior of Carrollian (including the normalization) \cite{Bagchi:2025vri} and celestial 2-point functions \cite{Raclariu:2021zjz, Pasterski:2021rjz}. This result agrees with that obtained in \cite{deGioia:2023cbd, Alday:2024yyj} by considering a boundary CFT$_3$ two-point function with bulk point kinematics. Interestingly, we see explicitly that in this simple case, the  integral localizes around the bulk point $r = 0$.

\section{AdS scattering bases from bulk-to-boundary propagators}
\label{sec:AdS-bases}

The Klein--Gordon inner products computed in Section \ref{sec:bulk-to-boundary-basis} allow us to construct alternative bases of solutions to the KG equation in AdS to the standard wavefunctions  \eqref{eq:AdS-wf} in terms of (anti-)time-ordered bulk-to-boundary propagators.  We first recall that the AdS propagators admit a decomposition in terms of the advanced and retarded constituents given by 
\begin{equation}
\label{eq:gp4}
G_{\Delta}^{+}(X; P) = \sum_{k = 0}^{\infty} \sum_{\ell = 0}^{\infty} \pi \sqrt{2 \ell + 1} (-1)^{k}\frac{2^{\Delta}\Gamma(\Delta + \ell + k)}{k! \Gamma(\Delta)\Gamma(\frac{3}{2} + \ell)}  \psi^{+}_{k \ell 0}(\tau - \tau_p, \rho, \Omega \cdot \Omega_p)
\end{equation}
and its complex conjugate $G^-(X; P)$. For positive $\Delta$, \eqref{eq:gp4} is manifestly a linear combination of the positive frequency wavefunctions \eqref{eq:AdS-wf} defined with respect to the boundary reference point $(\tau_p, \Omega_p)$. By virtue of orthogonality of the standard AdS wavefunctions, we showed in Section \eqref{sec:inner-prodG} that the Klein--Gordon inner products of \eqref{eq:Gpm-ip} and their complex conjugates on the bulk surface $\Sigma_{\tau_0}$ are given by
\begin{equation}
\label{eq:pm-ip}
   \begin{split}
        \langle G^{\pm}_{\Delta}(X;P_1), G^{\pm}_{\Delta}(X;P_2) \rangle_{\tau = \tau_0} &= \pm 2^{\Delta +1} \pi^{3/2} \frac{\Gamma(\Delta - \frac{1}{2})}{\Gamma(\Delta)} G_{\Delta}^{\pm}(P_1, P_2), \\
        \langle G^{\pm}_{\Delta}(X;P_1), G^{\mp}_{\Delta}(X;P_2) \rangle_{\tau = \tau_0} &= 0.
   \end{split}
\end{equation}
It follows that $G_{\Delta}^{\pm}(X,P)$ form a (non-orthogonal) basis of positive and negative frequency solutions to the Klein--Gordon equation. These basis elements are labeled by points on the boundary instead of the radial and angular momentum quantum numbers $(n, \ell, m)$. 

Note that the inner products \eqref{eq:pm-ip} are non-vanishing at non-coincident points and are proportional to (retarded/advanced) two-point functions in  CFT$_3$. The Lorentzian CFT two-point functions can be directly related to a higher-dimensional generalization of the BPZ inner product \cite{BELAVIN1984333}. Indeed, using orthogonality of the wavefunctions $\widehat{\psi}^{\pm}_{n \ell m}(\tau_i, \Omega_i)$ with respect to the boundary KG inner product\footnote{Note that the Fourier modes of boundary operators can be extracted using either orthogonality of $\widehat{\psi}^{\pm}_{n\ell m}$ with respect to the bulk KG inner product evaluated on the boundary, or from orthogonality of the same wavefunctions with respect to the boundary KG inner product evaluated on a time slice at large Euclidean times $\tau = \pm i\infty$.} one can extract from the RHS of \eqref{eq:pm-ip} an inner product among the incoming and outgoing Fourier modes of CFT$_3$ operators. An analogous construction involving the \textit{shadow} KG inner product in 4-dimensional Minkowski space (Mink$_4$) was shown to relate inner products of conformal primary wavefunctions to the BPZ inner product in (2d) celestial CFT \cite{Crawley:2021ivb}. Conversely, as we will show in Section \ref{sec:shadow-ip-AdS}, our computations in Section \ref{sec:shadow-ip-comp} suggest that the bulk-to-boundary propagators \eqref{eq:gp4} are delta-function normalizable with respect to the shadow inner product in AdS. 

\subsection{In/out bases from (anti-)time-ordered propagators}
\begin{figure}
    \centering
    \includegraphics[width=0.9\linewidth]{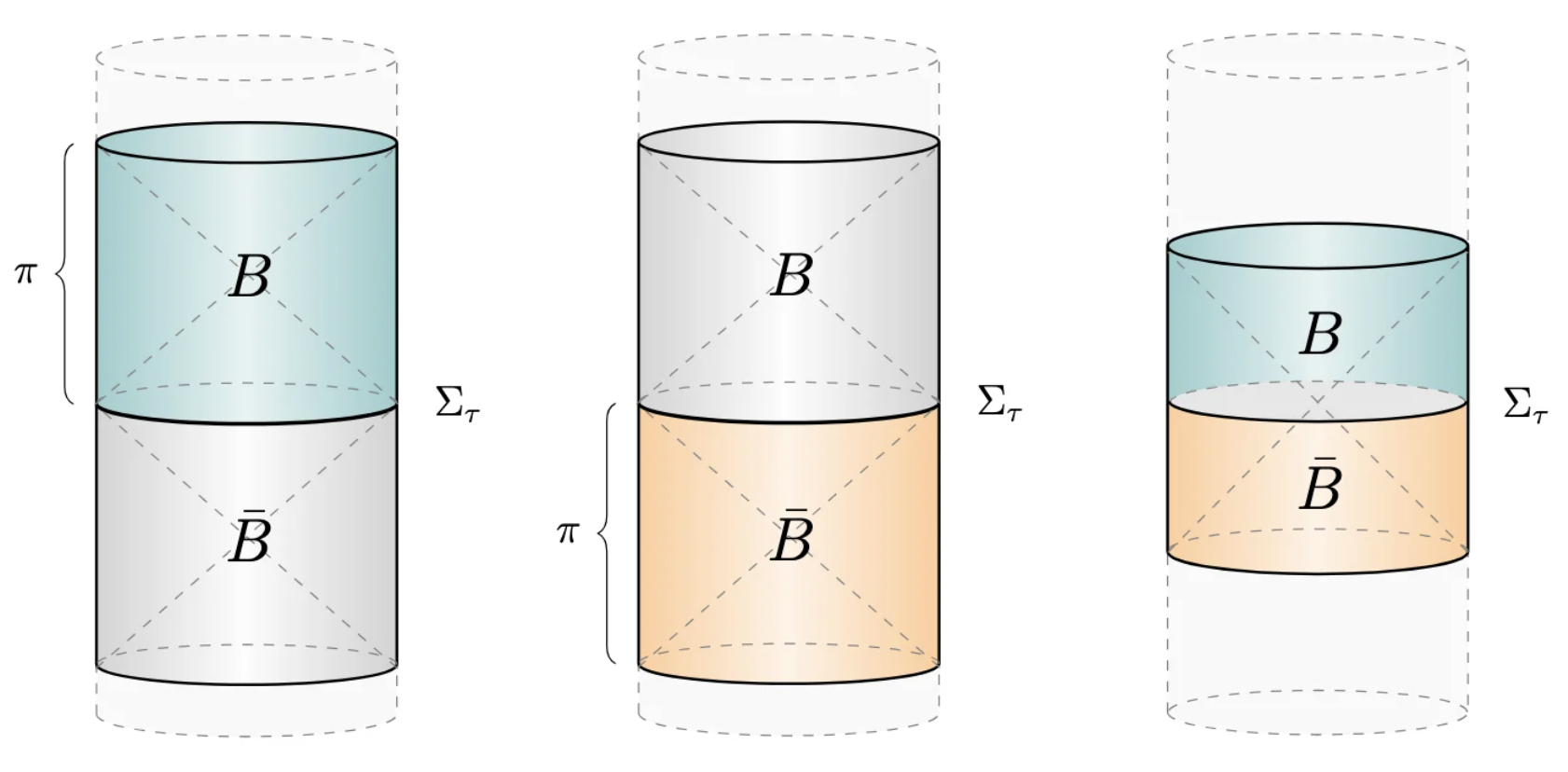}
    \caption{The normalizable component of a bulk field on $\Sigma_\tau$ in AdS can be reconstructed from boundary operators in region $B$ (left), $\bar{B}$ (middle) or a combination of $B$ and $\bar{B}$ (right) integrated against (anti-)time-ordered bulk-to-boundary propagators. The choice of time-ordering picks out the positive or negative energy components depending the choice of integration region. $B$ and $\bar{B}$ play the same role in AdS as $\mathscr{I}^+$ and $\mathscr{I}^-$ in flat space.}
    \label{fig:in-out}
\end{figure}

Using the definition of the (anti-)time-ordered bulk-to-boundary propagators 
\begin{equation}
\label{eq:to-ato}
\begin{split}
    G_{\Delta}^{\rm T.O.}(X, P) &=   G_{\Delta}^+(X; P) \Theta(\tau - \tau_p) +  G_{\Delta}^-(X;P) \Theta(\tau_p - \tau), \\
     G_{\Delta}^{\rm A.T.O.}(X, P) &=   G_{\Delta}^-(X; P) \Theta(\tau - \tau_p) +  G_{\Delta}^+(X;P) \Theta(\tau_p - \tau)
    \end{split}
\end{equation}
together with \eqref{eq:pm-ip}, we deduce that the (anti-)time-ordered propagators form a basis of (negative) positive frequency modes for global AdS provided that $\tau_p < \tau$ and vice versa for $\tau > \tau_p$. These bases can be used to define AdS analogs of incoming and outgoing scattering states in Minkowski space. We illustrate this in Figure \ref{fig:in-out}. As a result, these bulk-to-boundary propagators are the AdS counterparts of positive and negative energy plane waves, or rather, the superpositions thereof in \eqref{eq:Carroll-bases}, also known as Carrollian wavefunctions in Minkowski spacetime. Just like \eqref{eq:Carroll-bases}, the AdS bulk-to-boundary propagators do not diagonalize time translations, but rather boosts towards a point on the boundary. From a representation theoretic perspective, at fixed $\Delta$, $G^{\pm}_{\Delta}(X, P)$ play the same role for $\mathfrak{so}(3,2)$ as the Carrollian wavefunctions for the 4d Poincar\'e algebra.  We will elaborate on this point in Sections \ref{sec:boost-estates} and \ref{sec:flat-space-lim}.
 
Furthermore, we will show in Section \ref{sec:flat-space-lim} that the flat space limit $\ell, \Delta \rightarrow \infty$ relates time-ordered and anti-time-ordered bulk-to-boundary propagators with appropriate choices of boundary points to plane waves of either massive or massless particles. 
As a result, in this limit, \eqref{eq:pm-ip} become inner products of plane waves in flat space 
\begin{equation}
    \langle e^{\pm i k \cdot x}, e^{\pm i k' \cdot x} \rangle_{t = t_0} = \pm 2 (2\pi)^3 \delta^{(3)}(\vec{k} - \vec{k}'), \quad  \langle e^{\pm i k \cdot x}, e^{\mp i k' \cdot x} \rangle_{t = t_0} = 0.
\end{equation}
On the other hand, the flat space analog of the standard AdS wavefunctions \eqref{eq:AdS-wf}  can be shown to be related to Bessel functions. To see this one can use the well-known expansion of a plane wave in terms of spherical Bessel functions
\begin{equation}
    e^{ik r \Omega_k \cdot \Omega} = \sum_{\ell = 0}^{\infty} \sqrt{\frac{2\ell + 1}{4\pi} }i^{\ell} j_{\ell}(kr) Y_{\ell 0}(\Omega \cdot \Omega_k)
\end{equation}
and compare it to \eqref{eq:gp4}. 

An immediate application of this result is presented in Section \ref{sec:bulk-rec-AdS}, where we propose a new reconstruction formula for Lorentzian AdS bulk fields in terms of CFT boundary operators. Our formula generalizes the HKLL prescription, in that it allows for a free field at any point in AdS to be uniquely reconstructed from boundary operators. This is possible only because bulk-to-boundary propagators associated to specific boundary insertions and defined with appropriate time orderings (or $i\epsilon$ prescriptions)  span the positive and negative frequency subspaces of solutions to the Klein--Gordon equation in AdS. As a result, the bulk field can be uniquely reconstructed by analytic continuation from the regions where $X$ and $P$ are spacelike separated. 

The Klein--Gordon inner product is conserved upon deforming the bulk slice provided that no boundary insertions are crossed. Consequently, for appropriate configurations of boundary points, one can take $\Sigma$ to be a null plane without affecting the result \eqref{eq:pm-ip}. (As shown in Section \ref{sec:null-plane-ip}, additional delta function singularities appear if the boundary points lie on $\Sigma$.) This choice naturally relates to the prescription proposed in \cite{Alday:2024yyj} to separate boundary insertions leading to incoming and outgoing scattering states in a flat space limit. Consequently, our construction provides an explanation of why the naively different prescriptions proposed in \cite{deGioia:2024yne, Alday:2024yyj} lead to the same flat space amplitudes in a flat space limit.

\subsection{Shadow inner product}
\label{sec:shadow-ip-AdS}

Our analysis was motivated by trying to understand the precise bulk counterpart of the relation between CFT correlators and flat space scattering amplitudes and Carrollian/celestial correlators obtained using different representations and limits. In celestial CFT, the shadow transform has been argued in \cite{Pasterski:2017kqt,Fan:2021isc,Chang:2022jut,Banerjee:2024yir,Surubaru:2025qhs,Liu:2025dhh,Bhattacharyya:2025nfp} to remove some of the pathologies of correlation functions, in particular the distributional terms appearing at low points due to momentum conservation. These works suggest that, conversely, shadow transforms of standard CFT correlation functions will contain distributional terms, at least for low points. In this section we show this is indeed the case at the level of two-point functions by introducing a shadow KG inner product in AdS$_4$ (see \cite{Crawley:2021ivb} for a definition in Mink$_4$) and analyzing its properties. 

Our results in Section \ref{sec:shadow-ip-comp} allow us to deduce that the standard propagators \eqref{eq:gp4} admit a delta-function normalizable inner product with respect to the shadow Klein--Gordon inner product in AdS
\begin{equation}
    \llangle \Phi_1(X), \Phi_2(X) \rrangle \equiv i \int d\Sigma^{\mu} \left[\widetilde{\Phi}_1(X) \p_{\mu} \Phi_2(X) - \Phi_2(X) \p_{\mu} \widetilde{\Phi}_1(X) \right].
\end{equation}
This is obtained from the standard Klein--Gordon inner product \eqref{eq:KG-ip} by replacing complex conjugation by the shadow transform. In other words
\begin{equation}
     \llangle \Phi_1(X), \Phi_2(X) \rrangle = \langle \widetilde{\Phi}_1^*(X), \Phi_2(X)\rangle = -\langle \Phi_2^*(X), \widetilde{\Phi}_1(X)\rangle.
\end{equation}
It then follows from \eqref{eq:shadow-G-0} and \eqref{eq:shadow-bulk-ip} that 
\begin{equation}
\label{eq:shadow-ip-gpm}
\begin{split}
    \llangle G_{\Delta}^{\pm }(X;P_1), G_{\Delta}^{\pm}(X; P_2)\rrangle &= 0, \\
     \llangle G_{\Delta}^{\pm }(X;P_1), G_{\Delta}^{\mp}(X; P_2)\rrangle &= \langle G_{3-\Delta}^{\mp}(X;P_1), G_{\Delta}^{\mp}(X;P_2)\rangle = \mp \frac{(2\pi)^{3}  e^{\pm i\pi (\Delta - \frac{3}{2})}}{\Gamma(\Delta)\Gamma(3 - \Delta)} \delta(\Omega_{1}\cdot\Omega_{2}-1)\delta(\tau_{12}).
    \end{split}
\end{equation}
 These results are implied by the action of the shadow transform \eqref{eq:shadow-norm}  on $G^{\pm}_{\Delta}$, namely
\begin{equation}
    \widetilde{G^{\mp}_{\Delta}}(X;P) = G_{3 - \Delta}^{\mp}(X;P),
\end{equation}
the symmetries \eqref{eq:KG-sym} of the KG inner product, and \eqref{eq:to-sh}.
The inner products \eqref{eq:shadow-ip-gpm} assumed $|\tau_{12}| < \pi$, otherwise they receive additional contributions from light-like separated points on the boundary -- see \eqref{eq:shadow-ip}. In contrast to \eqref{eq:pm-ip}, for generic $\Delta$, the shadow inner product vanishes unless $\Omega_1 = \pm \Omega_2$.

As in \eqref{eq:to-ato}, one can construct time-ordered and anti-time-ordered shadow bulk-to-boundary propagators which once again form bases for either the positive or negative frequency solutions to the Klein--Gordon equation, depending on the location of the boundary insertion with respect to $\Sigma$. Specifically,
\begin{equation}
\begin{split}
    \llangle G_{\Delta}^{\rm T.O.}(X;P_1), G_{\Delta}^{\rm T.O.}(X;P_2) \rrangle &= \langle G_{3 - \Delta}^{\rm  T.O.}(X;P_1), G_{\Delta}^{\rm T.O.}(X;P_2)\rangle \\
    &=\frac{(2\pi)^{3} }{\Gamma(\Delta)\Gamma(3 - \Delta)} \rm \delta(\tau_{12})\delta(\Omega_{1}\cdot\Omega_{2}-1)\\
    &\times \left[e^{-i\pi(\Delta - \frac{3}{2})}\Theta(\tau_0 - \tau_1) \Theta(\tau_0 - \tau_2) -  e^{i\pi(\Delta - \frac{3}{2})}\Theta(\tau_1 - \tau_0) \Theta(\tau_2 - \tau_0)\right].
    \end{split}
\end{equation}
In the first equality we used that the shadow transform defined in \eqref{eq:shadow-G-0} maps a time-ordered bulk-to-boundary propagator to an anti-time-ordered one.

Consequently, one can choose to span the positive and negative energy subspaces of solutions to the KG equation in AdS with respectively $G_{\Delta}^{\rm T.O.}(X;P)$ and $G_{3 - \Delta}^{\rm A.T.O.}(X;P)$ provided that $\tau_p < \tau$, or equivalently, their complex conjugates for $\tau_p > \tau$.
Our results suggest that one can define generalized (Lorentzian) AdS Witten diagrams where the bulk integral is performed over the causal diamond anchored on $\Sigma$ and the bulk-to-boundary propagators are chosen appropriately depending on whether the boundary insertions are in $\bar{B}$ (corresponding to ``incoming'' states) or in $B$ (corresponding to ``outgoing'' states). We expect the observables constructed with the shadow basis to be closely related to scattering amplitudes in flat space.  We leave an exploration of this construction and its relation to previous works on extracting scattering amplitudes from CFT correlation functions \cite{Penedones:2010ue, Li:2021snj,Hijano:2019qmi, Hijano:2020szl, Komatsu:2020sag} to the future. The main take-home message is that these observables should contain no more or less information than the standard AdS Witten diagrams since the shadow transform provides a different boundary realization of the same bulk dynamics. We will see this explicitly in the next section, where we will show that a free field in AdS$_4$ with normalizable near-boundary fall-offs can be reconstructed in terms of either CFT$_3$ boundary operators or shadow transforms thereof.

\section{Bulk reconstruction of free fields in AdS, revisited}
\label{sec:bulk-rec-AdS}

The analysis in Sections \ref{sec:bulk-to-boundary-basis}, \ref{sec:AdS-bases} suggests that free fields in AdS$_4$ admit the following representation in terms of CFT$_3$ primary operators $\mathcal{O}_{\Delta}$
\begin{equation}
\label{eq:shadow-bulk-rec}
\begin{split}
     \Phi(\tau, \rho, \Omega) = \alpha \int_{B} d\tau_p d\Omega_p G^{\rm A.T.O.}_{3 - \Delta}(\tau, \rho, \Omega; \tau_p, \Omega_p) &\mathcal{O}_{\Delta}(\tau_p, \Omega_p) \\
     &+ \beta \int_{B} d\tau_p d\Omega_p G^{\rm T.O.}_{\Delta}(\tau, \rho, \Omega; \tau_p, \Omega_p) \widetilde{\mathcal{O}}_{3 - \Delta}^-(\tau_p, \Omega_p).
     \end{split}
\end{equation}
Here $B$ was defined in \eqref{eq:B-bB}, while the shadow operator $\widetilde{\mathcal{O}}^-_{3 - \Delta}$ is defined by 
\begin{equation}
\label{eq:shadow}
\begin{split}
     \widetilde{\mathcal{O}}^-_{3 - \Delta}(P) &\equiv \underbrace{\frac{e^{-i\pi(\Delta - \frac{3}{2})}}{\cos\pi\Delta}\frac{\Gamma( \Delta)}{\pi^{3/2}2^{\Delta +1}\Gamma(\Delta - \frac{3}{2})}}_{\mathcal{N}(\Delta)} \int_{\bar{B}} d^3 P' \frac{1}{(-P\cdot P' - i\epsilon)^{3-\Delta}} \mathcal{O}_{\Delta}(P'), \quad P \in B,
     \end{split}
\end{equation}
where $\bar{B}$ was defined in \eqref{eq:B-bB}. This definition is motivated by the transformation properties of bulk-to-boundary propagators under the shadow transform, which we worked out in Appendix \ref{sec:btb-shadow}. As a result, we can equivalently express \eqref{eq:shadow-bulk-rec} as 
\begin{equation}
\label{eq:phi-rec}
\begin{split}
    \Phi(\tau, \rho, \Omega) = \alpha \int_{B} d\tau_p d\Omega_p G^{\rm A.T.O.}_{3 - \Delta}(\tau, \rho, \Omega; \tau_p, \Omega_p) &\mathcal{O}_{\Delta}(\tau_p, \Omega_p) \\
    &+ \beta \int_{\bar{B}} d\tau_p d\Omega_p G^{\rm A.T.O.}_{3 - \Delta}(\tau, \rho, \Omega; \tau_p, \Omega_p) \mathcal{O}_{\Delta}(\tau_p, \Omega_p)\\
    \end{split}
\end{equation}
or 
\begin{equation}
\label{eq:phi-rec-1}
    \begin{split}
       \Phi(\tau, \rho, \Omega)  =  \alpha\int_{\bar{B}} d\tau_p d\Omega_p G^{\rm T.O.}_{\Delta}(\tau, \rho, \Omega; \tau_p, \Omega_p) &\widetilde{\mathcal{O}}_{3 - \Delta}^+(\tau_p, \Omega_p) \\
       &+ \beta \int_{B} d\tau_p d\Omega_p G^{\rm T.O.}_{\Delta}(\tau, \rho, \Omega; \tau_p, \Omega_p) \widetilde{\mathcal{O}}_{3 - \Delta}^-(\tau_p, \Omega_p),
    \end{split}
\end{equation}
where we defined
\begin{equation}
\label{eq:shadow-op}
    \widetilde{\mathcal{O}}^+_{3 - \Delta}(P) = \mathcal{N}^*(\Delta) \int_{B} d^3 P' \frac{1}{(-P\cdot P' - i\epsilon)^{3-\Delta}} \mathcal{O}_{\Delta}(P'), \quad P \in \bar{B} .
\end{equation}
We introduced the $\pm$ superscripts to distinguish between the positive/negative energy components of the boundary operators, as in \eqref{eq:extrapolate0}. We show in Appendix \ref{app:KA-AdS} that one obtains an equivalent formula to \eqref{eq:shadow-op} by integrating over $\bar{B}$ with a shadow kernel with the opposite $+i\epsilon$ prescription. Using these formulas, we can derive equivalent representations of the bulk field as integrals over boundary operators, involving either $\mathcal{O}_{\Delta}$ or its shadow, paired with bulk-to-boundary propagators with the appropriate dimension and time ordering, and integrated over the corresponding boundary interval of length $\Delta \tau_p = \pi$.

Assuming that $\Phi$ is Hermitian and that the boundary operators are given by the extrapolate dictionary \cite{Banks:1998dd,Harlow:2011ke}
\begin{equation}
\label{eq:extrapolate}
    \mathcal{O}_{\Delta}(\tau, \Omega) = \lim_{\rho \rightarrow \frac{\pi}{2} - \epsilon} \epsilon^{-\Delta} \Phi(\tau, \rho, \Omega)
\end{equation}
we find that 
\begin{equation}
    \alpha = \beta^*.
\end{equation}
This condition guarantees that $\Phi$ has normalizable near-boundary fall-off. To see this, we first express \eqref{eq:shadow} in terms of modes of the bulk field via \eqref{eq:extrapolate}. We find that
\begin{equation}
\begin{split}
    \widetilde{\mathcal{O}}^-_{3 - \Delta}(P) &= \lim_{\rho \rightarrow \frac{\pi}{2} - \epsilon} \epsilon^{\Delta - 3} \mathcal{N}(\Delta) \int_{\bar{B}} d\tau_p d\Omega_p G_{3 - \Delta}^{\rm A.T.O.}(X; P) \mathcal{O}_\Delta(P)\\
    &= \lim_{\rho \rightarrow \frac{\pi}{2} - \epsilon} \epsilon^{\Delta - 3} \mathcal{N}(\Delta) \sum_{n, \ell, m} \int_{\bar{B}} d\tau_p d\Omega_p G_{3 - \Delta}^{\rm A.T.O.}(X; P) \left( \widehat{\psi}_{n\ell m}^-(\tau_p, \Omega_p) a^{\dagger}_{n\ell m} + \widehat{\psi}^+_{n\ell m}(\tau_p, \Omega_p) a_{n \ell m} \right),
    \end{split}
\end{equation}
where $\widehat{\psi}_{n\ell m}$ were defined in \eqref{eq:hat-psi}.
The boundary integral can then be evaluated using the result \eqref{eq:ato} below,  
\begin{equation}
\label{eq:O-tilde-modes}
\begin{split}
    \widetilde{\mathcal{O}}^-_{3 - \Delta}(P) &= - \lim_{\rho \rightarrow \frac{\pi}{2} - \epsilon} \epsilon^{\Delta - 3} \mathcal{N}(\Delta) \frac{2^{4 - \Delta}\pi^{3/2} (-1)^{-\Delta + \frac{3}{2}} \Gamma(\frac{3}{2} - \Delta)\cos \pi \Delta}{\Gamma(3 - \Delta)} \sum_{n, \ell, m} \psi_{n \ell m}^-(\tau, \rho, \Omega) a_{n \ell m}^{\dagger} \\
    &= - \lim_{\epsilon \rightarrow 0} \epsilon^{2 \Delta - 3} e^{i\pi(3 - 2\Delta)} \frac{2^{3 - 2\Delta}\Gamma(\Delta) \Gamma(\frac{3}{2} - \Delta)}{\Gamma(3 - \Delta)\Gamma(\Delta - \frac{3}{2})} \sum_{n, \ell, m} \widehat{\psi}_{n \ell m}^-(\tau, \Omega) a_{n \ell m}^{\dagger}.
    \end{split}
\end{equation}
We hence see that $\widetilde{\mathcal{O}}_{3 - \Delta}^-$ is proportional to $\mathcal{O}^-_{\Delta}(P)$, up to a proportionality factor that depends on the boundary regulator. This dependence is consistent with scaling symmetry and ensures that \eqref{eq:phi-rec} is consistent with the mode expansion \eqref{eq:AdS-bulk-field}. 

One can obtain yet another equivalent expression to \eqref{eq:phi-rec} by trading $B$ for $\bar{B}$ (or vice versa) and complex conjugating the bulk-to-boundary propagator. For instance, 
\begin{equation}
\label{eq:exp-fin}
     \Phi(\tau, \rho, \Omega) =  \int_{B} d\tau_p d\Omega_p \left( \alpha G^{\rm A. T.O.}_{\Delta}(\tau, \rho, \Omega; \tau_p, \Omega_p) \widetilde{\mathcal{O}}_{3 - \Delta}^+(\tau_p, \Omega_p) + \beta G^{\rm T.O.}_{\Delta}(\tau, \rho, \Omega; \tau_p, \Omega_p) \widetilde{\mathcal{O}}_{3 - \Delta}^-(\tau_p, \Omega_p) \right).
\end{equation}
In the remainder of this section we prove that all of these decompositions are compatible with the extrapolate dictionary. The bulk inner products of bulk-to-boundary propagators computed in Section \ref{sec:bulk-to-boundary-basis} allow one to easily express the bulk field modes in terms of boundary operators, and vice versa.  

\subsection{Consistency with the extrapolate dictionary}

We use the mode expansion
\begin{equation}
\label{eq:shadow-ato-5}
    G_{3 - \Delta}^{\rm A.T.O.}(X,P) = G_{3 - \Delta}^-(X,P) \Theta(\tau - \tau_p) + G_{3 - \Delta}^+(X,P) \Theta(\tau_p - \tau),
\end{equation}
where 
\begin{equation}
    G_{3 - \Delta}^{\pm}(X,P) = \sum_{k = 0}^{\infty} \sum_{\ell = 0}^{\infty} \pi \sqrt{2 \ell + 1} (-1)^{\pm k} \frac{2^{3 - \Delta} \Gamma(3 - \Delta + \ell + k)}{k! \Gamma(3 - \Delta) \Gamma(\frac{3}{2} + \ell)} \psi_{k \ell 0}^{3-\Delta,\pm}(\tau - \tau_p,\rho, \Omega \cdot \Omega_p).
\end{equation}
Here $\psi^{\pm, 3 - \Delta}_{n\ell 0}$ are the highest/lowest weight wavefunctions \eqref{eq:AdS-wf} with $\Delta \rightarrow 3  - \Delta$. As a result we find
\begin{equation}
\label{eq:ato}
\begin{split}
     \int_{B} d\tau_p d\Omega_p G_{3 - \Delta}^{\rm A.T.O.}(\tau, \rho, \Omega; \tau_p, \Omega_p)& \widehat{\psi}^+_{n\ell m}(\tau_p, \Omega_p) =  \int_{B} d\tau_p d\Omega_p G_{3 - \Delta}^{+}(\tau, \rho, \Omega; \tau_p, \Omega_p) \widehat{\psi}^+_{n\ell m}(\tau_p, \Omega_p)\\
     &= - 2 \pi^{3/2} (-1)^{\Delta-\frac{3}{2}} \frac{2^{3 - \Delta} \Gamma(\frac{3}{2} - \Delta) \cos \pi \Delta}{\Gamma(3 - \Delta)} \\
     &\times \sin^{\ell} \rho \cos^{\Delta}\rho {}_2 F_1\left(-n, \ell + \Delta +n; \frac{3}{2} + \ell; \sin^2 \rho \right) Y_{\ell m}(\Omega)\\
     &= -2 \pi^{3/2} (-1)^{\Delta-\frac{3}{2}} \frac{2^{3 - \Delta} \Gamma(\frac{3}{2} - \Delta) \cos \pi \Delta}{\Gamma(3 - \Delta)} \psi_{n\ell m}^+(\tau, \rho, \Omega),
     \end{split}
\end{equation}
where we used 
\begin{equation}
    \int_{\tau}^{\pi + \tau} d \tau_p e^{-i\omega_n^{\Delta} \tau_p} e^{-i\omega^{3 - \Delta}_k(\tau - \tau_p)} = e^{-i\omega_k^{3-\Delta} \tau}\int_0^{\pi} d\tau_p e^{-i(\omega_n^{\Delta} - \omega_k^{3 - \Delta}) \tau_p} = \pi e^{-i\omega_n^{\Delta} \tau} \delta_{k, \Delta + n + \frac{\ell - \ell'}{2} - \frac{3}{2}},
\end{equation}
the identity
\begin{equation}
    \int d\Omega' Y_{\ell'  0}(\Omega \cdot \Omega') Y_{\ell  m}(\Omega') = \sqrt{\frac{4\pi}{2 \ell + 1}} \delta_{\ell \ell'} Y_{\ell m}(\Omega)
\end{equation}
derived in Appendix \ref{app:shadowO-O}, as well as the hypergeometric function identity \eqref{eq:hg-id-shadow}. 

Similarly, we find
\begin{equation}
     \int_{\bar{B}} d\tau_p d\Omega_p G_{3 - \Delta}^{\rm A.T.O.}(\tau, \rho, \Omega; \tau_p, \Omega_p) \widehat{\psi}^-_{n\ell m}(\tau_p, \Omega_p) = -2 \pi^{3/2} e^{i\pi (\frac{3}{2} - \Delta)} \frac{2^{3 - \Delta} \Gamma(\frac{3}{2} - \Delta) \cos \pi \Delta}{\Gamma(3 - \Delta)} \psi_{n\ell m}^-(\tau, \rho, \Omega), 
\end{equation}
while
\begin{equation}
     \int_{B} d\tau_p d\Omega_p G_{3 - \Delta}^{\rm A.T.O.}(\tau, \rho, \Omega; \tau_p, \Omega_p) \widehat{\psi}^-_{n\ell m}(\tau_p, \Omega_p)  =  \int_{\bar{B}} d\tau_p d\Omega_p G_{3 - \Delta}^{\rm A.T.O.}(\tau, \rho, \Omega; \tau_p, \Omega_p) \widehat{\psi}^+_{n\ell m}(\tau_p, \Omega_p) = 0.
\end{equation}
The formulas involving the time-ordered bulk-to-boundary propagators are obtained by complex conjugation. 

Substituting these formulas into  \eqref{eq:phi-rec}, and using the extrapolate dictionary \eqref{eq:extrapolate}, we find under the initial assumption that $\Delta - \frac{3}{2}\in \mathbb{N}$ -- assumed to hold for general $\Delta$,
\begin{equation}
\label{eq:first}
\begin{split}
    \Phi(\tau, \rho, \Omega) = -  \frac{2^{4 - \Delta}\pi^{3/2} \Gamma(\frac{3}{2} - \Delta) \cos\pi \Delta}{\Gamma(3 - \Delta)} \sum_{n, \ell, m} \Big(e^{i\pi(\Delta - \frac{3}{2})} \alpha &\psi_{n \ell m}^+(\tau, \rho, \Omega) a_{n\ell m} \\
    &+ e^{-i\pi(\Delta - \frac{3}{2})} \beta  \psi_{n \ell m}^-(\tau, \rho, \Omega) a^{\dagger}_{n\ell m} \Big).
\end{split}
\end{equation}
Consistency with the extrapolate dictionary then fixes 
\begin{equation}
\label{eq:a-b}
    \alpha = \beta^* = - e^{-i\pi(\Delta - \frac{3}{2})}\frac{\Gamma(3 - \Delta)}{2^{4 - \Delta} \pi^{3/2} \Gamma(\frac{3}{2} - \Delta) \cos \pi \Delta} \in \mathbb{R}.
\end{equation}

An identical computation to \eqref{eq:ato} yields
\begin{equation}
    \int_{\bar{B}} d\tau_p d\Omega_p G_{\Delta}^{\rm T.O.}(\tau, \rho, \Omega; \tau_p, \Omega_p) \widehat{\psi}^{+,3 - \Delta}_{n\ell m}(\tau_p, \Omega_p) = 2 \pi^{3/2} e^{-i\pi(\Delta - \frac{3}{2})} \frac{2^{\Delta} \Gamma(\Delta - \frac{3}{2}) \cos \pi \Delta}{\Gamma(\Delta)} \psi_{n\ell m}^{+,3 - \Delta}(\tau, \rho, \Omega)
\end{equation}
and
\begin{equation}
     \int_{B} d\tau_p d\Omega_p G_{\Delta}^{\rm T.O.}(\tau, \rho, \Omega; \tau_p, \Omega_p) \widehat{\psi}^{-,3 - \Delta}_{n\ell m}(\tau_p, \Omega_p) = 2 \pi^{3/2} e^{i\pi(\Delta -\frac{3}{2})} \frac{2^{\Delta} \Gamma(\Delta - \frac{3}{2}) \cos \pi \Delta}{\Gamma(\Delta)} \psi_{n\ell m}^{-,3 - \Delta}(\tau, \rho, \Omega).
\end{equation}
Furthermore, \eqref{eq:O-tilde-modes} can be rewritten in terms of $\widehat{\psi}^{-,3 - \Delta}_{n\ell m}(\tau', \Omega')$ by shifting the sum on $n$ and using \eqref{eq:hat-psi} as
\begin{equation}
\label{eq:shadow-O}
    \widetilde{\mathcal{O}}^-_{3 - \Delta}(P) = - e^{i\pi(3 - 2\Delta)}\frac{2^{3 - 2 \Delta} \Gamma(\Delta)\Gamma(\frac{3}{2} - \Delta)}{\Gamma(3 - \Delta)\Gamma(\Delta - \frac{3}{2})} \sum_{n = \Delta - \frac{3}{2}}^{\infty} \sum_{\ell = 0}^{\infty}\sum_{m = -\ell}^{\ell} \widehat{\psi}^{-, 3 - \Delta}_{n \ell m}(\tau_p, \Omega_p) a^{\dagger}_{n\ell m}.
\end{equation}
Together, these identities imply that \eqref{eq:phi-rec-1} can be expressed in terms of modes as
\begin{equation}
\begin{split}
    \Phi(\tau, \rho, \Omega) &= -\frac{2^{4 - \Delta} \pi^{3/2} \Gamma(\frac{3}{2} - \Delta)\cos \pi\Delta}{\Gamma(3 - \Delta)}\sum_{n = \Delta - \frac{3}{2}}^{\infty} \sum_{\ell = 0}^{\infty} \sum_{m = -\ell}^{\ell} \Big(e^{i\pi(\Delta - \frac{3}{2})}\alpha \psi^{+, 3 - \Delta}_{n\ell m}(\tau, \rho, \Omega) a_{n\ell m}\\
    &\hspace{240pt} + e^{-i\pi(\Delta - \frac{3}{2})} \beta \psi_{n \ell m}^{-, 3 - \Delta}(\tau, \rho, \Omega) a^{\dagger}_{n \ell m}\Big)\\
    &= - \frac{2^{4 - \Delta} \pi^{3/2} \Gamma(\frac{3}{2} - \Delta)\cos \pi\Delta}{\Gamma(3 - \Delta)}\sum_{n = 0}^{\infty}  \sum_{\ell = 0}^{\infty} \sum_{m = -\ell}^{\ell} \Big(e^{i\pi(\Delta - \frac{3}{2})} \alpha \psi^{+}_{n\ell m}(\tau, \rho, \Omega) a_{n\ell m} \\
    &\hspace{240pt} + e^{-i\pi(\Delta - \frac{3}{2})} \beta \psi_{n \ell m}^{-}(\tau, \rho, \Omega) a^{\dagger}_{n \ell m}\Big),
    \end{split}
\end{equation}
which again agrees with \eqref{eq:first} as it should. 

\subsection{Relation between bulk modes and boundary operators}
\label{sec:bulk-bdry-ads}

As an independent check of our results, we now show that one can use either the extrapolate dictionary \eqref{eq:extrapolate} or the bulk inner products computed in Section \eqref{sec:bulk-to-boundary-basis} together with the reconstruction formulas proposed in \eqref{eq:phi-rec}, \eqref{eq:phi-rec-1}, \eqref{eq:exp-fin}, to express the bulk field modes in terms of boundary operators.

One the one hand, we have
\begin{equation}
\begin{split}
\mathcal{O}_{\Delta}(\tau, \Omega) = \sum_{n, \ell, m}  &\frac{\Gamma(\frac{3}{2} - \Delta) \Gamma(\frac{3}{2} + \ell)}{\Gamma(\frac{3}{2} + \ell + n)\Gamma(\frac{3}{2} - \Delta - n)} \left[ e^{-i\omega_n \tau} Y_{\ell m}(\Omega) a_{n\ell m} + e^{i\omega_n \tau} Y_{\ell m}^*(\Omega) a_{n\ell m}^{\dagger} \right],\\
 \end{split}
\end{equation}
from which we infer, by using orthogonality of spherical harmonics and of the exponential, that
\begin{equation}
\label{eq:boundary-a}
    a^{\dagger}_{n\ell m} =  \frac{(-1)^{n}\Gamma(\frac{3}{2} + \ell + n)\Gamma(\Delta - \frac{1}{2}) }{\pi \Gamma(\frac{3}{2} + \ell)\Gamma(-\frac{1}{2} + \Delta + n)} \int_c^{\pi+c} d\tau\int d\Omega_p e^{-i\omega_n \tau} Y_{\ell m}(\Omega) \mathcal{O}_{\Delta}(\tau, \Omega).
\end{equation}
Note that one can pick out the modes by performing the boundary integral over any time interval of length $\pi$.
On the other hand, using the inner products computed in Section \ref{sec:bulk-to-boundary-basis}, one finds
\begin{equation}
\label{a-adagger}
\begin{split}
  a_{n\ell m}^{\dagger} &\equiv -\frac{1}{N_{n\ell}^{d=3}} \langle \psi^-_{n\ell m}(X), \Phi(X) \rangle_{\tau}, \\
  a_{n\ell m} &\equiv \frac{1}{N_{n\ell}^{d=3}} \langle \psi^+_{n\ell m}(X), \Phi(X) \rangle_{\tau},
  \end{split}
\end{equation}
which can be computed using the change of basis coefficients
\begin{equation}
\begin{split}
    \langle \psi^-_{n\ell m}(X), G_{\Delta}^{\rm T.O}(X;P)\rangle_{\tau} &=   - 2 \pi^{3/2} \frac{2^{\Delta} (-1)^n \Gamma(\Delta + \ell + n)}{n! \Gamma(\Delta) \Gamma(\frac{3}{2} + \ell)} N_{n\ell}^{d=3} e^{-i\omega_n\tau_p} Y_{\ell m}( \Omega_p) \Theta(\tau_p - \tau),\\
     \langle \psi^-_{n\ell m}(X), G^{\rm A.T.O.}_{3 - \Delta}(X;P)\rangle_{\tau} 
     &=  -2\pi^{3/2} (-1)^{-n-\Delta +\frac{3}{2}} 2^{3 - \Delta} \frac{\Gamma(n + \ell + \frac{3}{2})}{\Gamma(n + \Delta - \frac{1}{2})\Gamma(3-\Delta)\Gamma(\frac{3}{2} + \ell)} \\
     &\times N_{n\ell}^{d=3} e^{-i\omega_n \tau_p} Y_{\ell m}(\Omega_p) \Theta(\tau - \tau_p).\\
    \end{split}
\end{equation}
Applying these formulas to \eqref{eq:exp-fin}, we find
\begin{equation}
\label{eq:a-exp}
    a_{n \ell m}^{\dagger} = \gamma \int_{B} d\tau_p \int d\Omega_p   \widetilde{\mathcal{O}}^-_{3 - \Delta}(\tau_p, \Omega_p) e^{-i\omega_n \tau_p} Y_{\ell m}(\Omega_p) \Theta(\tau_p - \tau),
\end{equation}
where 
\begin{equation}
\begin{split}
    \gamma & = \beta 2 \pi^{3/2}  \frac{2^{\Delta} (-1)^n \Gamma(\Delta + \ell + n)}{n! \Gamma(\Delta) \Gamma(\frac{3}{2} + \ell)}.
 \end{split}   
\end{equation}
The expression for $a_{n\ell m}$ follows by complex conjugation. 

Furthermore, we show in Appendix \ref{app:shadowO-O} that
\begin{equation}
\begin{split}
& \int d\tau_p \int d\Omega_p   \widetilde{\mathcal{O}}_{3 - \Delta}(\tau_p, \Omega_p)  e^{-i\omega_n \tau_p} Y_{\ell m}(\Omega_p) \Theta(\tau_p - \tau) \\  &=\frac{ e^{-i\pi (\Delta - \frac{3}{2})}\Gamma(\Delta)}{ 2^{2\Delta -3} \Gamma(3 - \Delta)}  \int d\tau_{p'}\int d^2\Omega_{p'} \Theta(\tau - \tau_{p'}) \frac{\Gamma(\frac{3}{2} +\ell+n)\Gamma(1+n)}{\Gamma(-\frac{1}{2} + \Delta + n) \Gamma(\ell +\Delta + n)} e^{-i\omega^{\Delta}_n \tau_{p'}} Y_{\ell m}(\Omega_{p'}) \mathcal{O}_{\Delta}(P').
 \end{split}
\end{equation}
Substituting this into \eqref{eq:a-exp}, we find
\begin{equation}
\begin{split}
    a_{n\ell m}^{\dagger} &= 2 \beta e^{-i\pi (\Delta - \frac{3}{2})} (-1)^n \frac{2^{3 - \Delta}\pi^{3/2}}{\Gamma(3-\Delta)} \frac{\Gamma(\frac{3}{2} + \ell + n)}{ \Gamma(\frac{3}{2} + \ell)\Gamma(-\frac{1}{2} + \Delta + n)} \int_{\bar{B}} d\tau\int d\Omega_p e^{-i\omega_n \tau} Y_{\ell,m}(\Omega) \mathcal{O}_{\Delta}(\tau, \Omega)\\
    &= (-1)^n \frac{\Gamma(\frac{3}{2} + \ell + n)\Gamma(\Delta - \frac{1}{2})}{\pi \Gamma(\frac{3}{2} + \ell) \Gamma(\Delta + n - \frac{1}{2})} \int_{\bar{B}} d\tau\int d\Omega_p e^{-i\omega_n \tau} Y_{\ell,m}(\Omega) \mathcal{O}_{\Delta}(\tau, \Omega),
    \end{split}
\end{equation}
where we used \eqref{eq:a-b}. 
This agrees with \eqref{eq:boundary-a}, providing a non-trivial check of the validity of our mode expansions for bulk-to-boundary propagators and inner products thereof in Lorentzian AdS spacetimes.

\subsection{Kirchhoff--d'Adh\'emar formula in AdS}
\label{sec:AdS-KA}

In this section we use our bulk reconstruction formulas to derive an AdS$_4$ analogue of the Kirchhoff--d'Adh\'emar formula in 4d Minkowski spacetime. 

We start with the representation of an AdS free field in terms of CFT operators 
\begin{equation}
\Phi(X) = \int_{B} d\tau_p d\Omega_p \left( \alpha^* G^{\rm T.O.}_{\Delta}(\tau, \rho, \Omega; \tau_p, \Omega_p) \widetilde{\mathcal{O}}_{3 - \Delta}^-(\tau_p, \Omega_p) + \alpha G_{\Delta}^{\rm A.T.O}(\tau, \rho, \Omega; \tau_p, \Omega_p)\widetilde{\mathcal{O}}_{3 - \Delta}^{+}(\tau_p, \Omega_p)\right),
 \end{equation}
 where $\alpha$ is given in \eqref{eq:a-b}. Just as in flat space, one can invert this formula by taking the KG inner product of both sides with a bulk-to-boundary propagator $G_{\Delta}^{\rm T.O.}(X, P_1)$, with $P_1$ in $B$. Choosing the integration region to be $B$ and using \eqref{eq:boundary-ip}, \eqref{eq:bdr-to-ato}, we find
 \begin{equation}
 \begin{split}
     \langle G^{\rm T.O.}_{\Delta}(X; P_1), \Phi(X) \rangle \left. \right|_{B} &= -2^{\Delta + 1} \pi^{3/2}\alpha^* \frac{\Gamma(\Delta - \frac{1}{2})}{\Gamma(\Delta)} \int_B d^3P G_{\Delta}^{\rm T.O.}(P_1; P) \widetilde{\mathcal{O}}^-_{3 - \Delta}(P) \\
     &= -2^{\Delta + 1} \pi^{3/2} \frac{\Gamma(\Delta - \frac{1}{2})}{\Gamma(\Delta)}  \mathcal{O}^{\rm -}_{\Delta}(P_1),
     \end{split}
 \end{equation}
where we assumed that the shadow transform of the boundary-to-boundary propagator is given by a near-boundary limit of the Lorentzian shadow transform of the bulk-to-boundary propagator defined in Section \ref{sec:mode-exp-shadow}, and discarded the delta function contact term that appears, since $P_1$ was assumed to be in $B$. This assumption may need to be revisited in the future.
 
Similarly, 
\begin{equation}
 \begin{split}
     \langle G^{\rm A.T.O.}_{\Delta}(X; P_1), \Phi(X) \rangle\left. \right|_{B} &=  \frac{2^{\Delta + 1} \pi^{3/2} \alpha \Gamma(\Delta - \frac{1}{2})}{\Gamma(\Delta)} \int_B d^3P G_{\Delta}^{\rm A.T.O.}(P_1; P) \widetilde{\mathcal{O}}^{+}_{3 - \Delta}(P) \\
     &= 2^{\Delta + 1} \pi^{3/2} \frac{\Gamma(\Delta - \frac{1}{2})}{\Gamma(\Delta)}  \mathcal{O}^{+}_{\Delta}(P_1).
     \end{split}
 \end{equation}

 These representations of boundary operators in terms of boundary values of the bulk field can be substituted into the representation \eqref{eq:bulk-rec-intro} of the AdS field
\begin{equation}
    \begin{split}
     \Phi(X) &= \int_{B} d^3P \left( \alpha G_{3 - \Delta}^{\rm A.T.O}(X; P) \mathcal{O}_{\Delta}^{+}(P) + \alpha^* G_{3 - \Delta}^{\rm T.O}(X; P) \mathcal{O}_{\Delta}^{-}(P)\right)\\
     &= \mathcal{C}(\Delta) \langle G_{\Delta}^{\rm A. T.O.}(X; P'), \Phi(P') \rangle\left. \right|_{B} 
      - \mathcal{C}^*(\Delta) \langle G_{\Delta}^{\rm T.O.}(X; P'), \Phi(P') \rangle\left. \right|_{B},
     \end{split}
\end{equation}
where we defined
 \begin{equation}
     \mathcal{C}(\Delta) = \frac{\Gamma(\Delta)}{2^{\Delta + 1} \pi^{3/2}\Gamma(\Delta - \frac{1}{2})}.
 \end{equation}
The same procedure can be performed for the representation of the free field in the complementary region, using the identity
\begin{equation}
\langle G^{\rm T.O.}_\Delta(X;P_1),\Phi(X)\rangle_{\bar{B}} =-\langle G^{\rm A.T.O.}_\Delta(X;P_1),\Phi(X)\rangle_{B}.    
\end{equation}
This identity follows from \eqref{eq:boundary-ip} and \eqref{eq:bdr-to-ato}, recalling that the leading contributions to the boundary inner products of bulk-to-boundary propagators come from the delta function in \eqref{eq:boundary-exp}. 

 We finally can evaluate this formula in a coordinate system of our choice. For comparison to the flat space results in Appendix \ref{app:bulk-rec}, it will be convenient to work in the retarded coordinates \eqref{eq:AdS-retarded}. 
In these coordinates, taking $B$ to be a surface at $r_p = r_0$,~$r_0 \rightarrow \infty$,
\begin{equation}
     d\Sigma^\mu\partial_\mu= d\tau \,d\Omega r_p^2\left(-\partial_\tau +\frac{\ell^2+r_p^2}{\ell}\partial_{r_p}\right).
 \end{equation}
 Up to boundary terms, we find 
 \begin{equation}
 \label{eq:phi-G}
 \begin{split}
     \Phi(\tau, r, \Omega) &= 2i \int_{B} d\tau_p d\Omega_p ( \mathcal{C}(\Delta)G_{\Delta}^{\rm T.O.}(\tau, r, \Omega; \tau_p, \Omega_p) {r_p^2\left(-\partial_{\tau_p}+\frac{\ell^2+r_p^2}{\ell}\partial_{r_p}\right)} \Phi(\tau_p, r_p, \Omega_p)\left. \right|_{r_p = r_0} \\
     &- \mathcal{C}^*(\Delta) G_{\Delta}^{\rm A. T.O.}(\tau, r, \Omega; \tau_p, \Omega_p) r_p^2\left(-\partial_{\tau_p}+\frac{\ell^2+r_p^2}{\ell}\partial_{r_p}\right) \Phi(\tau_p,r_p, \Omega_p)  )\left. \right|_{r_p = r_0}\\
     &= 2i  \int_{B} d\tau_p d\Omega_p r_p^2 \left( \mathcal{C}(\Delta) \frac{\left(-\partial_{\tau_p}+\frac{\ell^2+r_p^2}{\ell}\partial_{r_p}\right) \Phi(\tau_p, r_p, \Omega_p)}{(r\cos(\tau - \tau_p)  + \ell \sin(\tau - \tau_p) - r \Omega \cdot \Omega_p + i\epsilon)^{\Delta}}  -  h.c.\right) \Bigg|_{r_p = r_0},
     \end{split}
 \end{equation}
 where in the last line we used \eqref{eq:G-null}. 

We next assume that $\Delta \in \mathbb{N}$  and write
 \begin{equation}
    G^{\rm T.O.}_{\Delta}(X; P) = \frac{1}{\Gamma(\Delta)}\left( \frac{1}{r} \Omega_p \cdot \p_{\Omega} \right)^{\Delta - 1} \frac{1}{(r\cos(\tau - \tau_p)  + \ell \sin(\tau - \tau_p) - r \Omega \cdot \Omega_p + i\epsilon)}.
 \end{equation}
 Substituting this into \eqref{eq:phi-G} and using the delta function identity
 \begin{equation}
 \begin{split}
     &\lim_{\epsilon \rightarrow 0}\left[\frac{1}{r \cos(\tau - \tau_p) + \ell \sin(\tau - \tau_p) - r\Omega \cdot \Omega_p + i\epsilon} -  \frac{1}{r \cos(\tau - \tau_p) + \ell \sin(\tau - \tau_p) - r\Omega \cdot \Omega_p - i\epsilon}\right]\\
     &= -2\pi i \delta(r \cos(\tau - \tau_p) + \ell \sin(\tau - \tau_p) - r\Omega \cdot \Omega_p ),
     \end{split}
 \end{equation}
 we find
 \begin{equation}
 \label{eq:phi}
     \begin{split}
         \Phi(\tau, r, \Omega) = 4\pi \frac{\mathcal{C}(\Delta) }{\Gamma(\Delta)}\int  d\Omega_p r_p^2 \left( \frac{1}{r} \Omega_p \cdot \p_{\Omega} \right)^{\Delta - 1} \int_B d\tau_p &\delta(r \cos(\tau - \tau_p) + \ell \sin(\tau - \tau_p) - r\Omega \cdot \Omega_p )\\
         &\times \left(-\partial_{\tau_p}+\frac{\ell^2+r_p^2}{\ell}\partial_{r_p}\right)\Phi(\tau_p,r_p,\Omega_p)\left. \right|_{r_p = r_0}.
     \end{split}
 \end{equation}
 The $\tau_p$ integral can now be evaluated by solving for $\tau_p$ in terms of $\tau, r, \Omega, \Omega_p$, namely
 \begin{equation}
     f(\tau_p) \equiv \frac{1}{2}(r-i\ell) e^{i(\tau - \tau_p)} +  \frac{1}{2}(r+i\ell) e^{-i(\tau - \tau_p)} - r \Omega \cdot \Omega_p = 0.
 \end{equation}
 The solutions to this equation are 
 \begin{equation}
 \label{eq:sol-quad}
     e^{i(\tau - \tau_p)} = \frac{i(\Omega\cdot \Omega_p r \pm \sqrt{r^2(\Omega\cdot \Omega_p)^2 - r^2 - \ell^2})}{\ell + ir} \implies \tau_p \equiv \tau_{\pm}.
 \end{equation}
In region $B$, $\sin(\tau - \tau_p) < 0$ and only one of the solutions is picked out by the $\tau_p$ integral. The bulk field is then determined in terms of its boundary value at $\tau_p = \tau_-$, 
 \begin{equation}
 \label{eq:KA-AdS}
 \begin{split}
      \Phi(\tau, r, \Omega) &=4\pi \frac{\mathcal{C}(\Delta) }{\Gamma(\Delta)}  \int  d\Omega_p \left( \frac{1}{r} \Omega_p \cdot \p_{\Omega} \right)^{\Delta - 1} \frac{1}{f'(\tau_i)}  r_p^2\left(-\partial_{\tau_p}+\frac{\ell^2+r_p^2}{\ell}\partial_{r_p}\right)\Phi(\tau_p, r_p, \Omega_p) \left. \right|_{\tau_p = \tau_-, r_p = r_0}. 
      \end{split}
 \end{equation}
 This is the AdS analog of the Kirchhoff--d'Adh\'emar formula in Minkowski space. 
 
We conclude this section by showing that this formula simplifies  in the flat space limit. This is given by $\tau = u/\ell$ and $\ell \rightarrow \infty$ keeping everything else fixed. This amounts to zooming near the lightcone through the origin. In this region, we expect AdS to be well approximated by Minkowski space. In this case, \eqref{eq:sol-quad} becomes
 \begin{equation}
     e^{i(u - u_p)/\ell} = i\ell^{-1} (\Omega \cdot \Omega_p \pm 1) r \mp 1 + \mathcal{O}(r^2). 
 \end{equation}
 This implies that 
 \begin{equation}
     \cos\left(\frac{u - u_p}{\ell}\right) = \mp 1 + \mathcal{O}(r^2), \quad \sin\left(\frac{u - u_p}{\ell}\right) = \frac{r}{\ell}(\Omega\cdot \Omega_p \pm 1).
 \end{equation}
 We see that only one of the solutions is allowed in the limit, in which case 
 \begin{equation}
     u_p = u - r (\Omega\cdot \Omega_p - 1).
 \end{equation}
 Consequently, \eqref{eq:phi} becomes
 \begin{equation}
 \begin{split}
     \Phi(u, r, \Omega) &= 4\pi \frac{\mathcal{C}(\Delta) }{\Gamma(\Delta)} \int d\Omega_p r_p^2  \left( \frac{1}{r} \Omega \cdot \p_{\Omega_p} \right)^{\Delta - 1} (-\p_{u_p} + 2 \p_{r_p}) \Phi\left(\frac{u_p}{\ell},r_p,\Omega_p\right)\left. \right|_{u_p = u - r(\Omega\cdot \Omega_p-1), r_p = r_0}\\
     &= 4\pi (-1)^{\Delta-1}\frac{\mathcal{C}(\Delta) }{\Gamma(\Delta)} \int d\Omega_p r_p^2 \p_{u_p}^{\Delta-1}  (-\p_{u_p} + 2 \p_{r_p}) \Phi\left(\frac{u_p}{\ell},r_p, \Omega_p\right)\left. \right|_{u_p = u - r(\Omega\cdot \Omega_p-1), r_p = r_0}.
     \end{split}
 \end{equation}
 The coefficient of the second term depends on whether $r_0/\ell$ is kept fixed or $r_0/\ell \rightarrow 0$ as $\ell \rightarrow \infty$. We have assumed the former, but either way, at large $r_0$, the second term is suppressed compared to the first, since $\Phi$ has normalizable asymptotics.  Further setting $\Delta = 1$, we find 
 \begin{equation}
     \Phi(u,r, \Omega) = -\frac{\ell}{\pi} \int d\Omega_p  \p_{u_p} \phi\left(\frac{u_p}{\ell}, \Omega_p\right)\Big|_{u_p = u - r(\Omega\cdot \Omega_p-1)},
 \end{equation}
 where we defined the boundary value of $\Phi$
 \begin{equation}
 \label{eq:bdry-operator}
     \phi(u_p, \Omega_p) = \lim_{r_{p} \rightarrow \infty} r_p \Phi(u_p, r_p, \Omega_p).
 \end{equation}

 This agrees up to a factor of $2$ with the Kirchhoff--Adh\'emar formula \eqref{eq:KA-Mink} in Mink$_4$. Note that from the perspective of the AdS extrapolate dictionary, \eqref{eq:bdry-operator} corresponds to a CFT$_3$ primary of dimension $\Delta = 1$, while from the perspective of the flat space dictionary, \eqref{eq:bdry-operator} defines an operator in a Carrollian field theory \cite{Bagchi:2025vri}. We will elaborate on this relation in Section \ref{sec:flat-space-lim}.

\section{$\mathfrak{so}(3,1)$ wavefunctions in AdS$_4$}
\label{sec:boost-estates}

In Section \ref{sec:bulk-to-boundary-basis} we have shown that bulk-to-boundary propagators can be expressed as linear combinations of the positive and negative energy wavefunctions that span the solution set of the KG equation in AdS$_4$. This implies that they are eigenstates of the $\mathfrak{so}(3,2)$ quadratic Casimir. However since they involve a sum over all radial modes, they are clearly neither lowest-weight states nor eigenstates of the dilatation generators. As we will point out below, they are instead eigenstates of the local AdS boost generators. Related observations appeared recently in \cite{Berenstein:2025tts}. 

The $\mathfrak{so}(3,2)$ lowest-weight wavefunction has lowest energy $\omega = \Delta$ (namely $n = \ell = 0$) and is defined by \cite{Kaplan1}
\be 
\label{eq:hw}
\begin{split}
K_i \left[ e^{-i\Delta \tau} \psi(\rho) \right] &= -i e^{i\tau} \left[ \Omega_i\left( \cos \rho \p_{\rho} + i\sin \rho \p_{\tau} \right) \right]\left[ e^{-i\Delta \tau} \psi(\rho) \right] = 0 \\
& \implies \left(\cos \rho \p_{\rho} + \Delta \sin \rho \right)\psi(\rho) = 0 \iff \psi(\rho) \propto (\cos \rho)^{\Delta},
\end{split}
\ee
where $K_i$ are given in \eqref{eq:AdS-Gens}.
This solution can be found more simply using the embedding space formalism by looking for eigenstates of  $D = -M_{04}$ that are annhilated by $K_i = X_i (\p_0 + i \p_4)$. These are easily seen to take the form
\begin{equation}
\label{eq:wf-Deltaestate}
\begin{split}
     \frac{1}{(X^0 + i X^4)^{\Delta}} = e^{-i\Delta \tau} (\cos\rho)^{\Delta} ,
     \end{split}
\end{equation}
in agreement with \eqref{eq:hw}. Equation \eqref{eq:wf-Deltaestate} automatically solves the wave equation, since it can be written as $(-P_0\cdot X)^{-\Delta}$ with $P_0 = (1,0,0,0,i)$ and $P_0^2 = 0$. A set of wavefunctions that span the positive energy irreducible representation of $\mathfrak{so}(3,2)$ is then $(-P_C\cdot X)^{-\Delta}$, where $P_C$ are null vectors in the embedding space, obtained from $P_0$ by an SO$(3,2)$ transformation. Since they are complex, $P_C$ do \textit{not} correspond to points on the AdS$_4$ boundary.

We can now construct counterparts of \ref{eq:wf-Deltaestate} that diagonalize AdS boosts. In AdS$_4$ these are generated by $K_i + P_i \propto M_{i0}$. Each of these generators is part of an $\mathfrak{sl}(2,\mathbb{R})$ subalgebra of $\mathfrak{so}(3,2)$ containing $D$. We can choose the Cartan subalgebra of $\mathfrak{so}(3,2)$ to be generated by $\{M_{30}, M_{12}\}$.\footnote{This Cartan subalgebra is related to the standard choice involving $D$ by a \textit{complex} inner automorphism. } A lowest weight state for this choice of Cartan subalgebra is given by the wavefunction
\begin{equation}
\label{eq:Boost-wf}
    \Psi_{\Delta}^0 \propto \frac{1}{(X^3 + X^0)^{\Delta}} .
\end{equation}
This solves the KG equation with $m^2 = \Delta (\Delta - 3)$, but in contrast to \eqref{eq:wf-Deltaestate}, it diagonalizes $M_{30}$ with eigenvalue $\Delta$. 
One can explicitly check that this wavefunction is annihilated by all $\mathfrak{so}(3,2)$ lowering operators. The resulting highest weight representation is then 
\begin{equation}
\label{eq:bulk-boundary}
    \Psi_{\Delta}(P; X) \propto \frac{1}{(-P\cdot X)^{\Delta}},
\end{equation}
where $P$ are embedding space null vectors that lie in the SO$(3,2)$ orbit of $P_0 = (-1,0,0,1,0)$. $P$ are now real and correspond to points on the AdS boundary and hence \eqref{eq:bulk-boundary} are bulk-to-boundary propagators. The precise relation between these representations and the positive/negative energy highest weight representations spanned by $\psi^+_{n\ell m}(X)$ was explicitly worked out in Sections \ref{sec:bulk-to-boundary-basis}-\ref{sec:bulk-rec-AdS}. 

The fact that bulk-to-boundary propagators are boost eigenstates in AdS is consistent with the observation in \cite{deGioia:2022fcn} that they become Carrollian wavefunctions in a flat space limit with fixed $\Delta$ \cite{deGioia:2022fcn}. Indeed, in the flat space limit, $M_{i0}$ become boosts in flat space. 
Specifically, taking \cite{Susskind:1998vk} $\tau = t \ell^{-1}, \rho = r \ell^{-1}$ with $\ell \rightarrow \infty$ and fixed $t, r$, the AdS$_4$ generators in \eqref{eq:AdS-Gens} become
\be
\label{eq:global-flat-space-gen}
\begin{split}
D &= i\ell \frac{\p}{\p t}, \quad M_{ij} = -i \left(\Omega_i \p_{\Omega_j} - \Omega_j \p_{\Omega_i} \right), \\
P_i &= i (1 - it \ell^{-1})\left[ \Omega_i\left( \ell \p_{r} - i r \p_{t} \right)  + \frac{\ell}{r} \nabla_i \right], \\
K_i &= -i(1+ it \ell^{-1})\left[ \Omega_i \left(\ell \p_{r} + i r\p_{t} \right) + \frac{\ell}{r} \nabla_i  \right].
\end{split}
\ee
Equivalently
\be 
\begin{split}
\frac{P_i + K_i}{2} &= \Omega_i\left( t\p_r + r \p_t \right) + \frac{t}{r} \nabla_i,\\
\frac{P_i - K_i}{2} &= \frac{\ell}{r} \nabla_i + i \Omega_i \ell \p_r.
\end{split}
\ee
which we recognize as boosts and translation generators in Mink$_4$.\footnote{Our combinations giving rise to boosts and translations are the opposite to those found in \cite{Berenstein:2025tts}. This follows from our different convention for $K$, namely $K_{\rm here} = - K_{\rm there}$.
}

The $M_{i0}$ generators together with the rotations $M_{ij}$ generate an $\mathfrak{so}(3,1)$ subalgebra of $\mathfrak{so}(3,2)$. The goal of this section is to construct the wavefunctions spanning irreducible representations of this subalgebra and therefore decompose  an $\mathfrak{so}(3,2)$ lowest weight representation into representations of $\mathfrak{so}(3,1)$. We will show in Section \ref{sec:flat-space-lim} that this construction parallels the decomposition of 4d Poincar\'e representations into representations of $\mathfrak{so}(3,1)$.

\subsection{Hyperbolic foliation}
\label{sec:hyper-fol}

The orbits of the Lorentz SO$(3,1)$ subgroup of SO$(3,2)$ that fixes the origin are hyperbolic AdS$_3$, dS$_3$ slices and the null cone through the origin. One way to identify the decomposition of $\mathfrak{so}(3,2)$ into representations of $\mathfrak{so}(3,1)$ is to solve the wave equation in a foliation of AdS$_4$ with (A)dS$_3$ and null slices. Reorganizing the solution to the wave equation in terms of solutions on the leaves will lead to the sought-after decomposition.

\subsubsection{$\mathfrak{so}(3,1)$ covariant basis in AdS$_4$}
\label{sec:boost-wf-hyp}

\begin{figure}
    \centering
    \begin{subfigure}[t]{0.45\linewidth}
    \includegraphics[width=\linewidth]{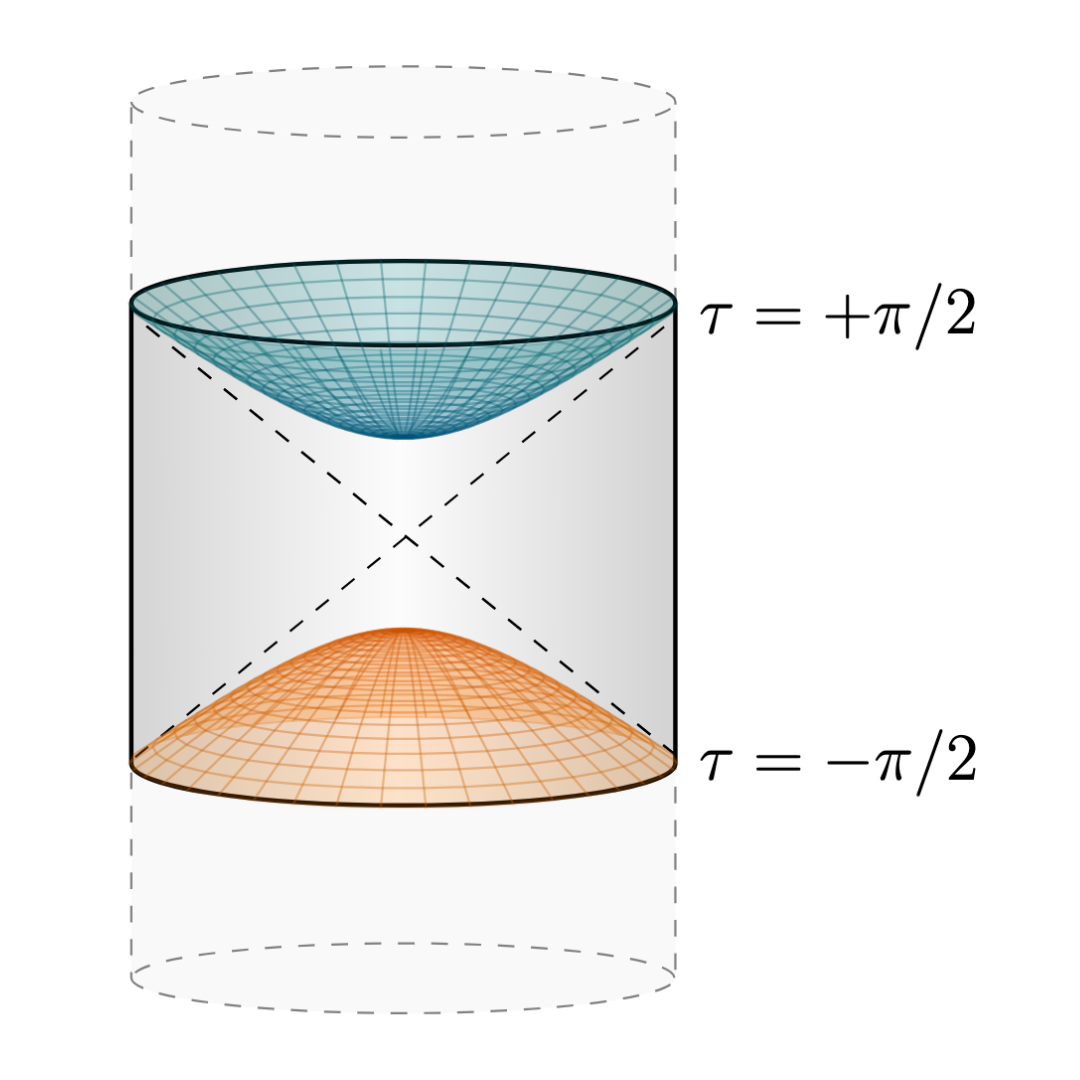}
    \subcaption{Euclidean AdS$_3$ slices foliating the future and past of the lightcone through the origin in Lorentzian AdS$_4$.}
    \label{fig:AdS-slices}
    \end{subfigure}\hfill
    \begin{subfigure}[t]{0.45\linewidth}
\includegraphics[width=\linewidth]{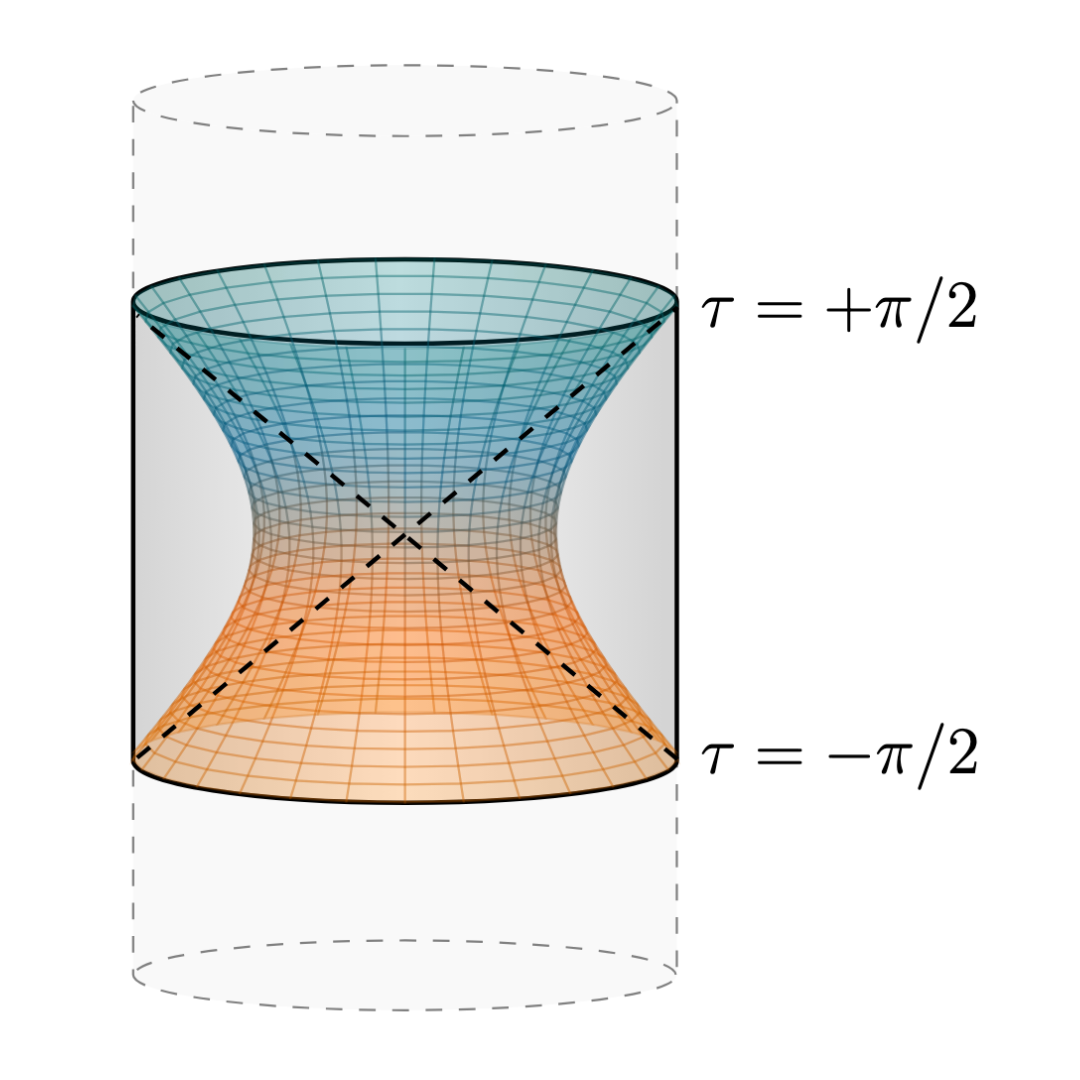}
    \subcaption{dS$_3$ surface foliating the exterior of the lightcone through the origin in Lorentzian AdS$_4$.}
    \label{fig:dS-slices}
    \end{subfigure}
    \caption{Foliations of Lorentzian AdS$_4$.}
    \label{fig:foliations}
\end{figure}

We start with the embedding \eqref{eq:AdS-emb} of AdS$_{4}$ as a hyperboloid in $\mathbb{R}^{2,3}$. 
Paralleling the construction of massive conformal primary wavefunctions in Mink$_{4}$ \cite{Pasterski:2017kqt}, we can foliate the future (+) and past (--) of the lightcone through the origin in AdS$_{4}$ with Euclidean AdS$_{3}$  (EAdS$_3$) slices of radii $\alpha$ using the following embedding
\begin{equation}
\label{eq:AdSd-emb}
     -(X^0)^2 + \sum_{i = 1}^3 (X^i)^2 = -\alpha^2, \quad X^{4} = \pm \sqrt{\ell^2 - \alpha^2}.
\end{equation}
It is straightforward to show that the AdS$_{4}$ metric then takes the form
\begin{equation}
\label{eq:massive-AdS-fol}
    ds^2_{4} = -\ell^2 \frac{d\alpha^2}{\ell^2 - \alpha^2} + \alpha^2 ds_{3}^2, 
\end{equation}
where $ds_{3}^2$ is the EAdS$_3$ metric.  In the limit $\ell^2 \rightarrow \infty$, the foliation \eqref{eq:massive-AdS-fol} reduces the one in the Milne regions of flat space \cite{Raclariu:2021zjz}.  The lightcone through the origin corresponds to $\alpha = 0$. Unlike in flat space, we encounter coordinate singularities as $\alpha = \pm \ell$. We illustrate this foliation in Figure \ref{fig:AdS-slices}.

Later on, we will relate our analysis to previous works in the flat space context, for which it will be useful to introduce Poincar\'e coordinates on the AdS$_3$ leaves of radius $\alpha$
\begin{equation}
\label{eq:Milne-AdS-flat}
\begin{split}
    X^0 &= \alpha \frac{1 + y^2 + z\bz}{2y}, \\
    X^1 &= \alpha \frac{z + \bz}{2y}, \quad X^2 = -i \alpha \frac{z - \bz}{2y}, \quad X^3 =  \alpha \frac{1 - y^2 - z \bz}{2y}.
    \end{split}
\end{equation}
In this case
\begin{equation}
    ds_3 = \frac{1}{y^2} \left(dy^2 + dz d\bz \right).
\end{equation}
 The region defined by the points spacelike separated from the origin can be foliated with dS$_{3}$ slices and is obtained by analytically continuing $\alpha \rightarrow i\alpha, ~~ y \rightarrow iy$. We illustrate these foliations in Figure \ref{fig:dS-slices}.  We will collectively denote the AdS$_3$ coordinates by $\hat{x}$.
 For simplicity, we set $\ell = 1$. Factors of $\ell$ can be restored in all formulas below by replacing $\alpha \rightarrow \alpha/\ell$ and rescaling the metric by $\ell^2$. The generalization to all $d$ is straightforward.

We now construct the AdS$_{4}$ analogs of massive conformal primary wavefunctions in Mink$_4$. These are defined by the condition that they transform in an irreducible representation of the Lorentz $\mathfrak{so}(3,1) \subset \mathfrak{so}(3,2)$. Recall that the Klein--Gordon equation in AdS$_{4}$ is an eigenvalue equation for the quadratic Casimir of $\mathfrak{so}(3,2)$. In the coordinates \eqref{eq:AdSd-emb}, the AdS$_{4}$ Laplacian decomposes as
\begin{equation}
    \Box_{AdS_{4}}  = \alpha^{-2} \Box_{AdS_{3}} + \alpha^{-1}\left((4\alpha^2 - 3) \p_{\alpha} + \alpha (\alpha^2 -1) \p_{\alpha}^2 \right).
\end{equation}
As a result, we can express the solutions to the KG equation in  AdS$_{4}$ as
\begin{equation}
\label{eq:AdS4-cpw}
    \Psi_{\Delta}(\alpha,\hat{x};\delta,\hat{q}) = f_{\Delta,\delta}(\alpha) G_{AdS_{3}}^{\delta}(\hat{x};\hat{q}),
\end{equation}
 where $f_{\Delta,\delta}(\alpha)$ satisfies the equation
\begin{equation}
\label{eq:massive-f}
  \left[ \alpha^{-2} \delta(\delta - 2)  + \alpha^{-1}\left((4\alpha^2 - 3) \p_{\alpha} + \alpha (\alpha^2 -1) \p_{\alpha}^2 \right) \right] f_{\Delta,\delta}(\alpha) = \Delta(\Delta - 3) f_{\Delta,\delta}(\alpha),
\end{equation}
while $G_{AdS_{3}}^{\delta}$ are bulk-to-boundary propagators on the AdS$_3$ leaves, namely
\begin{equation}
\label{eq:AdS3}
    \Box_{AdS_3} G^{\delta}_{AdS_{3}}(\hat{x};\hat{q}) = \delta(\delta - 2) G^{\delta}_{AdS_{3}}(\hat{x};\hat{q}).
\end{equation}
Here $\hat{q}$ is a vector in the $\mathbb{R}^{3,1}$ embedding space of AdS$_3$ satisfying $\hat{q}^2 = 0$, that labels points on the AdS$_3$ boundary.  The solutions to \eqref{eq:AdS3} can be shown to coincide with Fourier transforms of massive conformal primary wavefunctions in flat space \cite{Pasterski:2017kqt,Iacobacci:2022yjo,Sleight:2023ojm}. Just like in flat space, $\delta$ are not fixed. As we will see, in order to construct an irreducible representation of $\mathfrak{so}(3,2)$, we will have to sum/integrate over different values of $\delta$.  We next compute the functions $f_{\Delta,\delta}(\alpha)$. 

 Assuming $\Delta > \frac{3}{2}$ and $\delta >1$, the general solution to \eqref{eq:massive-f} is
\begin{equation}
\label{eq:fa}
\begin{split}
    f_{\Delta,\delta}(\alpha) &= c_2 i^{-\delta} \alpha^{-\delta} {}_2F_1\left[\frac{1}{2}(3 - \delta - \Delta), \frac{1}{2}(\Delta - \delta), 2 - \delta,\alpha^2 \right] \\
    &+ c_1 i^{\delta} \alpha^{\delta - 2} {}_2 F_1\left[\frac{1}{2}(1 + \delta - \Delta), \frac{1}{2}(\delta - 2 + \Delta), \delta, \alpha^2 \right].
    \end{split}
\end{equation}
These two independent solutions are related by a shadow transform $\delta \rightarrow 2 - \delta$ (as expected, since \eqref{eq:AdS3} coincides with the conformal Casimir equation of 2d CFT which is invariant under the shadow transform $\delta \rightarrow 2 - \delta$ \cite{Dolan:2000ut,Osborn:2012vt}). 

We would now like to find the conditions under which \eqref{eq:AdS4-cpw} form a complete set of solutions to the wave equation in AdS$_4$. 
We first observe that \eqref{eq:massive-f} can be put in Sturm-Liouville (SL) form 
\begin{equation}
\label{eq:SL}
  \mu^{-1}(\alpha) \left( - \frac{d}{d\alpha}\left[\mu(\alpha) P(\alpha) \frac{df_{\Delta,\delta}(\alpha)}{d\alpha}  \right] - R(\alpha)\mu(\alpha) f_{\Delta,\delta}(\alpha) \right) = 0,
\end{equation}
with
\begin{equation}
    P = \alpha^2 - 1, \quad Q = \frac{4\alpha^2 - 3}{\alpha} \quad R = -\Delta(\Delta - 3) +\alpha^{-2}\delta(\delta - 2), \quad \mu(\alpha) = \frac{1}{P} e^{\int \frac{Q}{P}d\alpha} = -\frac{\alpha^3}{ \sqrt{1 - \alpha^2}}.
\end{equation}
Equivalently, we have 
\begin{equation}
     \frac{d}{d\alpha}\left[\mu(\alpha) P(\alpha) \frac{df_{\Delta,\delta}(\alpha)}{d\alpha}  \right] -\Delta (\Delta - 3) \mu(\alpha) f_{\Delta,\delta}(\alpha) = \alpha^{-2} \mu(\alpha) \delta(2-\delta) f_{\Delta,\delta}(\alpha).
\end{equation}
The differential operator on the LHS of \eqref{eq:SL} is self-adjoint with respect to the inner product 
\begin{equation}
\label{eq:SLip}
    \langle f, g \rangle = \int_{0}^1 d\alpha \alpha^{-2} \mu(\alpha) f(\alpha) g(\alpha).
\end{equation}
and we restricted to $\alpha \in [0, 1)$ for which $\mu P > 0$ and $-\alpha^{-2} \mu(\alpha) > 0$ so that the SL problem is regular. Depending on the choice of boundary conditions imposed on \eqref{eq:fa}, we may get either a discrete or a continuous spectrum. A complete classification of allowed boundary conditions and resulting spectra is beyond the scope of this work. 

We instead will consider a particular choice of boundary conditions for which the wavefunctions \eqref{eq:AdS4-cpw} reduce to Mink$_4$ massive conformal primary wavefunctions in the flat space limit. As we will see, in this case the spectrum will be continuous with\footnote{Just as in \cite{Pasterski:2017kqt}, the $\lambda \in \mathbb{R}_-$ are obtained by a 2d shadow transform, and are therefore not linearly independent.} 
\begin{equation}
\label{eq:delta-ps}
    \delta = 1 + i\lambda, \quad \lambda \in \mathbb{R}_+ \implies \delta(2 - \delta) \geq 0.
\end{equation}
As a result, any massive solution to the scalar wave equation in AdS$_{4}$ admits a decomposition 
\begin{equation}
\label{eq:AdS-cpw-dec}
    \Phi(\alpha, \hat{x}) \supset \int_{\mathbb{R}} d\lambda \int d^{2} \hat{q} \Psi_{\Delta}(\alpha, \hat{x};\delta, \hat{q}) \mathcal{O}_{1 - i\lambda}(\hat{q})  = \int_{\mathbb{R}} d\lambda \int d^{2} \hat{q}  f_{\Delta,\delta}(\alpha)  G^{1 + i\lambda}_{AdS_3}(\hat{x};\hat{q}) \mathcal{O}_{1 - i\lambda}(\hat{q})
\end{equation}
in terms of unitary principal series representations of $\mathfrak{so}(3,1)$ and $f_{\Delta, \delta}(\alpha)$ given in \eqref{eq:fa} with specific non-vanishing $c_1$, $c_2$ that we will determine in the next section. The global wavefunction is obtained by summing over the contributions from all three regions -- see Figure \ref{fig:foliations}. 
Further imposing regularity at the origin, we expect to obtain a basis of positive and negative energy lowest/highest weight representations of $\mathfrak{so}(3,2)$ of $m^2 = \Delta (\Delta - 3)$. For concrete applications, it will be useful to explicitly evaluate the KG inner products  of AdS$_4$ bulk-to-boundary propagators and the wavefunctions \eqref{eq:AdS4-cpw}  which should yield the Clebsch--Gordan coefficients for the decomposition of $\mathfrak{so}(3,2)$ in terms of $\mathfrak{so}(3,1)$ representations. We leave this to future work. 

For later reference, we conclude this section by observing that we can also write
\begin{equation}
\label{eq:massive-boost}
    \Psi_{\Delta}( \alpha, \hat{x};\delta, \hat{q}) \propto \int_{EAdS_3} [dx'] G_{\delta}(\hat{x}';\hat{q}) B_{\Delta}(\hat{x}'; \alpha, \hat{x}),
\end{equation}
where $B_{\Delta}(\hat{x}';\alpha, \hat{x}) \propto (-\alpha \hat{x} \cdot \hat{x}')^{-\Delta}$ is a function of the AdS$_4$ invariant distance with one point restricted to the EAdS$_3$ hyperboloid of radius $\alpha = \ell$.\footnote{Note that $B_{\Delta}(\hat{x}'; \alpha, \hat{x})$ is \textit{not} the bulk-to-bulk propagator in AdS, so one may wonder why the Ansatz \eqref{eq:massive-boost} should be a solution to the wave equation in AdS$_4$. One can show that, despite $B_{\Delta}$ itself not being a solution, the integral in \eqref{eq:massive-boost} \textit{is}. To see this, we note that the violation of $B_{\Delta}$ being a solution is $\propto \frac{1}{(\hat{x}'\cdot \hat{x})^{\Delta + 2}}$ use the identity $\frac{1}{(\hat{x}\cdot \hat{x}')^{\Delta + 2}} \propto \p_{\hat{x}}^2 \frac{1}{(\hat{x}'\cdot \hat{x})^{\Delta}}\propto \frac{1}{\alpha^2} \p_{\hat{x}'}^2 \frac{1}{(\hat{x}'\cdot \hat{x})^{\Delta}} $, where $\p_{\hat{x}}^2$ is the ($\mathbb{R}^{1,3}$ embedding space version of the) Laplacian in AdS$_3$ and integrate by parts. This term then vanishes since $\hat{q}^2 = 0$. \label{footn:massive}} To show this, we multiply both sides by $B_{\Delta'}(\hat{x}_p;\alpha,\hat{x})$ and integrate over the future patch of AdS$_4$. Using orthogonality of $B_{\Delta}(\hat{x}';\alpha,\hat{x})$ and $B_{\Delta'}(\hat{x}_p;\alpha,\hat{x})$ for $\Delta' = 3 - \Delta$ \cite{Costa:2014kfa}, we expect this integral to be proportional to $G_{\delta}(\hat{x}_p;\hat{q})$.  We show this is indeed the case and evaluate the proportionality constant in Appendix \ref{app:so32-so31}. Similar integrals were evaluated in Minkowski space \cite{Iacobacci:2022yjo, Sleight:2023ojm}, where plane waves associated with massive particles play the same role as the bulk-to-bulk propagator here. This formula will be useful when constructing boost eigenstates in the null coordinates below.  \eqref{eq:massive-boost} is the $\mathfrak{so}(3,2)$ analog of the definition of massive conformal primary wavefunctions in Mink$_4$, while $\mathcal{O}_{1 - i\lambda}$ in \eqref{eq:AdS-cpw-dec} are analogous to the celestial CFT operators corresponding to massive particle states \cite{Pasterski:2016qvg}. 
 This follows by considering the flat space limits of the generators \eqref{eq:so32-hyp}, which we compute explicitly in Section \ref{sec:sym-hyp}. In Section \ref{sec:flat-lim-hyp}, we show that \eqref{eq:AdS4-cpw} become precisely the massive conformal primary wavefunctions constructed in \cite{Pasterski:2017kqt}.

\subsubsection{Symmetries}
\label{sec:sym-hyp}

By construction, the wavefunctions \eqref{eq:AdS4-cpw} transform in the principal series representation of $\mathfrak{so}(3,1)$. One can verify this explicitly by writing the $\mathfrak{so}(3,2)$ generators in the coordinates \eqref{eq:AdSd-emb}, \eqref{eq:Milne-AdS-flat} 
\begin{equation}
\label{eq:so32-hyp}
    \begin{split}
        M_{04} &= - i \sqrt{\ell^2-\alpha^2}\frac{ y \left(\left(\bz z-y^2+1\right) \p_{y} -2 y \left(\bz \p_{\bz}+z \p_z \right)\right)+\alpha  \left(\bz z+y^2+1\right)
   \p_{\alpha} }{2 \alpha  y}, \\
         M_{41} &= - i \sqrt{\ell^2 - \alpha^2}\frac{ \left(y \left((\bz +z) \p_y -2 y \left(\p_{\bz} + \p_z \right)\right)+\alpha  (\bz +z) \p_{\alpha}\right)}{2
   \alpha  y},\\
         M_{42} &= \sqrt{\ell^2 - \alpha^2}\frac{ \left(2 y^2 \p_{\bz} -2 y^2 \p_z +(\bz-z)
   \left(y \p_y +\alpha \p_{\alpha}\right)\right)}{2 \alpha  y},\\
         M_{43} &= i \sqrt{\ell^2 - \alpha^2}\frac{ \left(y \left(\left(\bz z-y^2-1\right) \p_y -2 y \left(\bz 
  \p_{\bz} +z \p_z \right)\right)+\alpha  \left(\bz  z+y^2-1\right)
   \p_{\alpha}\right)}{2 \alpha  y},\\
   \end{split}
   \end{equation}
   and
   \begin{equation}
   \label{eq:so31}
   \begin{split}
         M_{01} &= -\frac{1}{2} i \left(\left(\bz^2-y^2-1\right) \p_{\bz} +\left(z^2-y^2 -1\right) \p_z +y (\bz +z) \p_y \right),\\
         M_{02} &= \frac{1}{2} \left(\left(\bz^2+y^2+1\right) \p_{\bz}-\left(y^2+z^2+1\right) \p_z +y (\bz -z) \p_y \right),\\
         M_{03} &= -i \left(\bz\p_{\bz}+z \p_z +y \p_y \right),\\
         M_{12} &=  z \p_z -\bz \p_{\bz},\\
         M_{13} &= \frac{1}{2} i \left(\left(\bz^2-y^2+1\right)\p_{\bz} +\left(+z^2-y^2+1\right) \p_z +y (\bz +z) \p_y \right), \\
         M_{23} &=\frac{1}{2} \left(-\left(\bz^2+y^2-1\right) \p_{\bz}+\left(y^2+z^2-1\right) \p_z +y (z-\bz) \p_y \right).
    \end{split}
\end{equation}
Note that $M_{ij}$ and $M_{0i}$, $i = 1,2,3$ generate an $\mathfrak{so}(3,1)$ subalgebra which preserves the AdS$_3$ leaves.  

Applying the same reasoning around \eqref{eq:bulk-boundary} to the AdS$_3$ bulk-to-boundary propagator, we conclude that \eqref{eq:AdS4-cpw} with $\hat{q} = (1,0,0,1)$ diagonalize $M_{03}$ with eigenvalue $\delta$. For every fixed $\hat{q}$, $G_{AdS_3}^{\delta}$ diagonalizes a boost towards $\hat{q}$. Fixing $\delta$ to be given by \eqref{eq:delta-ps}, $G_{AdS_3}^{\delta}$ can be shown to span a principal series representation of $\mathfrak{so}(3,1)$ by varying $\hat{q}$ \cite{Pasterski:2017kqt, Sun:2021thf}.

$M_{0i}$ and $M_{ij}$ coincide with the Lorentz generators of Mink$_4$ \cite{Narayanan:2020amh}. Furthermore, restoring factors of $\ell$ by letting $\alpha \rightarrow \alpha/\ell$ and taking the flat space limit $\ell \rightarrow \infty$, we also find
\begin{equation}
 \lim_{\ell \rightarrow \infty}   \ell \sqrt{\ell^2 - \alpha^2 \ell^{-2}}  = \ell^2,
\end{equation}
which implies that $M_{\mu 4}$ become translation generators in Mink$_4$ upon rescaling $M_{\mu 4} \rightarrow \ell^{-2} M_{\mu 4}$. 

\subsubsection{Boundary conditions and flat space limit}
\label{sec:flat-lim-hyp}

We now show that the AdS wavefunctions \eqref{eq:AdS4-cpw} reduce precisely to the massive conformal primary wavefunctions in Mink$_4$ in the flat space limit for a certain choice of $c_1, c_2$.  

Restoring factors of the AdS radius $\ell,$ and recalling that the conformal dimensions are related to the mass of the AdS$_4$ fields via
\begin{equation}
    m^2 \ell^2 = \Delta( \Delta -3),
\end{equation}
 we see that we obtain a massive field of fixed $m$ in the flat space limit $\ell \rightarrow \infty$ by setting $\Delta \sim \ell \rightarrow \infty$. In this limit, each of the hypergeometric functions in \eqref{eq:fa} becomes proportional to a Bessel $I$ function. To see this, we use the series decomposition of the hypergeometric function
\begin{equation}
\begin{split}
    {}_2F_1\left( \frac{1+ \delta - \Delta}{2}, \frac{\delta - 2 + \Delta}{2},  \delta, \frac{\alpha^2}{\ell^2} \right) &= \sum_{n = 0}^{\infty} \frac{\Gamma(\frac{\delta+1 -\Delta}{2} + n)\Gamma(\frac{\delta - 2 + \Delta}{2} + n)\Gamma( \delta)}{\Gamma( \delta +n)\Gamma( \frac{\delta+1 - \Delta}{2})\Gamma(\frac{\delta - 2 + \Delta}{2})} \frac{\alpha^{2n}}{n! \ell^{2n}}
    \end{split}
\end{equation}
whose large $\ell$ limit is given by
\begin{equation}
    \begin{split}
   \lim_{\ell \sim \Delta \rightarrow \infty} {}_2F_1\left(\frac{\delta +1 - \Delta}{2}, \frac{\delta - 2 + \Delta}{2}, \delta, \frac{\alpha^2}{\ell^2} \right) &= \sum_{n = 0}^{\infty} \frac{\Gamma(\delta)}{\Gamma( \delta + n)} \frac{(-\Delta)^n \Delta^n}{2^{2n}}\frac{\alpha^{2n}}{n! \ell^{2n}}\\
  &= \sum_{n = 0}^{\infty}  \frac{(-1)^n \Gamma( \delta)}{\Gamma(\delta + n)}\frac{\alpha^{2n}}{2^{2n} n!} \\
  &= \Gamma(\delta)\left(\frac{2}{i\alpha} \right)^{\delta-1}I_{\delta-1}(i\alpha),
    \end{split}
\end{equation}
where $I_{\delta-1}$ is the modified Bessel function of first kind and in the last line we used its series representation
\begin{equation}
    I_{\lambda}(x) = \sum_{m = 0}^{\infty} \frac{1}{m! \Gamma(m + \lambda + 1)}\left(\frac{x}{2} \right)^{2m + \lambda}.
\end{equation}
As a result, in the flat space limit \eqref{eq:fa} reduces to
\begin{equation}
\label{eq:c12}
    \lim_{\ell \sim \Delta \rightarrow \infty} f_{\Delta,\delta}(\alpha) = c_1  (i\alpha)^{-1} 2^{\delta - 1} \Gamma(\delta) I_{\delta - 1}(i\alpha) +  c_2 2^{1 - \delta} (i\alpha)^{-1} \Gamma(2 - \delta) I_{1 - \delta}(i\alpha).
\end{equation}

In this limit, we expect that \eqref{eq:fa} becomes the Bessel function of the second kind (see for instance eqns. 3.16 - 3.19 in \cite{Sleight:2023ojm}), namely\footnote{This is obtained by parameterizing the Mink$_4$ coordinate in eq. 2.19 of \cite{Pasterski:2017kqt} according to \eqref{eq:Milne-AdS-flat}.}
\begin{equation}
    \lim_{\ell \sim \Delta \rightarrow \infty} f_{\Delta,\delta}(\alpha) \propto \alpha^{-1} K_{\delta - 1}(i \alpha).
\end{equation}
Using the relation between the Bessel functions of first and second kind
\begin{equation}
    K_{\lambda}(x) = \frac{\pi}{2} \frac{I_{-\lambda}(x) - I_{\lambda}(x)}{\sin(\lambda \pi)},
\end{equation}
this requirement fixes the ratio of coefficients $c_1$ and $c_2$ in \eqref{eq:c12} to
\begin{equation}
    \frac{c_1 2^{\delta - 1} \Gamma(\delta)}{c_2 2^{1 - \delta} \Gamma(2 - \delta)} = -1.
\end{equation}
The conformal primary wavefunctions are  completely determined by further demanding that they are appropriately normalized. 
We explicitly see from a bulk perspective how the flat space limit in AdS$_4$ implements the well-known In\"on\"u--Wigner contraction of $\mathfrak{so}(3,2)$ to 4d Poincar\'e. 

\subsection{Null foliation}
\label{sec:null-cpw-ads}

The foliation of AdS$_4$ introduced in Section \eqref{sec:hyper-fol} becomes degenerate at $\alpha = 0$. In this section, we would like to generalize the construction of $\mathfrak{so}(3,1)$ covariant wavefunctions to include points on the null cone through the origin.

To this end, we introduce a foliation of AdS$_4$ with null cones. We start with the lightcone through the origin which we parameterize by 
\begin{equation}
\label{eq:null-coords}
    X^0 = r,\quad X^i = r \Omega^i, \quad X^{4} = \ell.
\end{equation}
The foliation is defined by translating this cone with respect to global time. Such translations are implemented by a rotation in the $(X^0, X^{4})$ plane, namely
\begin{equation}
    \left(\begin{matrix}
        X^0\\
        X^{4}
    \end{matrix} \right) \rightarrow \left( \begin{matrix}
        \cos \tau & \sin \tau \\
        -\sin\tau & \cos \tau
    \end{matrix} \right)  \left(\begin{matrix}
        X^0\\
        X^{4}
    \end{matrix} \right).
\end{equation}
The result is
\begin{equation}
\label{eq:null-foliation}
    X^0 = \cos \tau r + \sin\tau \ell,\quad X^i = r \Omega^i, \quad X^{4} = -\sin\tau r + \cos\tau \ell.
\end{equation}
In this parameterization, the AdS$_{4}$ metric takes the form
\begin{equation}
\label{eq:AdS-retarded}
    ds_{AdS_{4}}^2 = -(\ell^2 + r^2) d\tau^2 - 2\ell d\tau dr + r^2 d\Omega_{2}^2.
\end{equation}
where $ r\in [0,\infty)$ is a radial parameter along the lightcones, $\tau$ (not to be confused with the AdS global time) plays the role of a retarded time and $d\Omega_2^2$ is the metric on $S^2$. Indeed, \eqref{eq:AdS-retarded} is nothing but the AdS$_{4}$ metric in retarded Bondi gauge \cite{Compere:2019bua}. Unlike \eqref{eq:Milne-AdS-flat} the Bondi coordinates are regular for all points away from $r = 0$, including points on the null cone through the origin. 

\subsubsection{$\mathfrak{so}(3,1)$ covariant basis in AdS$_4$}

As a first attempt to constructing boost eigenfunctions in the \eqref{eq:AdS-retarded} coordinates, we consider the null limit of the massive AdS conformal primary wavefunctions \eqref{eq:massive-boost}.  This is achieved by taking $y,~ \alpha \rightarrow 0$ for fixed
\begin{equation}
    r = \frac{\alpha}{y}. 
\end{equation}
One can check that in this limit, the metric \eqref{eq:massive-AdS-fol} becomes the  AdS metric in retarded Poincar\'e coordinates. 
Using the small $y$ expansion of
\begin{equation}
\label{eq:prop-null}
   \frac{1}{(-X\cdot X' + i\epsilon)^{\Delta}} = \frac{1}{\left( (r r' + \ell^2) \cos \tau + \ell(r' - r) \sin \tau - r r' \Omega \cdot \Omega' + i\epsilon \right)^{\Delta}}
\end{equation}
 (see eq. 4.2 of \cite{Pasterski:2017kqt})
we find\footnote{We write the contribution from the delta function term in the near boundary expansion of the AdS$_3$ propagator. The extra term arises from its shadow.}
\begin{equation}
\label{eq:limit-hyperbolic}
\begin{split}
    \lim_{\alpha, y \rightarrow 0} \Psi_{\Delta}(\delta, \hat{q}(z); \tau, r, w) &\propto \alpha^{-\delta} \int_0^{\infty} dr'  \frac{r'^{\delta - 1}}{(r' ( r \cos \tau + \ell \sin \tau - r \Omega \cdot \Omega_q) + \ell^2 \cos \tau - \ell r \sin \tau + i\epsilon)^{\Delta}} \\
    &+ \mathfrak{so}(3,1) ~~ {\rm shadow}.
    \end{split}
\end{equation}
Such integrals can be readily evaluated (see for instance \cite{deGioia:2024yne}). Upon renormalizing the divergent $\alpha^{-\delta}$ factor, we find 
\begin{equation}
\label{eq:psinull}
\begin{split}
    \Psi^{\rm null}_{\Delta}(\delta, \hat{q}; \tau, r, \Omega) &\propto \frac{1}{(r \cos \tau + \ell \sin \tau - r \Omega \cdot \Omega_q)^{\delta}} \int_0^{\infty} dr' \frac{r'^{\delta - 1}}{(r' + \ell^2 \cos \tau - \ell r \sin \tau + i\epsilon)^{\Delta}} \\
     &= \frac{\Gamma(\Delta - \delta)\Gamma(\delta)}{\Gamma(\Delta)}\frac{1}{(r \cos \tau + \ell \sin \tau - r \Omega \cdot \Omega_q)^{\delta}} \frac{1}{(\ell^2 \cos \tau - \ell r \sin \tau + i\epsilon)^{\Delta - \delta}} \\
     &\equiv \frac{\Gamma(\Delta - \delta)\Gamma(\delta)}{\Gamma(\Delta)} \widetilde{\Psi}_{\Delta}^{\rm null}(\delta, \hat{q}; \tau, r, \Omega).
     \end{split}
\end{equation}
By construction, these wavefunctions diagonalize boosts towards $\hat{q}$ (this is a point on the intersection of the null cone through the origin with the boundary) with boost weight $\delta$.
This can be checked explicitly using the explicit expressions for the $\mathfrak{so}(3,2)$ generators in retarded coordinates
\begin{equation}
\label{eq:boost-gens-ret}
    \begin{split}
        M_{04} &= -i\frac{\p}{\p \tau}, \\
         M_{4i} &= i (\ell \cos(\tau )- r \sin (\tau )) \left(\frac{1}{r} \nabla_i + \Omega_i \p_r \right) -i \Omega_i \cos (\tau ) \p_{\tau},\\
          M_{0i} &= i (\ell \sin (\tau )+r \cos (\tau )) \left(\frac{1}{r} \nabla_i + \Omega_i \p_r \right) -i \Omega_i \sin (\tau ) \p_{\tau},\\
         M_{ij} &= -i \left(\Omega_i \p_{\Omega_j} - \Omega_j \p_{\Omega_i} \right).
    \end{split}
\end{equation}

However, \eqref{eq:psinull} are \textit{not} solutions to the  wave equation in AdS$_4$. This is of course expected since in AdS \eqref{eq:prop-null} do not solve the wave equation.\footnote{Note that the wavefunctions with $\tau$ on the null cone through the origin are also not solutions. This case corresponds to taking $\alpha \rightarrow 0$ in footnote \ref{footn:massive}, which is singular and the argument breaks down. } We can construct a family of boost eigenstates that solve the KG equation in AdS by either replacing \eqref{eq:prop-null} for the bulk-to-bulk propagator in AdS \cite{Burgess:1984ti}, or by considering the linear combination 
\begin{equation}
\label{eq:Ansatz}
    \Phi_{\Delta}^{\rm null}(\delta,\hat{q}(z);\tau,r, w) = \sum_{n = 0}^{\infty} c_n(\Delta, \delta) \Psi_{\Delta+2n}^{\rm null}(\delta,\hat{q}(z);\tau,r, w),
\end{equation}
where the coefficients are determined recursively, using the fact that 
\begin{equation}
    \left[\Box_{AdS_4} - \frac{1}{\ell^2}\Delta(\Delta - 3) \right] \widetilde{\Psi}_{\Delta}^{\rm null} = -\frac{(\Delta - \delta)(\Delta + 1 -\delta) }{\ell^2(\ell \cos \tau - r \sin \tau)^2} \widetilde{\Psi}_{\Delta}^{\rm null} = -(\Delta - \delta)(\Delta + 1 -\delta) \widetilde{\Psi}_{\Delta + 2}^{\rm null}.
\end{equation}
The solution, whose derivation is presented in appendix \ref{app:null-boost-estate-solutions} is
\begin{equation}
\label{eq:final-sol}
\begin{split}
    \Phi_{\Delta}^{\rm null}(\delta, \hat{q}(z);\tau, r, w) &= \frac{\Gamma(\Delta - \delta)}{\Gamma(\Delta - \frac{1}{2})} {}_2F_1\left[\frac{1}{2}(\Delta - \delta), \frac{1}{2}(\Delta - \delta + 1); \Delta - \frac{1}{2}; \frac{1}{(\ell \cos\tau - r \sin \tau)^2} \right] \\
    &\times \widetilde{\Psi}_{\Delta}^{\rm null}(\delta, \hat{q}(z);\tau, r, w)\\
    &= \frac{\Gamma(\Delta)}{\Gamma(\Delta - \frac{1}{2})\Gamma(\delta)} {}_2F_1\left[\frac{1}{2}(\Delta - \delta), \frac{1}{2}(\Delta - \delta + 1); \Delta - \frac{1}{2}; \frac{1}{(\ell \cos\tau - r \sin \tau)^2} \right] \\
    &\times \Psi_{\Delta}^{\rm null}(\delta, \hat{q}(z);\tau, r, w).\\
    \end{split}
\end{equation}
One can show using \eqref{eq:boost-gens-ret} that $\Psi_{\Delta + 2n}^{\rm null}(\delta)$ are all boost eigenstates of eigenvalue $\delta$ and hence so is \eqref{eq:final-sol}.\footnote{Recall that $M_{43}$ is a boost towards $\hat{q} = (1,0,0,1)$.} As a result, $\Phi_{\Delta}^{\rm null}$ are again AdS$_4$ analogs of the Mink$_4$ conformal primary wavefunctions. 

We expect that for $\delta = 1 + i\lambda,~$ $\lambda \in \mathbb{R}$, \eqref{eq:final-sol} form yet another basis for $\mathfrak{so}(3,2)$ positive and negative energy representations. We show in the next section that, in the flat space limit with fixed $\Delta$, they becomes \textit{massless} conformal primary wavefunctions in Mink$_4$. 

\subsubsection{Boundary conditions and flat space limit}

We first notice that in the limit
\begin{equation}
\label{eq:limit-ret}
    \tau \rightarrow 0, ~~ \ell \rightarrow \infty, \quad  \tau \ell \equiv u ~~{\rm fixed},
\end{equation}
the AdS$_4$ metric \eqref{eq:AdS-retarded} becomes the Mink$_4$ metric in retarded coordinates. Furthermore, the $\mathfrak{so}(3,2)$ generators \eqref{eq:boost-gens-ret} become flat space Poincar\'e generators.
Here $u$ is to be identified with the retarded time $u = t - r$ in Mink$_4$. 

In the limit \eqref{eq:limit-ret}, we find
\begin{equation}
    \frac{1}{(\ell \cos \tau - r \sin \tau)^2} \rightarrow\frac{1}{\ell^2} \rightarrow 0,
\end{equation}
the hypergeometric function in \eqref{eq:final-sol} is set to 1, and 
\begin{equation}
    \Phi_{\Delta}^{\rm null}(\delta, \hat{q}(z), \tau,r, w) \rightarrow \frac{\Gamma(\Delta - \delta) }{\Gamma(\Delta - \frac{1}{2})} \frac{1}{(r + u - r \Omega \cdot \Omega_q + i\epsilon)^{\delta}}
\end{equation}
after renormalization. This is proportional to a conformal primary wavefunction of a \textit{massless} particle of boost weight $\delta$ in Minkowski space. The relation between AdS propagators  and massless conformal primary wavefunctions was first observed in \cite{deGioia:2022fcn} and later elaborated on \cite{Bagchi:2023fbj}. The result here arises instead from a limit of the  local Lorentz boost eigenstate \eqref{eq:final-sol} AdS$_4$ of dimension $\delta$. To the extent of our knowledge, this is a new construction of these wavefunctions.

\section{Flat space holography from AdS/CFT: the bulk picture}
\label{sec:flat-space-lim}

In this section we show that the AdS$_4$ bases constructed in Section \ref{sec:bulk-to-boundary-basis} reduce to the plane wave and conformal primary bases of Mink$_4$ in various limits. This explicitly realizes the contraction of $\mathfrak{so}(3,2)$ to 4d Poincar\'e at the level of the bulk Hilbert space of a free scalar in AdS$_4$. 
Furthermore, we show that the null bulk reconstruction formula \eqref{eq:bulk-reconstruction-null} localizes, in the flat space limit for fixed $\Delta$, to a conformal Carroll field whose modes are \textit{shadow} operators in the 3d CFT. We conclude by showing that the null inner product \eqref{eq:null} of AdS bulk-to-boundary propagators, as well as the inner product of of $G_{\Delta}$ with its shadow are solutions to the conformal Carroll Ward identities in 3d.

\subsection{Plane wave bases from flat space limit of bulk-to-boundary propagators} 
\label{sec:plane-waves}

In this section we show that the AdS bulk-to-boundary propagators become plane waves of massive and massless particles in appropriate limits. These limits were discussed from different perspectives in \cite{Hijano:2019qmi,Hijano:2020szl,Komatsu:2020sag, Li:2021snj}.
We will work in the global coordinates \eqref{eq:global-AdS} and consider the flat space limit (fixed $r, t$)
\be
\label{eq:global-flat}
\tau = t \ell^{-1}, \quad \rho  = r \ell^{-1}, \quad \ell \rightarrow \infty,
\ee
at the level of bulk-to-boundary with $\Delta \rightarrow \infty.$

We start with the mode expansion of the time-ordered propagator
\begin{equation}
\label{G-flat-exp}
    G_{\Delta}^{\rm T.O.}(X; P) = \sum_{k = 0}^{\infty} \sum_{\ell = 0}^{\infty}  \left[ b_{k\ell}^+\psi_{k\ell 0}^+(\tau - \tau_p, \rho, \Omega\cdot \Omega_p) \Theta(\tau - \tau_p) + b_{k\ell}^- \psi_{k\ell 0}^-(\tau - \tau_p, \rho, \Omega\cdot \Omega_p) \Theta(\tau_p - \tau) \right],
\end{equation}
with $b_{k\ell}^{\pm}$ given in \eqref{eq:bklto}. 
The flat space limit is given by $\ell \rightarrow \infty$ together with the coordinate rescaling \eqref{eq:global-flat}, keeping $m \equiv \Delta/\ell$ and the boundary insertions fixed. 
The goal is to show that in this limit, \eqref{G-flat-exp} reduces to a plane wave. Using the expansion of the Gamma function at large argument
\begin{equation}
\frac{\Gamma(x + n)}{\Gamma(x)} = x^n + \mathcal{O}(x^{n-1}),
\end{equation}
we find 
\begin{equation}
   b_{k\ell}^{\pm} f_{k\ell}(\rho) = \pi \sqrt{2\ell + 1}(-1)^k 2^{\Delta} \frac{\Delta^{k + \ell}}{k!\Gamma(\frac{3}{2} + \ell)} \sum_{n = 0}^k (-1)^n \left(\begin{matrix}
    k\\
    n
    \end{matrix}\right) \frac{\Delta^n\Gamma(\frac{3}{2} + \ell)}{\Gamma(\frac{3}{2} + \ell +  n)} \rho^{2n + \ell} + \cdots , \quad \Delta \rightarrow \infty, \rho \rightarrow 0.
\end{equation}
Switching the order of the $k$ and $n$ sums, and performing the sum on $k$, using
\begin{equation}
    \sum_{k = n}^{\infty} (-1)^k \frac{x^k}{k!}\left(\begin{matrix}
    k\\
    n
    \end{matrix}\right)  = (-1)^n\frac{x^n e^{-x}}{n!}
\end{equation}
with $x = \Delta e^{-2i(\tau - \tau_p)}$, we find for $\tau, \rho \rightarrow 0, \Delta \rightarrow \infty$
\begin{equation}
    G_{\Delta}^{+}(X;P) = \sum_{n = 0}^{\infty} \sum_{\ell = 0}^{\infty} 2^{\Delta}e^{-i\Delta (\tau -\tau_p)} e^{-\Delta (1 - 2i \tau)e^{2i\tau_p}} \frac{\Delta^{2n} e^{2in \tau_p}}{n!} \pi \sqrt{2 \ell + 1}  \frac{(\Delta e^{i\tau_p})^{\ell}}{\Gamma(\frac{3}{2} + \ell + n)} \rho^{2n + \ell} Y_{\ell 0}(\Omega \cdot \Omega_p).
\end{equation}
Setting $\Delta \tau = m t,~ \Delta \rho = m r$, and using the expansion of the spherical Bessel function
\begin{equation}
    j_\ell(x) = \sqrt{\frac{\pi}{2x}} \sum_{n = 0}^{\infty} \frac{(-1)^n}{n! \Gamma(n + \ell + \frac{3}{2})} \left( \frac{x}{2}\right)^{2n + \ell + \frac{1}{2}}
\end{equation}
we find
\begin{equation}
    G^{+}_{\Delta}(X; P) \rightarrow 2^{\Delta} e^{-\Delta e^{2i \tau_p}} e^{-i\Delta (\tau-\tau_p)} e^{2i m t e^{2i \tau_p}}\sum_{\ell = 0}^{\infty}  \sqrt{4\pi (2\ell + 1)} i^{\ell} j_{\ell}(-2 im r e^{i\tau_p}) Y_{\ell 0}(\Omega \cdot \Omega_p).
\end{equation}

We recognize the sum over $\ell$ as the expansion of the spatial component of a plane wave
\begin{equation}
    \sum_{\ell = 0}^{\infty}  \sqrt{4\pi (2\ell + 1)} i^{\ell} j_{\ell}(-2 im r e^{i\tau_p}) Y_{\ell 0}(\Omega \cdot \Omega_p) =  e^{-i r {\bf p} \cdot \Omega}
\end{equation}
with spatial momentum
\begin{equation}
    {\bf p} = 2 i m e^{i\tau_p} \Omega_p.
\end{equation}
Putting everything together, 
\begin{equation}
\label{eq:limitg+}
    G_{\Delta}^+(X; P) \rightarrow 2^{\Delta} e^{-\Delta e^{2i \tau_p}} e^{-i\Delta (\tau-\tau_p)}  e^{i p \cdot x},
\end{equation}
where $p$ is a complex $4$-momentum related to the boundary point $P$ by
\begin{equation}
\label{eq:p-flat}
    p = - 2m i e^{i\tau_p} ( - i e^{i\tau_p},  \Omega_p).
\end{equation}
We see that $p$ becomes real provided that
\begin{equation}
\label{eq:taup}
    \tau_p = \frac{t_p}{\ell} \pm \frac{\pi}{2} + i \hat{\tau}_p. 
\end{equation}

\begin{figure}
 \begin{center}
 \includegraphics[scale=0.35]{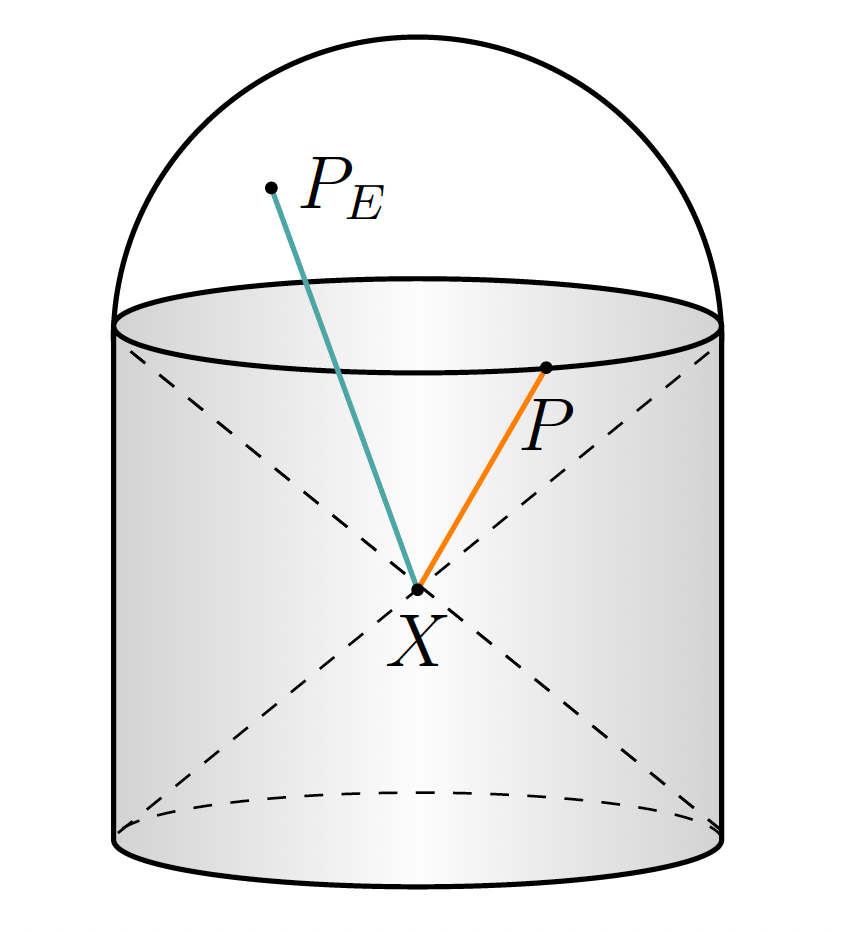}
 \caption{AdS-bulk-to-boundary propagators become proportional to massless plane waves in the flat space limit with $\Delta \rightarrow \infty$, provided that the boundary point ($P$) is approximately null separated from the bulk point. They become proportional to massive plane waves in the same limit, provided the boundary point $(P_E)$ is analytically continued to the Euclidean cap.}
 \label{eq:analytic-cont-mom}
 \end{center}
 \end{figure}

In the case where $\hat{\tau}_p = 0$, up to normalization (i.e. stripping off the coefficient including $e^{-i\Delta(\tau - \tau_p)}$ in \eqref{eq:limitg+}), we recover a massless plane wave of positive energy $\omega_p = m$. This is interesting because in previous works \cite{Penedones:2010ue, Komatsu:2020sag, Hijano:2020szl, Marotta:2024sce} it was found that the $\Delta \rightarrow \infty$ of CFT correlators in various representations gives rise to amplitudes involving massive particles in the flat space limit. Here we see that tuning the kinematics to the strip configurations (see eg.  \cite{Hijano:2020szl}) may also yield, after suitable renormalization, massless scattering amplitudes computed in the standard momentum space basis. The momenta with $\tau_{p} = \frac{t_p}{\ell}\pm \frac{\pi}{2}$ are antipodally related. 
 Massive plane waves are instead obtained for $\hat{\tau}_p \neq 0$. This case corresponds to taking the boundary point to lie on an Euclidean cap, as illustrated in Figure \ref{eq:analytic-cont-mom}.

 $G^{-}_{\Delta}$ gives a result of the same form as \eqref{eq:limitg+} with $\tau, \tau_p \rightarrow -\tau, -\tau_p$ and $p^0 \rightarrow -p^0$. From \eqref{eq:p-flat}, for the same choice of $\tau_p$ as in \eqref{eq:taup}, this corresponds, up to normalization, to a plane wave of negative energy. Substituting these results into the reconstruction formulae in Section \ref{sec:bulk-rec-AdS} we recover an expansion of the bulk field in terms of plane waves, which yields a precise map between CFT boundary operators and flat space creation and annihilation operators \eqref{eq:plane-waves}. \eqref{eq:p-flat} provides the map between points on the AdS boundary and flat space momenta. We will compute an explicit example in the next section. 

Using these results, it is straightforward to show that the inner products of of time-ordered and anti-time-ordered propagators computed in Section \ref{sec:AdS-bases} become delta-function normalized in the flat space limit. One can see this more directly using
\begin{equation}
   \lim_{\Delta \rightarrow \infty} \langle G_{\Delta}^{\pm}(P_1, X), G_{\Delta}^{\pm}(P_2, X)\rangle_{\tau} \propto \lim_{\Delta \rightarrow \infty} \frac{1}{(-P_1\cdot P_2)^{\Delta}}
\end{equation}
and using the Mellin representation of the boundary-to-boundary propagator which at large $\Delta$ localizes to 
\begin{equation}
     \lim_{\Delta \rightarrow \infty} \frac{1}{(-P_1\cdot P_2)^{\Delta}} \propto \delta(\tau_{12}) \delta^{(2)}(\Omega_1 - \Omega_2).
\end{equation}
The delta function in $\tau_{12}$ can be traded for a delta function in $p_{12}^0$ using \eqref{eq:p-flat}.
Keeping track of the time-ordering, we deduce that the bases of incoming and outgoing AdS modes constructed in Section \ref{sec:AdS-bases} reduce to positive and negative energy plane waves in the $\Delta \rightarrow \infty$ limit.

 If one instead fixes $\Delta$, it was shown in \cite{deGioia:2022fcn, Bagchi:2023fbj}, that the bulk-to-boundary propagators become time-ordered Carrollian wavefunctions, namely
 \begin{equation}
 \label{eq:Carroll}
    \mathcal{G}_{\Delta}^{+}(x;t_p,\hat{q}_p) =  \lim_{\ell \rightarrow \infty} G_{\Delta}^{\rm T.O.}(X; P) = \frac{1}{(t - t_p - r \Omega \cdot \Omega_p + i\epsilon)^{\Delta}},
 \end{equation}
 where we have chosen $\tau_p = \frac{t_p}{\ell} + \frac{\pi}{2}$. Together with the anti-time-ordered wavefunctions $\mathcal{G}_{\Delta}^{-}(x;t_p,\hat{q})$, they span the space of (positive and negative energy) solutions to the \textit{massless} KG equation in flat space (ie. a reducible Poincar\'e representation) for \textit{fixed} boost weight $\Delta = 1$ as $(t_p, \Omega_p)$ are varied. This is easy to see, as \eqref{eq:Carroll} are simply related to massless plane waves by the Fourier transform, 
 \begin{equation}
 \label{eq:flat-lim-pw}
     \int_{-\infty}^{\infty} dt_p e^{\pm i \omega t_p}  \mathcal{G}_{\Delta}^{\pm }(x;t_p,\hat{q}_p) =\frac{2\pi (\mp i)^{\Delta}}{\Gamma(\Delta)} \omega^{\Delta - 1} e^{\mp i\omega \hat{q}_p\cdot x},
 \end{equation}
 with the other combinations vanishing. Note that $\mathcal{G}_{\Delta}^+$ consists of negative frequency modes while $\mathcal{G}_{\Delta}^-$ consists of positive frequency modes. Given that $\tau_p > \tau$, this is in complete agreement with the AdS results obtained in Sections \ref{sec:AdS-bases} and \ref{sec:bulk-rec-AdS}. The relation between the modified Mellin transform introduced in \cite{Banerjee:2018gce} and the Carrollian wavefunctions \eqref{eq:Carroll} was derived in \cite{Bagchi:2023fbj}. We see here that these different prescriptions are directly related to flat space limits of (anti-)time-ordered bulk to boundary propagators in AdS and amount to a simple change of basis in the space of positive and negative frequency solutions to the KG equation.
Upon performing a further Mellin transform of \eqref{eq:Carroll} with respect to $t_p$, we obtain an $\mathfrak{so}(3,1)$ conformal primary wavefunction of dimension shifted by the Mellin weight \cite{Pasterski:2016qvg, Pasterski:2017kqt}.

\subsection{Conformal primary basis expansion of a free field in flat space from AdS}

In this section, we study the flat space limit of the AdS bulk reconstruction formulas proposed in Section \ref{sec:bulk-rec-AdS}. At fixed $\Delta$, these formulas reproduce, in the flat space limit, the Carrollian and conformal primary wavefunction decompositions of a free field in Mink$_4$. A byproduct of our analysis is a precise map between shadow transforms of local operators in a holographic CFT$_3$ and operators in Carrollian/celestial conformal field theories.

To easily compare with results in the flat space holography literature (see \cite{Raclariu:2021zjz, Pasterski:2021rjz, Donnay:2023mrd, Bagchi:2025vri} and references therein for reviews from different perspectives), it will be most convenient to work in the retarded coordinates \eqref{eq:AdS-retarded} in which the bulk-to-boundary propagator is given by
\begin{equation}
\label{eq:G-null}
     G_{\Delta}^{\rm T.O.}(P;X) = \frac{1}{\left( r \cos(\tau - \tau_p) + \ell \sin(\tau - \tau_p) -r \Omega \cdot \Omega_p +i\epsilon \right)^{\Delta}}.
\end{equation}
We recall  from \eqref{eq:AdS-retarded} that the flat space limit is given by
\begin{equation}
\label{eq:flat-lim-ret}
    \tau = \frac{u}{\ell}, \quad \ell \rightarrow \infty
\end{equation}
for fixed $u$. In this limit, the AdS$_4$ metric becomes the Mink$_4$ metric in retarded coordinates. Note that unlike the flat limit in the global AdS coordinates \eqref{eq:global-flat}, in the retarded coordinates, $r$ can be kept fixed.  

We start with the representation of the AdS field 
\begin{equation}
\label{eq:bulk-reconstruction-null}
    \Phi(X) = \int_{B} d^3P G^{\rm T.O.}_{\Delta}(P;X) \widetilde{\mathcal{O}}^-_{3 - \Delta}(P) + \int_{B} d\tau_p d\Omega_p G^{\rm  A.T.O.}_{\Delta}(P;X) \widetilde{\mathcal{O}}^{+}_{3 - \Delta}(P),
\end{equation}
which follows from \eqref{eq:shadow-bulk-rec}, and we absorbed the factors of $\alpha$ into the definition of $\widetilde{\mathcal{O}}^{\pm}_{3 - \Delta}$. We expect that in the flat space limit, the creation operators $(\widetilde{\mathcal{O}}^-_{3 - \Delta})$ will be paired with the negative energy plane waves while the annihilation operators $(\widetilde{\mathcal{O}}^{-\dagger}_{3 - \Delta}) = \widetilde{\mathcal{O}}_{3 - \Delta}^+$ will be paired with the positive energy ones. The time-orderings of the bulk-to-boundary propagators in our reconstruction formula are necessary to ensure this, as can be seen from \eqref{eq:Carroll} and \eqref{eq:flat-lim-pw}.

\begin{figure}
\begin{center}
\includegraphics[scale=0.35]{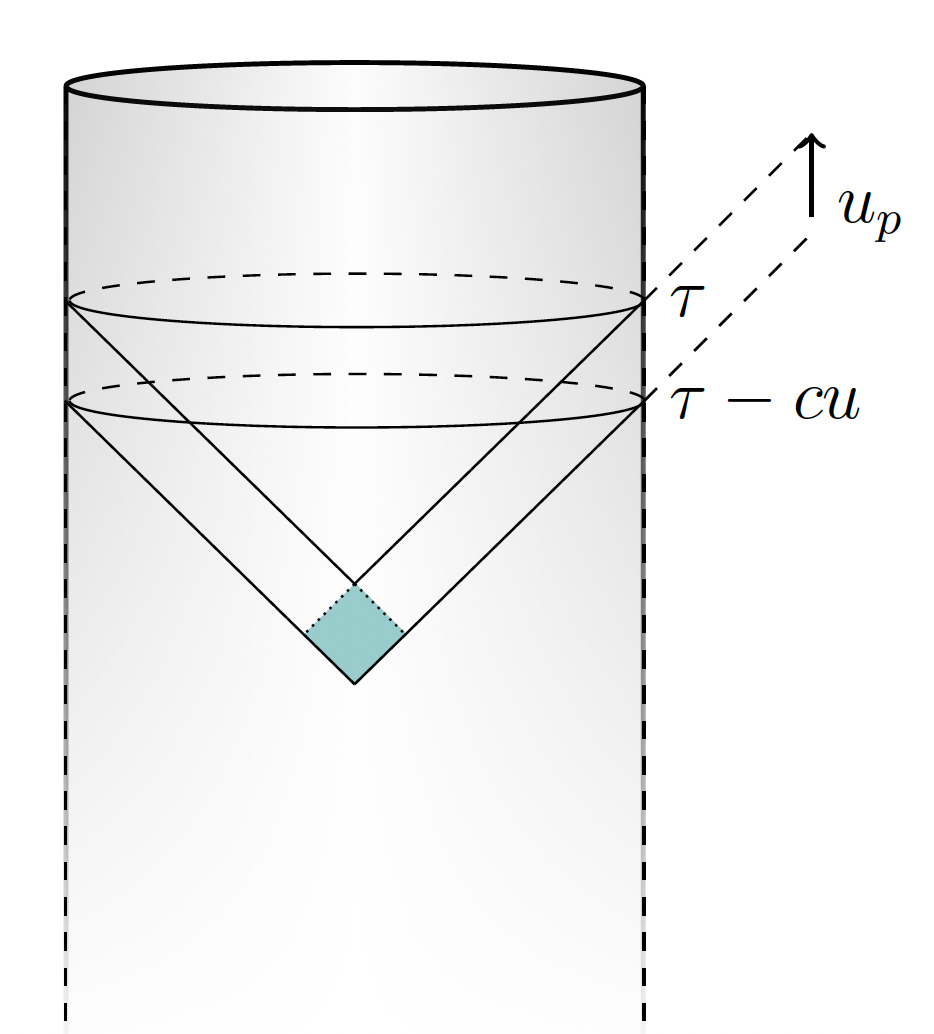}
\caption{Taking $B$ to be an infinitesimal strip around a constant time slice of the boundary allows for the field to be reconstructed in the central diamond. For fixed $\Delta$, our proposed AdS$_4$ reconstruction formulas reduce to the expansion of a free field in flat space in a basis of Carrollian wavefunctions. The modes in this expansion identify, upon quantization, operators in a 3d Carrollian field theory with shadow transforms of primary operators in CFT$_3$. }
\label{fig:bulk-from-boundary}
\end{center}
\end{figure}

Equation \eqref{eq:bulk-reconstruction-null} allows us to reconstruct the bulk scalar near the origin of the AdS from an integral over infinitesimal boundary regions. 
To see this, we take $B$ to be defined as in eq. \eqref{eq:B-bB} with $\Sigma_{\tau}$ at $\tau + c u_0$, $c = \frac{1}{\ell}$, and perform the change of variables
\begin{equation}
    \label{eq:boundary-retarded}
 \tau_p = \tau - c (u - u_p). 
\end{equation}
Then taking  $\ell \rightarrow \infty$ for fixed $r$ in \eqref{eq:G-null}, we find
\begin{equation}
\label{eq:Carroll-flat-lim}
\begin{split}
\Phi(\tau, r, \Omega) = \ell^{-1} \int_{u_0 + u}^{\infty}  du_p\int d\Omega_p \frac{1}{(u - u_p+ r(1-\Omega \cdot \Omega_p) + i\epsilon)^{\Delta}} \widetilde{\mathcal{O}}^-_{3 - \Delta}(\tau - \ell^{-1}(u - u_p), \Omega_p) + h.c.
\end{split}
\end{equation}
Further taking $u_0 \rightarrow -\infty$ for fixed $u$, we recognize the Carrollian expansion of a scalar field in flat space \eqref{eq:Carroll-exp}. Inside amplitudes, only the positive or negative frequency components will contribute, depending on whether the insertion is incoming or outgoing. 
Interestingly, one could directly obtain the result \eqref{eq:Carroll-flat-lim} by taking the integration regions in \eqref{eq:bulk-reconstruction-null} to be infinitesimal strips around the intersection of the lightcone with origin at $\tau$ with the boundary. We illustrate this in Figure \ref{fig:bulk-from-boundary}. Starting from \eqref{eq:bulk-reconstruction-null}, one can also recover the plane wave decomposition of a free field in Minkowski space by scaling both $\ell \rightarrow \infty$ and $\Delta \rightarrow \infty$, and using the behavior of the bulk-to-boundary propagator in this limit, derived in Section \ref{sec:plane-waves}. The localization of the bulk-to-boundary propagators in this limit is analogous to the localization of the HKLL kernel derived in \cite{Hijano:2019qmi,Hijano:2020szl}.

The formula \eqref{eq:Carroll-flat-lim} provides a precise  identification between shadow operators in CFT$_3$ and operators in a Carrollian field theory. To see this explicitly, we relabel
\begin{equation}
   \widetilde{\mathcal{O}}^-_{3 - \Delta}\left(\tau -\ell^{-1} (u-u_p), \Omega_p\right) \rightarrow \widetilde{\mathcal{O}}_{3 - \Delta}^-(u_p,\Omega_p)
\end{equation}
and take $r \rightarrow \infty$. In this case, we can apply the stationary phase formula \eqref{eq:stat-phase} to \eqref{eq:Carroll-flat-lim} which yields
\begin{equation}
\label{eq:AdS-field}
\begin{split}
     \lim_{r\rightarrow \infty}\Phi(\tau,r,\Omega) &= \frac{(-i)^{\Delta}}{\Gamma(\Delta)}\ell^{-1} \lim_{r\rightarrow \infty}\int_0^{\infty} d\omega \int_{-\infty}^{\infty} du_p \int d\Omega_p \omega^{\Delta - 1} e^{i\omega \left(r\left(1 - \Omega \cdot \Omega_p\right) + (u - u_p) + i\epsilon \right)} \widetilde{\mathcal{O}}^-_{3 - \Delta}(u_p,\Omega_p) + h.c.\\
     &= \frac{2\pi \ell^{-1}}{\Delta -1} \frac{1}{r} \int_{-\infty}^{\infty} du_p (u - u_p + i\epsilon)^{1 - \Delta} \widetilde{\mathcal{O}}^-_{3 - \Delta}(u_p,\Omega) + \mathcal{O}(r^{-2}) + h.c..
     \end{split}
\end{equation}
Taking the boundary integration region to be an infinitesimal strip gives rise to a bulk AdS field with $r^{-1}$ asymptotics, characteristic of a massless field in Minkowski space for any $\Delta$. 

We can further extract from \eqref{eq:AdS-field} an $\mathfrak{sl}(2,\mathbb{C})$ primary by setting 
\begin{equation}
\label{eq:AdS-center}
    \tau = \ell^{-1}u
\end{equation}
and taking a light transform with respect to $u$. \eqref{eq:AdS-center} corresponds to a bulk flat space limit -- \eqref{eq:flat-lim-ret}. In this case we schematically obtain
\begin{equation}
\label{eq:bulk-Carroll-boundary-operator}
  \lim_{r \rightarrow \infty} \int_{-\infty}^{\infty} du (u \pm i\epsilon)^{-\lambda} \Phi(u, r, \Omega) \propto \frac{\ell^{-1}}{r} \int_{-\infty}^{\infty} du_p  (u_p \pm i\epsilon)^{2 - \Delta - \lambda} \widetilde{\mathcal{O}}^{\pm}_{3 - \Delta}(u_p,\Omega) + \mathcal{O}(r^{-2}),
\end{equation}
 where the integral should be defined with an $i\epsilon$ prescription which can be determined using identities such as 
\begin{equation}
 \int_{-\infty}^{\infty} du (u \pm i\epsilon )^{-\lambda} (u - 1 \pm i\epsilon )^{1 - \Delta} = 0
\end{equation}
and
\begin{equation}
\label{eq:norm-carr}
     \int_{-\infty}^{\infty} du (u \mp i\epsilon )^{-\lambda} (u - 1 \pm i\epsilon )^{1 - \Delta} =  \mp 2i e^{\mp i \pi \Delta} \sin \pi \lambda B(1 - \lambda, \Delta + \lambda - 2).
\end{equation}
Equation \eqref{eq:bulk-Carroll-boundary-operator} provides a map between Carrollian bulk fields near the null cone through the origin of AdS  and \textit{shadows} of CFT$_3$ primary operators around equal time slices (or dimensional reductions thereof). Eq. \eqref{eq:bulk-Carroll-boundary-operator} allows one to infer the precise normalization relating CFT$_3$ and 3d Carrollian/2d celestial operators. The normalization naively implied by \eqref{eq:norm-carr} can be shown to agree with that computed in \cite{deGioia:2024yne, deGioia:2025mwt}. We note, however, that the change of variables used in passing from \eqref{eq:AdS-field} to \eqref{eq:bulk-Carroll-boundary-operator} introduces $u_p$ dependence in the $i\epsilon$ prescription, potentially leading to a different overall prefactor to \eqref{eq:norm-carr}. It will be interesting to see whether this can resolve the normalization mismatch encountered in \cite{deGioia:2024yne} when extracting conformally soft operators from light-transforms of CFT$_3$ operators via a Carrollian limit.

Note that we could have also started with the equivalent bulk reconstruction formula \eqref{eq:phi-rec} involving $\mathcal{O}_{\Delta}$. This would instead yield an identification of the form \eqref{eq:bulk-Carroll-boundary-operator} with $\Delta \rightarrow 3 - \Delta$. These results are consistent, as can be seen by recalling the relation \eqref{eq:O-tilde-modes} between primary and shadow modes. Provided one uses the correct $i\epsilon$ prescriptions, the difference between the flat space observables obtained starting from either representation  disappears upon recasting them in a conformal primary basis. This explains, for instance, why taking flat space limits of CFT$_3$ correlators involving primary operators \cite{deGioia:2025mwt} or their shadows \cite{deGioia:2023cbd} led to the same results upon transforming to the conformal primary basis. It will be interesting to revisit the construction of the celestial/Carrollian stress tensor \cite{Kapec:2016jld, Bagchi:2024gnn}, as well as the associated non-locality issues \cite{Banerjee:2022wht, Himwich:2025bza, Pranzetti:2025flv} in light of these observations. 

Finally, we note that one could have also considered a large-$r$ expansion of the reconstruction formula \eqref{eq:bulk-reconstruction-null}. The stationary phase approximation then identifies the boundary value of $\Phi$ with a CFT$_3$ scalar with $\Delta = \frac{3}{2}$.\footnote{For this value of $\Delta$, primary operators coincide with their shadows.} A light transform in $u$ with the appropriate dimension will again yield a conformal primary operator, but the precise normalization will differ from that in \eqref{eq:bulk-Carroll-boundary-operator}.

\subsection{Carrollian two-point functions from shadow inner products in AdS}
\label{sec:carroll}

In this section, we show that the shadow inner product defined in Section \ref{sec:shadow-ip-AdS} computes two point functions in the electric branch of a Carrollian field theory. 

We start with the formula
\begin{equation}
\label{eq:G-sing-ip}
\begin{split}
    \langle G^{\rm T.O.}_{\Delta}(P_1, X), G^{\rm T.O.}_{3 - \Delta}(P_2,X) \rangle_{\tau_0}  &=  \delta(\tau_1 -\tau_2)\delta^{(2)}(\Omega_1 - \Omega_2)\\
    &\times \left[N_{\Delta}\Theta(\tau_0 - \tau_1)\Theta(\tau_0 - \tau_2) - N_{\Delta}^*\Theta(\tau_1 - \tau_0)\Theta(\tau_2 - \tau_0) \right], \\
    N_{\Delta} &= (2\pi)^3 \frac{(-1)^{\Delta - \frac{3}{2}}}{\Gamma(\Delta)\Gamma(3 - \Delta)}.
    \end{split}
\end{equation}
This inner product is equivalent to the inner product of bulk to boundary propagators evaluated on a null plane -- see eq. \eqref{eq:null}. 
We now prove that \eqref{eq:G-sing-ip} is a solution to the conformal Carroll identities. 

The inner product \eqref{eq:G-sing-ip} is manifestly invariant under translations 
\begin{equation}
    \delta_{M_{r,s}}   \langle G^{\rm T.O.}_{\Delta}(P_1, X), G_{3-\Delta}^{\rm T.O.}(P_2,X) \rangle \propto -i \left( z_1^r \bz_1^{s} \p_{\tau_1} +z_2^r \bz_2^{s} \p_{\tau_2} \right) \delta(\tau_{12}) \delta^{(2)}(z_1 - z_2) = 0.
\end{equation}
On the support of the delta function in $z_{12}$, the non-trivial constraints from Lorentz invariance take the form \cite{Bagchi:2025vri}
\begin{equation}
    \begin{split}
        &\left[ (\Delta_1 + \Delta_2 - 2) + (\tau_1 - \tau_2) \p_{\tau_1} \right] \delta(\tau_{12})\delta^{(2)}(z_{12}) = 0, \\
       (z_1 + \bar{z}_1) &\left[ (\Delta_1 + \Delta_2 - 2) + (\tau_1 - \tau_2) \p_{\tau_1} \right] \delta(\tau_{12})\delta^{(2)}(z_{12}) = 0.
    \end{split}
\end{equation}
In our case $\Delta_1 + \Delta_2 = 3$ and furthermore
\begin{equation}
   ( \tau_1 - \tau_2) \p_{\tau_1}  \delta(\tau_{12}) = - \delta(\tau_{12}).
\end{equation}
 As a result, we find that the two-point function of a CFT$_3$ primary and its (3d) shadow is a natural solution to the conformal Carroll Ward identities. Note that no Carrollian limit had to be taken at the level of the boundary CFT. 
 
 While it is well known that Carrollian two point functions in the electric sector are distributional, the time-dependence is usually taken to be a power-law 
\begin{equation}
    \langle O_{\Delta_1}(u_1, z_1) O_{\Delta_2}(u_2, z_2)\rangle \propto f(u_{12}) \delta^{(2)}(z_{12}), \quad f(u_{12}) = \frac{1}{(u_{12} + i\epsilon)^{\Delta_1 + \Delta_2 - 2}}.
\end{equation}
In the case of \eqref{eq:G-sing-ip}, $\Delta_1 + \Delta_2 = 3$, hence
\begin{equation}
    f(u_{12}) = \frac{1}{u_{12} + i\epsilon}.
\end{equation}
Since real and imaginary parts of this function should solve the conformal Carroll Ward identities independently, we could have more directly concluded that \eqref{eq:G-sing-ip} is also consistent with the conformal Carroll Ward identities.

\section{Discussion}

This work started as an attempt to better understand the relation between AdS and flat space holography. While many aspects of this relation have been developed over the past decades from different perspectives, we found that the literature did not address a few elementary questions. The first one concerns the different approaches to quantizing fields and constructing observables in AdS vs. flat space. Quantum field theory in Minkowski spacetime starts by quantizing a free field, either canonically or via the path integral in Lorentzian signature. This is the first step in constructing asymptotic states and defining scattering amplitudes. 

On the other hand, the standard treatment in AdS  relies instead on observables such as the Euclidean path integral or AdS/Witten diagrams which are by default defined in Euclidean signature. The map between these quantities and Euclidean boundary CFT correlators is by now well understood \cite{Gubser:1998bc,Witten:1998qj, Hamilton:2006az}. However, the analytic continuation to the Lorentzian setting, while crucial in extracting the flat space S-matrix \cite{Penedones:2010ue, Komatsu:2020sag, vanRees:2022zmr, Marotta:2024sce}, remains quite subtle. 

Our first goal was to establish a precise and explicit relation between canonical quantization in Lorentzian AdS and flat spacetimes. In the elementary case of a free scalar field in AdS, we showed in Sections \ref{sec:bulk-to-boundary-basis} and \ref{sec:AdS-bases} that bulk-to-boundary propagators with appropriate time orderings provide a \textit{basis} for the positive and negative energy subspaces of solutions to the Klein--Gordon equation. The key ingredients to establishing this claim was the computation of \textit{bulk} inner products of bulk-to-boundary propagators which, to the best of our knowledge, appears in Section \ref{sec:bulk-to-boundary-basis} for the first time. We refer the reader to \cite{Costa:2014kfa} for complementary formulas for various conformal integrals (but not KG inner products) involving bulk-to-boundary propagators with dimensions on the principal series and \cite{Melton:2025ecj} for boundary inner products.  

This result allowed us to express a bulk field in AdS in terms of integrals of boundary operators against bulk-to-boundary propagators with the appropriate time orderings -- see for instance \eqref{eq:shadow-bulk-rec}. Our formula is similar in spirit to the HKLL proposal, however, it is more general, as it reconstructs the field at any bulk point in AdS. This is because the HKLL kernel is replaced by bulk-to-boundary propagators with specified time orderings, and hence one can uniquely analytically continue it to arbitrary points in AdS. For related discussions involving the reconstruction of a free field in Minkowski space from its representation in bulk subregions, see \cite{Chen:2023tvj, Chen:2024kuq}. Our formula passes a range of consistency checks. In particular, we show in Section \ref{sec:bulk-rec-AdS} that it is equivalent to the standard decomposition of a free field in AdS in terms of highest/lowest-weight wavefunctions and, consequently, it agrees with the AdS extrapolate dictionary. 
As a byproduct of this analysis, we obtained an AdS analog of the Kirchhoff--d'Adh\'emar formula in Minkowski space. In particular, we showed that a free AdS field of positive integer $\Delta$ can be reconstructed in terms of a dual boundary operator at a cut of the Lorentzian cylinder.

Our proposed bulk reconstruction formula is the AdS representation theoretic counterpart of the expansion of a free field in terms of Carrollian wavefunctions. In particular, time-ordered and anti-time-ordered bulk-to-boundary propagators with boundary points restricted to half the Lorentzian cylinder play the same role as the superpositions of positive and negative frequency plane waves defining the Carrollian wavefunctions \eqref{eq:Carroll-bases}. We showed in Section \ref{sec:flat-space-lim} that our formulas have well-defined flat space limits that yield plane waves of either massive or massless particles depending on the location of the boundary insertions, provided that $\Delta \rightarrow \infty$. On the other hand, at fixed $\Delta$ it was already observed in \cite{deGioia:2022fcn, Bagchi:2023fbj} that one obtains Carrollian wavefunctions. Our analysis clarifies the relation between these wavefunctions and representation theory: AdS$_4$ bulk-to-boundary propagators and Mink$_4$ Carrollian wavefunctions with the appropriate $i\epsilon$ prescriptions respectively span the positive and negative energy $\mathfrak{so}(3,2)$ and 4d Poincar\'e representations for fixed $\Delta$. 

In Section \ref{sec:boost-estates} we constructed the AdS analog of the decomposition of a free field in Mink$_4$ in terms of conformal primary wavefunctions. Specifically, we constructed wavefunctions for an $\mathfrak{so}(3,1)$ subalgebra of $\mathfrak{so}(3,2)$ for hyperbolic and null foliations of AdS$_4$ and showed that they reduce to conformal primary wavefunctions of massive and massless particles \cite{Pasterski:2016qvg,Pasterski:2017kqt} in Mink$_4$. 

Finally, in Section \ref{sec:flat-space-lim} we discussed some further applications of our results to flat space holography. In particular,  we clarified the relation between primary and shadow operators in CFT$_3$ and AdS fields in a bulk-point limit. This analysis allowed us to elaborate on the relation between Carrollian operators obtained via the extrapolate dictionary from these bulk fields and operators in a CFT$_3$. 

There are many interesting directions to be explored in the future. Easiest among these should be to generalize our analysis to higher spin fields. A first step in this direction will be to derive Lorentzian reconstruction formulas for gauge fields and gravitons in AdS. Equipped with these results, one should be able to define and compute Lorentzian counterparts of AdS-Witten diagrams. Upon incorporating interactions, these should match, at least order by order in perturbation theory, with  the analytic continuations of CFT correlators to the scattering regime \cite{Gary:2009ae,Penedones:2010ue,Raju:2012zr,Maldacena:2015iua}. Conversely, the orthogonality properties of the different wavefunctions with respect to the boundary KG inner product, should allow one to extract flat space observables directly from CFT correlators, by integrating each insertion against the appropriate kernel. Position and momentum space representations of these observables \cite{Susskind:1998vk,Hijano:2019qmi,Marotta:2024sce} should follow from our analysis after integrating each insertion against bulk-to-boundary propagators and Fourier transforms thereof with the appropriate time ordering. It is not immediately clear how flat space limit of Mellin CFT correlators \cite{Penedones:2010ue} arises from this perspective. 

It will be particularly interesting to study the decomposition of AdS observables in the basis of the $\mathfrak{so}(3,1)$ wavefunctions introduced in Section \ref{sec:boost-estates}. Since $\mathfrak{so}(3,1)$ is a common subalgebra of AdS$_{4}$, dS$_4$ and 4d flat spacetimes, one may expect their assembly into global observables to share some universal properties. We expect this decomposition to reveal the bulk counterparts of the $w_{1+\infty}$ algebras and their deformations found recently with different methods in the boundary CFTs \cite{Guevara:2021abz, Strominger:2021mtt, deGioia:2025mwt,Himwich:2025ekg,Strominger:2026yrh}. This problem will be addressed in future work. 

\section*{Acknowledgements}

We are grateful to Dionysios Anninos, Laurent Freidel, Damian Galante, Walker Melton, Maita Micol, Richard Myers, Romain Ruzziconi, Kostas Skenderis, Andrew Svesko, Tianli Wang and Yifei Zhao for helpful discussions. A.R. was in part supported by the Simons Foundation through the Emmy Noether Fellows Program at Perimeter Institute (1034867, Dittrich).
 We are grateful for the hospitality of Perimeter Institute where part of this work was carried out.

\appendix 

\section{Bulk reconstruction in higher-dimensional flat space}
\label{app:bulk-rec}

It is well known that massless free fields in 4d Minkowski space are completely determined by their value at a cut of $\mathscr{I}^{\pm}$ \cite{Penrose}. In this appendix we review this fact, keeping the spacetime dimension arbitrary. 

We would like to express the massless bulk field in $(d+1)$-dimensional Minkowski spacetime in terms of its value at $\mathscr{I}^{\pm}$. The boundary value is evaluated by taking the $r \rightarrow \infty$ limit of \eqref{eq:plane-waves} for fixed $u.$ Using the stationary phase approximation \eqref{eq:stat-phase}, 
\begin{equation}
    \int d^{d - 1}\Omega e^{\pm i \omega r (1-\hat{x} \cdot \Omega)} = i^{\frac{d - 1}{2}} \frac{(2\pi)^{\frac{d-1}{2}}}{(\pm \omega r)^{\frac{d-1}{2}}} = \frac{(2\pi)^{\frac{d-1}{2}}}{(\mp i \omega r)^{\frac{d-1}{2}}}
\end{equation}
we find
\begin{equation}
    \Phi(x) \left. \right|_{r\rightarrow \infty}= \frac{1}{2 (2\pi)^d}\int_0^{\infty} \frac{d\omega}{\omega} \omega^{d-1}  \left[ \frac{(2\pi)^{\frac{d-1}{2}}}{(i\omega r)^{\frac{d-1}{2}}} e^{-i\omega u} a(\omega\hat{x}) + \frac{(2\pi)^{\frac{d-1}{2}}}{(-i\omega r)^{\frac{d-1}{2}}}  e^{i\omega u} a^{\dagger}(\omega\hat{x})\right] + \mathcal{O}(r^{-\frac{d+1}{2}}).
\end{equation}
Consequently,
\begin{equation}
\begin{split}
   \int du e^{- i\omega' u}  \p_u \Phi(x) \left. \right|_{r\rightarrow \infty} &= i\frac{\omega'^{\frac{d - 1}{2} }}{2(2\pi)^{\frac{d - 1}{2}}r^{\frac{d-1}{2}}}  i^{\frac{d-1}{2}}a^{\dagger}(\omega' \hat{x}), \\
   \int du e^{i\omega' u}  \p_u \Phi(x) \left. \right|_{r\rightarrow \infty} &= -i \frac{\omega'^{\frac{d - 1}{2} }}{2(2\pi)^{\frac{d - 1}{2}}r^{\frac{d-1}{2}}}  i^{-\frac{d-1}{2}}a(\omega' \hat{x}). 
   \end{split}
\end{equation}
Substituting this formula back into the mode expansion \eqref{eq:plane-waves}, we obtain
\begin{equation}
\begin{split}
    \Phi(x) &= \frac{i}{(2\pi)^{\frac{d+1}{2}} } \int du \int_0^{\infty} d\omega \int d^{d-1}\Omega_p \omega^{\frac{d-3}{2} } \left[ i^{\frac{d - 1}{2}}e^{i(p\cdot x + \omega u)} \p_{u} \phi(u,\Omega_p) - i^{-\frac{d - 1}{2}} e^{-i(p\cdot x + \omega u)} \p_u\phi(u,\Omega_p)\right] + \cdots\\
    &= \frac{\Gamma(\frac{d-1}{2})}{(2\pi)^{\frac{d+1}{2}}} i^{d} \int du \int d^{d-1}\Omega_p\left(\frac{1}{\left(\hat{p}\cdot x + u + i\epsilon\right)^{\frac{d-1}{2}}} - \frac{(-1)^{d-1}}{\left(\hat{p}\cdot x + u - i\epsilon\right)^{\frac{d-1}{2}}} \right)\p_u\phi(u,\Omega_p) + \cdots,\\
    \end{split}
\end{equation}
where we defined
\begin{equation}
    \phi(u, \Omega) = \lim_{r\rightarrow \infty} r^{\frac{d-1}{2}} \Phi(x). 
\end{equation}
and $\cdots$ denote subleading corrections in $r^{-1}$.

We now see that for $d = 2k + 1$ (which corresponds to an even dimensional spacetime), we can write
\begin{equation}
\begin{split}
    \Phi(x) &= \frac{\Gamma(k)}{(2\pi)^{k+1} } i^{2k+1}\int du \int d^{d-1}\Omega_p \frac{(-1)^{k-1}}{\Gamma(k)}\p_u^{k-1}\left( \frac{1}{\hat{p}\cdot x + u + i\epsilon} - \frac{1}{\hat{p}\cdot x + u - i\epsilon}\right)\p_u\phi(u,\Omega_p), \\
    &= \frac{1}{(2\pi)^{k+1}} i^{2k+1} \int du \int d^{d-1}\Omega_p \left( \frac{1}{\hat{p}\cdot x + u + i\epsilon} -\frac{1}{\hat{p}\cdot x + u - i\epsilon}\right)\p_u^{k} \phi(u,\Omega_p), \\
    & = \frac{1}{(2\pi)^{k}}  i^{2k}\int d^{d-1}\Omega_p \p_u^{k} \phi(-\hat{p}\cdot x,\Omega_p),~~k \geq 1,
    \end{split}
\end{equation}
where in the last line we have used the delta function identity
\begin{equation}
    \lim_{\epsilon \rightarrow 0} \frac{\epsilon}{x^2 + \epsilon^2} = \pi \delta(x).
\end{equation}
Consequently, for all odd values of $d$, the value of the bulk field at a point $x$ can be reconstructed in terms of a derivative of the boundary field evaluated at a specific cut of $\mathscr{I}^+$. 
For the case of Mink$_{4}$, $d = 3~ (k = 1)$  we recover the Kirchhoff--d'Adh\'emar formula \cite{Donnay:2022wvx, Donnay:2023mrd}. 

In the case where $d = 2k$, the $u$ integrand has branch cuts and the $u$ integral fails to localize. We can see this explicitly by writing

\begin{equation}
\begin{split}
    \Phi(x) &= \frac{\Gamma(k-\frac{1}{2})}{(2\pi)^{k+\frac{1}{2}} } i^{2k} \int du \int d^{d-1}\Omega_p \sqrt{\pi}\frac{(-1)^{k-1}}{\Gamma(k - \frac{1}{2})}\p_u^{k-1}\left( \frac{1}{(\hat{p}\cdot x + u + i\epsilon)^{\frac{1}{2}}} +\frac{1}{(\hat{p}\cdot x + u - i\epsilon)^{\frac{1}{2}}}\right)\p_u\phi(u,\Omega_p), \\
    &= \frac{1}{\sqrt{2}(2\pi)^{k}} i^{2k} \int du \int d^{d-1}\Omega_p  \left( \frac{1}{(\hat{p}\cdot x + u + i\epsilon)^{\frac{1}{2}}} + \frac{1}{(\hat{p}\cdot x + u - i\epsilon)^{\frac{1}{2}}}\right)\p_u^k\phi(u,\Omega_p). \\
    \end{split}
\end{equation}
To proceed, we would need to know the analytic properties of $\p_u^k\phi$ as a function of $u$. It is nevertheless clear that the value of $\p_u^k\phi$ at a cut of $\mathscr{I}$ no longer determines the field at any bulk point. One instead needs to specify an infinite number of $\p_u$ derivatives of $\phi$ to reconstruct the bulk field. One can see this explicitly by considering a simple example in which $\phi(u, \Omega_p)$ admits a Taylor expansion in powers of $u$. This is presumably a simple consequence of the failure of the Huygen's principle in odd spacetime dimensions.

\section{Free scalars in AdS$_{d+1}$ and $\mathfrak{so}(d,2)$ representations}

In this appendix we review basic properties of free fields in AdS.

\subsection{Highest/lowest-weight representations of $\mathfrak{so}(d,2)$}
\label{app:highest-weight}

Particles in Lorentzian AdS$_{d+1}$ correspond to unitary lowest-weight representations of $\mathfrak{so}(d,2)$. States in these representations are labeled by their eigenvalues with respect to the Cartan subalgebra and the quadratic Casimir
\be 
\label{eq:Lorentz-Casimir}
\frac{1}{2}M_{AB} M^{AB} = \frac{1}{2}M_{ij}M^{ij} + \frac{1}{2}\left(P_i K_i + K_i P_i \right) + D^2.
\ee
Note that 
\be 
 \frac{1}{2} M_{ij} M^{ij} \equiv C_{\mathfrak{so}(d)}^2
\ee
is a quadratic Casimir of $\mathfrak{so}(d)$, while for fixed $i$ (no sum on $i$ below)
\be 
\frac{1}{2} \left(P_i K_i + K_i P_i \right) + D^2 \equiv C_{\mathfrak{sl}(2)}^{(i)}
\ee
is the quadratic Casimir of $\mathfrak{sl}(2,\mathbb{R})$. 

We define particles in (Lorentzian) AdS$_{d+1}$ as irreducible representations of $\mathfrak{so}(d,2)$. A special case of interest are gauge bosons which transform in the traceless, symmetric representation of $\mathfrak{so}(d)$, namely $[s, 0, \cdots, 0]$. In this case the eigenvalues of the quadratic Casimir are\footnote{For derivation, see for instance \cite{higher-d-sph-h}.} 
\be 
C^2_{\mathfrak{so}(d)} = -s (s+d - 2).
\ee
 The lowest weight representations are defined by
\be 
K_i |\Delta\rangle = 0, \quad \forall i
\ee
and hence
\be 
C_{\mathfrak{sl}(2)}^2 = \langle \Delta | \frac{1}{2}\left(P_i K_i + K_i P_i \right) + D^2| \Delta\rangle = -\Delta (d - \Delta) .
\ee

Putting everything together, we find the eigenvalues of the quadratic Casimir for $\mathfrak{so}(d,2)$ (in the case where the $\mathfrak{so}(d)$ reps are symmetric traceless)
\be 
\label{eq:sod2-Casimir}
C^2_{\mathfrak{so}(d,2)}(s, \Delta) = -s (s + d - 2) - \Delta (d - \Delta).
\ee
We conclude that gauge bosons in AdS are states in $\mathfrak{so}(2,d)$ representations 
\be 
|s; \vec{m} \rangle  \otimes |\Delta \rangle,
\ee
These representations can be realized on the space of functions on AdS$_{d+1}$ by introducing the $\mathfrak{so}(d)$ spherical harmonics 
$Y_{s, \vec{m}}$.

It will also be useful to understand the relation between \eqref{eq:Lorentz-Casimir} and the wave equation in embedding space. To this end, write
\be 
\begin{split}
 -M_{AB} M^{AB} &= \left( X_A \p_B - X_B \p_A \right) \left( X^A \p^B - X^B \p^A \right)\\
&= X^2 \p^2 + X\cdot \p - 2(d+2) X\cdot \p - 2 X_A X_B \p^A \p^B + X\cdot \p + X^2 \p^2\\
&= 2 X^2 \p^2 - 2(d+1) X\cdot \p - 2 X_A X_B \p^A \p^B\\
&= 2 X^2 \p^2 - 2 X\cdot \p \left( d + X \cdot \p \right)\\
&= -2 \ell^2 \p^2 - 2 X\cdot \p \left( d + X \cdot \p \right).
\end{split}
\ee
If we allow for $\ell = R$ to vary, which amounts to foliating $\mathbb{R}^{2,d}$ with AdS$_{d+1}$ slices of different curvature radii, we have
\be 
X\cdot \p = R \p_R
\ee
and 
\be 
\begin{split}
\frac{1}{2} M_{AB}M^{AB} &= R^2 \p^2 + R\p_R(d + R\p_R)\\
&=  R^2 \left(\p^2 + R^{-d-1}\p_R (R^{d+1}\p_R) \right).
\end{split}
\ee
When restricting to a constant $R$ hypersurface, 
\be 
  \frac{1}{2} M_{AB} M^{AB} = R^2 \p^2.
\ee
As a result, the AdS$_{d+1}$ Laplacian coincides with the quadratic Casimir \eqref{eq:Lorentz-Casimir} of $\mathfrak{so}(d,2)$. Combined with \eqref{eq:sod2-Casimir}, we conclude that the Klein--Gordon equation defines an $s = 0$ representation of $\mathfrak{so}(d,2)$ provided that 
\begin{equation}
    m^2 = -\Delta(d - \Delta).
\end{equation}

\subsection{Solving the Klein--Gordon equation in AdS}
\label{app:scalar}

For wavefunctions $\psi$ that diagonalize $D$ and $\Box_{d-1}$, the Klein--Gordon equation reduces to
\be 
\begin{split}
\left(-\Box_{AdS} + m^2\right) \psi &= 0 \implies\\
&-\psi'' + \frac{1-d}{\cos \rho \sin \rho} \psi' + \frac{\ell(\ell + d - 2)}{\sin^2\rho} \psi + \frac{m^2}{\cos^2\rho} \psi + \p_{\tau}^2 \psi = 0.
\end{split}
\ee
Further setting
\be 
\label{eq:full}
\psi = e^{i\omega \tau} f(\rho) Y_{\ell J}, \quad f = \sin^{\ell} \rho \cos^{\Delta} \rho F_{}(\rho), \quad m^2 = \Delta(\Delta - d),
\ee
we find that $F$ satisfies the following differential equation 
\be 
\label{eq:radial-eq}
(\ell + \Delta)^2 F(\rho) - 2 \left( \frac{-1 + d + \ell - \Delta}{\sin 2\rho} + (\ell + \Delta) \cot 2\rho\right)F'(\rho) - F''(\rho) - \omega^2 F(\rho) = 0.
\ee
This equation can be solved by first changing variables $z = \sin^2\rho$. In this case
\be 
\begin{split}
\frac{d}{d\rho} &= \sin 2 \rho \frac{d}{dz},\\
\frac{d^2}{d\rho^2} &= (2 - 4z) \frac{d}{dz} + 4 z(1 - z) \frac{d^2}{dz^2},
\end{split}
\ee
and \eqref{eq:radial-eq} simplifies to 
\be 
\left((\ell + \Delta)^2 - \omega^2 \right) F - 2(-1 + d + \ell - \Delta + (\ell + \Delta + 1)(1 - 2 z)) \frac{dF}{dz} - 4 z(1 - z) \frac{d^2F}{dz^2}.
\ee
This takes the form of a standard hypergeometric equation with
\be 
\begin{split}
a &= \frac{\ell + \Delta - \omega}{2} = -n, \\
b &= \frac{\Delta + \ell + \omega}{2} = \ell + \Delta + n, \\
c &= \frac{d}{2} + \ell.
\end{split}
\ee
The general solution to this equation is
\begin{equation}
\label{eq:global-radial}
    F(z) \propto  {}_2 F_1\left(a, b; c; z \right) +  i^{2 - d - 2\ell} r z^{\frac{1}{2}(2 - d - 2\ell)} {}_2 F_1\left(\tilde{a}, \tilde{b}; \tilde{c}; z \right),
\end{equation}
where $\tilde{a}, \tilde{b}, \tilde{c}$ are $a, b, c$ with $\ell \rightarrow - \ell + 2 - d$ and $r$ is an arbitrary constant. 

Demanding that the solution is regular at the origin as $z \rightarrow 0$ sets $r = 0$. 
For general $\omega$, the remaining hypergeometric function has both normalizable and non-normalizable fall-offs near the boundary. The non-normalizable mode is set to zero provided that $\omega$ is quantized as
\begin{equation}
    \omega\equiv \Delta + 2n + \ell.
\end{equation}
For these values of $\omega$, the solution reduces to
\be 
\label{eq:radial-wf}
F_{n\ell}(\rho) = {}_{2} F_1\left(a, b;c; \sin^2 \rho \right) =  {}_{2} F_1\left(-n , \ell + \Delta + n;\frac{d}{2} + \ell; \sin^2 \rho \right).
\ee
The conformal primary (lowest weight) solution/wavefunction is obtained by setting $n = 0$ in \eqref{eq:full}. 
For $d = 3$, we recover the expansion of a free field in AdS$_4$
\begin{equation}
\label{eq:AdS-free}
    \Phi(X) = \sum_{n, \ell, m} e^{-i(\Delta + 2n + \ell)\tau} \sin^{\ell} \rho \cos^{\Delta} \rho F_{n \ell}(\rho) Y_{\ell m}(\Omega) a_{n \ell m} + h.c. .
\end{equation}

\subsection{Null coordinates}
\label{app:null-foliation}

 In this appendix we consider a foliation \eqref{eq:null-foliation} of AdS$_{4}$ with null cones. We set $\ell = 1$.  The factors of $\ell$ can be restored by $r \rightarrow r/\ell$ and $g_{\mu\nu} \rightarrow \ell^2 g_{\mu\nu}$. In retarded coordinates, the metric is given by 
 \begin{equation}
     ds^2 = -(1+r^2) du^2 - 2 du dr + r^2 d\Omega_{d - 1}^2.
 \end{equation}
 These coordinates are related to the global AdS coordinates \eqref{eq:global-AdS} by
 \begin{equation}
 \label{eq:global-null}
     \sin^2 \rho = \frac{r^2}{1 + r^2}, \quad \tau = u + \arctan r.
 \end{equation}
  The generalization to any $d$ is straightforward. Our coordinates provide a global foliation of AdS with null slices, while those introduced in \cite{Alday:2024yyj} only cover a Poincar\'e patch. 
 
The AdS$_{4}$ Laplacian in the parameterization \eqref{eq:null-foliation} becomes
\begin{equation}
\label{eq:null-Laplacian}
\begin{split}
  \Box_{AdS_{4}} &=   \frac{1}{r^2}\left(\Box_{S^{2}} + r^2(1+r^2)\p_{r}^2 + 2r(1 + 2r^2) \p_{r} - 2r \p_{\tau} - 2 r^2 \p_{r}\p_{\tau} \right) \\
  &= \frac{1}{r^2}\left(\Box_{Mink_{4}} + 4r^3 \p_{r} + r^4 \p_{r}^2\right).
  \end{split}
\end{equation}
The AdS$_4$ wavefunctions in these coordinates are related to \eqref{eq:full} by the coordinate transform \eqref{eq:global-null}. We find
\begin{equation}
   \psi_{n\ell m}^+ = e^{-i\omega \tau} R(r) Y_{\ell m},
   \end{equation}
   where 
\begin{equation}
\label{eq:R}
\begin{split}
    R(r) &=  r^{-1 - \ell} e^{- i\omega \arctan(r)} (1 + r^2)^{\omega/2} \left[  c_1 \;{}_2F_1\left[ \frac{1}{2}(2 - \Delta - \ell + \omega), \frac{1}{2}(\Delta - 1 - \ell + \omega), \frac{1}{2} - \ell, -r^2 \right] \right. \\
    &\left. + c_2 \;r^{1 + 2\ell}\;{}_2F_1\left[\frac{1}{2}(3 - \Delta + \ell + \omega), \frac{1}{2}(\Delta + \ell + \omega), \frac{3}{2} + \ell, -r^2 \right] \right].
    \end{split}
\end{equation}
The hypergeometric functions appearing here are related to those in \eqref{eq:global-radial} by the identity
\begin{equation}
      {}_2 F_1(a, b; c; z) = (1 - z)^{-a} {}_2 F_1(a, c - b; c; \frac{z}{z - 1}).
\end{equation}
 The large $r$ expansion of \eqref{eq:R} takes the form
\begin{equation}
\begin{split}
    R(r) &\sim r^{-\Delta} \Gamma\left(\frac{3}{2} - \Delta\right) \left(\frac{c_1 \Gamma(\frac{1}{2} - \ell)}{\Gamma(\frac{1}{2}(2 - \Delta - \ell \pm  \omega))} + \frac{c_2\Gamma(\frac{3}{2} + \ell)}{\Gamma(\frac{1}{2}(3 - \Delta + \ell \pm  \omega))}  \right)\\
    &+ r^{-3+\Delta} \Gamma\left(-\frac{3}{2} +\Delta\right)\left(\frac{c_1 \Gamma(\frac{1}{2} - \ell)}{\Gamma(\frac{1}{2}( \Delta -1 - \ell \pm \omega))} + \frac{c_2\Gamma(\frac{3}{2} + \ell)}{\Gamma(\frac{1}{2}(\Delta + \ell \pm \omega))}  \right) + \cdots ,
    \end{split}
\end{equation}
where we introduced the notation
\begin{equation}
    \Gamma(x \pm \omega) \equiv \Gamma(x + \omega) \Gamma(x -\omega).
\end{equation}
Restoring the AdS radius, we see that $r \rightarrow 0$ limit corresponds to a flat space limit. Therefore demanding that the solutions are regular at the origin is equivalent to imposing that the flat space limit is well defined and hence $c_1 = 0$. We are left with a radial wavefunction that has both the normalizable and non-normalizable modes near the boundary \textit{unless}
\begin{equation}
    \omega = \Delta + \ell + 2n, \quad n \in \mathbb{N}. 
\end{equation}

\subsection{Rindler coordinates}
\label{app:Rindler}

We follow the notation of \cite{Sugishita:2022ldv}. We consider the embedding of a Rindler wedge of AdS$_4$ into $\mathbb{R}^{3,2}$
\begin{equation}
\begin{split}
X^{0} &= \xi_R \sinh t_R, \\
X^1 &= \xi_R \cosh t_R, \\
X^2 &= \sqrt{1 + \xi_R^2} \sinh \chi_R \cos\phi, \\
X^3 &= \sqrt{1 + \xi_R^2} \sinh \chi_R \sin \phi, \\
X^4 &= \sqrt{1 + \xi_R^2} \cosh\chi_R.
\end{split}
\end{equation}
The global coordinates $(\tau, \rho, \theta, \varphi)$ are related to the Rindler coordinates $(t, \xi, \chi, \phi)$ by
\begin{equation}
\label{eq:Rindler}
\begin{split}
\tan \tau &= \frac{X^0}{X^4} = \frac{\xi_R \sinh t_R}{\sqrt{1 + \xi_R^2} \cosh \chi_R}, \\
\frac{1}{\cos^2\rho} &= (X^0)^2 + (X^4)^2 = \xi_R^2 (\sinh t_R)^2 + (1 + \xi_R^2) \cosh^2\chi_R,\\
\tan \theta &= \frac{X^3}{X^1 \sin \phi} = \frac{\sqrt{1 + \xi_R^2} \sinh \chi_R}{\xi_R \cosh t_R }, \quad \varphi = \phi.
\end{split}
\end{equation}
One can proceed following \cite{Sugishita:2022ldv} to construct the Rindler wavefunctions. 

The analysis therein can be simplified by observing that
\begin{equation}
\eta_{AB} X^A X^B = \ell^2
\end{equation}
is preserved by the transformation
\begin{equation}
\label{eq:transform}
X^1 \rightarrow i X^4, \quad X^4 \rightarrow iX^1.
\end{equation}
Under this transformation, the two sets of coordinates will be instead related by the formal replacements
\begin{equation}
\label{eq:formal-ct}
\begin{split}
\frac{1}{\cos \rho} &= i\xi, \\
\tau &= -it, \\
\theta &= i \chi, \quad  \phi = \varphi.
\end{split}
\end{equation}
The equations of motion in AdS$_4$ are manifestly invariant under \eqref{eq:transform} and hence the solutions/wavefunctions in the Rindler region can be simply obtained from the global wavefunctions \eqref{eq:full},\eqref{eq:radial} by the complexified coordinate transformations \eqref{eq:formal-ct}. 

Note that neither the radial quantum number $n$ nor the angular momentum quantum number $\ell$ will be quantized anymore. In particular, the Laplacian eigenvalue equation on $S^2$ 
\begin{equation}
\nabla_{S^2} Y_{\ell,m} = -\ell (\ell + d - 2) Y_{\ell, m}
\end{equation}
gets replaced by the Laplacian on $H^2$
\begin{equation}
\nabla_{H^2} Y_{\lambda, m} = -\left(\lambda^2 + \frac{(d - 2)^2}{4}\right) Y_{\lambda,\mu}.
\end{equation}
As a result we have the identification
\begin{equation}
\lambda^2 = -\left(\ell^2 + \ell(d - 2) + \frac{(d - 2)^2}{4}\right) =  \left(\ell + \frac{d - 2}{2} \right)^2 \implies \lambda =\pm i \left(\ell + \frac{d - 2}{2}\right)
\end{equation}
or equivalently
\begin{equation}
\ell = \mp i \lambda  - \frac{d - 2}{2}. 
\end{equation}
For the radial quantum number we have
\begin{equation}
i\omega = \Delta + 2n + \ell , \quad n \in \mathbb{R}. 
\end{equation}
From here we deduce that 
\begin{equation}
n = \frac{i\omega - \Delta - \ell}{2}.
\end{equation}

The global positive energy solutions take the form
\begin{equation}
\label{eq:global-wf}
\varphi_G = e^{-i(\Delta + 2n + \ell)\tau} \sin^{\ell} \rho \cos^{\Delta} \rho {}_2 F_1 \left(-n, \ell + \Delta + n, \frac{d}{2} + \ell; \sin^2\rho \right) Y_{\ell,m}.
\end{equation}
In terms of the coordinates \ref{eq:formal-ct}, we have
\begin{equation}
\sin \rho = \frac{\sqrt{1 + \xi^2}}{\xi} \implies \sin^2 \rho = \frac{1 + \xi^2}{\xi^2}.
\end{equation}
We finally use the hypergeometric function identity
\begin{equation}
{}_2 F_1 \left(-n,a; a- b + 1, \frac{1}{1 - z} \right) = \frac{(1 - b)_n}{(a - b + 1)_n} (1  - z)^a {}_2 F_1\left(a,b;b-n;z \right)
\end{equation}
to write
\begin{equation}
\begin{split}
 {}_2 F_1 \left(-n, \ell + \Delta + n, \frac{d}{2} + \ell; \sin^2\rho \right) &= \frac{(\frac{d}{2} - \Delta - n)_n}{(\frac{d}{2} + \ell)_n} \sin^{-2(\ell + \Delta + n)}\rho\\
 &\times {}_2 F_1\left(\ell + \Delta + n, \Delta+n+1 - \frac{d}{2}; \Delta+ 1 - \frac{d}{2}; -\frac{\cos^2\rho}{\sin^2\rho} \right)\\
 &= \frac{\Gamma(\frac{d}{2} - \Delta)\Gamma(\frac{d}{2} + \ell)}{\Gamma(\frac{d}{2} -\Delta - n)\Gamma(\frac{d}{2} + \ell +n)} \sin^{-2(\ell + \Delta + n)}\rho\\
 &\times {}_2 F_1\left(\ell + \Delta + n, \Delta+n+1 - \frac{d}{2}; \Delta+ 1 - \frac{d}{2}; -\frac{\cos^2\rho}{\sin^2\rho} \right)\\
 &= \frac{\Gamma(\frac{d}{2} - \Delta)\Gamma(\frac{d}{2} + \ell)}{\Gamma(\frac{d}{2} -\Delta - n)\Gamma(\frac{d}{2} + \ell +n)} (1 + \xi^2)^{\frac{-i\omega - \Delta  - \ell}{2}} \xi^{i\omega + \Delta + \ell}\\
 &\times {}_2 F_1\left(\frac{i\omega + \Delta+\ell}{2},\frac{i\omega + \Delta - \ell}{2} + 1 - \frac{d}{2}; \Delta+ 1 - \frac{d}{2}; \frac{1}{1 + \xi^2} \right).\\
 \end{split}
\end{equation}
Introducing
\begin{equation}
\Delta - \frac{d}{2} = \nu,
\end{equation}
we see that
\begin{equation}
n = \frac{i\omega + i\lambda - \nu - 1}{2}.
\end{equation}
Substituting for $\ell$ in terms of $\lambda$, we obtain
\begin{equation}
\begin{split}
\varphi_G &\rightarrow  \frac{\Gamma(\frac{d}{2} - \Delta)\Gamma(- i\lambda + 1)}{\Gamma(\frac{d}{2} -\Delta - n)\Gamma( -i\lambda + 1 +n)}  e^{-i\omega t}  \xi^{-\ell}(i\xi)^{-\Delta}(1 + \xi^2)^{\frac{-i\omega - \Delta}{2}} \xi^{i\omega + \Delta + \ell}\\
&\times {}_2 F_1\left(\frac{i\omega -i\lambda +\nu + 1}{2},\frac{i\omega + i\lambda  + \nu+ 1}{2}; \nu + 1; \frac{1}{1 + \xi^2} \right)\\
&= i^{-\Delta}  \frac{\Gamma(-\nu)\Gamma(- i\lambda + 1)}{\Gamma(\frac{-i\omega - i\lambda - \nu + 1}{2})\Gamma(\frac{i\omega - i\lambda - \nu + 1}{2})}  e^{-i\omega t}  (1 + \xi^2)^{\frac{-i\omega - \Delta}{2}} \xi^{i\omega }\\
&\times {}_2 F_1\left(\frac{i\omega -i\lambda +\nu + 1}{2},\frac{i\omega + i\lambda  + \nu+ 1}{2}; \nu + 1; \frac{1}{1 + \xi^2} \right).\\
\end{split}
\end{equation}
These are precisely the Rindler wavefunctions constructed in \cite{Sugishita:2022ldv}.

It is interesting to compare the Rindler wavefunctions to the wavefunctions in null coordinates described in Appendix \ref{app:null-foliation}. The former are obtained from the latter by setting
\begin{equation}
    r^2 = -\xi^2 - 1 \implies r = i \sqrt{1 + \xi^2}.
\end{equation}
The Rindler horizon is at $\xi = 0$ or equivalently $r = i$. This singularity is manifest in the $c_1 = 0$ wavefunctions in \eqref{eq:R} through a divergence $\sim (\widetilde{r} - 1)^{-\omega/2}$ at $\widetilde{r} = 1$, after analytically continuing $r \rightarrow i \widetilde{r}$.

\section{Inner products of AdS bulk-to-boundary propagators}

In this appendix we decompose the (anti-)time-ordered bulk-to-boundary propagators in terms of the positive and negative energy wavefunctions \eqref{eq:AdS-wf}. This allows us to compute inner products thereof on bulk hypersurfaces in AdS and its boundary. 

\subsection{Mode expansion of $G_{\Delta}^{\pm}$}
\label{eq:btb-exp}

The coefficients in the expansion of the bulk-to-boundary propagator  \eqref{eq:scalar-btb} in terms of Legendre polynomials are obtained by evaluating 
\begin{equation}
\begin{split}
c_{\ell} = \int_0^{\pi} d\theta \sin \theta & \frac{(\cos \rho)^{\Delta}}{\left(\cos(\tau - \tau_p) - \sin \rho \cos\theta \right)^{\Delta}}P_{\ell}(\cos\theta) \\
&= \int_{-1}^{1} dx  \frac{(\cos \rho)^{\Delta}}{\left(\cos(\tau - \tau_p) - \sin \rho x \right)^{\Delta}} \frac{1}{2^{\ell} \ell!} \frac{d^{\ell}}{dx^{\ell}} \left[(x^2 - 1)^{\ell} \right]\\
&= \int_{-1}^{1} dx (-1)^{\ell} \frac{\Gamma(\Delta + \ell)}{ \Gamma(\Delta)} \frac{(\cos\rho)^{\Delta} (\sin\rho)^{\ell}}{\left(\cos(\tau - \tau_p) - \sin \rho x \right)^{\Delta + \ell}} \frac{(x^2 - 1)^{\ell}}{2^{\ell} \ell!}\\
&= \sqrt{\pi} \frac{\Gamma(\Delta + \ell)}{2^{\ell}  \ell! \Gamma(\Delta)} \frac{\Gamma(1 + \ell)}{\Gamma(\frac{3}{2} + \ell)} (\cos(\tau - \tau_p))^{-\Delta - \ell} (\cos\rho)^{\Delta} (\sin\rho)^{\ell}\\
&\times {}_2F_1\left(\frac{\Delta + \ell}{2}, \frac{1 + \Delta + \ell}{2}; \frac{3}{2} + \ell; \frac{\sin^2 \rho}{\cos^2(\tau - \tau_p)} \right).
\end{split}
\end{equation}
Formally, this result holds only for $|\sin\rho| \leq |\cos(\tau - \tau_p)|$ since the infinite sum defining the hypergeometric function diverges otherwise. Later, when we compute the bulk inner products of bulk-to-boundary propagators, we will have to integrate over values of $\rho$ that violate this inequality. We haven't proven that exchanging the summation with the integration is allowed outside the domain of convergence of the sum, but the results and multiple consistency checks that we performed suggest that this holds. We hope it can be formally justified by analytic continuation in the future. 

To proceed, we use the definition of the hypergeometric function
\begin{equation}
{}_2F_1\left(a, b; c; z \right) = \sum_{n = 0}^{\infty} \frac{\Gamma(a + n) \Gamma(b + n) \Gamma(c)}{n! \Gamma(a) \Gamma(b) \Gamma(c + n)} z^n,
\end{equation}
and expand
\begin{equation}
(\cos(\tau  - \tau_p) )^{-\Delta - \ell - 2n} =  e^{-i(\tau - \tau_p)(\Delta + \ell + 2n)} 2^{\Delta + \ell + 2n} \sum_{k = 0}^{\infty} \left(\begin{matrix}
-\Delta - \ell - 2n\\
k
\end{matrix} \right) e^{-2ik(\tau - \tau_p)}.
\end{equation}
Then
\begin{equation}
\begin{split}
c_{\ell} &= \sum_{n, k = 0}^{\infty} \sqrt{\pi}  \frac{2^{\Delta + 2n}\Gamma(\Delta + \ell)\Gamma(1 + \ell)\Gamma(\frac{\Delta + \ell}{2} + n) \Gamma(\frac{1 + \Delta + \ell}{2} + n) \Gamma(\frac{3}{2} + \ell) \Gamma(-\Delta - \ell - 2n + 1)}{n! \ell! \Gamma(\Delta)\Gamma(\frac{3}{2} + \ell)\Gamma(\frac{\Delta + \ell}{2}) \Gamma(\frac{1 + \Delta + \ell}{2}) \Gamma(\frac{3}{2} + \ell + n) \Gamma(k + 1)\Gamma(-\Delta - \ell - 2n + 1 - k)}\\
&\times (\cos\rho)^{\Delta} (\sin\rho)^{\ell + 2n} e^{-i(\tau - \tau_p)(\Delta + \ell + 2n + 2k)}\\
&= \sum_{n, k = 0}^{\infty} \sqrt{\pi} (-1)^{ k}  \frac{2^{\Delta} \Gamma(\Delta + \ell + 2n + k)}{n! \Gamma(\Delta)\Gamma(\frac{3}{2} + \ell + n) \Gamma(k + 1)} (\cos\rho)^{\Delta} (\sin\rho)^{\ell + 2n} e^{-i(\tau - \tau_p)(\Delta + \ell + 2n + 2k)}
\end{split}
\end{equation}
where we have used the Euler identity 
\begin{equation}
\Gamma(z) \Gamma(1 - z) = \frac{\pi}{\sin\pi z}
\end{equation}
and the  Legendre duplication formula 
\begin{equation}
\Gamma\left(\frac{1}{2} + z\right) \Gamma(z) = \sqrt{\pi} 2^{1 - 2z}\Gamma(2z)
\end{equation}
to simplify the coefficient. 
We now let $k \rightarrow k - n$ and perform the sum on $n$ which ranges from $0$ to $k$ after switching the order of sums. Using
\begin{equation}
\sum_{n = 0}^k (-1)^n \frac{\Gamma(\Delta + \ell + k + n)}{n! (k - n)! \Gamma(\frac{3}{2} + \ell + n)} (\sin \rho)^{2n} = \frac{\Gamma(\Delta + \ell + k)}{\Gamma(\frac{3}{2} + \ell) k!} {}_2 F_1\left(-k, \Delta + \ell + n; \frac{3}{2} + \ell; \sin^2 \rho\right)
\end{equation}
we obtain
\begin{equation}
c_{\ell} = \sum_{k = 0}^{\infty} \sqrt{\pi}  (-1)^{k}\frac{2^{\Delta}\Gamma(\Delta + \ell + k)}{k! \Gamma(\Delta)\Gamma(\frac{3}{2} + \ell)} (\cos\rho)^{\Delta} (\sin\rho)^{\ell} e^{-i(\tau - \tau_p)(\Delta + \ell + 2k)} {}_2 F_1\left(-k, \Delta + \ell + n; \frac{3}{2} + \ell; \sin^2 \rho\right).
\end{equation}

\subsubsection{Agreement with previous work}
\label{eq:Giddings}

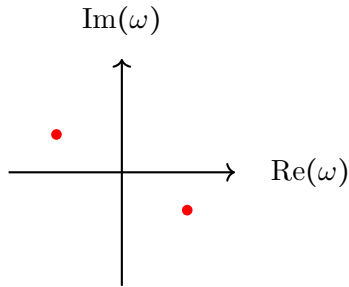
\begin{figure}[h]
\begin{center}
\begin{tikzpicture}
\draw [->,black, thick] (-1.5, 0) -- (1.5,0);
\draw [->,black, thick] (0, -1.5) -- (0, 1.5);
\fill[red]  (-0.866,0.5) circle[radius=2pt];
\fill[red]  (0.866,-0.5) circle[radius=2pt];
\node at (2.5,0){$\text{Re}(\omega)$};
\node at (0,2) {$\text{Im}(\omega)$};
\end{tikzpicture}
\caption{Location of poles in the integrand of \eqref{eq:giddings-btb}.}
\label{fig:poles}
\end{center}
\end{figure}

 In this appendix we compare the result \eqref{eq:time-ordered-btb} obtained by direct mode expansion with the representation of the AdS bulk-to-boundary propagator stated in eq. (A.31) of \cite{Giddings:1999jq},
\begin{equation}
\label{eq:giddings-btb}
K_{B\partial}(b,x')=2\nu R^{d-1}\int \frac{d\omega}{2\pi} \sum_{nl\vec{m}}e^{i\omega(\tau-\tau')} \frac{k_{nl}Y_{l\vec{m}}^*(\hat{e})\phi_{nl\vec{m}}(\vec{x}')}{\omega_{nl}^2-\omega^2-i\epsilon}.
\end{equation}
Here 
\begin{equation}
\label{eq:k}
k_{nl}=(-1)^nA_{nl} \frac{\Gamma(n+\nu+1)}{n!\Gamma(\nu+1)}, 
\end{equation}
\begin{equation}
\label{eq:A}
A_{nl}^2=\frac{2 \omega_{nl}}{R^{d-1}} \frac{n!\Gamma(n+2h_{+}+l)}{\Gamma\left( n+l+\frac{d}{2} \right)\Gamma(n+\nu+1)}
\end{equation}
and
\begin{equation}
    h_+ = \frac{d + 2\nu}{4} = \frac{\Delta}{2}, \quad \nu =  \Delta - \frac{d}{2}.
\end{equation}
Furthermore, $\omega_{n\ell} > 0$ and is given in \eqref{eq:omega}. 
The poles are located at $\omega=\pm\omega_{nl}\mp i\epsilon$, as shown in Figure \ref{fig:poles}. 

For $\tau-\tau'>0$ we take an upper half closed contour, picking up the pole at $\omega=-\omega_{nl}$, while for $\tau-\tau'<0$ we take the lower half closed contour, picking up the pole at $\omega=\omega_{nl}$.
After integration, we get 
\begin{equation}
K_{B\partial}(b,x')= i\nu R^{d-1}\sum_{nl\vec{m}} \frac{k_{nl}}{\omega_{nl}}Y_{l\vec{m}}^*(\hat{e})\phi_{nl\vec{m}}(\vec{x}')\left( e^{i\omega_{nl}(\tau-\tau')} \Theta(\tau'-\tau)+e^{-i\omega_{nl}(\tau-\tau')} \Theta(\tau-\tau')\right).
\end{equation}
Setting $d = 3$ and using the definition of $\phi_{nl m}$ (eq. A.18 of \cite{Giddings:1999jq}), we find
\begin{equation}
\label{eq:Giddings-btb-semi}
\begin{split}
    K_{B \partial}(b,x') &= i\sum_{n, \ell, m} (-1)^n \frac{\Gamma(n + \Delta + \ell)}{ {\color{red}\sqrt{\omega_{nl}}}\Gamma(\Delta - \frac{3}{2})\Gamma(n + \ell + \frac{3}{2})} Y_{\ell,m}^*(\hat{e}) Y_{\ell,m}(\vec{e}') P_n^{\ell + \frac{1}{2}, \nu}(1 - 2 \sin^2\rho)\\
    &\times \left( e^{i\omega_{nl}(\tau-\tau')} \Theta(\tau'-\tau)+e^{-i\omega_{nl}(\tau-\tau')} \Theta(\tau-\tau')\right),
    \end{split}
\end{equation}
where $P_n^{\ell + \frac{1}{2},\nu}$ are Jacobi polynomials, which can be also written as 
\begin{equation}
     P_{n}^{\ell+\frac{1}{2},\nu}(1 - 2 \sin^2 \rho) \equiv \left(\begin{matrix}
        n + \ell + \frac{1}{2}\\
        n
    \end{matrix} \right){}_2F_1\left(-n, n + \ell  + \Delta; \ell + \frac{3}{2}; \sin^2\rho \right).
\end{equation}
As a result, \eqref{eq:Giddings-btb-semi} simplifies to 
\begin{equation}
\begin{split}
    K_{B \partial}(b,x') &= i\sum_{n, \ell} (-1)^n \frac{\Gamma(n + \Delta + \ell)}{ {\color{red}\sqrt{\omega_{nl}}}n! \Gamma(\Delta - \frac{3}{2}) \Gamma(\ell + \frac{3}{2})} \frac{2\ell + 1}{4\pi} P_{\ell}(\vec{e}\cdot \vec{e}') {}_2F_1\left(-n, n + \ell  + \Delta; \ell + \frac{3}{2}; \sin^2\rho \right)\\
    &\times \left( e^{i\omega_{nl}(\tau-\tau')} \Theta(\tau'-\tau)+e^{-i\omega_{nl}(\tau-\tau')} \Theta(\tau-\tau')\right)
    \end{split}
\end{equation}
where we used \eqref{eq:LP-SH} to perform the sum on $m$. This formula agrees up to the factors of $\sqrt{\omega_{n\ell}}$ highlighted in red and a difference in normalization of bulk-to-boundary propagator -- see eq. A.33 of \cite{Giddings:1999qu} -- with the mode expansion of the (time-ordered) bulk-to-boundary propagator \eqref{eq:time-ordered-btb}. We suspect that this factor arises due to an inconsistency in the normalization of the wavefunctions $\phi_{n\ell m}$ and the definitions \eqref{eq:k}-\eqref{eq:A} in \cite{Giddings:1999jq}. 

\subsection{Bulk inner product of AdS wavefunctions}
\label{sec:AdS-prop}

In this Appendix we prove \eqref{eq:wf-ip-main}. Using the definitions \eqref{eq:AdS-wf} of the wavefunctions, we find
\begin{equation}
\begin{split}
\langle \psi^\pm_{n\ell 0}(\tau - \tau_1, \rho, \Omega \cdot \Omega_1), & \psi^{\pm}_{k \ell' 0}(\tau - \tau_2, \rho, \Omega \cdot \Omega_2)\rangle = \pm e^{\pm i\omega_n(\tau - \tau_1)} e^{\mp i\omega_k(\tau - \tau_2)} \\
&\times \int_0^{\pi/2} d\rho (\tan\rho)^{2} \int_{S^{2}} d \Omega \left[(\omega_k + \omega_{n}) f_{n\ell}(\rho) f_{k\ell'}(\rho) Y_{\ell 0}^*(\Omega \cdot \Omega_1) Y_{\ell' 0}(\Omega \cdot \Omega_2) \right].
\end{split}
\end{equation}
We now use the addition theorem for spherical harmonics 
\begin{equation}
\label{eq:LP-SH}
P_{\ell}(\Omega \cdot \Omega_i) = \frac{4\pi}{2\ell + 1}\sum_{m = -\ell}^{\ell} Y_{\ell m}(\Omega_i) Y^*_{\ell m}(\Omega), \quad \forall ~~\Omega, \Omega_i \in \mathbb{R}^2, \quad \Omega^2 = \Omega^2_i = 1
\end{equation}
to evaluate the angular integral. We find
\begin{equation}
\label{eq:Y-prod-int}
\begin{split}
 \int_{S^{2}} d \Omega Y_{\ell 0}^*(\Omega \cdot \Omega_1) Y_{\ell' 0}(\Omega \cdot \Omega_2) &= \sqrt{(2\ell + 1)(2\ell' + 1)} (4\pi)^{-1}  \int_{S^{2}} d \Omega  P^*_{\ell}(\Omega \cdot \Omega_1)P_{\ell'}(\Omega \cdot \Omega_2)\\
 &= \frac{4\pi}{\sqrt{(2\ell + 1) (2\ell' + 1)}} \sum_{m = -\ell}^{\ell} \sum_{m' = -\ell'}^{\ell'} \int_{S^{2}} d \Omega Y^*_{\ell m}(\Omega_1) Y_{\ell m}(\Omega) Y_{\ell' m'}(\Omega_2) Y^*_{\ell' m'}(\Omega)\\
 &= \frac{4\pi}{2\ell + 1} \sum_{m = -\ell}^{\ell} Y^*_{\ell m} (\Omega_1) Y_{\ell m}(\Omega_2) \delta_{\ell \ell'} = P_{\ell}(\Omega_1 \cdot \Omega_2) \delta_{\ell \ell'}\\
 &= \sqrt{\frac{4\pi}{2\ell + 1}} Y_{\ell 0}(\Omega_1 \cdot \Omega_2) \delta_{\ell\ell'},
 \end{split}
\end{equation}
where we used orthogonality of spherical harmonics. Furthermore, the radial integral can be computed directly using \eqref{eq:Jacobi-ortho} and \eqref{eq:wf-ip-fin}. As a result, we find 
\begin{equation}
\begin{split}
\langle \psi^\pm_{n\ell 0}(\tau - \tau_1, \rho, \Omega \cdot \Omega_1), & \psi^\pm_{k \ell' 0}(\tau - \tau_2, \rho, \Omega \cdot \Omega_2)\rangle =   \pm\sqrt{\frac{4\pi}{2\ell + 1}} N_{n \ell}^{d=3} ~ e^{\mp i\omega_n (\tau_1 - \tau_2)}Y_{\ell 0}(\Omega_1 \cdot \Omega_2)  \delta_{nk} \delta_{\ell \ell'}.
\end{split}
\end{equation}

\subsection{Shadows of bulk-to-boundary propagators}
\label{sec:btb-shadow}

In this appendix we compute the shadow transform of the bulk to boundary propagator. Given the definition of the shadow transform of boundary operators, we expect 
\begin{equation}
\label{eq:shadow-ie}
   \widetilde{G}^{\pm i\epsilon}_{3 - \Delta}(X;P)  \propto \int_I d^3 P' \frac{1}{(-P\cdot P' \pm i\epsilon)^{3 - \Delta}} G^{\rm T.O.}_{\Delta}(X;P') \propto G_{3 - \Delta}(X;P).
\end{equation}
These shadow transforms can be computed using the mode expansions \eqref{eq:time-ordered-btb}, \eqref{eq:mode-exp-ato-btb}. They take a simple form consistent with the expectation \eqref{eq:shadow-ie} in the case where $P$ and $P'$ lie on opposite sides of $\Sigma_{\tau}$, where $\tau$ is the time of the bulk point at which $G_{\Delta}^{\rm T.O.}(X;P)$ is evaluated. Furthermore, we will see that in this case the result is only non-vanishing in the case where the shadow kernel is defined with the $-i\epsilon$ prescription.

To this end, we use the mode expansions of the bulk-to-boundary and boundary-to-boundary propagators
\begin{equation}
\label{eq:to-prop}
    G_{\Delta}^{\rm T.O.}(X;P) = \sum_{k = 0}^{\infty} \sum_{\ell = 0}^{\infty}  \Big[b_{k\ell}^+ \psi^{+}_{k \ell 0}(\tau - \tau_{p}, \rho, \Omega \cdot \Omega_{p})\Theta(\tau-\tau_{p})+ b_{k\ell}^-\psi^{-}_{k \ell 0}(\tau - \tau_{p}, \rho, \Omega \cdot \Omega_{p})\Theta(\tau_{p}-\tau) \Big]
\end{equation}
and 
\begin{equation}
\label{eq:bdry-bdry}
    G_{\Delta}^{\rm T.O.}(P';P) = \sum_{k = 0}^{\infty} \sum_{\ell = 0}^{\infty} \Big[b_{k\ell}^+ \widehat{\psi}^{+}_{k \ell 0}(\tau_{p'} - \tau_p, \Omega_{p'} \cdot \Omega_p)\Theta(\tau_{p'}-\tau_{p})+ b_{k\ell}^-\widehat{\psi}^{-}_{k \ell 0}(\tau_{p'} - \tau_p, \Omega_{p'} \cdot \Omega_p)\Theta(\tau_{p}-\tau_{p'}) \Big],
\end{equation}
where 
\begin{equation}
    b_{k\ell}^{\pm} = \pi \sqrt{2 \ell + 1} (-1)^{\pm k}\frac{2^{\Delta}\Gamma(\Delta + \ell + k)}{k! \Gamma(\Delta)\Gamma(\frac{3}{2} + \ell)}
\end{equation}
and $\widehat{\psi}_{k\ell 0}$ are the boundary values of the wavefunctions given in \eqref{eq:hat-psi}. They can be simplified using the Gauss formula
\begin{equation}
    {}_2F_1\left(-n,\Delta + \ell  + n, \frac{3}{2} + \ell; 1\right) = \frac{\Gamma(\frac{3}{2} + \ell)\Gamma(\frac{3}{2} - \Delta)}{\Gamma(\frac{3}{2} + \ell + n)\Gamma(\frac{3}{2} - \Delta - n)}.
\end{equation}
Note that \eqref{eq:bdry-bdry} and its complex conjugate define the shadow kernels with the $i\epsilon$ and $-i\epsilon$ prescriptions respectively.

We start with the $+i \epsilon$ case and consider a shadow integration region $I = \bar{B}$ defined by $-\pi + \tau < \tau_{p'} < \tau$, in which case only the first term in \eqref{eq:to-prop} contributes. Further taking $\tau_p > \tau$ we see that only the second term in \eqref{eq:bdry-bdry} contributes on the support of $\Theta(\tau - \tau_{p'})$ (since $\tau_p > \tau > \tau_{p'}$). As a result, we find 
\begin{equation}
\begin{split}
   \widetilde{G}^{+i\epsilon}_{3 - \Delta}(X;P) &=\sum_{k,n = 0}^{\infty} \sum_{\ell,\ell' = 0}^{\infty} \int_{-\pi + \tau}^{\tau} d\tau_{p'} \int d\Omega_{p'} \pi^2 \sqrt{(2 \ell + 1)(2\ell' + 1)}(-1)^{k-n}\frac{\Gamma(\Delta + \ell + k)}{k! \Gamma(\Delta)\Gamma(\frac{3}{2} + \ell)}  \\
   &\times \frac{2^{3}\Gamma(3 - \Delta + \ell' + n)}{n! \Gamma(3-\Delta)\Gamma(\frac{3}{2} + \ell')}\frac{\Gamma(\frac{3}{2} + \ell')\Gamma(-\frac{3}{2} + \Delta )}{\Gamma(\frac{3}{2} + \ell' + n)\Gamma(-\frac{3}{2} + \Delta - n)} e^{-i(\omega_k^\Delta\tau+\omega_n^{3-\Delta}\tau_p)} e^{i(\omega_k^{\Delta} + \omega_n^{3 - \Delta})\tau_{p'}}\\&\times\Theta(\tau_p - \tau)
    (\cos\rho)^{\Delta} (\sin \rho)^{\ell} {}_{2}F_1(-k,\Delta + \ell + k, \frac{3}{2} + \ell; \sin^2\rho) Y_{\ell 0}(\Omega\cdot \Omega_{p'
   }) Y_{\ell' 0}(\Omega_{p'}\cdot \Omega_{p}).
   \end{split}
\end{equation}
We can now do the $\Omega_{p'}$ integral using \eqref{eq:Y-prod-int} and the $\tau_{p'}$ integral, which sets 
\begin{equation}
    \omega_n^{3-\Delta} = -\omega_k^{\Delta} \implies \frac{3}{2} + n + k + \frac{\ell}{2} + \frac{\ell'}{2} = 0.
\end{equation}
This condition is never obeyed, so we see that the  $+i\epsilon$ prescription in the shadow Kernel \eqref{eq:shadow-ie} leads to a vanishing result. It is easy to see that the same holds for the shadow integration region $B$ defined by $\tau < \tau_{p'} < \tau + \pi$, provided that $\tau_p < \tau$.

If we instead use the anti-time-ordered $i\epsilon$ prescription in the definition of the shadow transform, or equivalently, the complex conjugate of the shadow kernel, we get a non-vanishing result, as we now show. Evaluating the shadow in $\bar{B}$, the $\tau_{p'}$ integral now gives
\begin{equation}
    \omega_n^{3 - \Delta} = \omega_k^{\Delta} \implies  \Delta - \frac{3}{2} + k - n + \frac{\ell}{2} - \frac{\ell'}{2} = 0.
\end{equation}
We obtain, provided that $\Delta - \frac{3}{2} \in \mathbb{Z}$ and $\Delta - \frac{3}{2} \leq 0$,
\begin{equation}
\label{eq:shadow-integral}
    \begin{split}
         \widetilde{G}^{-i\epsilon}_{3 - \Delta}(X;P) &= \sum_{n = 0}^{\infty} \sum_{\ell = 0}^{\infty}\frac{2^2 \Gamma(\Delta - \frac{3}{2})}{\Gamma(\Delta)\Gamma(3 - \Delta)} \frac{\Gamma(3 - \Delta + \ell + n)}{n! \Gamma(\frac{3}{2} + \ell)} \frac{ 4 \pi^{3} \pi^{1/2} \sqrt{2 \ell + 1} (-1)^{\frac{3}{2} - \Delta}}{\Gamma(n - \Delta + \frac{5}{2})\Gamma(\Delta - n - \frac{3}{2})}  Y_{\ell 0}(\Omega \cdot \Omega_p)\\
         &\times (\cos \rho)^{3 - \Delta} (\sin\rho)^{\ell} {}_{2}F_1(-n,3 - \Delta + \ell + n, \frac{3}{2} + \ell; \sin^2\rho)  e^{-i\omega_n^{3 - \Delta}(\tau - \tau_p)}\Theta(\tau_p - \tau).
    \end{split}
\end{equation}
In the last line we used the hypergeometric identity
\begin{equation}
\label{eq:hg-id-shadow}
    {}_2 F_1(a,b;c;z) = (1 - z)^{c - a - b} {}_2F_1(c - a, c - b;c;z) .
\end{equation}
We recognize on the RHS of \eqref{eq:shadow-integral} the expansion of $G_{3 - \Delta}^{\rm A. T.O.}(P;X)$, 
\begin{equation}
    \widetilde{G}^{-i\epsilon}_{3- \Delta}(X;P) = e^{i\pi(\frac{3}{2}-\Delta)}   \pi^{3/2} 2^{\Delta + 1}\cos(\pi  \Delta) \frac{\Gamma(\Delta - \frac{3}{2})}{\Gamma(\Delta)} G_{3 - \Delta}^{\rm A.T.O.}(X;P), \quad \tau_p > \tau.
\end{equation}

To summarize, we found that 
\begin{equation}
\label{eq:final-shadow}
\int_{\bar{B}} d^3 P' \frac{1}{(-P\cdot P' - i\epsilon)^{3 - \Delta}} G^{\rm T.O.}_{\Delta}(X;P')  = (-1)^{\frac{3}{2} - \Delta}   \pi^{3/2} 2^{\Delta+1} \cos\pi  \Delta \frac{\Gamma(\Delta - \frac{3}{2})}{\Gamma( \Delta)} G_{3 - \Delta}^{\rm A.T.O.}(X;P), \quad P \in B.
\end{equation}
This shadow transform maps a time-ordered propagator with $\tau_p < \tau$ to an anti-time-ordered propagator with $\tau_p > \tau$. A similar result is obtained by taking the shadow integration region to be $B$ and $P \in \bar{B}$. 
In this case, we find
\begin{equation}
    \label{eq:final-shadow-bb}
\int_{B} d^3 P' \frac{1}{(-P\cdot P' - i\epsilon)^{3 - \Delta}} G^{\rm T.O.}_{\Delta}(X;P')  = (-1)^{-\frac{3}{2} +\Delta}   \pi^{3/2} 2^{\Delta+1} \cos\pi  \Delta \frac{\Gamma(\Delta - \frac{3}{2})}{\Gamma( \Delta)} G_{3 - \Delta}^{\rm A.T.O.}(X;P), \quad P \in \bar{B}.
\end{equation}

\subsection{Inner product of AdS wavefunctions and their shadow}
\label{app:shadow}

We start by evaluating \eqref{eq:shadow-int-eval}
\begin{equation}
\begin{split}
&\frac{1}{4} \int_{-1}^1 dx (1 - x)^{\alpha}  2^{-\alpha} \frac{n! \alpha!}{(n + \alpha)!}\frac{k! \alpha!}{(k + \alpha)!} P_n^{(\alpha, \beta)}(x) P_k^{(\alpha, -\beta)}(x) \\
&=  \frac{1}{4} \int_{-1}^1 dx (1 - x)^{\alpha} (1 + x)^{\beta}  2^{-\alpha - \beta} \frac{n! \alpha!}{(n + \alpha)!}\frac{k! \alpha!}{(k + \alpha)!} P_n^{(\alpha, \beta)}(x) \frac{\Gamma(k + \alpha + 1)}{\Gamma(k + \alpha + 1 - \beta)} \frac{(k - \beta)!}{k!} P_{k - \beta}^{( \alpha, \beta)}(x)\\
&=  \frac{1}{4}  2^{-\alpha - \beta} \frac{n! \alpha!}{(n + \alpha)!}\frac{k! \alpha!}{(k + \alpha)!} \dfrac{2^{\alpha + \beta + 1}}{\alpha + \beta + 2n + 1} \delta_{n, k - \beta},
\end{split}
\end{equation}
where $\alpha=\ell+\frac{1}{2}$ and $\beta=\Delta-\frac{3}{2}$.
Now we note that 
\begin{equation}
\p_{\tau} e^{-i(3 - \Delta + 2k + \ell)\tau} \delta_{n,  k - \Delta + \frac{3}{2}} = -i\underbrace{(\Delta + 2n + \ell)}_{\alpha + \beta + 2n + 1} e^{i(\Delta + 2n + \ell)\tau} \delta_{n,  k - \Delta + \frac{3}{2}}
\end{equation}
which implies that the bulk inner products of the highest weight wavefunctions and their shadow are given by
\begin{equation}
\begin{split}
\langle \psi^{\pm}_{n\ell 0}(\tau - \tau_1, \rho, \Omega\cdot \Omega_1), \psi^{\pm, 3 - \Delta}_{k\ell' 0}(\tau - \tau_2, \rho, \Omega\cdot \Omega_2) \rangle &=  \pm \frac{n! \alpha!}{(n + \alpha)!}\frac{k! \alpha!}{(k + \alpha)!}, \\
&\times e^{\mp i\omega_n(\tau_1-\tau_2)}\sqrt{\frac{4\pi}{2\ell+1}}Y_{\ell0}(\Omega_1\cdot\Omega_2)  \delta_{n, k - \beta} \delta_{\ell \ell'}\\
\langle \psi^{\pm}_{n\ell 0}(\tau - \tau_1, \rho, \Omega\cdot \Omega_1), \psi^{\mp, 3 - \Delta}_{k\ell' 0}(\tau - \tau_2, \rho, \Omega\cdot \Omega_2)\rangle &=  0.
\end{split}
\end{equation}

Using this result together with the definition \eqref{eq:time-ordered-btb}, we can now compute
\begin{equation}
\label{eq:G-Gsh}
\begin{split}
&\langle G_{\Delta}^{\rm T.O.}(X;P_1), G_{3 - \Delta}^{\rm T.O.}(X;P_2)\rangle = \sum_{n,k = 0}^{\infty} \sum_{\ell = 0}^{\infty} 2 \pi^{5/2} \sqrt{2 \ell + 1}  \frac{2^{3} \Gamma(\Delta  + \ell + n) \Gamma(3 - \Delta + \ell + k)}{n! k! \Gamma(\Delta)\Gamma(3 - \Delta) \Gamma(\frac{3}{2} + \ell)^2} \delta_{n, k - \Delta + \frac{3}{2}} \\
&\times  \frac{n! \alpha!}{(n + \alpha)!}\frac{k! \alpha!}{(k + \alpha)!}  \left[ (-1)^{-n+k} e^{-i\omega_n\tau_{12}} \Theta(\tau - \tau_1)\Theta(\tau - \tau_2) -  (-1)^{n-k} e^{i\omega_n\tau_{12}} \Theta(\tau_1 - \tau)\Theta(\tau_2 - \tau) \right] Y_{\ell0}(\Omega_1\cdot \Omega_2)\\
&=\sum_{n, \ell = 0}^{\infty} 2 \pi^{5/2} \sqrt{2 \ell + 1} \frac{2^{3} \Gamma(\Delta  + \ell + n) \Gamma(\frac{3}{2}  + \ell + n)}{\Gamma(\Delta)\Gamma(3 - \Delta) \Gamma(\frac{3}{2} + \ell)^2} \frac{(\ell + \frac{1}{2})!}{(n + \ell + \frac{1}{2})!}\frac{(\ell + \frac{1}{2})!}{(n+ \Delta - 1 + \ell)!} \\
&\times  \left[ (-1)^{\Delta - \frac{3}{2}} e^{-i\omega_n\tau_{12}} \Theta(\tau - \tau_1)\Theta(\tau - \tau_2) -  (-1)^{-\Delta + \frac{3}{2}} e^{i\omega_n\tau_{12}} \Theta(\tau_1 - \tau)\Theta(\tau_2 - \tau) \right] Y_{\ell 0}(\Omega_1\cdot \Omega_2)\\
&=\sum_{n, \ell = 0}^{\infty} \frac{2^4 \pi^{5/2} \sqrt{2 \ell + 1} }{\Gamma(\Delta)\Gamma(3 - \Delta)} \\
&\times \left[(-1)^{\Delta - \frac{3}{2}} e^{-i\omega_n\tau_{12}} \Theta(\tau - \tau_1)\Theta(\tau - \tau_2) -  (-1)^{-\Delta + \frac{3}{2}} e^{i\omega_n\tau_{12}} \Theta(\tau_1 - \tau)\Theta(\tau_2 - \tau) \right] Y_{\ell 0}(\Omega_1\cdot \Omega_2),
\end{split}
\end{equation}
where we assumed $\Delta - \frac{3}{2} \in \mathbb{Z}, \Delta - \frac{3}{2} > 0$. We note that in the last line the coefficients canceled precisely and recall that
\begin{equation}
\omega_n = \Delta + 2n + \ell.
\end{equation}

 \subsection{Near-boundary expansion of bulk-to-boundary propagator}
 \label{app:delta-id}

 In this Appendix we derive the identity \eqref{eq:boundary-exp}. To this end, we use the integral representation
 \begin{equation}
\label{eq:Mellin-tog}
G^{\rm T.O.}_{\Delta}(X;P) = \frac{i^{-\Delta}}{\Gamma(\Delta)} \int_0^{\infty} d\alpha \alpha^{\Delta - 1} e^{i \alpha(-P\cdot X + i\epsilon)}. 
\end{equation}
Using the parameterization \eqref{eq:global-AdS} and expanding near the boundary by setting $\rho = \frac{\pi}{2} - \varepsilon$, we find
\begin{equation}
   G^{\rm T.O.}_{\Delta}(X;P) \sim \frac{i^{-\Delta}}{\Gamma(\Delta)} \int_0^{\infty} d\alpha \alpha^{\Delta - 1} e^{i \frac{\alpha}{\varepsilon}(\cos(\tau - \tau_p) - \Omega \cdot \Omega_p(1 - \frac{\varepsilon^2}{2}) + i\epsilon)}, \quad \varepsilon \rightarrow 0.
\end{equation}
The distributional contributions to this propagator can be evaluated by integrating it against a test function over $\tau,~ \Omega$ and using the stationary phase approximation \eqref{eq:stat-phase} with
\begin{equation}
\label{eq:ourf}
f(x) = \cos(\tau - \tau_p) - \Omega \cdot \Omega_p.
\end{equation}

Writing 
\begin{equation}
\Omega \cdot \Omega_p = \pm \frac{2 - \left(\Omega \mp \Omega_p \right)^2}{2},
\end{equation}
we find that the critical points of \eqref{eq:ourf} are given by
\begin{equation}
\begin{split}
\sin(\tau - \tau_p) &= 0 \implies \tau = \tau_p + k \pi,\\
\Omega &= (-1)^k \Omega_p.
\end{split}
\end{equation}
Furthermore, 
\begin{equation}
{\rm det}\left({\rm Hess}(f(x_0)) \right) = (-1)^k, \quad {\rm sgn}\left({\rm Hess}(f(x_0)) \right) = (-1)^k.
\end{equation}
As a result, we find (setting $\epsilon = 0$ at the end while keeping $\varepsilon$ finite)
\begin{equation}
\label{eq:nb-exp}
\begin{split}
    G_{\Delta}^{\rm T.O.}(X; P) &\sim \frac{i^{-\Delta}}{\Gamma(\Delta)} \sum_{k \in \mathbb{Z}} \int_0^{\infty} d\alpha \alpha^{\Delta - 1} e^{i\alpha \frac{\varepsilon}{2}{(-1)^k} - \alpha \frac{\epsilon}{\varepsilon}} e^{\frac{i\pi}{4} (-1)^k} \left( \frac{2\pi \varepsilon}{\alpha} \right)^{3/2} \delta(\tau - \tau_p + k \pi) \delta^{(2)}(\Omega - \Omega_p^k) + \cdots\\
    &= i^{-\frac{3}{2} }\frac{\Gamma(\Delta - \frac{3}{2})}{\Gamma(\Delta)} \varepsilon^{3 - \Delta} 2^{\Delta} \pi^{3/2} \sum_{k \in \mathbb{Z}} i^{(-1)^k \frac{1}{2}} e^{i\pi k(\frac{3}{2}-\Delta)} \delta(\tau - \tau_p + k \pi) \delta^{(2)}(\Omega - \Omega_p^k) + \cdots,
    \end{split}
\end{equation}
where $\cdots$ denote subleading contributions in the small $\epsilon$ expansion and $\Omega^k_p \equiv (-1)^k \Omega_p$. For $k = 0$, we recover precisely \eqref{eq:boundary-exp}.

\subsection{Boundary inner product of bulk-to-boundary propagators}
\label{app:boundary-ip}
In this Appendix we use \eqref{eq:nb-exp} to evaluate the boundary inner product of bulk-to-boundary propagators \eqref{eq:boundary-ip} and verify that Stokes' theorem \eqref{eq:boundary-stokes} is obeyed. We take $\Sigma$ to be the AdS boundary in the definition of the Klein-Gordon inner product \eqref{eq:KG-ip}, in which case
\begin{equation}
\label{eq:measure}
\begin{split}
    d\Sigma^{\mu}\p_{\mu} &= \frac{1}{\epsilon^2} d\tau d\Omega \p_{\epsilon}.
    \end{split}
\end{equation}
 This boundary inner product can then be performed using \eqref{eq:boundary-exp}. 
 
 The leading terms in the two series lead to $4$ contributions to the inner product, only three of which are non-vanishing (for $\Delta > 1$) at the boundary. The finite contribution comes from the cross terms where the delta function in $G_{\Delta}(X; P_1)$ multiplies the power-law term in $G_{\Delta}(X; P_2)$ and vice versa. For simplicity, we assume that only $P_1$ is contained in the boundary integration region.  We find
\begin{equation}
\begin{split}
    \langle G^{\rm T.O.}_{\Delta}(X,P_1), G^{\rm T.O.}_{\Delta}(X,P_2)\rangle_{\rho = \frac{\pi}{2} - \epsilon} &= i \frac{1}{\epsilon^2} \left(i \frac{\Gamma(\Delta - \frac{3}{2})}{\Gamma(\Delta)} \epsilon^{3 - \Delta} 2^{\Delta} \pi^{3/2} \Delta \epsilon^{\Delta - 1} G_{\Delta}^{\rm T.O.}(P_1, P_2) \right.\\
    &\left. - i \frac{\Gamma(\Delta - \frac{3}{2})}{\Gamma(\Delta)} \epsilon^{3 - \Delta} 2^{\Delta} \pi^{3/2} (3 - \Delta)\epsilon^{2 - \Delta} \epsilon^{\Delta} G_{\Delta}^{\rm T.O.}(P_1, P_2) \right)\\
    &= -2^{\Delta + 1} \frac{\Gamma(\Delta - \frac{1}{2})}{\Gamma(\Delta)} \pi^{3/2}G_{\Delta}^{\rm T.O.}(P_1, P_2) 
    \end{split}
\end{equation}
and
\begin{equation}
    \langle G^{\rm A.T.O.}_{\Delta}(X,P_1), G^{\rm A.T.O.}_{\Delta}(X,P_2)\rangle_{\rho = \frac{\pi}{2} - \epsilon} = 2^{\Delta + 1} \frac{\Gamma(\Delta - \frac{1}{2})}{\Gamma(\Delta)} \pi^{3/2}G_{\Delta}^{\rm A.T.O.}(P_1, P_2) .
\end{equation}

The result in the case when $P_2$ is contained in the boundary integration region instead turns out to be the complex conjugate of the RHS above. When both $P_1$ and $P_2$ are contained in the integration region, we find
\begin{equation}
\label{eq:final-bip}
\begin{split}
\langle G^{\rm T.O.}_{\Delta}(X,P_1), G^{\rm T.O.}_{\Delta}(X,P_2)\rangle_{\rho = \frac{\pi}{2}} &= -2^{\Delta + 1} \pi^{3/2}  \frac{\Gamma(\Delta - \frac{1}{2})}{\Gamma(\Delta)} \left( G^{\rm T.O.}_{\Delta}(P_1, P_2) + G^{\rm A.T.O.}_{\Delta}(P_1, P_2)\right)\\
&= - 2^{\Delta + 1} \pi^{3/2}  \frac{\Gamma(\Delta - \frac{1}{2})}{\Gamma(\Delta)} \left( G^{\rm+}_{\Delta}(P_1, P_2) + G^{\rm -}_{\Delta}(P_1, P_2)\right).
\end{split}
\end{equation}
Similarly, one obtains
\begin{equation}
     \langle G_{\Delta}^{\rm A.T.O.}(X; P_1), G^{\rm T.O.}_{\Delta}(X; P_2) \rangle \left. \right|_{\rho = ct.} 
     = 2^{\Delta + 1}\pi^{3/2} \frac{\Gamma(\Delta - \frac{1}{2})}{\Gamma(\Delta)}G_{\Delta}^{\rm T.O.}(P_1, P_2)
\end{equation}
when $P_1$ is in the integration region, and
\begin{equation}
     \langle G_{\Delta}^{\rm A.T.O.}(X; P_1), G^{\rm T.O.}_{\Delta}(X; P_2) \rangle \left. \right|_{\rho = ct.} 
     = -2^{\Delta + 1}\pi^{3/2} \frac{\Gamma(\Delta - \frac{1}{2})}{\Gamma(\Delta)}G_{\Delta}^{\rm T.O.}(P_1, P_2)
\end{equation}
when only $P_2$ is in the integration region, such that the inner product vanishes when both $P_1$ and $P_2$ are contained in the integration region. The last line of \eqref{eq:final-bip} follows upon recalling the relation between the different propagators, which we review for completeness below.

We start by recalling the relation between time-ordered propagators and Wightman functions in Lorentzian CFT (a clear summary in AdS can be found in an appendix of \cite{Sleight:2019hfp}). Wightman functions are obtained from Euclidean correlators by analytic continuation. They are defined for fixed operator orderings by
 \begin{equation}
     \label{eq:Wightman}
     \begin{split}
    \mathcal{W}_{12} &=  \langle \phi(t_1 - \frac{i\epsilon}{2}, \vec{x}_1) \phi(t_2 + \frac{i\epsilon}{2}, \vec{x}_2) \rangle = \frac{1}{\left( -t_{12}^2 + |\vec{x}_{12}|^2 + i\epsilon \sgn(t_{12}) \right)^{\Delta}}, \quad \epsilon > 0, \\
    \mathcal{W}_{21} &=   \langle \phi(t_2 - \frac{i\epsilon}{2}, \vec{x}_1) \phi(t_1 + \frac{i\epsilon}{2}, \vec{x}_2) \rangle = \frac{1}{\left( -t_{12}^2 + |\vec{x}_{12}|^2 - i\epsilon \sgn(t_{12}) \right)^{\Delta}} = \mathcal{W}_{12}^*, \quad \epsilon > 0. \\
     \end{split}
 \end{equation}
Time ordered correlators are obtained from Wightman functions by the following formula
\begin{equation}
\label{eq:time-ord-corr}
    \langle \mathcal{T} \phi_1 \phi_2\rangle = \mathcal{W}_{12} \Theta(t_{12}) + \mathcal{W}_{21} \Theta(t_{21}). 
\end{equation}
We see that Wightman functions are analytic, while time-ordered correlators are not. 

The bulk-to-boundary propagators $G^{\pm}_{\Delta}(X;P)$ are defined in analogy with \eqref{eq:Wightman} where $x_{12}^2 \rightarrow -X\cdot P$, while the time-ordered bulk-to-boundary propagator is the counterpart of \eqref{eq:time-ord-corr}. As a result, $G^{\pm}$ are free of delta function singularities in the limit as $X$ approaches the boundary \cite{H:2024cfo}. 
One way to see this is to write the Wightman functions as
\begin{equation}
\label{eq:Wightman-cc}
    \mathcal{W}_{12} = \frac{1}{\left( -t_{12}^2 + |\vec{x}_{12}|^2 + i\epsilon  \right)^{\Delta}} \Theta(t_{12}) +  \frac{1}{\left( -t_{12}^2 + |\vec{x}_{12}|^2 - i\epsilon  \right)^{\Delta}} \Theta(t_{21}) = \mathcal{W}_{21}^*,
\end{equation}
which are real in the limit as $t_1 \rightarrow t_2$. As shown in \eqref{eq:boundary-exp}, delta function contact terms appear as imaginary contributions in the near-boundary expansion of the constituents, and are hence absent above. 
The time-ordered correlator, on the other hand, receives a contribution from the discontinuity
\begin{equation}
    \langle \mathcal{T} \phi_1 \phi_2 \rangle \supset {\rm Im}\left( \frac{1}{(-t_{12}^2 + |\vec{x}_{12}|^2 + i\epsilon)^{\Delta}} \right) \propto \delta(t_{12}) \delta(\vec{x}_{12}).
\end{equation}
The last line of \eqref{eq:final-bip} follows from \eqref{eq:time-ord-corr} and \eqref{eq:Wightman-cc}. 

 We finally point out that the overleading contributions to the boundary inner product (i.e. the terms arising from products of delta functions and scaling as $\epsilon^{3 - 2\Delta}$) vanish.

\subsection{Near-boundary contact terms from mode expansion}
\label{sec:delta-function-modes}

In this appendix we show how the delta function term in \eqref{eq:boundary-exp} arises from the mode-expansion of the bulk-to-boundary propagator. We first note, that the mode expansion \eqref{eq:time-ordered-btb} for points $\tau$ close to $\tau_p$ can be equivalently written as
\begin{equation}
\label{eq:time-ord}
\begin{split}
    G^{\rm T.O.}_{\Delta}(X; P) &= \sum_{k = -\infty}^{\infty} \sum_{\ell = 0}^{\infty} \pi \sqrt{2\ell + 1} (-1)^k \frac{2^{\Delta} \Gamma(\Delta + \ell + k)}{k! \Gamma(\Delta) \Gamma(\frac{3}{2} + \ell)} e^{-i\omega_k(\tau - \tau_p) - i\omega_k \epsilon \sgn (\tau - \tau_p)} Y_{\ell,m}(\Omega\cdot \Omega_p) \\
    &\times \sin^{\ell}\rho \cos^{\Delta} \rho {}_2F_1\left(-k,\ell + \Delta + k; \frac{3}{2} + \ell; \sin^2\rho \right).
    \end{split}
\end{equation}
One can show by direct computation that the near-boundary expansion of the hypergeometric function takes the form
\begin{equation}
\label{eq:f21}
\begin{split}
    &{}_2F_{1}\left(-n,\ell + \Delta + n; \frac{3}{2} + \ell; \sin^2\rho \right) \left. \right|_{\rho = \frac{\pi}{2} - \epsilon}  \\
    &= \epsilon^{3-2\Delta} \frac{\Gamma(\Delta - \frac{3}{2})\Gamma(\frac{3}{2} + \ell)}{\Gamma(-n)\Gamma(\Delta + \ell + n)}  + \cdots + \frac{\Gamma(\frac{3}{2} - \Delta)\Gamma(\frac{3}{2} + \ell)}{\Gamma(\frac{3}{2} - \Delta - n)\Gamma(\frac{3}{2} + \ell + n)} + \cdots 
    \end{split}
\end{equation}

Mode-by-mode (that is for \textit{fixed} $n \in \mathbb{N}$), the first series vanishes and we recover the statement that the standard AdS wavefunctions \eqref{eq:AdS-wf} are normalizable and span representations of $\mathfrak{so}(3,2)$. We now show that it is precisely this naively vanishing series that is responsible for the delta function contact terms in the near-boundary expansion of the bulk-to-boundary propagator. To this end, it will be convenient to not yet set the first term above to $0$.  We will adopt the regulation scheme $n \rightarrow n +i\epsilon$ and take $\epsilon \rightarrow 0$ at the end.

Setting $\tau \simeq \tau_p$ in definition of the time-ordered bulk-to-boundary propagator
\begin{equation}
\label{eq:series-btb}
\begin{split}
G^{\rm T.O.}(X;P) \supset - \epsilon^{3 - \Delta} \frac{2^{\Delta}\Gamma(\Delta - \frac{3}{2})}{\Gamma(\Delta)} \pi^{1/2} &\sum_{k, \ell = 0}^{\infty} (-1)^k \frac{\sin\pi (k+ i\epsilon)}{\pi} \left[e^{-i\omega_k(\tau - \tau_p - i\epsilon)} \Theta(\tau - \tau_p) \right. \\
&\left. + e^{i\omega_k(\tau - \tau_p+i\epsilon)} \Theta(\tau_p - \tau)\right] \sqrt{\pi(2 \ell + 1)} Y_{\ell,0}(\Omega \cdot \Omega_p),
\end{split}
\end{equation}
we find 
\begin{equation}
\label{eq:delta-gto-modes}
\begin{split}
    G^{\rm T.O.}(X;P)  &\supset - \epsilon^{3 - \Delta} \frac{2^{\Delta}\Gamma(\Delta - \frac{3}{2})}{\Gamma(\Delta)} \pi^{1/2} \sum_{ \ell = 0}^{\infty}  \frac{i\epsilon}{2 i} \frac{1}{ - i\epsilon(\tau - \tau_p)} (\tau - \tau_p) \delta(\tau - \tau_p) \\
 &\times \sqrt{\pi(2 \ell + 1)} Y_{\ell,0}(\Omega \cdot \Omega_p) =   -i\epsilon^{3 - \Delta} \frac{2^{\Delta-1}\Gamma(\Delta - \frac{3}{2})}{\Gamma(\Delta)} \pi^{1/2} \delta(\tau - \tau_p) \delta(\Omega\cdot \Omega_p - 1).
 \end{split}
\end{equation}
Using 
\begin{equation}
    \delta(\Omega \cdot  \Omega_p - 1) = 2\pi \delta^{(2)}(\Omega - \Omega_p),
\end{equation}
we find agreement with the near-boundary expansion of the bulk-to-boundary propagator stated in \eqref{eq:boundary-exp}. However, we remark that the overall coefficient appears to depend on the regularization scheme used. We leave a better understanding of this point to future work.

The retarded and advanced propagators contain no discontinuity as $\tau \rightarrow \tau_p$ and hence no delta functions in the near-boundary expansion.

\newpage
\section{Details of AdS bulk reconstruction}
\label{app:AdS-bulk-reconstruction}

In this appendix we present some consistency checks on the bulk reconstruction formulae proposed in Section \ref{sec:bulk-rec-AdS}.

\subsection{Relation between bulk modes and boundary operators}

\label{app:KA-AdS}

Using the schematic formulas
\begin{equation}
\begin{split}
    G^{\rm A.T.O}(\tau; \tau_p) &\sim \psi^-(\tau - \tau_p) \Theta_{\bar{B}} + \psi^{+}(\tau-\tau_p) \Theta_{B}, \\
     G^{\rm T.O}(\tau; \tau_p) &\sim \psi^-(\tau-\tau_p) \Theta_{B} + \psi^{+}(\tau-\tau_p) \Theta_{\bar{B}}, \\
     \mathcal{O}(\tau_p) &\sim \psi^+(\tau_p) a + \psi^-(\tau_p) a^{\dagger}
    \end{split}
\end{equation}
where $\Theta_{\bar{B}}$ is supported in $\bar{B}$ and $\Theta_{B}$ is supported in $B$, we find
\begin{equation}
\label{eq:O-}
\begin{split}
    \widetilde{\mathcal{O}}_{3-\Delta}^-(P) &= \mathcal{N}(\Delta)\int_{\bar{B}} dP' \frac{1}{(-P\cdot P' - i\epsilon)^{3 - \Delta}} \mathcal{O}_{\Delta}(P'), \quad P \in B \\
    &= \mathcal{N}(\Delta) \int_{B} dP' \frac{1}{(-P\cdot P' + i\epsilon)^{3-\Delta}}  \mathcal{O}_{\Delta}(P'), \quad P \in \bar{B}, \quad \sim a^{\dagger}
    \end{split}
\end{equation}
and 
\begin{equation}
\label{eq:O+}
\begin{split}
    \widetilde{\mathcal{O}}_{3- \Delta}^+(P) &= \mathcal{N}^*(\Delta)\int_{B} dP' \frac{1}{(-P\cdot P' - i\epsilon)^{3 - \Delta}} \mathcal{O}_{\Delta}(P') ,\quad P \in \bar{B}\\
    &= \mathcal{N}^*(\Delta) \int_{\bar{B}} dP' \frac{1}{(-P\cdot P' + i\epsilon)^{3-\Delta}}  \mathcal{O}_{\Delta}(P'), \quad P \in B \quad \sim a,
    \end{split}
\end{equation}
where 
\begin{equation}
    \mathcal{N}(\Delta) = \frac{e^{-i\pi(\Delta - \frac{3}{2})}\Gamma(\Delta)}{\cos(\pi \Delta) \pi^{3/2} 2^{\Delta + 1} \Gamma(\Delta - \frac{3}{2})}.
\end{equation}
Note that
\begin{equation}
    \widetilde{\mathcal{O}}^{-}_{3 - \Delta}(P) =  \widetilde{\mathcal{O}}_{3- \Delta}^{+\dagger}(P).
\end{equation}

The normalization ensures that the shadow is an involution since
\begin{equation}
   \int_B dP' \int_{\bar{B}} dP'' \frac{1}{(-P\cdot P' + i\epsilon)^{\Delta}} \frac{1}{(-P'\cdot P'' - i\epsilon)^{3 - \Delta}} \mathcal{O}_{\Delta}(P'')  = \frac{1}{\mathcal{N}(\Delta) \mathcal{N}(3 - \Delta)}  \mathcal{O}_{\Delta}(P) .
\end{equation}
It will also be useful to observe that 
\begin{equation}
    \alpha = \mathcal{N}^{*}(3 - \Delta).
\end{equation}
We summarize this discussion in Table \ref{table:presc}.

\begin{table}[h]
\centering
\begin{tabular}{@{}ccc@{}}
\toprule
& $G^{\rm T.O.}_\Delta$  & $G^{\rm A.T.O.}_\Delta$    \\ \midrule
$B \;(\tau_p>\tau)$       & $\psi^-$ & $\psi^+$ \\ \midrule
$\bar{B}\;(\tau>\tau_p)$ & $\psi^+$ & $\psi^-$ \\ \bottomrule
\end{tabular}
\quad\quad
\begin{tabular}{@{}ccc@{}}
\toprule
& $+i\epsilon$  & $-i\epsilon$    \\ \midrule
$B \;(\tau_p>\tau)$       & $\widetilde{\mathcal{O}}^+_{3-\Delta}$ & $\widetilde{\mathcal{O}}^-_{3-\Delta}$  \\ \midrule
$\bar{B}\;(\tau>\tau_p)$ & $\widetilde{\mathcal{O}}^-_{3-\Delta}$ & $\widetilde{\mathcal{O}}^+_{3-\Delta}$ \\ \bottomrule
\end{tabular}
\caption{The left table shows the type (positive/negative energy) standard AdS wavefunction that appears in the decomposition of $G_{\Delta}^{\rm (A).T.O.}$ upon restriction to regions $B$ or $\bar{B}$. The right table shows the regions where $\widetilde{\mathcal{O}}^{\pm}_{3 - \Delta}$ are defined. They are integrated against shadow kernels with the shown $i\epsilon$ prescription over the \textit{complementary} region. }
\label{table:presc}
\end{table}
These formulas, together with the inner products of bulk-to-boundary propagators derived in section \ref{sec:bulk-to-boundary-basis}, can be used to obtain the Kirchhoff--d'Adh\'emar formula \eqref{eq:KA-AdS}.

\subsection{Mode expansions of shadow operators}
\label{app:shadowO-O}

In this appendix we derive the relation \eqref{eq:shadow-O} between modes of a primary operator $\mathcal{O}$ and its shadow. We start with the definition \eqref{eq:O-}, \eqref{eq:O+} of $\widetilde{\mathcal{O}}_{3 - \Delta}$ in $B$ and apply the integral transform
\begin{equation}
    \int d\tau_p \int d^2\Omega_p e^{-i\omega_n \tau_p} Y_{\ell m}(\Omega_p) \Theta(\tau_p - \tau)
\end{equation}
to both sides. Noting that the $\tau_p$ integral projects out the $\widetilde{\mathcal{O}}^+_{3 - \Delta}$ component and using the mode expansion \eqref{eq:shadow-ato-5} of the shadow kernel which follows from \eqref{eq:bdry-bdry}, we find
\begin{equation}
\begin{split}
    & \int d\tau_p \int d^2 \Omega_p e^{-i\omega_n^{\Delta} \tau_p} Y_{\ell m}(\Omega_p) \Theta(\tau_p - \tau) \widetilde{\mathcal{O}}_{3 - \Delta}(P) = \frac{e^{-i\pi (\Delta - \frac{3}{2})}}{\cos\pi\Delta}\frac{\Gamma(\Delta)}{\pi^{3/2} 2^{\Delta + 2} \Gamma(\Delta - \frac{3}{2})} \\
    &\times \sum_{k, \ell'}^{\infty} \int d\tau_{p'}\int d^2\Omega_{p'} \int d\tau_p \int d^2 \Omega_p e^{-i\omega_n^{\Delta} \tau_p} Y_{\ell m}(\Omega_p)\Theta(\tau_p - \tau)\Theta(\tau - \tau_{p'}) \pi \sqrt{2\ell' + 1} \\
    &\times 2 \frac{2^{3 - \Delta} \Gamma(3 - \Delta + \ell' + k)\Gamma(\frac{5}{2} - \Delta + k)}{\Gamma(k + 1)\Gamma(3 - \Delta)\Gamma(\frac{3}{2} + \ell' + k)\Gamma(\frac{5}{2} - \Delta)} e^{i\omega^{3 - \Delta}_k(\tau_p - \tau_{p'})} Y_{\ell' 0}(\Omega_p\cdot \Omega_{p'}) \mathcal{O}_{\Delta}(P')\\
  &= \frac{4 \pi e^{-i\pi (\Delta - \frac{3}{2})} \Gamma(\Delta)}{ 2^{\Delta+2}{\cos \pi\Delta} \Gamma(\Delta - \frac{3}{2})\Gamma(\frac{5}{2} - \Delta)\Gamma(3 - \Delta)}  \int d\tau_{p'}\int d^2\Omega_{p'} \Theta(\tau - \tau_{p'}) \frac{2^{3 - \Delta} \Gamma(\frac{3}{2} +\ell+n)\Gamma(1+n)}{\Gamma(-\frac{1}{2} + \Delta + n) \Gamma(\ell +\Delta + n)} \\
  &\times e^{-i\omega^{\Delta}_n \tau_{p'}} Y_{\ell m}(\Omega_{p'}) \mathcal{O}_{\Delta}(P').
    \end{split}
\end{equation}
The $\Omega_p$ integral was performed using the identity
\begin{equation}
\label{eq:sph-ortho}
    \int d^2 \Omega Y_{\ell m}(\Omega) Y_{\ell' 0}(\Omega \cdot \Omega_p) = \frac{\sqrt{4\pi}}{\sqrt{2\ell' + 1}} \delta_{\ell \ell'} Y_{\ell' m}(\Omega_p),
\end{equation}
and we assumed $\Delta-\frac{3}{2}\in\mathbb{Z}$. For completeness, we include the proof of \eqref{eq:sph-ortho}:
\begin{align}
\int d\Omega'Y_{\ell0}(\Omega\cdot\Omega')Y_{\ell'm}(\Omega')&=\sqrt{ \frac{2\ell+1}{4\pi} }\int d\Omega'P_{\ell}(\Omega\cdot\Omega')Y_{\ell'm}(\Omega')= \\
&=\sum_{m'=-\ell}^\ell \sqrt{ \frac{4\pi}{2\ell+1} }\int d\Omega'Y_{\ell m'}(\Omega)Y^*_{\ell m'}(\Omega')Y_{\ell'm}(\Omega')= \\
&=\sqrt{ \frac{4\pi}{2\ell+1} }Y_{\ell m}(\Omega)\delta_{\ell \ell'},
\end{align}
where in the first line we used the relation between $Y_{\ell 0}$ and the Legendre polynomials, in the second line we used the addition theorem \eqref{eq:LP-SH}, and in the last line we used orthogonality of spherical harmonics.

\subsection{Agreement with the extrapolate dictionary}

According to the extrapolate dictionary,
\begin{align}
\mathcal{O}_{\Delta}(\tau,\Omega)&=\lim_{ \rho \to \frac{\pi}{2} }(\cos \rho)^{-\Delta}\;\Phi(\tau,\rho,\Omega)= \sum_{n\ell m}\left(\widehat{\psi}_{n\ell m}^+(\tau,\Omega)\,a^{\dagger}_{n\ell m}+\widehat{\psi}_{n\ell m}^-(\tau,\Omega)\,a_{n\ell m}\right) \\
&= \sum_{n\ell m}\frac{\Gamma\left( \ell+\frac{3}{2} \right)\Gamma\left( \frac{3}{2}-\Delta \right)}{\Gamma\left( n+\ell+\frac{3}{2} \right)\Gamma\left( \frac{3}{2}-\Delta-n \right)} \left(e^{-i\omega_{n}\tau}a^{\dagger}_{n\ell m}Y_{\ell m}(\Omega)+e^{i\omega_{n}\tau}a_{n\ell m}Y^*_{\ell m}(\Omega)\right).
\end{align}
This expression can be substituted into the various reconstruction formulae proposed in Section \ref{sec:bulk-rec-AdS} to show that they all yield the standard mode decomposition \eqref{eq:AdS-bulk-field} in terms of highest/lowest weight wavefunctions. 

Using the mode expansion for the bulk-to-boundary propagator, we find
\begin{align}
\Phi(\tau,&\rho,\Omega)\supset\int_{I}d\tau'd\Omega'\,G^{{\rm (A.)T.O.}}_{3-\Delta}(\tau,\rho,\Omega;\tau',\Omega')\,\mathcal{O}_{\Delta}(\tau',\Omega')= \\
&=\sum_{k,\ell}\sum_{n,\ell',m}\pi \sqrt{ 2\ell+1 } \frac{2^{3-\Delta}\Gamma(3-\Delta+\ell+k)}{k!\Gamma(3-\Delta)\Gamma\left( \ell+\frac{3}{2} \right)}\cdot \frac{\Gamma\left( \ell'+\frac{3}{2} \right)\Gamma\left( \frac{3}{2}-\Delta \right)}{\Gamma\left( n+\ell'+\frac{3}{2} \right)\Gamma\left( \frac{3}{2}-\Delta-n \right)} \\
&\times (\sin\rho)^\ell(\cos \rho)^{3-\Delta} {}_{2}F_{1}\left( -k,\ell+3-\Delta+k;\ell+\frac{3}{2};\sin^2\rho \right) \\
&\times \int_{I} d\tau'd\Omega' Y_{\ell0}(\Omega\cdot\Omega')\left((-1)^{\pm k} e^{\mp i\omega_{k}(\tau-\tau')}\Theta(\tau-\tau')+ (-1)^{\mp k} e^{\pm i\omega_{k}(\tau-\tau')}\Theta(\tau'-\tau)\right) \\
&\times \left( e^{-i\omega_{n}\tau'}a^{\dagger}_{n\ell' m}Y_{\ell'm}(\Omega')+ e^{i\omega_{n}\tau'}a_{n\ell' m}Y^*_{\ell'm}(\Omega')\right),
\end{align}
where the plus minus signs correspond to particular time orderings; upper signs correspond to the time-ordered bulk-to-boundary propagator and the lower signs correspond to the anti-time-ordered bulk-to-boundary propagator.
The non-vanishing integrals are those that set $\omega_{n}=\omega_{k}$, namely, $\int_{0}^{\pi} d\tau' e^{2 i(n - k)\tau'}= \pi \delta_{nk}$. 

To summarize, we find for the different integration regions

\begin{align}
\int_{\bar{B}} d^3P' G^{\rm T.O.}_{3 - \Delta}(\tau, \rho, \Omega; P') \mathcal{O}_{\Delta}(P') =\sum_{n\ell m}\pi^{3/2} e^{i\pi (\Delta-1/2)} \cos( \pi\Delta )\frac{2^{4-\Delta}\Gamma\left( \frac{3}{2}-\Delta \right)}{\Gamma(3-\Delta)} \psi^+_{n\ell m}(\tau,\rho,\Omega)a^{\dagger}_{n\ell m},
\end{align}
\begin{align}
\int_{\bar{B}} d^3P' G^{\rm A.T.O.}_{3 - \Delta}(\tau, \rho, \Omega; P') \mathcal{O}_{\Delta}(P') =\sum_{n\ell m}\pi^{3/2} e^{- i\pi (\Delta-1/2)} \cos( \pi\Delta )\frac{2^{4-\Delta}\Gamma\left( \frac{3}{2}-\Delta \right)}{\Gamma(3-\Delta)} \psi^-_{n\ell m}(\tau,\rho,\Omega)a_{n\ell m},
\end{align}
\begin{align}
\int_{B} d^3P' G^{\rm T.O.}_{3 - \Delta}(\tau, \rho, \Omega; P') \mathcal{O}_{\Delta}(P') =\sum_{n\ell m}\pi^{3/2} e^{-i\pi (\Delta-1/2)} \cos( \pi\Delta )\frac{2^{4-\Delta}\Gamma\left( \frac{3}{2}-\Delta \right)}{\Gamma(3-\Delta)} \psi^-_{n\ell m}(\tau,\rho,\Omega)a_{n\ell m}
\end{align}
and 
\begin{align}
\int_B d^3P' G^{\rm A. T.O.}_{3 - \Delta}(\tau, \rho, \Omega; P') \mathcal{O}_{\Delta}(P') =\sum_{n\ell m}\pi^{3/2} e^{i\pi (\Delta-1/2)} \cos( \pi\Delta )\frac{2^{4-\Delta}\Gamma\left( \frac{3}{2}-\Delta \right)}{\Gamma(3-\Delta)} \psi^+_{n\ell m}(\tau,\rho,\Omega)a^{\dagger}_{n\ell m}.
\end{align}

\newpage
\section{Decomposition of $\mathfrak{so}(3,2)$ wavefunctions into $\mathfrak{so}(3,1)$ components}

In this appendix we provide alternative derivations of the $\mathfrak{so}(3,1)$ wavefunctions constructed in Sections \ref{sec:hyper-fol} and \ref{sec:null-cpw-ads}. 

\subsection{Hyperbolic foliation}
\label{app:so32-so31}

Following the discussion in Section \ref{sec:boost-wf-hyp}, we evaluate the integral
\begin{equation}
   I =  \int \frac{d\alpha \alpha^3}{\sqrt{\ell^2 - \alpha^2}} \int_{AdS_3} d^3\hat{x} G_{\Delta'}(X;\hat{x}) f(\alpha) G_{\delta}^{\rm AdS_3}(\hat{x}; \hat{q}),
\end{equation}
where $x^2 = -\ell^2$ is a point on the hyperboloid in the future of the lightcone through the origin (AdS$_4^+$) of radius $\alpha = \ell$. We will set $\ell = 1$ for simplicity. Parameterizing $X$ with the hyperbolic coordinates \eqref{eq:AdSd-emb}, \eqref{eq:Milne-AdS-flat}, we find
\begin{equation}
    G_{\Delta'}(X;\hat{x}) = \left(\frac{2 y_p y}{\alpha (y^2 + y_p^2 + |z - z_p|^2)} \right)^{\Delta'}  = \alpha^{-\Delta'} \frac{1}{(-\hat{x}_p\cdot \hat{x})^{\Delta'}}, \quad \hat{x}, \hat{x}_p \in \mathbb{R}^{3,1}, \quad \quad \hat{x}^2 = \hat{x}_p^2 = -1.
\end{equation}
We will also need the Mellin-Barnes representation of the hypergeometric function
\begin{equation}
\begin{split}
    {}_2F_1\left[\frac{1}{2}(3 - \delta - \Delta), \frac{1}{2}(\Delta - \delta), 2 - \delta,\alpha^2 \right] &= \frac{ \Gamma(2 - \delta)}{\Gamma(\frac{1}{2}(3 - \delta - \Delta))\Gamma(\frac{1}{2}(\Delta - \delta))}\\
    &\times \frac{1}{2\pi i} \int_{-i\infty}^{i\infty} ds \frac{\Gamma(-s)\Gamma(\frac{1}{2}(3 - \delta - \Delta) + s) \Gamma( \frac{1}{2}(\Delta - \delta) + s)}{ \Gamma(2 - \delta + s)}(-\alpha^2)^s,
    \end{split}
\end{equation}
as well as the formula for the bulk-to-boundary propagator in AdS$_3$
\begin{equation}
   G_{\delta}^{\rm AdS_3}(\hat{x}; \hat{q}) = \frac{1}{(-\hat{x}\cdot \hat{q})^{\delta}}, \quad \hat{q} \in \mathbb{R}^{3,1}, \quad \hat{q}^2 = 0.
\end{equation}

Introducing Schwinger parameters, we find
\begin{equation}
\begin{split}
    I &= \frac{i^{-\delta - \Delta'}}{\Gamma(\Delta') \Gamma(\delta)}\int_{AdS_4^+} d^4X \int_0^{\infty} du \int_0^{\infty} dt \int_0^{\infty} dv u^{\Delta' - 1} t^{\delta - 1}   e^{iu\alpha(-\hat{x}\cdot \hat{x}_p + i\epsilon) + it\alpha(-\hat{x}\cdot \hat{q} + i\epsilon)}e^{i\alpha^2 v}\\
   & \times \frac{1}{2\pi i}\frac{ \Gamma(2 - \delta)}{\Gamma(\frac{1}{2}(3 - \delta - \Delta))\Gamma(\frac{1}{2}(\Delta - \delta))}  \int ds v^{-s-1} i^{-s}\frac{\Gamma(\frac{1}{2}(3 - \delta - \Delta) + s) \Gamma( \frac{1}{2}(\Delta - \delta) + s)}{ \Gamma(2 - \delta + s)}.
    \end{split}
\end{equation}
We next evaluate the integral over AdS$_4^+$. We recall from \eqref{eq:massive-AdS-fol} that the measure is 
\begin{equation}
\begin{split}
  \int_{AdS_4}  d^4X &=  \int_0^{1} \frac{d\alpha}{\sqrt{1 - \alpha^2}} \alpha^3 \int_{AdS_3} d^3\hat{x} =  \int_0^{1} \frac{d\alpha}{\sqrt{1 - \alpha^2}} \alpha^3 \int_{\mathbb{R}_4} d^4\hat{x} \delta(\hat{x}^2 + 1) \\
  &=\frac{1}{2\pi}  \int_0^{1} \frac{d\alpha}{\sqrt{1 - \alpha^2}} \alpha^3 \int_{\mathbb{R}_4} d^4\hat{x} \int dz e^{iz(\hat{x}^2+ 1)}.
  \end{split}
\end{equation}
The integral over $\hat{x}^2$ is then Gaussian and yields
\begin{equation}
\begin{split}
    I &=\frac{1}{2\pi} \frac{i^{-\delta - \Delta'}}{\Gamma(\Delta') \Gamma(\delta)}\int_0^{1} \frac{d\alpha \alpha^3}{\sqrt{1 - \alpha^2}}\int_0^{\infty} du \int_0^{\infty} dt \int_0^{\infty} dv u^{\Delta' - 1} t^{\delta - 1}  \int_{-\infty}^{\infty} dz \left(\frac{\pi}{z} \right)^2 e^{-\frac{i\alpha^2}{4z}(u \hat{x}_p + t \hat{q})^2 + iz} e^{i\alpha^2 v}\\
   & \times \frac{1}{2\pi i}\frac{ \Gamma(2 - \delta)}{\Gamma(\frac{1}{2}(3 - \delta - \Delta))\Gamma(\frac{1}{2}(\Delta - \delta))}  \int ds v^{-s-1} i^{-s}\frac{\Gamma(\frac{1}{2}(3 - \delta - \Delta) + s) \Gamma( \frac{1}{2}(\Delta - \delta) + s)}{ \Gamma(2 - \delta + s)}.
   \end{split}
\end{equation}

We observe that
\begin{equation}
    (u\hat{x}_p + t \hat{q})^2 = - u^2 + 2 ut \hat{x}_p\cdot \hat{q}
\end{equation}
and further evaluate the $t$ integral 
\begin{equation}
  \begin{split}
    I &= \frac{1}{2\pi}\frac{i^{-\Delta'}}{\Gamma(\Delta')}\int_0^{1} \frac{d\alpha \alpha^3}{\sqrt{1 - \alpha^2}}\int_0^{\infty} du \int_0^{\infty} dv u^{\Delta' - 1} \int_{-\infty}^{\infty} dz \left(\frac{\pi}{z} \right)^2 e^{\frac{i\alpha^2}{4z} u^2 + iz} e^{i\alpha^2 v} \frac{1}{(-\hat{x}_p \cdot \hat{q})^{\delta}} \left(\frac{\alpha^2 u}{2 z} \right)^{-\delta}\\
   & \times \frac{1}{2\pi i}\frac{ \Gamma(2 - \delta)}{\Gamma(\frac{1}{2}(3 - \delta - \Delta))\Gamma(\frac{1}{2}(\Delta - \delta))}  \int ds v^{-s-1} i^{-s}\frac{\Gamma(\frac{1}{2}(3 - \delta - \Delta) + s) \Gamma( \frac{1}{2}(\Delta - \delta) + s)}{ \Gamma(2 - \delta + s)}.
   \end{split}
\end{equation}
As a result, the $u, v$ and $z$ integrals factorize. 
They can be expressed in terms of Mellin integrals of the exponential, namely
\begin{equation}
    \int_0^{\infty} du u^{\Delta' - \delta - 1} e^{\frac{i\alpha^2 u^2}{4z}} = 2^{\Delta' - \delta - 1} \Gamma\left(\frac{\Delta' - \delta}{2}\right) \left( -\frac{i\alpha^2}{z} \right)^{\frac{\delta - \Delta'}{2}},
\end{equation}
\begin{equation}
    \int_0^{\infty} dv v^{-s-1} e^{i\alpha^2 v} = \Gamma(-s) i^{-s} \alpha^{2s}
\end{equation}
and 
\begin{equation}
    \int_{-\infty}^{\infty} dz (z+i\epsilon)^{\frac{\Delta' + \delta}{2} - 2} e^{iz} = \Gamma\left(\frac{\Delta' + \delta}{2} - 1\right) 2 \sin \left[\pi\left(\frac{\Delta' + \delta}{2} -1\right)\right] i^{1 - \frac{\Delta' + \delta}{2}}.
\end{equation}
Putting everything together we find 
\begin{equation}
\label{eq:final-intE}
    \begin{split}
         I ~&\propto~ 2^{\Delta' -2} i  \frac{i^{ -\Delta' - \delta}}{\Gamma(\Delta')}\int_0^{\ell^2} \frac{d\alpha \alpha^3}{\sqrt{1 - \alpha^2}}  \frac{1}{(-\hat{x}_p \cdot \hat{q})^{\delta}} 
  \frac{ \Gamma(2 - \delta)\Gamma(\frac{\delta + \Delta'}{2} - 1)}{\Gamma(\frac{1}{2}(3 - \delta - \Delta'))} \sin \left[\pi\left(\frac{\delta + \Delta'}{2} - 1\right)\right] 
\frac{\Gamma(\frac{1}{2}(\Delta' - \delta))}{\Gamma(\frac{1}{2}(\Delta - \delta))}\\ &\times \int ds (-1)^s (\alpha^2)^{s +\frac{-\delta - \Delta'}{2}} \frac{\Gamma(-s)\Gamma(\frac{1}{2}(3 - \delta - \Delta) + s) \Gamma( \frac{1}{2}(\Delta - \delta) + s)}{ \Gamma(2 - \delta + s)}\\
   &= 2^{\Delta'-2} \pi i  \frac{i^{-\Delta' - \delta}}{\Gamma(\Delta')} \frac{1}{(-\hat{x}_p \cdot \hat{q})^{\delta}} 
  \frac{ \Gamma(2 - \delta)}{\Gamma(\frac{1}{2}(3 - \delta - \Delta))\Gamma(2 - \frac{\delta + \Delta'}{2})}  \frac{\Gamma(\frac{1}{2}(\Delta' - \delta))}{\Gamma(\frac{1}{2}(\Delta - \delta))}\\ &\times \int ds (-1)^s \frac{\Gamma(-s)\Gamma(\frac{1}{2}(3 - \delta - \Delta) + s) \Gamma( \frac{1}{2}(\Delta - \delta) + s)}{ \Gamma(2 - \delta + s)}\frac{\sqrt{\pi}\Gamma(2 + s - \frac{\delta + \Delta'}{2})}{2\Gamma(\frac{5}{2} + s - \frac{\Delta' + \delta}{2})}\\
   &= -2^{\Delta'-2} \pi^{5/2} \frac{i^{-\Delta' - \delta}}{\Gamma(\Delta')} \frac{1}{(-\hat{x}_p \cdot \hat{q})^{\delta}} 
 \frac{\Gamma(\frac{1}{2}(\Delta' - \delta))}{\Gamma(\frac{5 - \Delta' - \delta}{2})}\\
   &\times {}_{3}F_2\left[ \frac{1}{2}(3 - \delta - \Delta), \frac{1}{2}(\Delta - \delta), 2 -\frac{\delta + \Delta'}{2}; 2 - \delta, \frac{5}{2} - \frac{\Delta' + \delta}{2},1 \right].
    \end{split}
\end{equation}

Interestingly, the hypergeometric function is Saalschutzian and simplifies, provided that one of the first 3 arguments of the hypergeometric function is a negative integer. This case may be relevant when considering the decomposition of $\mathfrak{so}(3,2)$ representations into integer dimension $\mathfrak{sl}(2,\mathbb{C})$ ones. In this case \eqref{eq:final-intE} simplifies to
\begin{equation}
\begin{split}
   I=-2^{\Delta'-2} \pi^{3/2} i^{-\Delta' - \delta}\cos\left(\frac{\pi(\delta+\Delta')}{2}\right)\frac{\Gamma(2-\delta)\Gamma\left(\frac{\Delta'-\Delta}{2}\right)\Gamma\left(\frac{\Delta'+\Delta-3}{2}\right)}{\Gamma\left(2-\frac{\Delta+\delta}{2}\right)\Gamma\left(\frac{\Delta-\delta+1}{2}\right)\Gamma(\Delta')} \frac{1}{(-\hat{x}_p \cdot \hat{q})^{\delta}}.
   \end{split}
\end{equation}

\subsection{Null foliation}
\label{app:null-boost-estate-solutions}

We start with the recursion relation
\begin{equation}
    \left[\Box_{AdS_4} - \frac{1}{\ell^2}(\Delta+2n)(\Delta +2n - 3) \right] \widetilde{\Psi}_{\Delta+2n}^{\rm null} = -(\Delta + 2n - \delta)(\Delta+ 2n + 1 -\delta)  \widetilde{\Psi}_{\Delta + 2n+ 2}^{\rm null}, 
\end{equation}
which can be rewritten as
\begin{equation}
     \left[\Box_{AdS_4} - \frac{1}{\ell^2}\Delta(\Delta - 3) - \frac{1}{\ell^2}2n(2\Delta + 2n -3) \right] \widetilde{\Psi}_{\Delta+2n}^{\rm null} = -(\Delta + 2n - \delta)(\Delta+ 2n + 1 -\delta) \widetilde{\Psi}_{\Delta + 2n+ 2}^{\rm null}.
\end{equation}
Demanding that \eqref{eq:Ansatz} is a solution to the scalar wave equation in AdS$_4$ implies that 
\begin{equation}
    \sum_{n = 0}^{\infty}\left( 2n(2\Delta + 2n - 3)c_n \widetilde{\Psi}_{\Delta + 2n} - \ell^2 (\Delta + 2n - \delta)(\Delta+ 2n + 1 -\delta) c_{n}  \widetilde{\Psi}_{\Delta + 2n+ 2}^{\rm null} \right) = 0.
\end{equation}
Noting that the $n = 0$ term vanishes in the first term and shifting $n \rightarrow n + 1$, we find
\begin{equation}
\begin{split}
    \sum_{n = 0}^{\infty}&\left( 2(n+1)(2\Delta + 2n - 1)c_{n+1}  \right. \left. - \ell^2 (\Delta + 2n - \delta)(\Delta+ 2n + 1 -\delta) c_{n} \right)\widetilde{\Psi}_{\Delta + 2n+ 2}^{\rm null}  = 0.
\end{split}
\end{equation}
As a result we find the following recursion relation
\begin{equation}
    c_{n+1} = \ell^2 \frac{(\Delta + 2n - \delta)(\Delta+ 2n + 1 -\delta)}{2(n+1)(2\Delta + 2n - 1)}c_{n}.
\end{equation}
The solution is 
\begin{equation}
    c_n = \ell^{2n} \frac{\Gamma(\Delta + 2n - \delta)}{2^{2n} \Gamma(n + 1) \Gamma(\Delta + n - \frac{1}{2})}.
\end{equation}

We conclude that the KG equation admits a family of solutions
\begin{equation}
\label{eq:null-solution}
\begin{split}
    \Phi_{\Delta}^{\rm null}(\delta, \hat{q}(z);\tau, r, w) &= \frac{\Gamma(\Delta - \delta)}{\Gamma(\Delta - \frac{1}{2})} {}_2F_1\left[\frac{1}{2}(\Delta - \delta), \frac{1}{2}(\Delta - \delta + 1); \Delta - \frac{1}{2}; \frac{1}{(\ell \cos\tau - r \sin \tau)^2} \right] \\
    &\times \widetilde{\Psi}_{\Delta}^{\rm null}(\delta, \hat{q}(z);\tau, r, w).
    \end{split}
\end{equation}
all of which diagonalize boosts towards $\hat{q}$ with eigenvalue $\delta$. 
One can also check that replacing \eqref{eq:prop-null} by the AdS$_4$ bulk-to-bulk propagator \cite{Burgess:1984ti} (cf. eq. 8 there and note their definitions $\beta = \Delta$ and $1 + \sigma = -X \cdot X'$ in AdS radius units)
\begin{equation}
    G_{bulk-bulk}(X, X') \propto \frac{1}{(-X \cdot X')^{\Delta}} {}_2 F_1\left[\frac{\Delta}{2},\frac{\Delta+1}{2}; \Delta - \frac{1}{2}; \frac{1}{(-X \cdot X')^2}\right],
\end{equation}
using the definition of the hypergeometric function and performing the integrals in  \eqref{eq:limit-hyperbolic} term by term again yields \eqref{eq:null-solution}.

\bibliographystyle{utphys}
\bibliography{references}

@article{jacobipol,
      title={Orthogonality of Jacobi polynomials with general parameters}, 
      author={A. B. J. Kuijlaars and A. Martinez-Finkelshtein and R. Orive},
      year={2003},
      eprint={math/0301037},
      archivePrefix={arXiv},
      primaryClass={math.CA},
      url={https://arxiv.org/abs/math/0301037}, 
}

@article{Aharony:1999ti,
    author = "Aharony, Ofer and Gubser, Steven S. and Maldacena, Juan Martin and Ooguri, Hirosi and Oz, Yaron",
    title = "{Large N field theories, string theory and gravity}",
    eprint = "hep-th/9905111",
    archivePrefix = "arXiv",
    reportNumber = "CERN-TH-99-122, HUTP-99-A027, LBNL-43113, RU-99-18, UCB-PTH-99-16, LBL-43113",
    doi = "10.1016/S0370-1573(99)00083-6",
    journal = "Phys. Rept.",
    volume = "323",
    pages = "183--386",
    year = "2000"
}

@article{Witten:1998qj,
    author = "Witten, Edward",
    title = "{Anti de Sitter space and holography}",
    eprint = "hep-th/9802150",
    archivePrefix = "arXiv",
    reportNumber = "IASSNS-HEP-98-15",
    doi = "10.4310/ATMP.1998.v2.n2.a2",
    journal = "Adv. Theor. Math. Phys.",
    volume = "2",
    pages = "253--291",
    year = "1998"
}

@article{Gubser:1998bc,
    author = "Gubser, S. S. and Klebanov, Igor R. and Polyakov, Alexander M.",
    title = "{Gauge theory correlators from noncritical string theory}",
    eprint = "hep-th/9802109",
    archivePrefix = "arXiv",
    reportNumber = "PUPT-1767",
    doi = "10.1016/S0370-2693(98)00377-3",
    journal = "Phys. Lett. B",
    volume = "428",
    pages = "105--114",
    year = "1998"
}

@article{Maldacena:1997re,
    author = "Maldacena, Juan Martin",
    title = "{The Large $N$ limit of superconformal field theories and supergravity}",
    eprint = "hep-th/9711200",
    archivePrefix = "arXiv",
    reportNumber = "HUTP-97-A097, HUTP-98-A097",
    doi = "10.4310/ATMP.1998.v2.n2.a1",
    journal = "Adv. Theor. Math. Phys.",
    volume = "2",
    pages = "231--252",
    year = "1998"
}

@article{Aharony:2008ug,
    author = "Aharony, Ofer and Bergman, Oren and Jafferis, Daniel Louis and Maldacena, Juan",
    title = "{N=6 superconformal Chern-Simons-matter theories, M2-branes and their gravity duals}",
    eprint = "0806.1218",
    archivePrefix = "arXiv",
    primaryClass = "hep-th",
    reportNumber = "WIS-12-08-JUN-DPP",
    doi = "10.1088/1126-6708/2008/10/091",
    journal = "JHEP",
    volume = "10",
    pages = "091",
    year = "2008"
}

@article{deHaro:2000vlm,
    author = "de Haro, Sebastian and Solodukhin, Sergey N. and Skenderis, Kostas",
    title = "{Holographic reconstruction of space-time and renormalization in the AdS / CFT correspondence}",
    eprint = "hep-th/0002230",
    archivePrefix = "arXiv",
    reportNumber = "SPIN-2000-05, ITP-UU-00-03, PUTP-1921",
    doi = "10.1007/s002200100381",
    journal = "Commun. Math. Phys.",
    volume = "217",
    pages = "595--622",
    year = "2001"
}

@article{Skenderis:2002wp,
    author = "Skenderis, Kostas",
    editor = "de Wit, B. and Vandoren, S.",
    title = "{Lecture notes on holographic renormalization}",
    eprint = "hep-th/0209067",
    archivePrefix = "arXiv",
    reportNumber = "PUTP-2047",
    doi = "10.1088/0264-9381/19/22/306",
    journal = "Class. Quant. Grav.",
    volume = "19",
    pages = "5849--5876",
    year = "2002"
}

@article{Ryu:2006bv,
    author = "Ryu, Shinsei and Takayanagi, Tadashi",
    title = "{Holographic derivation of entanglement entropy from AdS/CFT}",
    eprint = "hep-th/0603001",
    archivePrefix = "arXiv",
    reportNumber = "NSF-KITP-06-11, NSF-KITP-06-11",
    doi = "10.1103/PhysRevLett.96.181602",
    journal = "Phys. Rev. Lett.",
    volume = "96",
    pages = "181602",
    year = "2006"
}

@article{Maldacena:2002vr,
    author = "Maldacena, Juan Martin",
    title = "{Non-Gaussian features of primordial fluctuations in single field inflationary models}",
    eprint = "astro-ph/0210603",
    archivePrefix = "arXiv",
    doi = "10.1088/1126-6708/2003/05/013",
    journal = "JHEP",
    volume = "05",
    pages = "013",
    year = "2003"
}

@article{Burgess:1984ti,
    author = "Burgess, C. P. and Lutken, C. A.",
    title = "{Propagators and Effective Potentials in Anti-de Sitter Space}",
    reportNumber = "UTTG-29-84",
    doi = "10.1016/0370-2693(85)91415-7",
    journal = "Phys. Lett. B",
    volume = "153",
    pages = "137--141",
    year = "1985"
}

@article{Sleight:2019hfp,
    author = "Sleight, Charlotte and Taronna, Massimo",
    title = "{Bootstrapping Inflationary Correlators in Mellin Space}",
    eprint = "1907.01143",
    archivePrefix = "arXiv",
    primaryClass = "hep-th",
    reportNumber = "PUPT-2590",
    doi = "10.1007/JHEP02(2020)098",
    journal = "JHEP",
    volume = "02",
    pages = "098",
    year = "2020"
}

@article{H:2024cfo,
    author = "H, Anupam A. and Athira, P. V. and Paul, Priyadarshi and Raju, Suvrat",
    title = "{Interacting Fields at Spatial Infinity}",
    eprint = "2405.20326",
    archivePrefix = "arXiv",
    primaryClass = "hep-th",
    month = "5",
    year = "2024"
}

@article{Giddings:1999qu,
    author = "Giddings, Steven B.",
    title = "{The Boundary S matrix and the AdS to CFT dictionary}",
    eprint = "hep-th/9903048",
    archivePrefix = "arXiv",
    doi = "10.1103/PhysRevLett.83.2707",
    journal = "Phys. Rev. Lett.",
    volume = "83",
    pages = "2707--2710",
    year = "1999"
}

@article{Giddings:1999jq,
    author = "Giddings, Steven B.",
    title = "{Flat space scattering and bulk locality in the AdS/CFT correspondence}",
    eprint = "hep-th/9907129",
    archivePrefix = "arXiv",
    doi = "10.1103/PhysRevD.61.106008",
    journal = "Phys. Rev. D",
    volume = "61",
    pages = "106008",
    year = "2000"
}

@article{Gary:2009ae,
    author = "Gary, Mirah and Giddings, Steven B. and Penedones, Joao",
    title = "{Local bulk S-matrix elements and CFT singularities}",
    eprint = "0903.4437",
    archivePrefix = "arXiv",
    primaryClass = "hep-th",
    reportNumber = "CERN-PH-TH-2009-035, NSF-KITP-09-35",
    doi = "10.1103/PhysRevD.80.085005",
    journal = "Phys. Rev. D",
    volume = "80",
    pages = "085005",
    year = "2009"
}

@article{Penedones:2010ue,
    author = "Penedones, Joao",
    title = "{Writing CFT correlation functions as AdS scattering amplitudes}",
    eprint = "1011.1485",
    archivePrefix = "arXiv",
    primaryClass = "hep-th",
    doi = "10.1007/JHEP03(2011)025",
    journal = "JHEP",
    volume = "03",
    pages = "025",
    year = "2011"
}

@article{Komatsu:2020sag,
    author = "Komatsu, Shota and Paulos, Miguel F. and Van Rees, Balt C. and Zhao, Xiang",
    title = "{Landau diagrams in AdS and S-matrices from conformal correlators}",
    eprint = "2007.13745",
    archivePrefix = "arXiv",
    primaryClass = "hep-th",
    reportNumber = "CPHT-RR119.122020",
    doi = "10.1007/JHEP11(2020)046",
    journal = "JHEP",
    volume = "11",
    pages = "046",
    year = "2020"
}

@article{Li:2021snj,
    author = "Li, Yue-Zhou",
    title = "{Notes on flat-space limit of AdS/CFT}",
    eprint = "2106.04606",
    archivePrefix = "arXiv",
    primaryClass = "hep-th",
    doi = "10.1007/JHEP09(2021)027",
    journal = "JHEP",
    volume = "09",
    pages = "027",
    year = "2021"
}

@article{vanRees:2022zmr,
    author = "van Rees, Balt C. and Zhao, Xiang",
    title = "{Quantum Field Theory in AdS Space instead of Lehmann-Symanzik-Zimmerman Axioms}",
    eprint = "2210.15683",
    archivePrefix = "arXiv",
    primaryClass = "hep-th",
    doi = "10.1103/PhysRevLett.130.191601",
    journal = "Phys. Rev. Lett.",
    volume = "130",
    number = "19",
    pages = "191601",
    year = "2023"
}

@article{Hijano:2020szl,
    author = "Hijano, Eliot and Neuenfeld, Dominik",
    title = "{Soft photon theorems from CFT Ward identites in the flat limit of AdS/CFT}",
    eprint = "2005.03667",
    archivePrefix = "arXiv",
    primaryClass = "hep-th",
    doi = "10.1007/JHEP11(2020)009",
    journal = "JHEP",
    volume = "11",
    pages = "009",
    year = "2020"
}

@article{Hijano:2019qmi,
    author = "Hijano, Eliot",
    title = "{Flat space physics from AdS/CFT}",
    eprint = "1905.02729",
    archivePrefix = "arXiv",
    primaryClass = "hep-th",
    doi = "10.1007/JHEP07(2019)132",
    journal = "JHEP",
    volume = "07",
    pages = "132",
    year = "2019"
}

@article{Hamilton:2005ju,
    author = "Hamilton, Alex and Kabat, Daniel N. and Lifschytz, Gilad and Lowe, David A.",
    title = "{Local bulk operators in AdS/CFT: A Boundary view of horizons and locality}",
    eprint = "hep-th/0506118",
    archivePrefix = "arXiv",
    reportNumber = "BROWN-HET-1448, CU-TP-1130",
    doi = "10.1103/PhysRevD.73.086003",
    journal = "Phys. Rev. D",
    volume = "73",
    pages = "086003",
    year = "2006"
}

@article{Hamilton:2006fh,
    author = "Hamilton, Alex and Kabat, Daniel N. and Lifschytz, Gilad and Lowe, David A.",
    title = "{Local bulk operators in AdS/CFT: A Holographic description of the black hole interior}",
    eprint = "hep-th/0612053",
    archivePrefix = "arXiv",
    reportNumber = "CU-TP-1162",
    doi = "10.1103/PhysRevD.75.106001",
    journal = "Phys. Rev. D",
    volume = "75",
    pages = "106001",
    year = "2007",
    note = "[Erratum: Phys.Rev.D 75, 129902 (2007)]"
}

@article{Hamilton:2006az,
    author = "Hamilton, Alex and Kabat, Daniel N. and Lifschytz, Gilad and Lowe, David A.",
    title = "{Holographic representation of local bulk operators}",
    eprint = "hep-th/0606141",
    archivePrefix = "arXiv",
    reportNumber = "CU-TP-1149",
    doi = "10.1103/PhysRevD.74.066009",
    journal = "Phys. Rev. D",
    volume = "74",
    pages = "066009",
    year = "2006"
}

@article{Kajuri:2020vxf,
    author = "Kajuri, Nirmalya",
    title = "{Lectures on Bulk Reconstruction}",
    eprint = "2003.00587",
    archivePrefix = "arXiv",
    primaryClass = "hep-th",
    doi = "10.21468/SciPostPhysLectNotes.22",
    journal = "SciPost Phys. Lect. Notes",
    volume = "22",
    pages = "1",
    year = "2021"
}

@article{Barnich:2010eb,
    author = "Barnich, Glenn and Troessaert, Cedric",
    title = "{Aspects of the BMS/CFT correspondence}",
    eprint = "1001.1541",
    archivePrefix = "arXiv",
    primaryClass = "hep-th",
    reportNumber = "ULB-TH-09-28",
    doi = "10.1007/JHEP05(2010)062",
    journal = "JHEP",
    volume = "05",
    pages = "062",
    year = "2010"
}

@article{Bagchi:2010zz,
    author = "Bagchi, Arjun",
    title = "{Correspondence between Asymptotically Flat Spacetimes and Nonrelativistic Conformal Field Theories}",
    eprint = "1006.3354",
    archivePrefix = "arXiv",
    primaryClass = "hep-th",
    doi = "10.1103/PhysRevLett.105.171601",
    journal = "Phys. Rev. Lett.",
    volume = "105",
    pages = "171601",
    year = "2010"
}

@article{Fitzpatrick:2011jn,
    author = "Fitzpatrick, A. Liam and Kaplan, Jared",
    title = "{Scattering States in AdS/CFT}",
    eprint = "1104.2597",
    archivePrefix = "arXiv",
    primaryClass = "hep-th",
    reportNumber = "SLAC-PUB-14507",
    month = "4",
    year = "2011"
}

@article{Strominger:2001pn,
    author = "Strominger, Andrew",
    title = "{The dS / CFT correspondence}",
    eprint = "hep-th/0106113",
    archivePrefix = "arXiv",
    doi = "10.1088/1126-6708/2001/10/034",
    journal = "JHEP",
    volume = "10",
    pages = "034",
    year = "2001"
}

@article{Anninos:2011ui,
    author = "Anninos, Dionysios and Hartman, Thomas and Strominger, Andrew",
    title = "{Higher Spin Realization of the dS/CFT Correspondence}",
    eprint = "1108.5735",
    archivePrefix = "arXiv",
    primaryClass = "hep-th",
    doi = "10.1088/1361-6382/34/1/015009",
    journal = "Class. Quant. Grav.",
    volume = "34",
    number = "1",
    pages = "015009",
    year = "2017"
}

@article{Alday:2024yyj,
    author = "Alday, Luis F. and Nocchi, Maria and Ruzziconi, Romain and Yelleshpur Srikant, Akshay",
    title = "{Carrollian amplitudes from holographic correlators}",
    eprint = "2406.19343",
    archivePrefix = "arXiv",
    primaryClass = "hep-th",
    doi = "10.1007/JHEP03(2025)158",
    journal = "JHEP",
    volume = "03",
    pages = "158",
    year = "2025"
}

@article{Narayanan:2020amh,
    author = "Narayanan, Sruthi A.",
    title = "{Massive Celestial Fermions}",
    eprint = "2009.03883",
    archivePrefix = "arXiv",
    primaryClass = "hep-th",
    doi = "10.1007/JHEP12(2020)074",
    journal = "JHEP",
    volume = "12",
    pages = "074",
    year = "2020"
}

@article{deBoer:2003vf,
    author = "de Boer, Jan and Solodukhin, Sergey N.",
    title = "{A Holographic reduction of Minkowski space-time}",
    eprint = "hep-th/0303006",
    archivePrefix = "arXiv",
    reportNumber = "ITFA-2003-11",
    doi = "10.1016/S0550-3213(03)00494-2",
    journal = "Nucl. Phys. B",
    volume = "665",
    pages = "545--593",
    year = "2003"
}

@article{Pasterski:2017kqt,
    author = "Pasterski, Sabrina and Shao, Shu-Heng",
    title = "{Conformal basis for flat space amplitudes}",
    eprint = "1705.01027",
    archivePrefix = "arXiv",
    primaryClass = "hep-th",
    doi = "10.1103/PhysRevD.96.065022",
    journal = "Phys. Rev. D",
    volume = "96",
    number = "6",
    pages = "065022",
    year = "2017"
}

@article{Pasterski:2016qvg,
    author = "Pasterski, Sabrina and Shao, Shu-Heng and Strominger, Andrew",
    title = "{Flat Space Amplitudes and Conformal Symmetry of the Celestial Sphere}",
    eprint = "1701.00049",
    archivePrefix = "arXiv",
    primaryClass = "hep-th",
    doi = "10.1103/PhysRevD.96.065026",
    journal = "Phys. Rev. D",
    volume = "96",
    number = "6",
    pages = "065026",
    year = "2017"
}

@article{Raclariu:2021zjz,
    author = "Raclariu, Ana-Maria",
    title = "{Lectures on Celestial Holography}",
    eprint = "2107.02075",
    archivePrefix = "arXiv",
    primaryClass = "hep-th",
    month = "7",
    year = "2021"
}

@article{Crawley:2021ivb,
    author = "Crawley, Erin and Miller, Noah and Narayanan, Sruthi A. and Strominger, Andrew",
    title = "{State-operator correspondence in celestial conformal field theory}",
    eprint = "2105.00331",
    archivePrefix = "arXiv",
    primaryClass = "hep-th",
    doi = "10.1007/JHEP09(2021)132",
    journal = "JHEP",
    volume = "09",
    pages = "132",
    year = "2021"
}

@article{deGioia:2025mwt,
    author = "de Gioia, Leonardo Pipolo and Raclariu, Ana-Maria",
    title = "{Infinite towers of 2d symmetry algebras from Carrollian limit of 3d CFT}",
    eprint = "2508.19981",
    archivePrefix = "arXiv",
    primaryClass = "hep-th",
    month = "8",
    year = "2025"
}

@article{deGioia:2023cbd,
    author = "de Gioia, Leonardo Pipolo and Raclariu, Ana-Maria",
    title = "{Celestial sector in CFT: Conformally soft symmetries}",
    eprint = "2303.10037",
    archivePrefix = "arXiv",
    primaryClass = "hep-th",
    doi = "10.21468/SciPostPhys.17.1.002",
    journal = "SciPost Phys.",
    volume = "17",
    number = "1",
    pages = "002",
    year = "2024"
}

@article{deGioia:2022fcn,
    author = "de Gioia, Leonardo Pipolo and Raclariu, Ana-Maria",
    title = "{Eikonal approximation in celestial CFT}",
    eprint = "2206.10547",
    archivePrefix = "arXiv",
    primaryClass = "hep-th",
    doi = "10.1007/JHEP03(2023)030",
    journal = "JHEP",
    volume = "03",
    pages = "030",
    year = "2023"
}

@article{Jorstad:2023ajr,
    author = "J{\o}rstad, Eivind and Pasterski, Sabrina and Sharma, Atul",
    title = "{Equating extrapolate dictionaries for massless scattering}",
    eprint = "2310.02186",
    archivePrefix = "arXiv",
    primaryClass = "hep-th",
    doi = "10.1007/JHEP02(2024)228",
    journal = "JHEP",
    volume = "02",
    pages = "228",
    year = "2024"
}

@article{Banerjee:2024yir,
    author = "Banerjee, Shamik",
    title = "{Boundary operators in asymptotically flat space-time}",
    eprint = "2406.06690",
    archivePrefix = "arXiv",
    primaryClass = "hep-th",
    doi = "10.1007/JHEP05(2025)033",
    journal = "JHEP",
    volume = "05",
    pages = "033",
    year = "2025"
}

@article{Osborn:2012vt,
    author = "Osborn, H.",
    title = "{Conformal Blocks for Arbitrary Spins in Two Dimensions}",
    eprint = "1205.1941",
    archivePrefix = "arXiv",
    primaryClass = "hep-th",
    reportNumber = "DAMTP-12-37",
    doi = "10.1016/j.physletb.2012.09.045",
    journal = "Phys. Lett. B",
    volume = "718",
    pages = "169--172",
    year = "2012"
}

@article{Dolan:2000ut,
    author = "Dolan, F. A. and Osborn, H.",
    title = "{Conformal four point functions and the operator product expansion}",
    eprint = "hep-th/0011040",
    archivePrefix = "arXiv",
    reportNumber = "DAMTP-2000-125",
    doi = "10.1016/S0550-3213(01)00013-X",
    journal = "Nucl. Phys. B",
    volume = "599",
    pages = "459--496",
    year = "2001"
}

@article{Maldacena:2015iua,
    author = "Maldacena, Juan and Simmons-Duffin, David and Zhiboedov, Alexander",
    title = "{Looking for a bulk point}",
    eprint = "1509.03612",
    archivePrefix = "arXiv",
    primaryClass = "hep-th",
    doi = "10.1007/JHEP01(2017)013",
    journal = "JHEP",
    volume = "01",
    pages = "013",
    year = "2017"
}

@article{deGioia:2024yne,
    author = "de Gioia, Leonardo Pipolo and Raclariu, Ana-Maria",
    title = "{Celestial amplitudes from conformal correlators with bulk-point kinematics}",
    eprint = "2405.07972",
    archivePrefix = "arXiv",
    primaryClass = "hep-th",
    month = "5",
    year = "2024"
}

@article{Bagchi:2023fbj,
    author = "Bagchi, Arjun and Dhivakar, Prateksh and Dutta, Sudipta",
    title = "{AdS Witten diagrams to Carrollian correlators}",
    eprint = "2303.07388",
    archivePrefix = "arXiv",
    primaryClass = "hep-th",
    doi = "10.1007/JHEP04(2023)135",
    journal = "JHEP",
    volume = "04",
    pages = "135",
    year = "2023"
}

@article{Lipstein:2025jfj,
    author = "Lipstein, Arthur and Ruzziconi, Romain and Yelleshpur Srikant, Akshay",
    title = "{Towards a flat space Carrollian hologram from AdS$_{4}$/CFT$_{3}$}",
    eprint = "2504.10291",
    archivePrefix = "arXiv",
    primaryClass = "hep-th",
    doi = "10.1007/JHEP06(2025)073",
    journal = "JHEP",
    volume = "06",
    pages = "073",
    year = "2025"
}

@article{Bagchi:2016bcd,
    author = "Bagchi, Arjun and Basu, Rudranil and Kakkar, Ashish and Mehra, Aditya",
    title = "{Flat Holography: Aspects of the dual field theory}",
    eprint = "1609.06203",
    archivePrefix = "arXiv",
    primaryClass = "hep-th",
    doi = "10.1007/JHEP12(2016)147",
    journal = "JHEP",
    volume = "12",
    pages = "147",
    year = "2016"
}

@article{Compere:2019bua,
    author = "Comp{\`e}re, Geoffrey and Fiorucci, Adrien and Ruzziconi, Romain",
    title = "{The $\Lambda$-BMS$_4$ group of dS$_4$ and new boundary conditions for AdS$_4$}",
    eprint = "1905.00971",
    archivePrefix = "arXiv",
    primaryClass = "gr-qc",
    doi = "10.1088/1361-6382/ab3d4b",
    journal = "Class. Quant. Grav.",
    volume = "36",
    number = "19",
    pages = "195017",
    year = "2019",
    note = "[Erratum: Class.Quant.Grav. 38, 229501 (2021)]"
}

@article{Xiao:2014uea,
    author = "Xiao, Xiao",
    title = "{Holographic representation of local operators in de sitter space}",
    eprint = "1402.7080",
    archivePrefix = "arXiv",
    primaryClass = "hep-th",
    doi = "10.1103/PhysRevD.90.024061",
    journal = "Phys. Rev. D",
    volume = "90",
    number = "2",
    pages = "024061",
    year = "2014"
}

@article{Bzowski:2023nef,
    author = "Bzowski, Adam and McFadden, Paul and Skenderis, Kostas",
    title = "{Renormalisation of IR divergences and holography in de Sitter}",
    eprint = "2312.17316",
    archivePrefix = "arXiv",
    primaryClass = "hep-th",
    doi = "10.1007/JHEP05(2024)053",
    journal = "JHEP",
    volume = "05",
    pages = "053",
    year = "2024"
}

@article{Parikh:2012kg,
    author = "Parikh, Maulik and Samantray, Prasant",
    title = "{Rindler-AdS/CFT}",
    eprint = "1211.7370",
    archivePrefix = "arXiv",
    primaryClass = "hep-th",
    doi = "10.1007/JHEP10(2018)129",
    journal = "JHEP",
    volume = "10",
    pages = "129",
    year = "2018"
}

@article{Sugishita:2022ldv,
    author = "Sugishita, Sotaro and Terashima, Seiji",
    title = "{Rindler bulk reconstruction and subregion duality in AdS/CFT}",
    eprint = "2207.06455",
    archivePrefix = "arXiv",
    primaryClass = "hep-th",
    reportNumber = "YITP-22-70",
    doi = "10.1007/JHEP11(2022)041",
    journal = "JHEP",
    volume = "11",
    pages = "041",
    year = "2022"
}

@article{Bagchi:2025vri,
    author = "Bagchi, Arjun and Banerjee, Aritra and Dhivakar, Prateksh and Mondal, Saikat and Shukla, Ashish",
    title = "{The Carrollian Kaleidoscope}",
    eprint = "2506.16164",
    archivePrefix = "arXiv",
    primaryClass = "hep-th",
    month = "6",
    year = "2025"
}

@article{Pasterski:2021rjz,
    author = "Pasterski, Sabrina",
    title = "{Lectures on celestial amplitudes}",
    eprint = "2108.04801",
    archivePrefix = "arXiv",
    primaryClass = "hep-th",
    doi = "10.1140/epjc/s10052-021-09846-7",
    journal = "Eur. Phys. J. C",
    volume = "81",
    number = "12",
    pages = "1062",
    year = "2021"
}

@article{Navarro:2025,
    author = "Navarro, Núria and Raclariu, Ana-Maria",
    title = "{Canonical quantization in Lorentzian AdS spacetimes and its flat space limit}",
    eprint = "to appear"
}

@inproceedings{Penedones:2016voo,
    author = "Penedones, Joao",
    title = "{TASI lectures on AdS/CFT.}",
    booktitle = "{Theoretical Advanced Study Institute in Elementary Particle Physics}: {New Frontiers in Fields and Strings}",
    eprint = "1608.04948",
    archivePrefix = "arXiv",
    primaryClass = "hep-th",
    doi = "10.1142/9789813149441_0002",
    pages = "75--136",
    year = "2017"
}

@article{Kaplan1,
    author = "Kaplan, Jared",
    title = "{Lectures on AdS/CFT from the Bottom Up}"
}

@article{Melton:2025ecj,
    author = "Melton, Walker and Strominger, Andrew and Wang, Tianli",
    title = "{Quantum Fields on Time-Periodic AdS$_3/\mathbb{Z}$}",
    eprint = "2510.15036",
    archivePrefix = "arXiv",
    primaryClass = "hep-th",
    month = "10",
    year = "2025"
}

@article{Harlow:2011ke,
    author = "Harlow, Daniel and Stanford, Douglas",
    title = "{Operator Dictionaries and Wave Functions in AdS/CFT and dS/CFT}",
    eprint = "1104.2621",
    archivePrefix = "arXiv",
    primaryClass = "hep-th",
    reportNumber = "SU-ITP-11-22",
    month = "4",
    year = "2011"
}

@article{Banks:1998dd,
    author = "Banks, Tom and Douglas, Michael R. and Horowitz, Gary T. and Martinec, Emil J.",
    title = "{AdS dynamics from conformal field theory}",
    eprint = "hep-th/9808016",
    archivePrefix = "arXiv",
    reportNumber = "NSF-ITP-98-082, EFI-98-30",
    month = "8",
    year = "1998"
}

@article{Papadodimas:2012aq,
    author = "Papadodimas, Kyriakos and Raju, Suvrat",
    title = "{An Infalling Observer in AdS/CFT}",
    eprint = "1211.6767",
    archivePrefix = "arXiv",
    primaryClass = "hep-th",
    reportNumber = "HRI-ST-1107, ICTS-2012-10",
    doi = "10.1007/JHEP10(2013)212",
    journal = "JHEP",
    volume = "10",
    pages = "212",
    year = "2013"
}

@article{Harlow:2018fse,
    author = "Harlow, Daniel",
    title = "{TASI Lectures on the Emergence of Bulk Physics in AdS/CFT}",
    eprint = "1802.01040",
    archivePrefix = "arXiv",
    primaryClass = "hep-th",
    doi = "10.22323/1.305.0002",
    journal = "PoS",
    volume = "TASI2017",
    pages = "002",
    year = "2018"
}

@article{cite-key,
	Author = {Newman, Ezra T. },
	Da = {1976/01/01},
	Doi = {10.1007/BF00762018},
	Id = {Newman1976},
	Isbn = {1572-9532},
	Journal = {General Relativity and Gravitation},
	Number = {1},
	Pages = {107--111},
	Title = {Heaven and its properties},
	Ty = {JOUR},
	Url = {https://doi.org/10.1007/BF00762018},
	Volume = {7},
	Year = {1976}}

@article{Berenstein:2025tts,
    author = "Berenstein, David and Simon, Joan",
    title = "{Aspects of the bulk flat space limit in AdS/CFT}",
    eprint = "2510.23697",
    archivePrefix = "arXiv",
    primaryClass = "hep-th",
    month = "10",
    year = "2025"
}

@article{Costa:2014kfa,
    author = "Costa, Miguel S. and Gon{\c{c}}alves, Vasco and Penedones, Jo{\~a}o",
    title = "{Spinning AdS Propagators}",
    eprint = "1404.5625",
    archivePrefix = "arXiv",
    primaryClass = "hep-th",
    doi = "10.1007/JHEP09(2014)064",
    journal = "JHEP",
    volume = "09",
    pages = "064",
    year = "2014"
}

@article{BELAVIN1984333,
title = {Infinite conformal symmetry in two-dimensional quantum field theory},
journal = {Nuclear Physics B},
volume = {241},
number = {2},
pages = {333-380},
year = {1984},
issn = {0550-3213},
doi = {https://doi.org/10.1016/0550-3213(84)90052-X},
url = {https://www.sciencedirect.com/science/article/pii/055032138490052X},
author = {A.A. Belavin and A.M. Polyakov and A.B. Zamolodchikov},
}

@article{Simmons-Duffin:2012juh,
    author = "Simmons-Duffin, David",
    title = "{Projectors, Shadows, and Conformal Blocks}",
    eprint = "1204.3894",
    archivePrefix = "arXiv",
    primaryClass = "hep-th",
    doi = "10.1007/JHEP04(2014)146",
    journal = "JHEP",
    volume = "04",
    pages = "146",
    year = "2014"
}

@article{Himwich:2025bza,
    author = "Himwich, Elizabeth and Pate, Monica",
    title = "{Shadow celestial operator product expansions}",
    eprint = "2505.09499",
    archivePrefix = "arXiv",
    primaryClass = "hep-th",
    doi = "10.1007/JHEP11(2025)139",
    journal = "JHEP",
    volume = "11",
    pages = "139",
    year = "2025"
}

@article{Guevara:2021abz,
    author = "Guevara, Alfredo and Himwich, Elizabeth and Pate, Monica and Strominger, Andrew",
    title = "{Holographic symmetry algebras for gauge theory and gravity}",
    eprint = "2103.03961",
    archivePrefix = "arXiv",
    primaryClass = "hep-th",
    doi = "10.1007/JHEP11(2021)152",
    journal = "JHEP",
    volume = "11",
    pages = "152",
    year = "2021"
}

@article{Strominger:2021mtt,
    author = "Strominger, Andrew",
    title = "{$w_{1+\infty}$ Algebra and the Celestial Sphere: Infinite Towers of Soft Graviton, Photon, and Gluon Symmetries}",
    eprint = "2105.14346",
    archivePrefix = "arXiv",
    primaryClass = "hep-th",
    doi = "10.1103/PhysRevLett.127.221601",
    journal = "Phys. Rev. Lett.",
    volume = "127",
    number = "22",
    pages = "221601",
    year = "2021"
}

@article{Banerjee:2022wht,
    author = "Banerjee, Shamik and Pasterski, Sabrina",
    title = "{Revisiting the shadow stress tensor in celestial CFT}",
    eprint = "2212.00257",
    archivePrefix = "arXiv",
    primaryClass = "hep-th",
    doi = "10.1007/JHEP04(2023)118",
    journal = "JHEP",
    volume = "04",
    pages = "118",
    year = "2023"
}

@article{Chen:2023tvj,
    author = "Chen, Hong Zhe and Myers, Robert C. and Raclariu, Ana-Maria",
    title = "{Entanglement, soft modes, and celestial holography}",
    eprint = "2308.12341",
    archivePrefix = "arXiv",
    primaryClass = "hep-th",
    doi = "10.1103/PhysRevD.109.L121702",
    journal = "Phys. Rev. D",
    volume = "109",
    number = "12",
    pages = "L121702",
    year = "2024"
}

@article{Chen:2024kuq,
    author = "Chen, Hong Zhe and Myers, Robert C. and Raclariu, Ana-Maria",
    title = "{Entanglement, soft modes, and celestial CFT}",
    eprint = "2403.13913",
    archivePrefix = "arXiv",
    primaryClass = "hep-th",
    doi = "10.1007/JHEP04(2025)074",
    journal = "JHEP",
    volume = "04",
    pages = "074",
    year = "2025"
}

@article{Donnay:2022wvx,
    author = "Donnay, Laura and Fiorucci, Adrien and Herfray, Yannick and Ruzziconi, Romain",
    title = "{Bridging Carrollian and celestial holography}",
    eprint = "2212.12553",
    archivePrefix = "arXiv",
    primaryClass = "hep-th",
    doi = "10.1103/PhysRevD.107.126027",
    journal = "Phys. Rev. D",
    volume = "107",
    number = "12",
    pages = "126027",
    year = "2023"
}

@article{He:2014laa,
    author = "He, Temple and Lysov, Vyacheslav and Mitra, Prahar and Strominger, Andrew",
    title = "{BMS supertranslations and Weinberg{\textquoteright}s soft graviton theorem}",
    eprint = "1401.7026",
    archivePrefix = "arXiv",
    primaryClass = "hep-th",
    doi = "10.1007/JHEP05(2015)151",
    journal = "JHEP",
    volume = "05",
    pages = "151",
    year = "2015"
}

@article{Kapec:2016jld,
    author = "Kapec, Daniel and Mitra, Prahar and Raclariu, Ana-Maria and Strominger, Andrew",
    title = "{2D Stress Tensor for 4D Gravity}",
    eprint = "1609.00282",
    archivePrefix = "arXiv",
    primaryClass = "hep-th",
    doi = "10.1103/PhysRevLett.119.121601",
    journal = "Phys. Rev. Lett.",
    volume = "119",
    number = "12",
    pages = "121601",
    year = "2017"
}

@article{Iacobacci:2022yjo,
    author = "Iacobacci, Lorenzo and Sleight, Charlotte and Taronna, Massimo",
    title = "{From celestial correlators to AdS, and back}",
    eprint = "2208.01629",
    archivePrefix = "arXiv",
    primaryClass = "hep-th",
    doi = "10.1007/JHEP06(2023)053",
    journal = "JHEP",
    volume = "06",
    pages = "053",
    year = "2023"
}

@article{Sleight:2023ojm,
    author = "Sleight, Charlotte and Taronna, Massimo",
    title = "{Celestial Holography Revisited}",
    eprint = "2301.01810",
    archivePrefix = "arXiv",
    primaryClass = "hep-th",
    doi = "10.1103/PhysRevLett.133.241601",
    journal = "Phys. Rev. Lett.",
    volume = "133",
    number = "24",
    pages = "241601",
    year = "2024"
}

@article{Almheiri:2014lwa,
    author = "Almheiri, Ahmed and Dong, Xi and Harlow, Daniel",
    title = "{Bulk Locality and Quantum Error Correction in AdS/CFT}",
    eprint = "1411.7041",
    archivePrefix = "arXiv",
    primaryClass = "hep-th",
    reportNumber = "SU-ITP-14-30, SU-ITP-14/30",
    doi = "10.1007/JHEP04(2015)163",
    journal = "JHEP",
    volume = "04",
    pages = "163",
    year = "2015"
}

@article{Dong:2016eik,
    author = "Dong, Xi and Harlow, Daniel and Wall, Aron C.",
    title = "{Reconstruction of Bulk Operators within the Entanglement Wedge in Gauge-Gravity Duality}",
    eprint = "1601.05416",
    archivePrefix = "arXiv",
    primaryClass = "hep-th",
    reportNumber = "NSF-KITP-16-005, NSF-KITP-16-005",
    doi = "10.1103/PhysRevLett.117.021601",
    journal = "Phys. Rev. Lett.",
    volume = "117",
    number = "2",
    pages = "021601",
    year = "2016"
}

@article{Caron-Huot:2025she,
    author = "Caron-Huot, Simon and Chakravarty, Joydeep and Namjou, Keivan",
    title = "{Boundary imprint of bulk causality}",
    eprint = "2501.13182",
    archivePrefix = "arXiv",
    primaryClass = "hep-th",
    doi = "10.1007/JHEP07(2025)076",
    journal = "JHEP",
    volume = "07",
    pages = "076",
    year = "2025"
}

@article{Caron-Huot:2025hmk,
    author = "Caron-Huot, Simon and Chakravarty, Joydeep and Namjou, Keivan",
    title = "{Looking at bulk points in general geometries}",
    eprint = "2502.14963",
    archivePrefix = "arXiv",
    primaryClass = "hep-th",
    doi = "10.1007/JHEP06(2025)197",
    journal = "JHEP",
    volume = "06",
    pages = "197",
    year = "2025"
}

@article{Jain:2023fxc,
    author = "Jain, Diksha and Kundu, Suman and Minwalla, Shiraz and Parrikar, Onkar and Prabhu, Siddharth G. and Shrivastava, Pushkal",
    title = "{The S-matrix and boundary correlators in flat space}",
    eprint = "2311.03443",
    archivePrefix = "arXiv",
    primaryClass = "hep-th",
    month = "11",
    year = "2023"
}

@article{Kim:2023qbl,
    author = "Kim, Seolhwa and Kraus, Per and Monten, Ruben and Myers, Richard M.",
    title = "{S-matrix path integral approach to symmetries and soft theorems}",
    eprint = "2307.12368",
    archivePrefix = "arXiv",
    primaryClass = "hep-th",
    doi = "10.1007/JHEP10(2023)036",
    journal = "JHEP",
    volume = "10",
    pages = "036",
    year = "2023"
}

@article{Donnay:2023mrd,
    author = "Donnay, Laura",
    title = "{Celestial holography: An asymptotic symmetry perspective}",
    eprint = "2310.12922",
    archivePrefix = "arXiv",
    primaryClass = "hep-th",
    doi = "10.1016/j.physrep.2024.04.003",
    journal = "Phys. Rept.",
    volume = "1073",
    pages = "1--41",
    year = "2024"
}

@article{Anninos:2023epi,
    author = "Anninos, Dionysios and Galante, Dami{\'a}n A. and Maneerat, Chawakorn",
    title = "{Gravitational observatories}",
    eprint = "2310.08648",
    archivePrefix = "arXiv",
    primaryClass = "hep-th",
    doi = "10.1007/JHEP12(2023)024",
    journal = "JHEP",
    volume = "12",
    pages = "024",
    year = "2023"
}

@article{Anninos:2024wpy,
    author = "Anninos, Dionysios and Galante, Dami{\'a}n A. and Maneerat, Chawakorn",
    title = "{Cosmological observatories}",
    eprint = "2402.04305",
    archivePrefix = "arXiv",
    primaryClass = "hep-th",
    doi = "10.1088/1361-6382/ad5824",
    journal = "Class. Quant. Grav.",
    volume = "41",
    number = "16",
    pages = "165009",
    year = "2024"
}

@article{Anninos:2024xhc,
    author = "Anninos, Dionysios and Arias, Ra{\'u}l and Galante, Dami{\'a}n A. and Maneerat, Chawakorn",
    title = "{Gravitational observatories in AdS$_{4}$}",
    eprint = "2412.16305",
    archivePrefix = "arXiv",
    primaryClass = "hep-th",
    doi = "10.1007/JHEP07(2025)234",
    journal = "JHEP",
    volume = "07",
    pages = "234",
    year = "2025"
}

@article{Anninos:2025zgr,
    author = "Anninos, Dionysios and Galante, Dami{\'a}n A. and Georgescu, Silvia and Maneerat, Chawakorn and Svesko, Andrew",
    title = "{The Stretched Horizon Limit}",
    eprint = "2512.16738",
    archivePrefix = "arXiv",
    primaryClass = "hep-th",
    month = "12",
    year = "2025"
}

@article{Silverstein:2024xnr,
    author = "Silverstein, Eva and Torroba, Gonzalo",
    title = "{Timelike-bounded dS$_{4}$ holography from a solvable sector of the T$^{2}$ deformation}",
    eprint = "2409.08709",
    archivePrefix = "arXiv",
    primaryClass = "hep-th",
    doi = "10.1007/JHEP03(2025)156",
    journal = "JHEP",
    volume = "03",
    pages = "156",
    year = "2025"
}

@article{McFadden:2009fg,
    author = "McFadden, Paul and Skenderis, Kostas",
    title = "{Holography for Cosmology}",
    eprint = "0907.5542",
    archivePrefix = "arXiv",
    primaryClass = "hep-th",
    reportNumber = "ITF-22",
    doi = "10.1103/PhysRevD.81.021301",
    journal = "Phys. Rev. D",
    volume = "81",
    pages = "021301",
    year = "2010"
}

@article{Baumann:2022jpr,
    author = "Baumann, Daniel and Green, Daniel and Joyce, Austin and Pajer, Enrico and Pimentel, Guilherme L. and Sleight, Charlotte and Taronna, Massimo",
    title = "{Snowmass White Paper: The Cosmological Bootstrap}",
    eprint = "2203.08121",
    archivePrefix = "arXiv",
    primaryClass = "hep-th",
    doi = "10.21468/SciPostPhysCommRep.1",
    journal = "SciPost Phys. Comm. Rep.",
    volume = "2024",
    pages = "1",
    year = "2024"
}

@article{Baumann:2020dch,
    author = "Baumann, Daniel and Duaso Pueyo, Carlos and Joyce, Austin and Lee, Hayden and Pimentel, Guilherme L.",
    title = "{The Cosmological Bootstrap: Spinning Correlators from Symmetries and Factorization}",
    eprint = "2005.04234",
    archivePrefix = "arXiv",
    primaryClass = "hep-th",
    doi = "10.21468/SciPostPhys.11.3.071",
    journal = "SciPost Phys.",
    volume = "11",
    pages = "071",
    year = "2021"
}

@article{Anninos:2012qw,
    author = "Anninos, Dionysios",
    title = "{De Sitter Musings}",
    eprint = "1205.3855",
    archivePrefix = "arXiv",
    primaryClass = "hep-th",
    doi = "10.1142/S0217751X1230013X",
    journal = "Int. J. Mod. Phys. A",
    volume = "27",
    pages = "1230013",
    year = "2012"
}

@article{Anninos:2011af,
    author = "Anninos, Dionysios and Hartnoll, Sean A. and Hofman, Diego M.",
    title = "{Static Patch Solipsism: Conformal Symmetry of the de Sitter Worldline}",
    eprint = "1109.4942",
    archivePrefix = "arXiv",
    primaryClass = "hep-th",
    doi = "10.1088/0264-9381/29/7/075002",
    journal = "Class. Quant. Grav.",
    volume = "29",
    pages = "075002",
    year = "2012"
}

@article{PhysRevD.15.2738,
  title = {Cosmological event horizons, thermodynamics, and particle creation},
  author = {Gibbons, G. W. and Hawking, S. W.},
  journal = {Phys. Rev. D},
  volume = {15},
  issue = {10},
  pages = {2738--2751},
  numpages = {0},
  year = {1977},
  month = {May},
  publisher = {American Physical Society},
  doi = {10.1103/PhysRevD.15.2738},
  url = {https://link.aps.org/doi/10.1103/PhysRevD.15.2738}
}

@article{Kapec:2014opa,
    author = "Kapec, Daniel and Lysov, Vyacheslav and Pasterski, Sabrina and Strominger, Andrew",
    title = "{Semiclassical Virasoro symmetry of the quantum gravity $ \mathcal{S}$-matrix}",
    eprint = "1406.3312",
    archivePrefix = "arXiv",
    primaryClass = "hep-th",
    doi = "10.1007/JHEP08(2014)058",
    journal = "JHEP",
    volume = "08",
    pages = "058",
    year = "2014"
}

@article{Freidel:2021ytz,
    author = "Freidel, Laurent and Pranzetti, Daniele and Raclariu, Ana-Maria",
    title = "{Higher spin dynamics in gravity and w1+{\ensuremath{\infty}} celestial symmetries}",
    eprint = "2112.15573",
    archivePrefix = "arXiv",
    primaryClass = "hep-th",
    doi = "10.1103/PhysRevD.106.086013",
    journal = "Phys. Rev. D",
    volume = "106",
    number = "8",
    pages = "086013",
    year = "2022"
}

@article{Adamo:2021lrv,
    author = "Adamo, Tim and Mason, Lionel and Sharma, Atul",
    title = "{Celestial $w_{1+\infty}$ Symmetries from Twistor Space}",
    eprint = "2110.06066",
    archivePrefix = "arXiv",
    primaryClass = "hep-th",
    doi = "10.3842/SIGMA.2022.016",
    journal = "SIGMA",
    volume = "18",
    pages = "016",
    year = "2022"
}

@article{Bittleston:2024efo,
    author = "Bittleston, Roland and Costello, Kevin and Zeng, Keyou",
    title = "{Self-Dual Gauge Theory from the Top Down}",
    eprint = "2412.02680",
    archivePrefix = "arXiv",
    primaryClass = "hep-th",
    month = "12",
    year = "2024"
}

@article{Fernandez:2024qnu,
    author = "Fern{\'a}ndez, V{\'\i}ctor E. and Paquette, Natalie M.",
    title = "{Associativity is enough: an all-orders 2d chiral algebra for 4d form factors}",
    eprint = "2412.17168",
    archivePrefix = "arXiv",
    primaryClass = "hep-th",
    doi = "10.1088/1361-6382/ae0089",
    journal = "Class. Quant. Grav.",
    volume = "42",
    number = "18",
    pages = "185005",
    year = "2025"
}

@article{Charanya:2026pnh,
    author = "Charanya, Jaazib and Morales, Anthony and Paquette, Natalie M.",
    title = "{Form factors of $\mathscr{N}=4$ self-dual Yang-Mills from the chiral algebra bootstrap}",
    eprint = "2604.21015",
    archivePrefix = "arXiv",
    primaryClass = "hep-th",
    month = "4",
    year = "2026"
}

@article{Susskind:1998vk,
    author = "Susskind, Leonard",
    editor = "Burgess, C. P. and Myers, Robert C.",
    title = "{Holography in the flat space limit}",
    eprint = "hep-th/9901079",
    archivePrefix = "arXiv",
    doi = "10.1063/1.1301570",
    journal = "AIP Conf. Proc.",
    volume = "493",
    number = "1",
    pages = "98--112",
    year = "1999"
}

@article{Marotta:2024sce,
    author = "Marotta, Raffaele and Skenderis, Kostas and Verma, Mritunjay",
    title = "{Flat space spinning massive amplitudes from momentum space CFT}",
    eprint = "2406.06447",
    archivePrefix = "arXiv",
    primaryClass = "hep-th",
    doi = "10.1007/JHEP08(2024)226",
    journal = "JHEP",
    volume = "08",
    pages = "226",
    year = "2024"
}

@article{Raju:2012zr,
    author = "Raju, Suvrat",
    title = "{New Recursion Relations and a Flat Space Limit for AdS/CFT Correlators}",
    eprint = "1201.6449",
    archivePrefix = "arXiv",
    primaryClass = "hep-th",
    reportNumber = "HRI-ST-1201",
    doi = "10.1103/PhysRevD.85.126009",
    journal = "Phys. Rev. D",
    volume = "85",
    pages = "126009",
    year = "2012"
}

@article{Fitzpatrick:2011ia,
    author = "Fitzpatrick, A. Liam and Kaplan, Jared and Penedones, Joao and Raju, Suvrat and van Rees, Balt C.",
    title = "{A Natural Language for AdS/CFT Correlators}",
    eprint = "1107.1499",
    archivePrefix = "arXiv",
    primaryClass = "hep-th",
    reportNumber = "SLAC-PUB-14506, HRI-ST-1107",
    doi = "10.1007/JHEP11(2011)095",
    journal = "JHEP",
    volume = "11",
    pages = "095",
    year = "2011"
}

@article{Melton:2023bjw,
    author = "Melton, Walker and Sharma, Atul and Strominger, Andrew",
    title = "{Celestial leaf amplitudes}",
    eprint = "2312.07820",
    archivePrefix = "arXiv",
    primaryClass = "hep-th",
    doi = "10.1007/JHEP07(2024)132",
    journal = "JHEP",
    volume = "07",
    pages = "132",
    year = "2024"
}

@article{Penrose,
	Author = {Penrose, R. },
	Doi = {10.1007/BF00756234},
	Id = {Penrose1980},
	Isbn = {1572-9532},
	Journal = {General Relativity and Gravitation},
	Number = {3},
	Pages = {225--264},
	Title = {Golden Oldie: Null Hypersurface Initial Data for Classical Fields of Arbitrary Spin and for General Relativity},
	Ty = {JOUR},
	Url = {https://doi.org/10.1007/BF00756234},
	Volume = {12},
	Year = {1980}}

@article{Banerjee:2018gce,
    author = "Banerjee, Shamik",
    title = "{Null Infinity and Unitary Representation of The Poincare Group}",
    eprint = "1801.10171",
    archivePrefix = "arXiv",
    primaryClass = "hep-th",
    doi = "10.1007/JHEP01(2019)205",
    journal = "JHEP",
    volume = "01",
    pages = "205",
    year = "2019"
}

@book{Weinberg:1995mt,
    author = "Weinberg, Steven",
    title = "{The Quantum theory of fields. Vol. 1: Foundations}",
    doi = "10.1017/CBO9781139644167",
    isbn = "978-0-521-67053-1, 978-0-511-25204-4",
    publisher = "Cambridge University Press",
    month = "6",
    year = "2005"
}

@article{Newman:1976gc,
    author = "Newman, E. T.",
    title = "{Heaven and Its Properties}",
    doi = "10.1007/BF00762018",
    journal = "Gen. Rel. Grav.",
    volume = "7",
    pages = "107--111",
    year = "1976"
}

@article{Freidel:2022skz,
    author = "Freidel, Laurent and Pranzetti, Daniele and Raclariu, Ana-Maria",
    title = "{A discrete basis for celestial holography}",
    eprint = "2212.12469",
    archivePrefix = "arXiv",
    primaryClass = "hep-th",
    doi = "10.1007/JHEP02(2024)176",
    journal = "JHEP",
    volume = "02",
    pages = "176",
    year = "2024"
}

@article{Donnay:2022sdg,
    author = "Donnay, Laura and Pasterski, Sabrina and Puhm, Andrea",
    title = "{Goldilocks modes and the three scattering bases}",
    eprint = "2202.11127",
    archivePrefix = "arXiv",
    primaryClass = "hep-th",
    reportNumber = "CPHT-RR004.012022",
    doi = "10.1007/JHEP06(2022)124",
    journal = "JHEP",
    volume = "06",
    pages = "124",
    year = "2022"
}

@article{Mitra:2024ugt,
    author = "Mitra, Prahar",
    title = "{Celestial Conformal Primaries in Effective Field Theories}",
    eprint = "2402.09256",
    archivePrefix = "arXiv",
    primaryClass = "hep-th",
    month = "2",
    year = "2024"
}

@article{Pano:2023slc,
    author = "Pano, Yorgo and Puhm, Andrea and Trevisani, Emilio",
    title = "{Symmetries in Celestial CFT$_{d}$}",
    eprint = "2302.10222",
    archivePrefix = "arXiv",
    primaryClass = "hep-th",
    reportNumber = "CPHT-RR071.122022",
    doi = "10.1007/JHEP07(2023)076",
    journal = "JHEP",
    volume = "07",
    pages = "076",
    year = "2023"
}

@article{Kapec:2021eug,
    author = "Kapec, Daniel and Mitra, Prahar",
    title = "{Shadows and soft exchange in celestial CFT}",
    eprint = "2109.00073",
    archivePrefix = "arXiv",
    primaryClass = "hep-th",
    doi = "10.1103/PhysRevD.105.026009",
    journal = "Phys. Rev. D",
    volume = "105",
    number = "2",
    pages = "026009",
    year = "2022"
}

@article{Bagchi:2023cen,
    author = "Bagchi, Arjun and Dhivakar, Prateksh and Dutta, Sudipta",
    title = "{Holography in flat spacetimes: the case for Carroll}",
    eprint = "2311.11246",
    archivePrefix = "arXiv",
    primaryClass = "hep-th",
    doi = "10.1007/JHEP08(2024)144",
    journal = "JHEP",
    volume = "08",
    pages = "144",
    year = "2024"
}

@article{Taylor:2023ajd,
    author = "Taylor, Tomasz R. and Zhu, Bin",
    title = "{w1+{\ensuremath{\infty}} Algebra with a Cosmological Constant and the Celestial Sphere}",
    eprint = "2312.00876",
    archivePrefix = "arXiv",
    primaryClass = "hep-th",
    doi = "10.1103/PhysRevLett.132.221602",
    journal = "Phys. Rev. Lett.",
    volume = "132",
    number = "22",
    pages = "221602",
    year = "2024"
}

@article{higher-d-sph-h,
author = {Gallier, Jean},
year = {2009},
month = {02},
pages = {},
title = {Notes on Spherical Harmonics and Linear Representations of Lie Groups}
}

@article{Sun:2021thf,
    author = "Sun, Zimo",
    title = "{A note on the representations of SO(1,d + 1)}",
    eprint = "2111.04591",
    archivePrefix = "arXiv",
    primaryClass = "hep-th",
    doi = "10.1142/S0129055X24300073",
    journal = "Rev. Math. Phys.",
    volume = "37",
    number = "01",
    pages = "2430007",
    year = "2025"
}

@article{Pranzetti:2025flv,
    author = "Pranzetti, Daniele and Salluce, Domenico Giuseppe",
    title = "{Double-soft limit and celestial shadow OPE from charge bracket}",
    eprint = "2510.26520",
    archivePrefix = "arXiv",
    primaryClass = "hep-th",
    doi = "10.1007/JHEP03(2026)096",
    journal = "JHEP",
    volume = "03",
    pages = "096",
    year = "2026"
}

@article{Bagchi:2024gnn,
    author = "Bagchi, Arjun and Dhivakar, Prateksh and Dutta, Sudipta",
    title = "{3D stress tensor for gravity in 4D flat spacetime}",
    eprint = "2408.05494",
    archivePrefix = "arXiv",
    primaryClass = "hep-th",
    doi = "10.1103/fmvk-vrdz",
    journal = "Phys. Rev. D",
    volume = "113",
    number = "6",
    pages = "066013",
    year = "2026"
}

@article{Himwich:2025ekg,
    author = "Himwich, Elizabeth and Pate, Monica",
    title = "{Light-ray Operators and the ${\rm w}_{1+\infty}$ Algebra}",
    eprint = "2512.18973",
    archivePrefix = "arXiv",
    primaryClass = "hep-th",
    month = "12",
    year = "2025"
}

@article{Strominger:2026yrh,
    author = "Strominger, Andrew and Wei, Hongji",
    title = "{EVERY CFT$_3$ HAS AN $\mathcal{L}_\Lambda$ $w_{1+\infty}$ SYMMETRY}",
    eprint = "2603.26459",
    archivePrefix = "arXiv",
    primaryClass = "hep-th",
    month = "3",
    year = "2026"
}

@article{Kraus:2024gso,
    author = "Kraus, Per and Myers, Richard M.",
    title = "{Carrollian partition functions and the flat limit of AdS}",
    eprint = "2407.13668",
    archivePrefix = "arXiv",
    primaryClass = "hep-th",
    doi = "10.1007/JHEP01(2025)183",
    journal = "JHEP",
    volume = "01",
    pages = "183",
    year = "2025"
}

@article{Ammon:2025avo,
    author = "Ammon, Martin and Capone, Federico and Sieling, Christoph",
    title = "{Flat Holography {\&} Holographic Renormalization: Scalar Field}",
    eprint = "2512.14818",
    archivePrefix = "arXiv",
    primaryClass = "hep-th",
    month = "12",
    year = "2025"
}

@article{Jafferis:2015del,
    author = "Jafferis, Daniel L. and Lewkowycz, Aitor and Maldacena, Juan and Suh, S. Josephine",
    title = "{Relative entropy equals bulk relative entropy}",
    eprint = "1512.06431",
    archivePrefix = "arXiv",
    primaryClass = "hep-th",
    reportNumber = "NSF-KITP-15-162, NSF-KITP-15-162",
    doi = "10.1007/JHEP06(2016)004",
    journal = "JHEP",
    volume = "06",
    pages = "004",
    year = "2016"
}

@article{Bhattacharyya:2025nfp,
    author = "Bhattacharyya, Arpan and Ghosh, Saptaswa and Pal, Sounak",
    title = "{The sky remembers everything: Celestial amplitude, shadow and OPE in quadratic EFT of gravity}",
    eprint = "2505.02899",
    archivePrefix = "arXiv",
    primaryClass = "hep-th",
    doi = "10.21468/SciPostPhys.19.2.041",
    journal = "SciPost Phys.",
    volume = "19",
    number = "2",
    pages = "041",
    year = "2025"
}

@article{Liu:2025dhh,
    author = "Liu, Reiko and Ma, Wen-Jie",
    title = "{Amplitude from crossing-symmetric celestial OPE}",
    eprint = "2503.21512",
    archivePrefix = "arXiv",
    primaryClass = "hep-th",
    month = "3",
    year = "2025"
}

@article{Surubaru:2025qhs,
    author = "Surubaru, Iustin and Zhu, Bin",
    title = "{Conformal blocks from celestial graviton amplitudes}",
    eprint = "2501.05805",
    archivePrefix = "arXiv",
    primaryClass = "hep-th",
    doi = "10.1007/JHEP06(2025)174",
    journal = "JHEP",
    volume = "06",
    pages = "174",
    year = "2025"
}

@article{Chang:2022jut,
    author = "Chang, Chi-Ming and Cui, Wei and Ma, Wen-Jie and Shu, Hongfei and Zou, Hao",
    title = "{Shadow celestial amplitudes}",
    eprint = "2210.04725",
    archivePrefix = "arXiv",
    primaryClass = "hep-th",
    doi = "10.1007/JHEP02(2023)017",
    journal = "JHEP",
    volume = "02",
    pages = "017",
    year = "2023"
}

@article{Fan:2021isc,
    author = "Fan, Wei and Fotopoulos, Angelos and Stieberger, Stephan and Taylor, Tomasz R. and Zhu, Bin",
    title = "{Conformal blocks from celestial gluon amplitudes}",
    eprint = "2103.04420",
    archivePrefix = "arXiv",
    primaryClass = "hep-th",
    doi = "10.1007/JHEP05(2021)170",
    journal = "JHEP",
    volume = "05",
    pages = "170",
    year = "2021"
}

\end{document}